\documentclass[letterpaper]{article}

\usepackage[round]{natbib}
\usepackage{authblk}

\usepackage[margin=1.25in]{geometry}
\usepackage{algorithm}
\usepackage{algorithmic}
\usepackage{xspace}
\newcommand{\scp}{SC\xspace}
\newcommand{\scpl}{Submodular Cover\xspace}
\newcommand{\smp}{KCSM\xspace}
\newcommand{\smpl}{Knapsack Constrained Submodular Maximization\xspace}

\newcommand{\definef}{$f:2^U\to\mathbb{R}_{\geq 0}$\xspace}
\newcommand{\definew}{$w:U\to\mathbb{R}_{\geq 0}$\xspace}

\newcommand{\multi}{\textsc{Multi}\xspace}
\newcommand{\single}{\textsc{Single}\xspace}
\newcommand{\bimax}{\textsc{SingleMax}\xspace}
\newcommand{\uncons}{\textsc{UnconsMax}$_{\gamma}$\xspace}
\newcommand{\stream}{\textsc{Stream}\xspace}

\newcommand{\defineepsilon}{$\epsilon\in(0,1)$\xspace}
\newcommand{\smallpb}{2/\epsilon}
\newcommand{\js}{\{1,...,\smallpb\}}
\newcommand{\jsb}{\{0,...,\smallpb\}}
\newcommand{\Si}[1]{S_{ #1 }}
\newcommand{\opta}{X_1}
\newcommand{\optb}{X_2}
\newcommand{\guess}{\overline{OPT}}
\newcommand{\guesssingle}{\sigma}

\newcommand{\ratio}{(1+\epsilon)(4/\epsilon^2+1)}
\newcommand{\maxsize}{2\guess/\epsilon}
\newcommand{\gain}{w(x)\epsilon\tau/(2\guess)}

\newcommand{\argmax}{\text{argmax}}
\newcommand{\argmin}{\text{argmin}}

\usepackage{amsmath}
\usepackage{amssymb}
\usepackage{amsthm}
\usepackage{booktabs}
\theoremstyle{theorem}
\newtheorem{theorem}{Theorem}
\newtheorem{lemma}{Lemma}
\newtheorem{claim}{Claim}
\newtheorem{definition}{Definition}
\newtheorem{problem}{Problem}

\usepackage{enumitem}
\usepackage[dvipsnames]{xcolor}
\usepackage{tikz}
\usetikzlibrary{arrows.meta}
\usetikzlibrary{decorations.pathreplacing}
\usepackage{subfigure}
\begin{document}

\title{Scalable Bicriteria Algorithms for Non-Monotone Submodular Cover}
\author[1]{Victoria G. Crawford}
\affil[1]{vcrawford@tamu.edu, Department of Computer Science \& Engineering, Texas A\&M University}
\date{}

\maketitle

\begin{abstract}
  In this paper, we consider the optimization problem \scpl (\scp), which is to find a minimum cost subset of a ground set $U$ such that the value of a submodular function $f$ is above a threshold $\tau$. In contrast to most existing work on \scp, it is not assumed that $f$ is monotone. Two bicriteria approximation algorithms are presented for \scp that, for input parameter $0 < \epsilon < 1$, give $O( 1 / \epsilon^2 )$ ratio to the optimal cost and ensures the function $f$ is at least $\tau(1 - \epsilon)/2$. A lower bound shows that under the value query model shows that no polynomial-time algorithm can ensure that $f$ is larger than $\tau/2$.
  %\ratioallb bicriteria
  %approximation guarantee to the problem.
  Further, the algorithms presented are scalable to large data sets, processing the ground set in a stream.
%	Impossibility results are presented that show under the value query model these
%	algorithms are in a sense optimal.
	Similar algorithms developed for \scp also work for the related
	optimization problem of \smpl (\smp).
	%ensuring a \bimaxratiob bicriteria approximation guarantee.
	Finally, the algorithms are demonstrated to be effective in experiments
	involving graph cut and data summarization functions.
\end{abstract}

\section{Introduction}
Submodular set functions arise in many applications such as
summarization of data sets \citep{tschiatschek2014learning},
feature selection in machine learning \citep{das2018approximate},
cut functions in graphs \citep{balkanski2018non},
viral marketing in a social network
\citep{kempe2003maximizing}, and many others.
Intuitively, submodularity describes a diminishing returns property of set functions. Formally,
let $f:2^U\to\mathbb{R}$ be defined over subsets of a universe $U$ of size $n$.
Then $f$ is \textit{submodular}
if for all $A\subseteq B\subseteq U$
and $x\notin B$, $f(A\cup\{x\})-f(A) \geq f(B\cup\{x\})-f(B)$.
Moreover, $f$ is \textit{monotone} if
for all $A\subseteq B\subseteq U$, $f(A)\leq f(B)$.
% As a result of submodular functions wide applicability, much work has focused on
% developing approximation algorithms for NP-hard
% combinatorial optimization problems involving submodular functions.
In this paper, the \scpl problem (\scp) is considered.
%\scp is defined as follows.
\begin{problem}[\scp]
Define submodular \definef over subsets of the universe $U$ of size $n$,
and non-negative cost function \definew.
Given threshold
$\tau\leq \max\{f(X):X\subseteq U\}$, find
%\begin{align*}
    $$\argmin_{X\subseteq U}\left\{\sum_{x\in X}w(x): f(X)\geq\tau\right\}.$$
%\end{align*}
Let $OPT$ refer to the cost of the optimal solution.
\end{problem}

The \scp formulation captures applications where we wish to
ensure that a submodular function $f$ is sufficiently high, while minimizing cost. When $f$ is monotone, \scp
has been studied extensively, \textit{e.g.} \citet{wolsey1982,mirzasoleiman2015,mirzasoleiman2016,norouzi2016efficient}.
To the best of our knowledge, the only work to consider \scp with non-monotone $f$ is \citet{iyer2013}, discussed in Section \ref{section:relatedwork} below.
Therefore, non-monotone \scp is relatively unexplored, despite such problems arising in learning applications.

For example, non-monotone submodular functions frequently arise in data summarization
tasks as a measure of how effectively
a subset $X\subseteq U$ summarizes a data set $U$
\citep{gillenwater2012near,tschiatschek2014learning} .
Data summarization can then be formulated as an instance of \scp where we seek to
pick the summary of minimum memory (e.g. if $U$ is images then the cost $w$ may be
the size of each image) that reaches a constant factor of the maximum value of $f$
(i.e. $\tau=\xi\max_{X\subseteq U}f(X)$).
As a second example, non-monotone, submodular revenue functions
may be formulated on a social network \citep{hartline2008optimal,balkanski2018non}.
In this context, the \scp problem asks to guarantee a certain amount of revenue with minimum cost.
%Analogous formulations can be made for applications such as
%cut functions on a graph \citep{balkanski2018non}.
%While the related submodular maximization problem \citep{nemhauser1978analysis} and the
%monotone,
%submodular cover problem \citep{wolsey1982} have both received significant attention,
%\scp is relatively unexplored.

The above examples demonstrate that applications of non-monotone submodular functions require algorithms that are able to run on very large data sets.
For example, summarization of massive data sets,
or revenue problems involving huge social networks.
Therefore the ability of algorithms developed for \scp to scale to massive data sets
is of utmost importance.
Properties of an algorithm can be used in order to determine how well it
will scale to large data sets include:
1) The number of queries the algorithm makes to $f$ because it is
assumed that this is the main bottleneck as far as time complexity;
2) Whether the algorithm can process $U$ in a stream, making a single or
few passes through $U$, while requiring low memory because $U$ may be much too large
to fit into memory at once.
In many applications, the cost function $w$ can be interpreted as
the size in memory of an element of $U$.
The simplest example is where $w$ is uniformly 1 \citep{norouzi2016efficient}.
Another possibility is the data summarization application discussed above.

\textbf{Contributions.}
In this paper, scalable bicritera approximation algorithms
for \scp are developed.
It is proven in Theorem \ref{thm:impossible} below that
one cannot find a feasible solution to \scp in polynomially many queries to $f$ assuming the
  value oracle model. In particular, we cannot guarantee a solution $X$ to an instance
  of \scp such that $f(X)> \tau/2$ in general.
  This result motivates the development of \textit{bicriteria approximation algorithms}
  for \scp.
An $(\alpha,\beta)$-bicriteria approximation algorithm
for \scp produces a solution $X\subseteq U$ such that $w(X)\leq\alpha OPT$
and $f(X)\geq\beta\tau$;
%  A $(\sigma, \rho)$-bicriteria algorithm for \scp finds a set $S$ such that $w( S ) \le \sigma OPT$ and $f(S) \ge \rho \tau$.

The algorithm \multi is presented in Section \ref{section:streamingmulti},
  which is an approximation algorithm with a bicriteria approximation guarantee of
  $((1+\epsilon)(4/\epsilon^2+1),(1-\epsilon)/2)$
  for \scp.
  %where $\gamma$ is the approximation ratio of an unconstrained submodular
  %maximization algorithm used as a subroutine.
  %$\gamma$ can reach $1/2$ by using the algorithm of
  %\citet{buchbinder2018deterministic}. Therefore,
  The guarantee of \multi on the
  constraint nearly matches the impossibility result stated in Theorem \ref{thm:impossible} and
  so is optimal in that sense.
  The total number of queries \multi makes to $f$ is $O(n\ln(n))$.
  If $w$ is interpreted as the cost to store an element,
  \multi is a scalable algorithm for \scp in terms of memory:
  \multi takes at most $\mathcal{O}(\ln(OPT))$ passes through $U$ in an arbitrary order,
  while storing elements of total cost at most $\mathcal{O}(OPT)$.

The algorithm \single is presented in Section \ref{section:streamingsingle},
  which takes a single pass
  through $U$ in an arbitrary order and has the same bicriteria approximation
  guarantee as \multi.
  However, \single does not have a bound on the total cost of stored elements
  relative to $OPT$, but instead has a \textit{competitive bound} on the memory.
  %Instead, \single has a bound on the total weight of stored elements relative to
  %the cost of the optimal solution
  %restricted to the first $i$ elements read in, provided it exists.
  %Once the $i$th element of the universe has been read in by \single,
  %the total weight of all elements stored at one time
  %at most $\ratio \ln(2OPT_i/(\epsilon\tau\xi))/\ln(1+\epsilon)OPT_i$
  %from that point on, where $\xi$ is instance dependent.
  The total number of queries \single makes to $f$ is $\mathcal{O}(n^2)$.
  Further,
  a more scalable version of \single with total number of queries
  $\mathcal{O}(n\ln(n))$ is possible, but results in worse approximation guarantees.

The algorithm \bimax is proposed in Section \ref{section:bimax}
  for the related problem \smpl (\smp) \citep{nemhauser1978analysis}.
  \bimax takes a single pass through $U$ in an arbitrary order and has a
  bicriteria approximation guarantee of
  $((1- \epsilon)/2 ,(1+\epsilon)(4/\epsilon^2+1))$.
  Because \bimax only returns an approximately feasible solution, \bimax has a
  better approximation guarantee (nearly $1/2$)
  compared to all existing approximation algorithms
  for \smp; \citet{gharan2011submodular} showed
  no approximation ratio better than $0.491$ is achievable with polynomially many queries to $f$ if a feasible solution must be obtained.
  The total cost of all stored elements at once is at most
  $\mathcal{O}(\kappa\ln(\kappa))$, where $\kappa$
  is the knapsack constraint, and
  \bimax makes a total of at most $\mathcal{O}(n\ln(\kappa))$
  queries to $f$.

\multi and \single are empirically evaluated in
  %on instances of \scp with cut and diverse data summarization objectives in
  Section \ref{section:streamexperiments}.
  \multi and \single are demonstrated to be able to run on large data sets,
  using relatively little memory.
  Further, \multi and \single outcompete alternative approaches to solving
  \scp in terms of solution quality, as well as total number of queries.
%\end{itemize}

\textbf{Definitions and Notation.}
The following definitions and notation are used throughout the paper. 1) The notation \scp$(U,f,w,\tau)$ is used to refer to an instance of \scp with universe $U$, submodular function $f$, weight function $w$, and threshold $\tau$. 2) $\Delta f(X,x)=f(X\cup\{x\})-f(X)$.
3) $w(X)=\sum_{x\in X}w(x)$ where \definew;
4) $w_{min} = \min\{w(x): x\in U\}$ and $w_{max}= \max\{w(x): x\in U\}$.

%%% Local Variables:
%%% mode: latex
%%% TeX-master: "../paper.tex"
%%% End:

\section{Related Work}
\label{section:relatedwork}
The related optimization problem Unconstrained Submodular Maximization (USM)
is simply to find a subset of $U$ that maximizes $f$.
USM cannot be approximated in polynomially many queries of $f$
better than $1/2$ assuming the value query
model \citep{feige2011maximizing}.
A number of approximation algorithms have been proposed for USM
\citep{feige2011maximizing,buchbinder2015tight,buchbinder2018deterministic}.
%One approach is via local search algorithms \citep{feige2011maximizing} but this method
%is relatively slow.
Notably,
\citet{buchbinder2018deterministic} introduced an algorithm that gives a
$1/2-\epsilon$ guarantee in $O(n/\epsilon)$ time.
%On the other hand,
%\citet{buchbinder2015tight} is a randomized $1/2$-approximation in linear time.
%A very simple algorithm is to return a random set, which as been proven to be a
%randomized $1/4$-approximation in constant time \citep{feige2011maximizing}.
%\cite{chen2019unconstrained} introduced a constant adaptivity algorithm with a
%$1/2-\epsilon$ randomized guarantee, based on the multilinear extension.
%Independently, the same result was found by \cite{ene2018parallel}.
%All of the unconstrained submodular maximization algorithms require
%the entire ground set to be stored in memory.
%To the best of our knowledge,
%none of the above algorithms for unconstrained submodular cover have been shown to
%give an approximation guarantee for \scp. For some of them, it is easy to see that
%they do not give a non-trivial approximation guarantee for \scp: Both the local
%search algorithm of \cite{feige2011maximizing} and the double greedy algorithm of
%\cite{buchbinder2015tight} can return solutions that have $n$ times the cost of
%that of the optimal.

The special case of \scp where $f$ is monotone has been considered in a number of
works \citep{wolsey1982,wan2010,mirzasoleiman2015,mirzasoleiman2016,crawford2019submodular}.
A classic result is that the standard greedy algorithm produces a logarithmic
approximation guarantee \citep{wolsey1982}.
%In particular, the greedy algorithm produces a
%$1+\ln(\alpha/\beta)$-approximate solution, where $\alpha$ and $\beta$ are instance
%dependent parameters \citep{wolsey1982}.
%In addition, a slightly modified greedy algorithm produces a
%$(\ln(1/\epsilon),1-\epsilon)$-bicriteria approximation ratio.
However, the greedy algorithm does not have any non-trivial approximation guarantee for
\scp if monotonicity is not assumed.
%It appears that for the more general problem \scp,
%entirely new types of algorithms must be developed in order
%to deal with non-monotonicity.
In addition, monotone \scp has been studied previously in
a streaming-like setting \citep{norouzi2016efficient}.
%\citeauthor{norouzi2016efficient} showed that any single pass streaming algorithm
%using sublinear in $n$ memory will fail to provide any non-trivial
%approximation guarantees.
If an upper bound $\epsilon M$ on $OPT$ is given,
the streaming algorithm of \citeauthor{norouzi2016efficient}
makes a single pass through $U$ in an arbitrary order and returns a
$(2/\epsilon,1-\epsilon)$-bicriteria approximate solution,
storing a maximum of $M$ elements,
and making at most $\mathcal{O}(nM)$ evaluations of $f$.

To the best of our knowledge,
\citet{iyer2013} is the only other work to consider \scp where there is no assumption
of monotonicity.
\citeauthor{iyer2013} proposes a method of converting
algorithms for \smp to ones for \scp.
Because there is a long line of work on \smp
\citep{gupta2010constrained,buchbinder2014submodular,buchbinder2017comparing},
especially if the cost is uniform, this introduces many possible algorithms.
In particular, given an $(\alpha,\beta)$ bicriteria approximation algorithm for \smp,
the approach of
\citeauthor{iyer2013} produces a $((1+\epsilon)\beta,\alpha)$-bicriteria
approximation algorithm for \scp with uniform cost.
However, this method is limited:
Even assuming a cardinality constraint,
the current best approximation algorithm for \smp
has an approximation ratio of
$\alpha=0.385$ \citep{buchbinder2016constrained}, and it is impossible under the value query
model in order to get a better approximation ratio than 0.491
in polynomially many queries of $f$ \citep{gharan2011submodular}.
$\alpha$ would need to get arbitrarily close to 1/2 in order to outperform
\multi and \single.

A number of algorithms have been proposed for \smp with uniform cost
in the streaming setting \citep{chakrabarti2015submodular,alaluf2020optimal}.
Similar to \multi and \single, the
algorithm of \citeauthor{alaluf2020optimal}
stores arriving elements from the stream in
$O(1/\epsilon)$ disjoint sets, and once the stream is complete an offline algorithm
is run on their union (but not an algorithm for USM as we will propose here).
See the appendix for a more thorough comparison with
\citeauthor{alaluf2020optimal}.
%because as elements arrive from the stream they either discarded
%or stored in $O(1/\epsilon)$ disjoint sets, and once all elements have arrived
%an offline algorithm for CCSM is run on the union of the sets.
%The algorithm of \citeauthor{alaluf2020optimal} takes a single pass through the
%universe, uses $O(\kappa/\epsilon^2)$ memory where $\kappa$ is the cardinality
%constraint, and produces a $\alpha/(1+\alpha)-\epsilon$-approximate solution where
%$\alpha$ is the approximation guarantee of the offline algorithm.
%This yields a 0.2779 approximation
%guarantee if using the state-of-the-art offline CCSM algorithm.

\section{Algorithms and Theoretical Guarantees}
\label{section:streamingtheoretical}
In this section, two approximation algorithms are proposed for \scp:
\multi and \single.
\multi is presented in Section \ref{section:streamingmulti}, and \single is
presented in Section \ref{section:streamingsingle}.
%The main difference between the two algorithms is that
%the former takes multiple passes through the universe $U$ when selecting
%its solution while the latter takes a single pass.
At the core of both of these
algorithms is the subroutine \stream, which is presented in Section \ref{section:stream}.

Before presenting the algorithms,
we first consider the limitations in finding feasible solutions to \scp.
\scp is related to USM:
In particular, %if $\tau=\max_{X\subseteq U}f(X)$ then any
any $(\alpha,\beta)$-bicriteria
approximation algorithm for \scp can be converted into a $\beta$ approximation algorithm for
USM.
Because USM
cannot be approximated in polynomial time better than $1/2$, assuming the value query
model \citep{feige2011maximizing}, it is not possible to develop an
$(\alpha,\beta)$-bicriteria approximation algorithm for \scp such that
$\beta > 1/2$.
This is formalized in Theorem \ref{thm:impossible}, the proof of which can
be found in the appendix.

\begin{theorem}
\label{thm:impossible}
For any $\epsilon > 0$, there are instances of
\scp where $f$ is assumed to be symmetric
such that there is no (adaptive, possibly randomized) algorithm
using fewer than $\Omega(\ln(1+\epsilon)e^{\epsilon^2n}/\ln(n))$ queries
that always finds a solution of
expected $f$ value at least $(1/2+\epsilon)\tau$.
\end{theorem}

%As discussed in Section \ref{section:relatedwork}, \citet{iyer2013} proposed
%solving \scp using algorithms for CSM as a subroutine.
%However, even for the cardinality case, we cannot reach arbitrarily close to
%$\beta=1/2$ in polynomial time using this method since
%it is impossible under the value query
%model in order to get a better approximation ratio than 0.491
%for CSM \cite{gharan2011submodular},
%and the current best approximation algorithm for CSM has an approximation ratio of
%0.385 \citep{buchbinder2016constrained}.

This leads us to the question of whether
$(\alpha,\beta)$-bicriteria approximation algorithms can be developed for \scp
where the best case $\beta=1/2$ is achieved or nearly achieved.
In this section, we answer this question affirmatively
by presenting the bicriteria approximation
algorithms \multi and \single, both of which can get arbitrarily close to a feasibility
guarantee of $\beta=1/2$.

\begin{subsection}{The Algorithm \stream}
\label{section:stream}
\begin{figure}[t]
\centering
\begin{tikzpicture}[scale=0.25]

\draw (17,0) ellipse (3cm and 4cm);
\draw [cyan, fill=cyan] (18,-0.5) circle [radius=0.25];
\draw [cyan, fill=cyan] (19,1.5) circle [radius=0.25];
\draw [cyan, fill=cyan] (17.5,1) circle [radius=0.25];
\draw [cyan, fill=cyan] (16,-3) circle [radius=0.25];
\draw [cyan, fill=cyan] (15.5,-1.5) circle [radius=0.25];
\draw (17,-5.5) node {\small$S_{2/\epsilon}$};

\draw (10,0) ellipse (3cm and 4cm);
\draw [cyan, fill=cyan] (9,2) circle [radius=0.25];
\draw [cyan, fill=cyan] (7.5,0.5) circle [radius=0.25];
\draw [cyan, fill=cyan] (12.5,1.5) circle [radius=0.25];
\draw [cyan, fill=cyan] (10.5,-1) circle [radius=0.25];
\draw [cyan, fill=cyan] (10,-3) circle [radius=0.25];
\draw [cyan, fill=cyan] (8.5,-1.5) circle [radius=0.25];
\draw (10,-5.5) node {\small$S_{2/\epsilon-1}$};

\draw (0,0) ellipse (3cm and 4cm);
\draw [cyan, fill=cyan] (-1,-2) circle [radius=0.25];
\draw [cyan, fill=cyan] (-2.5,0) circle [radius=0.25];
\draw [cyan, fill=cyan] (2.5,-0.5) circle [radius=0.25];
\draw [cyan, fill=cyan] (0,-1) circle [radius=0.25];
\draw [cyan, fill=cyan] (0,-3) circle [radius=0.25];
\draw [cyan, fill=cyan] (-1.5,-1.5) circle [radius=0.25];
\draw (0,-5.5) node {\small$S_1$};

\draw [dotted, thick] (6,0) to (4,0);

\draw [WildStrawberry,thick,-{Latex[length=2mm,width=2mm]}] (24,0) to (21,0);
\draw [cyan, fill=cyan] (25,0) circle [radius=0.25];
\draw [cyan, fill=cyan] (26,0) circle [radius=0.25];
\draw [cyan, fill=cyan] (27,0) circle [radius=0.25];
\draw [cyan, fill=cyan] (28,0) circle [radius=0.25];
\draw [dotted, thick] (30.5,0) to (29.5,0);
\draw node at (27,-1.5) {\small Stream of $U$};

\end{tikzpicture}
\caption{An illustration of \stream. Each blue dot represents an element of $U$.
As elements of $U$ arrive,
and each element is either discarded or stored in at most one of
the disjoint sets $S_1,...,S_{2/\epsilon}$.
%An element $u$ is stored
%in a set $S_j$ if both of the following are true:
%(i) $u$ has sufficiently low cost;
%(ii) adding $u$ is sufficiently beneficial to increasing
%the $f$ value of $S_j$.
%$\Delta f(S_j,u)\geq w(u)\epsilon \tau/(2 \guess)$.
%If no such $S_j$ exists, $u$ is discarded.
%If the total cost of one of the disjoint sets gets too high
%before all elements of $U$ have been read in, then \stream stops reading in elements.
At the completion of reading elements from the stream,
\stream runs \uncons on the union of the disjoint sets.}
\label{fig:stream}
\end{figure}
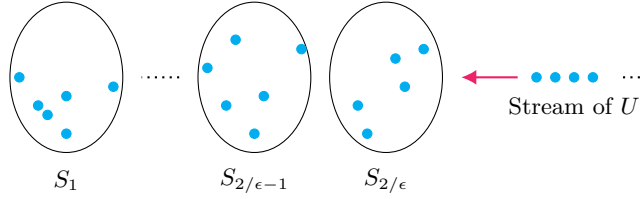

\begin{algorithm}[tb]
\caption{\stream}\label{alg:stream}
\textbf{Input}: $\guess$, $\epsilon$\\
\textbf{Output}: $S\subseteq U$
\begin{algorithmic}[1]
  \STATE $S_1\gets\emptyset,...,S_{\smallpb}\gets\emptyset$
  \FOR {$u$ received from stream such that $w(u)\leq\guess$}\label{line:loop}
    \IF {$\exists j$ s.t.
         $\Delta f(S_j,u)/w(u)\geq \epsilon \tau/(2 \guess)$}\label{line:add}
         \STATE $S_j\gets S_j\cup\{u\}$
         \IF{$w(S_j) > 2 \guess/\epsilon$}\label{line:full}
         \STATE \textbf{break}
         \ENDIF
    \ENDIF
  \ENDFOR
  \STATE $S_0\gets$\uncons$\left(\cup_{j=1}^{\smallpb}S_j\right)$\label{line:uncons}
  \STATE \textbf{return} $\argmax\{f(S_0),...,f(S_{\smallpb})\}$\label{line:complete}
  %\If {$\max\{f(S_0),...,f(S_{\smallpb})\}\geq \alpha(1-\epsilon)\tau$}\label{line:end}
  %  \State \textbf{return} $\argmax\{f(S_0),...,f(S_{\smallpb})\}$
  %\EndIf

\end{algorithmic}
\end{algorithm}

The subroutine \stream is a key subroutine of
both \multi and \single.
\stream takes a single pass through the universe $U$, choosing to store or
discard each element, filtering $U$ down to a set of much smaller
total cost.
An illustration of \stream is presented in Figure \ref{fig:stream}, and
pseudocode for \stream is presented in Algorithm \ref{alg:stream}.

\stream takes as input
$\epsilon\in (0,1)$, and a
guess of the cost of the optimal solution, $\guess$.
It will become clear how $\guess$ is chosen when \multi and \single are presented.
\stream makes a single pass
through the universe $U$ in a stream of arbitrary order, and stores
elements of total cost at most $(4/\epsilon^2 + 1)\guess$.
The stored elements are organized into $2/\epsilon$ disjoint sets,
$S_1,...,S_{2/\epsilon}$.
An element $u$ is stored
in at most one set $S_j$ if both of the following are true:
(i) $u$ has sufficiently low cost;
(ii) adding $u$ is sufficiently beneficial to increasing
the $f$ value of $S_j$.
%$\Delta f(S_j,u)\geq w(u)\epsilon \tau/(2 \guess)$.
If no such $S_j$ exists, $u$ is discarded.
If the total cost of one of the disjoint sets goes over $2\guess/\epsilon$
before all elements of $U$ have been read in, then \stream stops reading in elements.
Once reading from the stream is complete,
\stream runs \uncons on the union of the disjoint sets to get the set $S_0$,
where \uncons is an algorithm for Unconstrained Submodular Maximization \citep{feige2011maximizing}
with a
deterministic $\gamma$ approximation ratio.
\stream returns as its solution $\argmax\{f(S_0),...,f(S_{\smallpb})\}$.

\begin{subsubsection}{Theoretical Guarantees of \stream}
Several useful theoretical properties of \stream
are presented in Lemmas \ref{lemma:streamweight} and \ref{lemma:streamratio} below.

\begin{lemma}
  \label{lemma:streamweight}
  \stream has the following properties:
  (i) $w(S)\leq( 4/\epsilon^2+1)\guess$ where $S$ is the returned solution;
  (ii) The total cost of all
    elements stored at once is at most $(4/\epsilon^2+1)\guess$;
  (iii) The total number of queries to $f$ is at most
    %\begin{align*}
      $2n/\epsilon+\mathcal{T}((4/\epsilon^2+1)\guess/w_{min})$,
    %\end{align*}
    where $\mathcal{T}(m)$ is the number of queries to $f$ of \uncons on
    an input set of size $m$.
\end{lemma}
\begin{proof}
  Consider the state of \stream at the beginning of an iteration of the for loop
  on Line \ref{line:loop}, when an element $x$ has been read in but not yet added
  to any of the sets $S_1,...,S_{2/\epsilon}$.
  Then the if statement on Line \ref{line:full} ensures that
  for all $i\in\{1,...,2/\epsilon\}$, $w(S_i)\leq 2\guess/\epsilon$.
  At the end of the iteration, $u$ has been added to at most a single set $S_j$,
  and the condition that $w(u)\leq\guess$ on Line \ref{line:loop} to add $u$ ensures that
  $w(S_j)\leq (2/\epsilon+1)\guess$.
  Further $S_0$ is a subset of $\cup_{i=1}^{2/\epsilon}S_i$.
  Therefore, at any point in \stream before Line \ref{line:uncons},
  \begin{align*}
    w\left(\cup S_{i=0}^{2/\epsilon}\right) &\leq \sum_{i=0}^{2/\epsilon}w(S_i) \\
    &\leq (4/\epsilon^2+1)\guess.
  \end{align*}
  Therefore the bound on the total weight at any point in \stream
  stated in Lemma \ref{lemma:streamweight} (i) holds. Because the
  solution returned by \stream is a subset of $\cup S_{i=1}^{2/\epsilon}$, the
  bound on its weight stated in Lemma \ref{lemma:streamweight} (ii)
  is the same as the bound on its total memory.

  As each element arrives in the stream, \stream makes at most $2/\epsilon$ queries.
  Once the stream has been read in, \stream runs \uncons on  $\cup S_{i=1}^{2/\epsilon}$
  which is of total cost at most $(4/\epsilon^2+1)\guess$, and therefore at
  most $(4/\epsilon^2+1)\guess/w_{min}$ elements. The number of queries stated in
  Lemma \ref{lemma:streamweight} (iii) is then proven.
\end{proof}

\begin{lemma}
  \label{lemma:streamratio}
  Suppose that \stream is run with $\guess\geq OPT$.
  Let $S$ be the set returned by \stream.
  Then $f(S)\geq\gamma (1-\epsilon)\tau$.
\end{lemma}
\begin{proof}
  The loop on Line \ref{line:loop} of \stream completes in one of two ways:
  (i) The if statement on Line \ref{line:full} has been satisfied;
  or (ii) All of the elements of $U$ have been read from the stream, and Line
  \ref{line:full} was never satisfied.
  The proof of Lemma \ref{lemma:streamratio} is broken up into each of these two
  events.

  First suppose event (i) above occurs.
  Then at the completion of the loop there exists some $r\in\{1,...,\smallpb\}$ such that
  $w(\Si{r})\geq\maxsize$. Let $\Si{r}(\ell)$ be $\Si{r}$ after the $\ell th$ element
  was added to it and $\Si{r}(0)=\emptyset$. Then at the completion of \stream
  \begin{align*}
    f(\Si{r}) &\overset{(a)}{\geq} f(\Si{r})-f(\emptyset) \\
    &= \sum_{\ell=1}^{|\Si{r}|}\left(f(\Si{r}(\ell))-f(\Si{r}(\ell-1))\right) \\
    &\overset{(b)}{\geq} \sum_{x\in \Si{r}}\gain \\
    &= w(\Si{r})\epsilon\tau/(2\guess) \\
    &\overset{(c)}{\geq} \tau
  \end{align*}
  where (a) is because $f(\emptyset)\geq 0$; (b) is by the condition on Line \ref{line:add}; and
  (c) is by the assumption that $w(\Si{r})\geq\maxsize$.
  Therefore at the completion of \stream
  $\max\{f(\Si{0}),...,f(\Si{\smallpb})\}\geq f(\Si{r})\geq \tau.$

  Now suppose that event (ii) above occurs.
  Then at the end of \stream, $w(\Si{j})<\maxsize$ for all $j\in\js$.
  For this case, we need the following claim which is proven in the appendix,
  and is based on a result from \citet{feige2011maximizing}
  which is stated as Lemma \ref{lemma:expected} in the appendix.
  \begin{claim}
    \label{claim:mono}
    Let $A_1,...,A_m\subseteq U$ be disjoint, and $B\subseteq U$.
    Then there exists $i\in\{1,...,m\}$ such that
    $f(A_i\cup B)\geq (1-1/m)f(B)$.
  \end{claim}
  Let $S^*$ be an optimal solution to the instance of \scp.
  By Claim \ref{claim:mono}, there exists $t\in\{1,...,\smallpb\}$ such that
  $(1-\epsilon/2)\tau\leq f(S^*\cup S_t)$.
  Define $\opta=S^*\cap (\cup_{i=1}^{\smallpb}S_i)$ and
  $\optb=S^*\setminus \opta$.
  Then,
  \begin{align}
    (1-\epsilon/2)\tau &\leq f(S^*\cup S_t) \nonumber \\
    &= f(\opta\cup S_t) + f(S^*\cup S_t) - f(\opta\cup S_t) \nonumber\\
    &\overset{(a)}{\leq} f(\opta\cup S_t) +\sum_{x\in \optb}\Delta f(\opta\cup S_t, x)\nonumber\\
    &\overset{(b)}{\leq} f(\opta\cup S_t) +\sum_{x\in \optb}\Delta f(S_t, x)\label{eqn:1}
  \end{align}
  where (a) and (b) are both due to submodularity. In addition,
  \begin{align}
    \sum_{x\in \optb}\Delta f(S_t, x) &\overset{(a)}{<} \sum_{x\in \optb}\gain \nonumber \\
    &= w(\optb)\epsilon\tau/(2\guess)\nonumber \\
    &\overset{(b)}{\leq} w(\optb)\epsilon\tau/(2OPT)\nonumber \\
    &\overset{(c)}{\leq} \epsilon \tau/2\label{eqn:327}
  \end{align}
  where (a) is by submodularity and the condition on Line \ref{line:add};
  (b) is because $\guess\geq OPT$;
  (c) is because $\optb\subseteq S^*$ implies that $w(\optb)\leq OPT$.
  Then by combining Inequalities \ref{eqn:1} and \ref{eqn:327}, we have that
  \begin{align*}
    (1-\epsilon)\tau &\leq f(\opta\cup S_t)\\
    &\overset{(a)}{\leq}\max_{Y\subseteq \cup_{i=1}^{\smallpb}S_i}f(Y)\\
    &\overset{(b)}{\leq} \frac{1}{\gamma}f(\Si{0})
  \end{align*}
  where (a) is because $\opta\cup S_t\subseteq \cup_{i=1}^{\smallpb}S_i$;
  (b) is because $\Si{0}$ is an $\gamma$-approximate maximum of $f$ over
  $\cup_{i=1}^{\smallpb}\Si{i}$.
  Therefore $\max\{f(\Si{0}),...,f(\Si{\smallpb})\}\geq f(\Si{0})\geq \gamma(1-\epsilon)\tau.$
\end{proof}

As presented in Lemma \ref{lemma:streamweight} (iii), the number of queries
\stream makes to $f$ depends on the run time of \uncons.
%Assuming that $\guess/w_{min}$ is $\mathcal{O}(n)$ and
If a linear time algorithm is used for USM such as that of
\citet{buchbinder2018deterministic}, then
the overall number of queries to $f$ that \stream makes is linear.
In addition, the guarantee of Lemma \ref{lemma:streamratio}
assume that \uncons has a deterministic approximation ratio.
Alternatively, randomized approximation algorithms for \uncons
can be run $O(\ln(n))$ times in order to ensure their guarantee holds
with high probability. This possibility is described in more detail in the
appendix.

\end{subsubsection}
\end{subsection}

\begin{subsection}{The Algorithm \multi}
\label{section:streamingmulti}
We now present the constant factor bicriteria approximation algorithm \multi for \scp,
which takes $\mathcal{O}(\ln(OPT))$ passes through the universe $U$,
stores elements of total cost $\mathcal{O}(OPT)$ at once, and makes
$O(n\ln(OPT))$ queries to $f$ if a linear time algorithm is used for USM as a subroutine
\citep{Buchbinder2018}.

\multi works by sequentially running \stream for increasingly large guesses of $OPT$,
each guess corresponding to a pass through the universe $U$.
By the time that a guess is an upper bound for $OPT$,
Lemma \ref{lemma:streamratio} implies
that \multi has found a solution $S$ such that $f(S)\geq \gamma(1-\epsilon)\tau$,
where $\gamma$ is the approximation ratio of the algorithm used for USM, and
then \multi exits.
\multi does not make guesses that are much higher than $OPT$, and as a result
of Lemma \ref{lemma:streamweight} (ii)
the total cost of elements stored at one time is low.
An illustration of \multi is presented in Figure \ref{fig:multi}, and
pseudocode for \multi is presented in Algorithm \ref{alg:multi}.

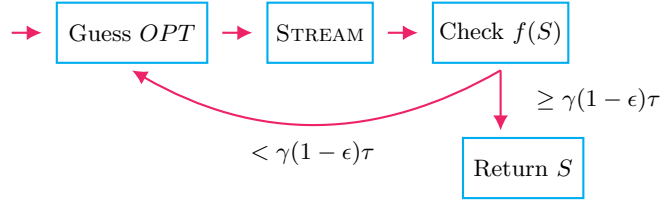
\begin{figure}[t]
\centering
\begin{tikzpicture}[scale=0.4]

\draw [WildStrawberry,thick,-{Latex[length=2mm,width=2mm]}] (14,0) to (15,0);

\node at (18,0) {\small Guess $OPT$};
\draw [cyan, thick] (15.5,-1) rectangle (20.5,1);
\draw [WildStrawberry,thick,-{Latex[length=2mm,width=2mm]}] (21,0) to (22,0);

\node at (24.25,0) {\small\stream};
\draw [cyan, thick] (22.5,-1) rectangle (26,1);
\draw [WildStrawberry,thick,-{Latex[length=2mm,width=2mm]}] (26.5,0) to (27.5,0);

\node at (30.25,0) {\small Check $f(S)$};
\draw [cyan, thick] (28,-1) rectangle (32.5,1);

\draw [WildStrawberry,thick,-{Latex[length=2mm,width=2mm]}] (30.25,-1.25) to [out=-150,in=-30] (18,-1.25);
\node at (24,-4) {\small$< \gamma(1-\epsilon)\tau$};

\draw [WildStrawberry,thick,-{Latex[length=2mm,width=2mm]}] (30.25,-1.25) to (30.25,-3.25);
\node at (33.5,-2.25) {\small$\geq\gamma(1-\epsilon)\tau$};

\node at (31,-4.5) {\small Return $S$};
\draw [cyan, thick] (29,-5.5) rectangle (33,-3.5);

\end{tikzpicture}
\caption{An illustration of the algorithm \multi. \multi makes increasingly large
guesses for $OPT$, and runs an instance of \stream for each guess.
Once \stream returns a
solution with sufficiently high $f$ value, \multi returns this solution.}
\label{fig:multi}
\end{figure}

\begin{algorithm}[tb]
\caption{\multi}\label{alg:multi}
\textbf{Input}: $\epsilon$\\
\textbf{Output}: $S\subseteq U$
\begin{algorithmic}[1]
\STATE{$\guess\gets w_{min}$}
 \WHILE{\textbf{true}}\label{line:guess}%\Comment{Iterate through smallest to largest
  \STATE $S\gets\stream(\guess,\epsilon)$
  \IF {$f(S)\geq \gamma(1-\epsilon)\tau$}\label{line:end}
    \STATE \textbf{return} $S$
  \ENDIF
    \STATE{$\guess\gets (1+\epsilon)\guess$}
 \ENDWHILE

\end{algorithmic}
\end{algorithm}

\multi takes as input a parameter $\epsilon\in (0,1)$.
\multi makes a sequence of runs of \stream with increasing guesses of $OPT$.
The first guess is $\guess=w_{min}$ (notice that
$w_{min}$ can be computed in a preliminary pass).
At the end of each run of \stream,
the solution $S$ returned by \stream is tested as to whether
$f(S)\geq\gamma(1-\epsilon)\tau$.
If $f(S)\geq\gamma(1-\epsilon)\tau$, then $S$ is returned and \multi terminates.
Otherwise, the guess is increased by a multiplicative factor of $1+\epsilon$ and
\stream is run again.

\begin{subsubsection}{Theoretical Guarantees of \multi}
We now present the theoretical guarantees of \multi in Theorem \ref{theorem:multi}.
%The theoretical guarantees of \multi are stated in Theorem \ref{theorem:multi}.
%Assuming that $\guess/w_{min}$ is $\mathcal{O}(n)$ and a $O(n)$ algorithm is
%used for USM \citep{Buchbinder2018},
%the overall number of queries to $f$ that \multi makes (item (iv)) can
%be $\mathcal{O}(n\ln(n))$.
\begin{theorem}
  \label{theorem:multi}
  Suppose that \multi is run for an instance of \scp.
  Then \multi:
  \vspace{-1em}
  \begin{itemize}[noitemsep]
    \item[(i)] Returns $S$ such that $f(S)\geq\gamma(1-\epsilon)\tau$
    and $w(S)\leq \ratio OPT$;
    \item[(ii)] Makes at most $\ln(OPT/w_{min})/\ln(1+\epsilon)$ passes through $U$;
    \item[(iii)] The total cost of all elements stored at once is at most $\ratio OPT$;
    \item[(iv)] Makes at most
      $$\small\frac{\ln(OPT/w_{min})}{\ln(1+\epsilon)}
        \left(\frac{2n}{\epsilon}+\mathcal{T}\left(\frac{(1+\epsilon)(4/\epsilon^2+1)OPT}{w_{min}}\right)\right)$$
      queries of $f$,
      where $\mathcal{T}(m)$ is the number of queries to $f$ of \uncons on an
      input set of size $m$.
  \end{itemize}
\end{theorem}
\begin{proof}
  Define $q\in\mathbb{Z}_{> 0}$ to be the unique value where
  \begin{align}
    (1+\epsilon)^{q-1}w_{min} < OPT \leq (1+\epsilon)^{q}w_{min}. \label{eqn:defineq}
  \end{align}
  By Lemma \ref{lemma:streamratio}, if the loop on Line \ref{line:guess} reaches
  $OPT=(1+\epsilon)^qw_{min}$, \stream will return a set $S$ that satisfies
  $f(S)\geq\gamma(1-\epsilon)\tau$. Then the if statement on Line \ref{line:end}
  will be satisfied, and \multi will terminate with solution $S$.
  Further, by Lemma \ref{lemma:streamweight},
  $w(S)\leq (4/\epsilon^2+1)(1+\epsilon)^{q}w_{min}\leq
  (4/\epsilon^2+1)(1+\epsilon)OPT$. Therefore item (i) is proven.

  Each iteration of the loop on Line \ref{line:guess} corresponds to one pass through $U$.
  Since the loop on Line \ref{line:guess} stops before or once $\kappa$ reaches
  $(1+\epsilon)^qw_{min}$ (as explained above), there are at most
  $\ln(OPT/w_{min})/\ln(1+\epsilon)$ passes through $U$.
  Therefore item (ii) is proven.

  Over the course of \multi, $\kappa$ increases from $w_{min}$ to
  $(1+\epsilon)^qw_{min}$ (as explained above). Further, each iteration of the
  for loop on Line \ref{line:guess} stores elements only needed in the corresponding
  call of \stream.
  By Lemma \ref{lemma:streamweight}, therefore
  the total weight of all elements stored at once
  is at most $(4/\epsilon^2+1)(1+\epsilon)^{q}w_{min}\leq
  (4/\epsilon^2+1)(1+\epsilon)OPT$.
  Therefore item (iii) is proven.

  Item (iv) is a result of Lemma \ref{lemma:streamweight}, and the fact that
  \stream is run at most
  $\ln(OPT/w_{min})/\ln(1+\epsilon)$ times with $\guess\leq (1+\epsilon)OPT$.
  \end{proof}

\end{subsubsection}
\end{subsection}

\begin{subsection}{The Algorithm \single}
\label{section:streamingsingle}
We now present the constant factor bicriteria approximation algorithm \single for \scp,
which takes a single pass through the universe $U$.
\single is more useful for applications where $U$ is received in a stream but is never
stored, making multiple passes impossible.
\single provides the same bicriteria approximation guarantees as \multi, but
has weaker guarantees on the total cost of elements stored and runtime.

\single works in a similar way to \multi, except
\single essentially runs the instances of \stream for each guess of $OPT$ in
parallel instead of sequentially. In the single pass setting, new difficulties arise
because without seeing the entire ground set $U$ it is difficult to determine a
useful upper bound on $OPT$ (see the appendix for further discussion on this problem).
Since the guess of $OPT$ determines the upper limit on
the total cost of elements stored by \stream at once (see Lemma \ref{lemma:streamweight}),
too large of a guess of $OPT$ may result in too high of total cost of elements stored.
In Theorem \ref{theorem:single}, we instead
present a competitive bound on the total cost of elements stored at once by \single,
where the cost is bounded
relative to the optimal solution if the instance of \scp were
restricted only to the elements of the universe received from the stream so far.

%Abstractly, \single works in a similar way to \multi except instead of sequentially
%calling \stream, \single runs a modified version of \stream, \streampar, in parallel for many
%guesses of $OPT$.
\single takes as input $\epsilon\in(0,1)$, and an upper bound on $OPT$, $B$.
\single runs instances of a modified version of \stream in parallel.
The modified version of
\stream runs \uncons on $\cup S_i$ after each element
is received from the stream, instead of after the entire stream has been read in.
Each instance of \stream therefore is associated with a set at any point of time
that is the output of \uncons, which we will call its \textit{best solution}.
A lower bound on the guesses of $OPT$, $L$, is updated lazily as \single runs.
In particular, if element $u$ arrives from the stream such that
$f(u)/w(u)>\epsilon\tau/(2 L)$, then $L$ is set to be $\epsilon\tau w(u)/(2f(u))$.
The guesses of $OPT$ are
$\{(1+\epsilon)^i: i\in\mathbb{Z}, L\leq (1+\epsilon)^i \leq B\}$.
An upper bound on the guesses of $OPT$, $B$, is initially given as an input.
In the case of streaming algorithms for monotone \scp,
an upper bound is also assumed as input \citep{norouzi2016efficient}.
$B$ is updated to be the smallest guess of $OPT$ for which the corresponding
parallel instance of stream has a best solution $X$ such
that $f(X)\geq\gamma(1-\epsilon)\tau$.
Once \single has read in $U$ from the stream,
the best solution of the instance of \stream
corresponding to $B$ is returned as a solution.
Pseudocode for \single is given in the appendix.

\begin{subsubsection}{Theoretical Guarantees of \single}
The theoretical guarantees of \single are presented in Theorem \ref{theorem:single}.
Items (iv) and (v) are \textit{competitive} guarantees
%on the maximum total cost stored at one time and the number of queries to $f$
in the sense that
they are with respect to the optimal solution over the set of elements seen so far
in the stream.
In contrast, items (ii) and (iii) are guarantees in terms of the input upper bound
on $OPT$, $B$, which could be quite bad.
Items (iv) and (v) are stronger than items (ii) and (iii), provided
the conditions of the former are met.

Relative to \multi, \single makes many more calls to \uncons.
Therefore if a linear time algorithm is used for \uncons, \single
potentially makes $\Omega(n^2)$ total queries to $f$.
However, a randomly chosen set is a constant time randomized approximation algorithm for USM
\citep{feige2011maximizing}, but the approximation guarantee is only 1/4.
Despite this, such an algorithm for \uncons may be practical if we seek to minimize runtime.
The proof of Theorem \ref{theorem:single} can be found in the appendix.
\begin{theorem}
  \label{theorem:single}
  Suppose that \single is run for an instance of \scp,
  and input $B\geq OPT$.
  Define the following two functions:
  $$m(x)=(1+\epsilon)(4/\epsilon^2+1)\frac{x\ln\left(2x/(\epsilon\tau\xi)\right)}{\ln(1+\epsilon)}$$
  $$q(x)=\frac{\ln\left(2x/(\epsilon\tau\xi)\right)}{\ln(1+\epsilon)}\left(\frac{2}{\epsilon}+\frac{\mathcal{T}((4/\epsilon^2+1)x)}{w_{min}}\right)$$
  where $\xi = \min_{u\in U}w(u)/f(\{u\})$, and
  $\mathcal{T}(m)$ is the number of queries of \uncons on an input set of size $m$.
  Then, \single:
  \vspace{-1em}
  \begin{itemize}[noitemsep]
    \item[(i)] Returns a set $S$ such that
      $f(S)\geq\gamma(1-\epsilon)\tau$ and $w(S) \leq \ratio OPT$;
    \item[(ii)] The total cost of all elements stored at once is at
      most $m(B)$;
    \item[(iii)] Makes at most $q(B)$
    queries of $f$ per arriving element of the stream.
  \end{itemize}
  Let $u_1,...,u_n$ be the order that the elements of $U$ arrive in, and
  $U_i=\{u_1,...,u_i\}$.
  If the instance $SC(U_i,f,w,\tau)$ is feasible and has optimal cost $OPT_i$,
  then from the end of the $i$th iteration of the loop in \single onwards:
  \begin{itemize}[noitemsep]
    \item[(iv)] The total cost of all elements stored at once is
      at most $m(OPT_i)$;
    \item[(v)] At most $q(OPT_i)$
    queries of $f$ are made per arriving element of the stream.
  \end{itemize}
\end{theorem}

\end{subsubsection}
\end{subsection}

\begin{subsection}{The Algorithm \bimax}
\label{section:bimax}
A related optimization problem to \scp is \smpl (\smp), defined as follows:
\begin{problem}[\smp]
Define submodular \definef over subsets of the universe $U$ of size $n$,
and non-negative cost function \definew.
Given budget $\kappa$, find
%\begin{align*}
    $$\small\argmax_{X\subseteq U}\left\{f(X): \sum_{x\in X}w(x)\leq\kappa\right\}.$$
%\end{align*}
%Let $OPT$ refer to the cost of the optimal solution.
\end{problem}
A bicriteria approximation algorithm that uses \stream as a subroutine,
\bimax,
can also be used for \smp in order to get an arbitrarily close to $1/2$
approximation guarantee (but not necessarily feasible).
%The guarantee of \bimax does not contradict existing impossibility results for
%\smp that were discussed in Section \ref{section:relatedwork},
%because \bimax is a bicriteria guarantee and does not return a feasible solution.
A description of \bimax and its theoretical guarantees are presented in this
section, but details are relegated to the appendix.
Note that the algorithm described here is not the same as converting algorithms
for \scp to ones for \smp as described by \citet{iyer2013}.

In \smp, the $f$ value of the optimal solution is unknown.
\bimax runs a variant of \stream in parallel for guesses of the $f$ value
of the optimal solution.
Because the cost of the optimal solution is known to be at most
$\kappa$ in \smp, the total stored cost
at once for every instance of \stream is bounded by $(4/\epsilon^2+1)\kappa$
(see Lemma \ref{lemma:streamweight}).
For this reason, we avoid difficulties of having too high
total stored cost as we did in \single.
\bimax lazily keeps track of an upper and lower bound
for the $f$ value of the optimal solution as elements arrive from the stream in
a similar manner as the lower bound $L$
was updated in \single.
%\uncons is only run for each instance of \stream
%once reading from the stream is complete.
Pseudocode for \bimax can be found in the appendix.

\subsubsection{Theoretical Guarantees of \bimax}
We now present the theoretical guarantees of the algorithm \bimax for \smp.
The proof of Theorem \ref{theorem:bimax} can be found in the appendix.
\begin{theorem}
  \label{theorem:bimax}
  Suppose that \bimax is run for an instance of \smp:
  Then:
  \vspace{-1em}
  \begin{itemize}[noitemsep]
    \item[(i)] The set $S$ returned by \bimax satisfies $f(S)\geq\gamma(1-\epsilon)OPT$
    and $w(S)\leq \ratio\kappa$;
    \item[(ii)] The total cost of all elements needing to be stored at once is at most
    $(4/\epsilon^2+1)\ln(2\kappa/(w_{min}\epsilon))/\ln(1+\epsilon)\kappa$;
    \item[(iii)] And at most
    $\ln(2\kappa/(w_{min}\epsilon))(2n/\epsilon + \mathcal{T}((4/\epsilon^2+1)\kappa))/\ln(1+\epsilon)$
    queries of $f$ are made in total
    where $\mathcal{T}(m)$ is the number of queries of \uncons on an input set of size $m$.
  \end{itemize}
\end{theorem}

\end{subsection}

\section{Experimental Results}
\label{section:streamexperiments}
In this section, the algorithms \single and \multi are evaluated
on instances of non-monotone submodular cover involving
diverse summarization \citep{tschiatschek2014learning} and
graph cut \citep{balkanski2018non} functions.
Additional experiments can be found in the appendix.

\begin{subsection}{Experimental Setup}
  \label{section:expsetup}
  The experimental setup is briefly described here, additional details can be
  found in the appendix.
  The graph cut instances presented are on the
  ca-AstroPh ($n=18772$, 198110 edges)
  networks from the SNAP large network collection \citep{snapnets}.
  The cost of each element is uniformly set as 1.
  The diverse summarization instances are on
  a subset of tagged webpages from the delicious.com website
  \citep{soleimani2016semi} ($n=50000$).
  The costs of the websites from the delicious.com website are uniform.
  Experiments involving non-uniform cost can be found in the appendix.

  \multi and \single require an algorithm for USM as a subroutine.
  \multi uses repeated runs of the randomized double greedy algorithm
  of \cite{buchbinder2015tight}, which results in $\mathcal{O}(n\ln(n))$ total
  queries of $f$.
  \single repeatedly samples a random
  set \citep{feige2011maximizing}, which also results in $\mathcal{O}(n\ln(n))$
  total queries of $f$.
  An experimental comparison of using different USM algorithms as subroutines for
  \multi and \single can be found in the appendix.

  The method proposed by \citet{iyer2013}, described in Section \ref{section:relatedwork},
  is used in order to convert algorithms for \smp with uniform cost
  to algorithms for \scp for comparison
  to \multi and \single.
  The first algorithm compared is the stochastic greedy algorithm
  of \citet{buchbinder2017comparing} with parameter 0.1, which results in a nearly
  %$(1+\delta,1/e-\epsilon)$-bicriteria
  $(1+\delta,1/e)$-bicriteria
  approximation algorithm for \scp in
  %$\mathcal{O}(n\ln(n)\ln(1/\epsilon)/(\ln(1+\delta)\epsilon^2))$ queries of $f$.
  $\mathcal{O}(n\ln(n)/\ln(1+\delta))$ queries of $f$.
  The second algorithm compared is the streaming algorithm of
  \citet{alaluf2019optimal} with stochastic greedy algorithm as its offline subroutine, which
  results in a nearly $(1+\delta,0.27)$-bicriteria approximation algorithm for \scp with
  at most $\mathcal{O}(\ln(n)/\ln(1+\delta))$ passes through $U$,
  storing elements of total cost at most $\mathcal{O}(OPT)$ at a time, and making
  $\mathcal{O}(n\ln(n)/\ln(1+\delta))$ queries of $f$.
\end{subsection}

\begin{subsection}{Experimental Results}
  \label{section:expresults}
  \begin{figure*}[t!]
  \centering
  %\hspace{-1em}
  %\subfigure[amazon] {
  %  \includegraphics[width=0.24\textwidth]{sections/figures/amazon_epsf.pdf}
  %  \label{amazon_epsf}
  %}
  \hspace{-1em}
  \subfigure[delicious50k $f$] {
    \includegraphics[width=0.24\textwidth]{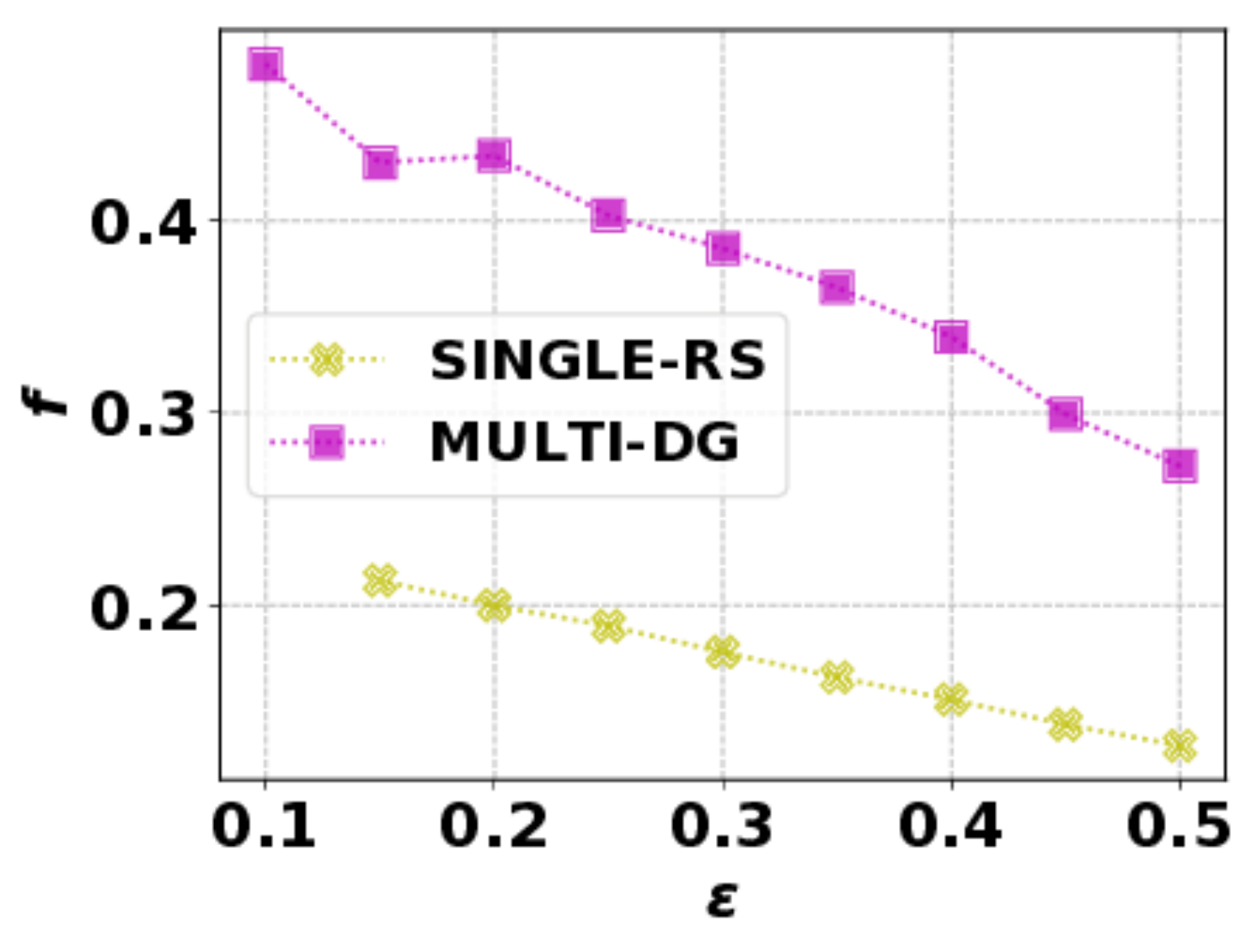}
    \label{delicious_epsf}
  }
  %\hspace{-1em}
  %\subfigure[amazon] {
  %  \includegraphics[width=0.24\textwidth]{sections/figures/amazon_epsc.pdf}
  %  \label{amazon_epsc}
  %}
  \hspace{-1em}
  \subfigure[delicious50k $c$] {
    \includegraphics[width=0.24\textwidth]{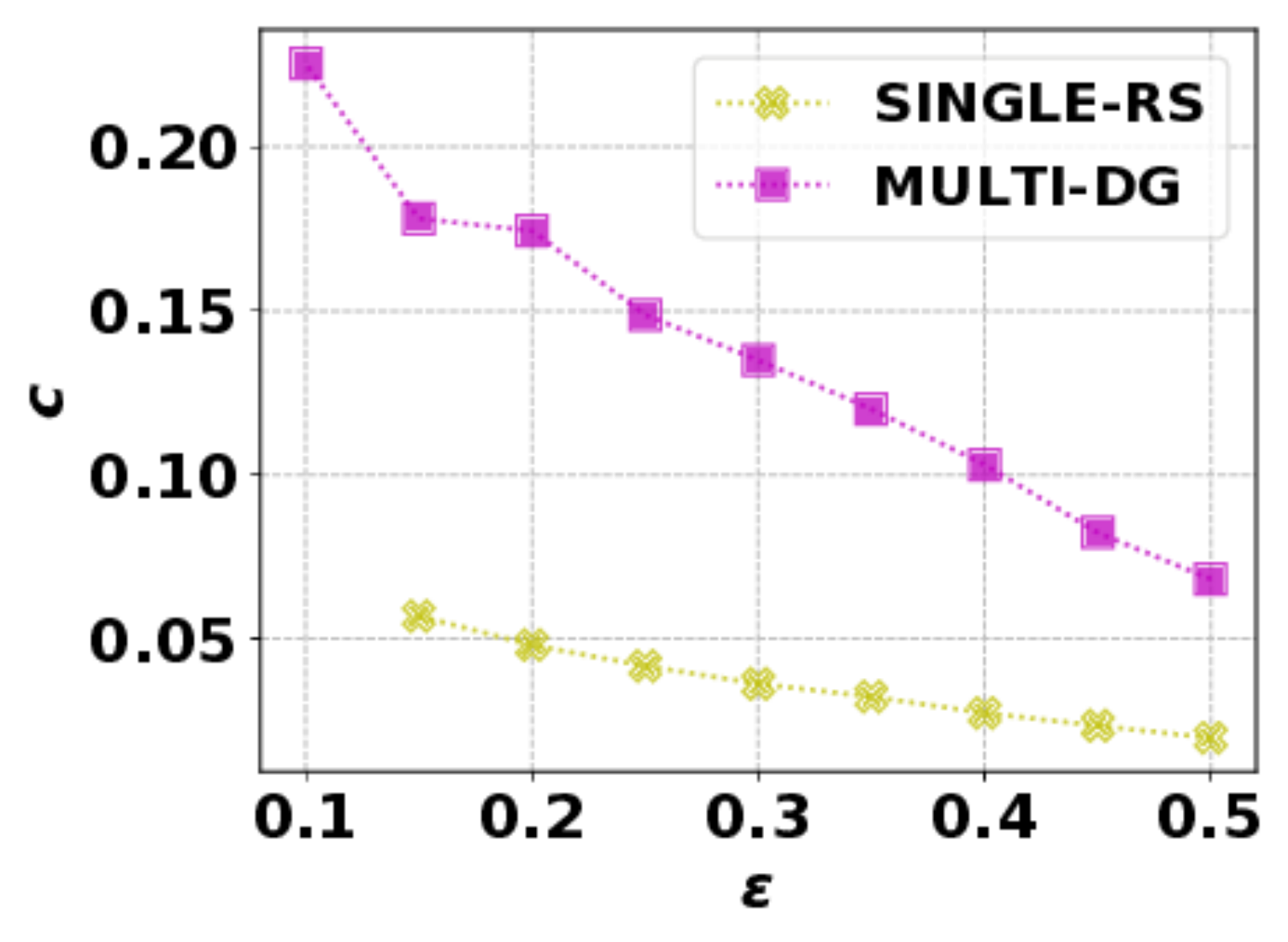}
    \label{delicious_epsc}
  }
  %\hspace{-1em}
  %\subfigure[amazon] {
  %  \includegraphics[width=0.24\textwidth]{sections/figures/amazon_epsq.pdf}
  %  \label{amazon_epsq}
  %}
  \hspace{-1em}
  \subfigure[delicious50k queries] {
    \includegraphics[width=0.24\textwidth]{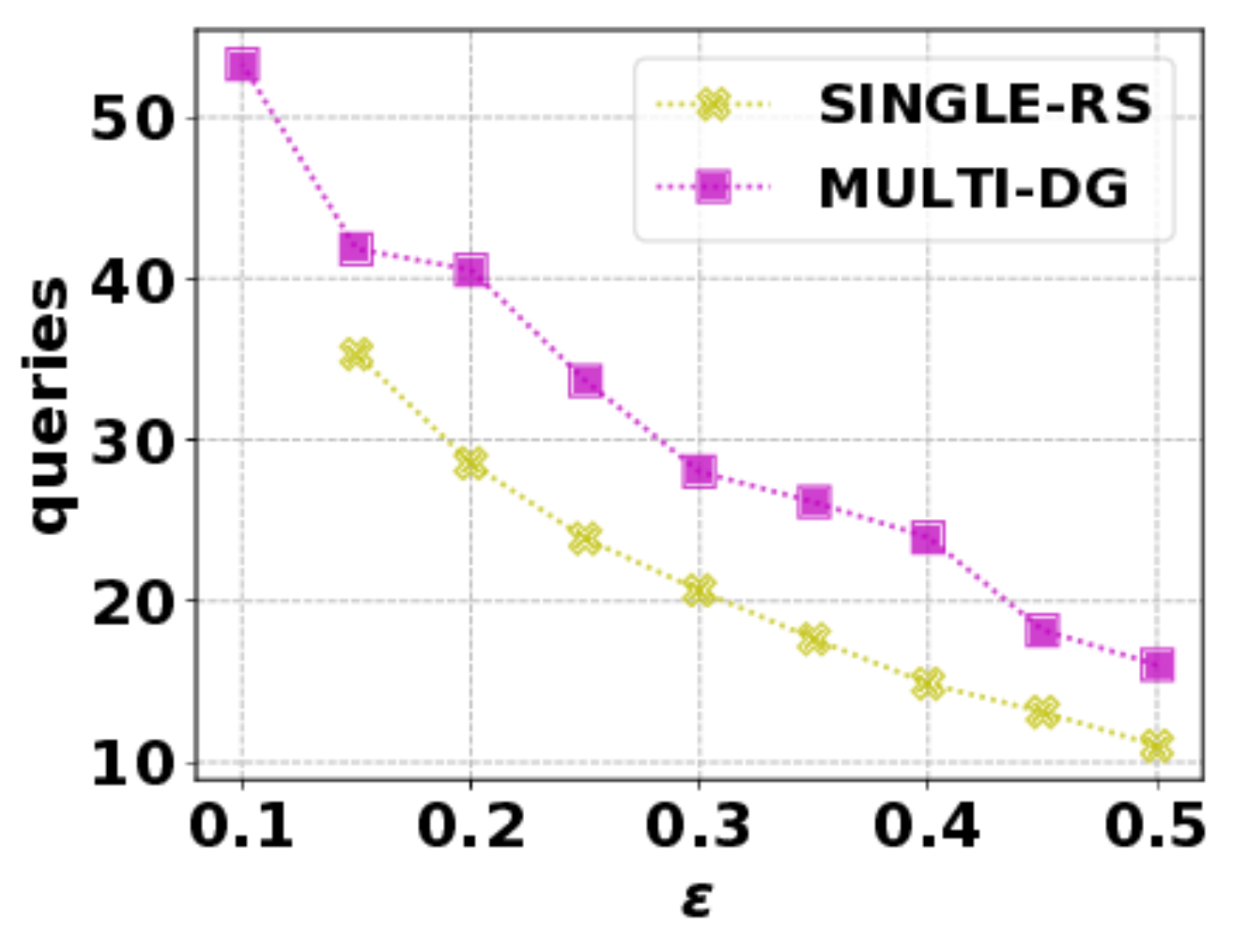}
    \label{delicious_epsq}
  }
  %\hspace{-1em}
  %\subfigure[amazon] {
  %  \includegraphics[width=0.24\textwidth]{sections/figures/amazon_epsmem.pdf}
  %  \label{amazon_epsmem}
  %}
  \hspace{-1em}
  \subfigure[delicious50k memory] {
    \includegraphics[width=0.24\textwidth]{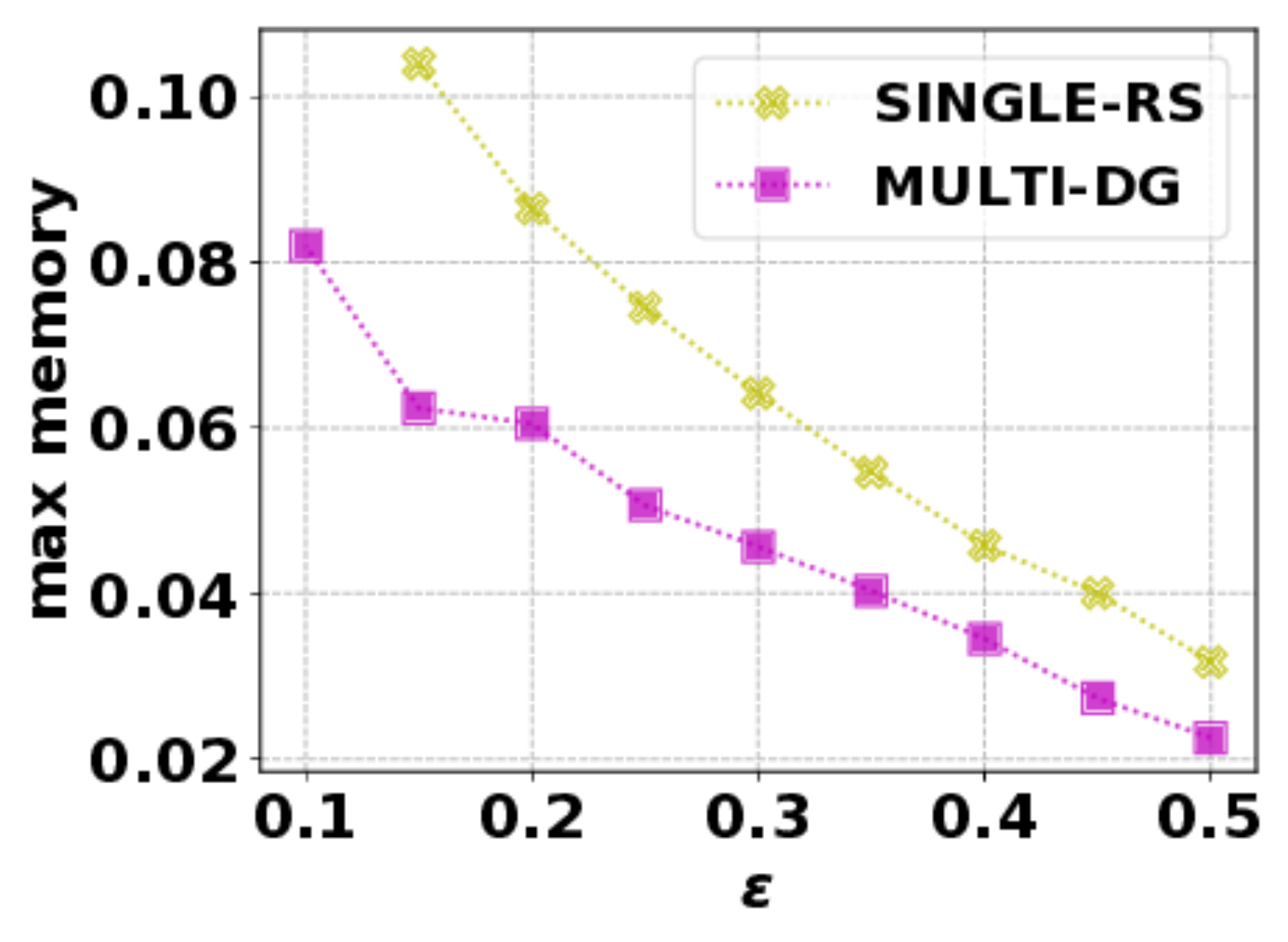}
    \label{delicious_epsmem}
  }
  \hspace{-1em}
  \subfigure[astro $f$] {
    \includegraphics[width=0.24\textwidth]{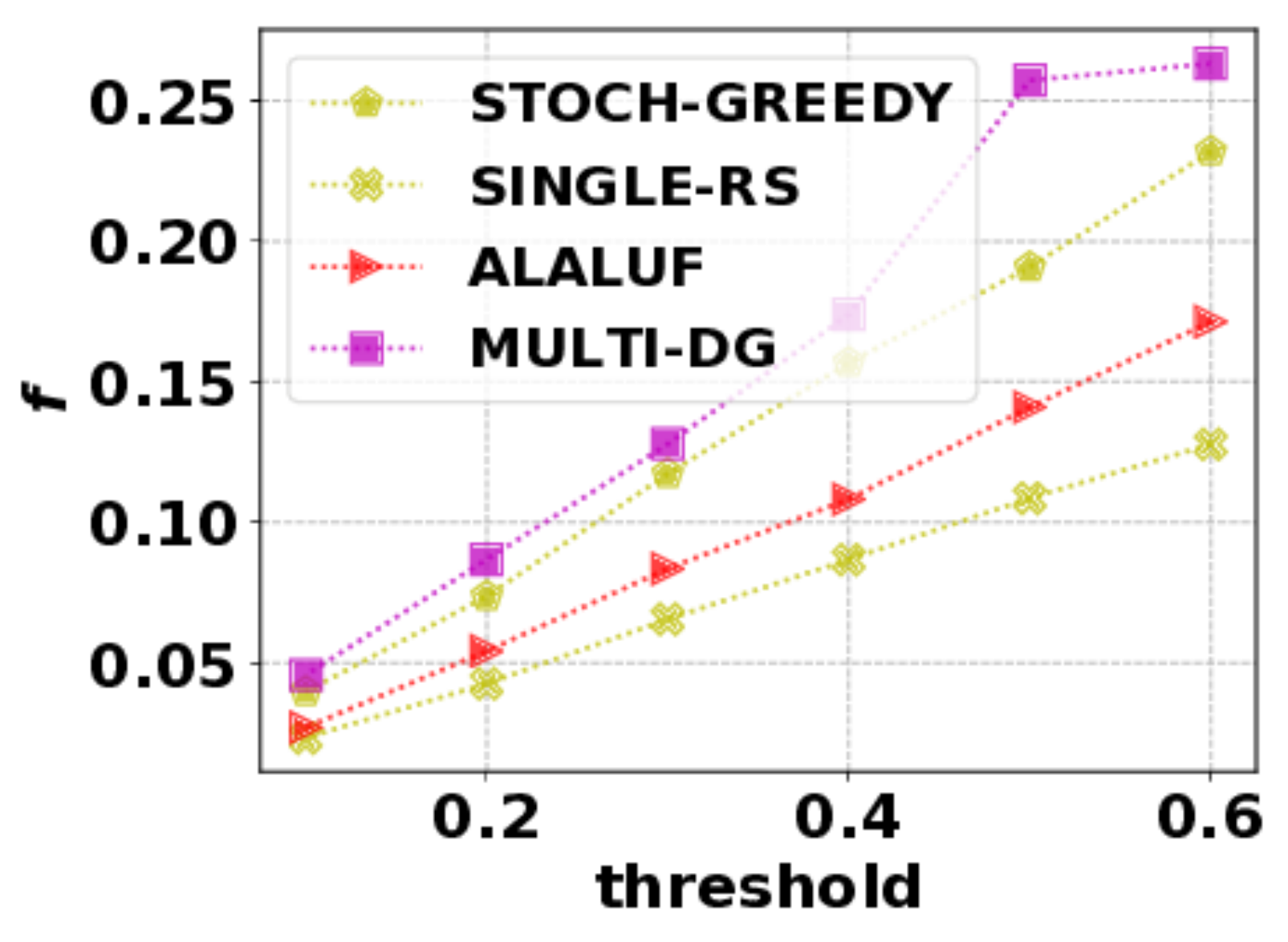}
    \label{astro_tauf}
  }
  %\hspace{-1em}
  %\subfigure[corel $f$] {
  %  \includegraphics[width=0.24\textwidth]{sections/figures/corel_tauf.pdf}
  %  \label{corel_tauf}
  %}
  \hspace{-1em}
  \subfigure[astro $c$] {
    \includegraphics[width=0.24\textwidth]{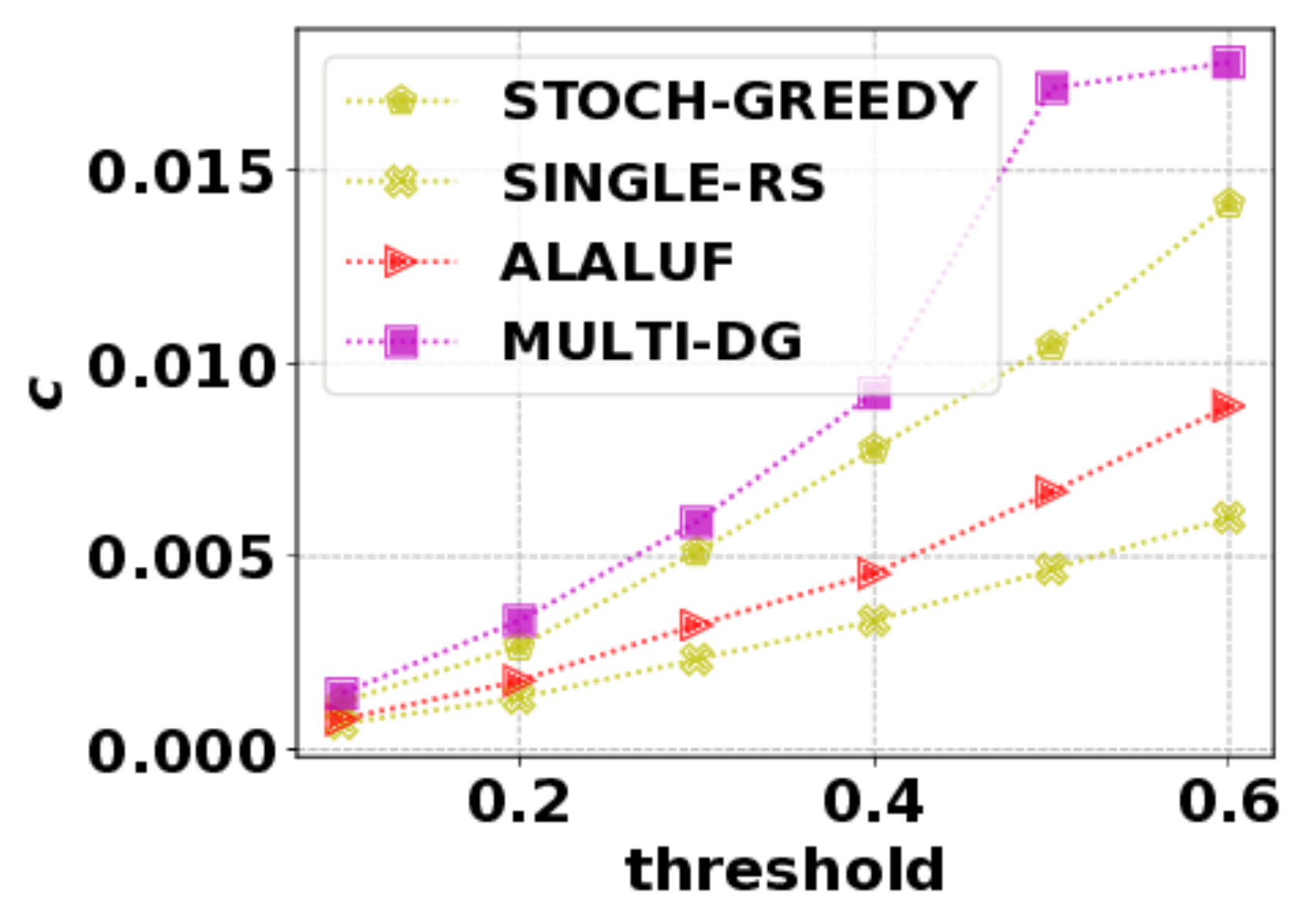}
    \label{astro_tauc}
  }
  %\hspace{-1em}
  %\subfigure[corel $c$] {
  %  \includegraphics[width=0.24\textwidth]{sections/figures/corel_tauc.pdf}
  %  \label{corel_tauc}
  %}
  \hspace{-1em}
  \subfigure[astro queries] {
    \includegraphics[width=0.24\textwidth]{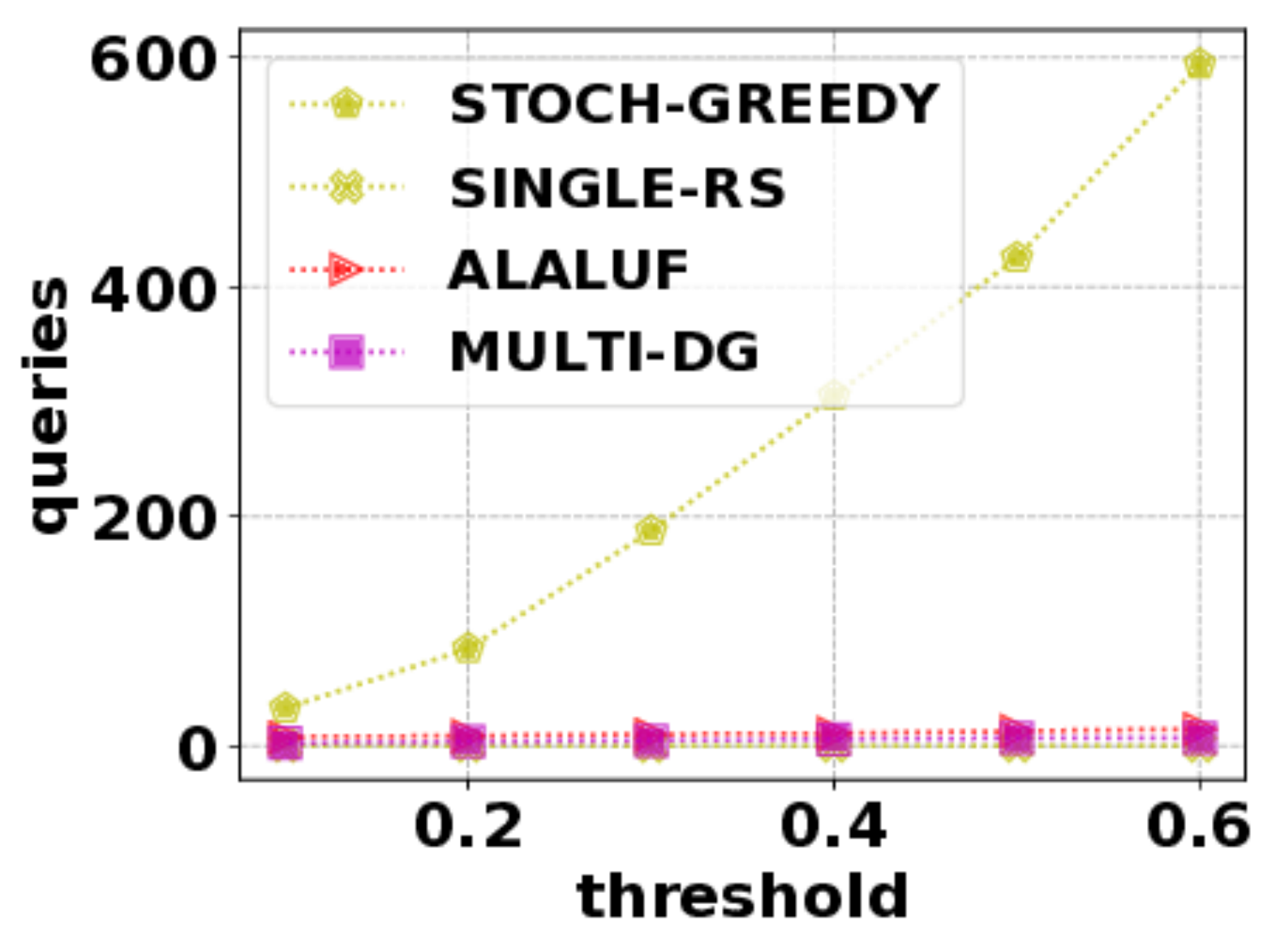}
    \label{astro_tauq}
  }
  %\hspace{-1em}
  %\subfigure[corel queries] {
  %  \includegraphics[width=0.24\textwidth]{sections/figures/corel_tauq.pdf}
  %  \label{corel_tauq}
  %}
  \hspace{-1em}
  \subfigure[astro memory] {
    \includegraphics[width=0.24\textwidth]{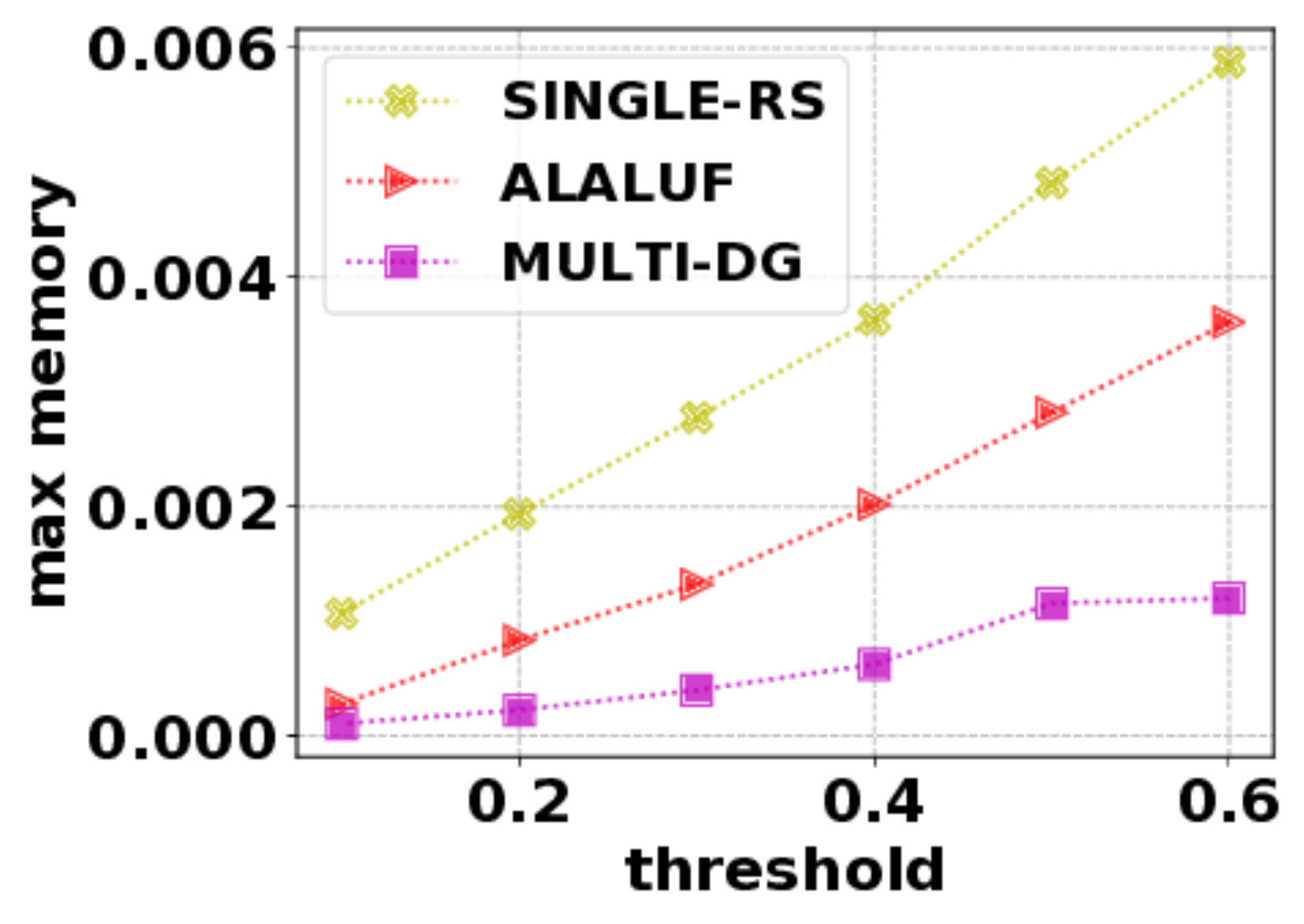}
    \label{astro_taumem}
  }
  %\hspace{-1em}
  %\subfigure[corel memory] {
  %  \includegraphics[width=0.24\textwidth]{sections/figures/corel_taumem.pdf}
  %  \label{corel_taumem}
  %}
  \caption{
  The experimental results of running the algorithms on instances of
  diverse data summarization on the
  %Corel5k (``corel'') and
  delicious (``delicious50k'') dataset,
  and instances of graph cut on the ca-AstroPh (``astro'')
  %and com-Amazon (``amazon'')
  dataset.
  \multi using the double greedy algorithm of \citet{buchbinder2015tight}
  is referred to as ``MULTI-DG''. \single using random sets as described by
  \citet{feige2011maximizing} is referred to as ``SINGLE-RS''.
  The \scp algorithms using the stochastic greedy algorithm of
  \citet{buchbinder2017comparing} and the streaming algorithm of \citet{alaluf2019optimal}
  are referred to as ``STOCH-GREEDY'' and ``ALALUF'' respectively.
  All $x$ and $y$ axes are normalized as described in Section \ref{section:expresults}.
  }
  \label{fig:shortexp}
\end{figure*}

  The experimental results are presented in Figure \ref{fig:shortexp}.
  In every experiment, the double greedy
  USM algorithm of \citet{buchbinder2015tight} (``DG'') is
  initially run as a baseline comparison.
  Let the cost, $f$ value, and number of queries of DG
  be $c_0$, $f_0$, and $q_0$ respectively.
  For all of the plots,
  the $f$ values on the y-axis are normalized by $f_0$, the $c$ values by $c_0$,
  the number of queries by $q_0$, and the max memory by $n$.
  Notice that the total cost in memory at one time of the algorithm is $n$.

  In the first set of experiments, \multi and \single are run on the
  delicious50k dataset with input $\tau=f_0$ and varying $\epsilon$
  (Figures \ref{delicious_epsf} to \ref{delicious_epsmem}).
  In Figure \ref{delicious_epsf}, one can see that
  the $f$ values of the
  solutions returned by each algorithm are close to their theoretical bounds.
  This is because the guaranteed lower bound on $f$ of $\gamma(1-\epsilon)\tau$
  (where $\gamma$ is 1/2 in \multi and 1/4 in \single)
  is used explicitly in both \multi and \single in order to decide which solutions
  to choose.
  In addition, one can see in Figures \ref{delicious_epsc} and \ref{delicious_epsmem}
  that \multi and \single substantially improve on the baseline DG when it comes to the
  solution cost $c$, as well as the maximum cost of all elements held in memory at once.
  On the other hand, \multi and \single make more queries to
  $f$ compared to DG as demonstrated in Figure \ref{delicious_epsq}.
  \single having lower $f$ and $c$ values compared to \multi is a result of
  the different subroutines used to solve USM; if they used the same one their
  solutions would be expected to be about the same. This is demonstrated in the
  experiments in the appendix.
  As seen in Figure \ref{delicious_epsmem},
  \single has higher maximum total cost of all elements held in memory at once than
  \multi, which is because \single may have instances of \stream corresponding to
  too high of guesses of $OPT$ as described in Section \ref{section:streamingsingle}.
  \single makes few queries to $f$ relative to \multi, but this is because it is
  using a constant time subroutine for USM, if \single used DG it would make significantly
  more queries to $f$ compared to \multi, as shown in the experiments
  in the appendix.

  In the second set of experiments, \multi and \single are run on the ca-AstroPh
  dataset with varying input $\tau$
  (Figures \ref{astro_tauf} to \ref{astro_taumem}).
  Figure \ref{astro_tauf} demonstrates that
  \multi reaches the highest $f$
  values, followed by the cover algorithm using the
  stochastic greedy algorithm of \citep{buchbinder2017comparing}
  (``SG''), then the cover algorithm using the streaming algorithm of
  \citet{alaluf2019optimal} (``AL''),
  and then finally \single.
  This is the order of their theoretical guarantees on $f$, with the exception
  of AL and \single being reversed.
  In addition, the algorithms all perform close to their worst case theoretical guarantees
  on $f$ as seen in the first set of experiments.
  Again, both these patterns are to be expected because
  the algorithms all explicitly choose their
  solutions based on their lower bound.
  Heuristic versions of the algorithms where the lower bound is assumed to be the
  same for all of them are compared in the appendix.
  In Figure \ref{astro_tauc}, one can see that
  \multi tends to produce a solution of higher cost compared to
  SG and AL, which is in line with their theoretical guarantees.
  However, \single produces the lowest cost solution of all despite having the same
  theoretical bound on cost as \multi.
  In Figure \ref{astro_tauq}, \multi and \single make relatively low total number of
  queries to $f$.
  %SG make excessively high queries to $f$ as $\tau$ increases.
  %While this may be because of constant factors in their runtimes,
  %another explanation is that SG tends to be more conservative with the
  %number of elements chosen into a solution compared to \multi and \single since SG
  %uses a cardinality constraint and has a strict budget,
  %resulting in many more guesses of
  %$OPT$ before getting one that yields sufficiently high $f$ value.
  SG is not included in Figure \ref{astro_taumem} since it requires the entire ground set in memory.
  Figure \ref{astro_taumem} demonstrates that
  \multi has the lowest maximum total stored cost in memory at once, and
  despite having the highest \single is reasonably close to the other two.
  %However, \single is the only algorithm
  %that take a single pass through $U$, and so the extra memory may be worth it
  %if multiple passes are not feasible.
  %In addition, recall that \single does not have a bound on its stored cost compared to $OPT$
  %(see Theorem \ref{theorem:single}), but despite this maintains a reasonably
  %low memory compared to the other algorithms.
\end{subsection}

\clearpage
\bibliography{etc/paper}
\clearpage
%\section{Appendix}
%In this section, additional content is presented that is omitted
%from the main paper due to space considerations.

\section{Additional Content to Section \ref{section:relatedwork}}
\label{section:appendixrelatedwork}
In this section, we include a more thorough comparison with the algorithm of
\cite{alaluf2020optimal}.
\citeauthor{alaluf2020optimal} studied the problem of non-monotone submodular maximization
subject to a cardinality constraint $\kappa$, which is the special case of \smp where
the cost $w$ is uniform, in the streaming setting.
Their algorithm uses an offline $\alpha$-approximation algorithm for non-monotone submodular
maximization as a subroutine, and finds an $\alpha/(1+\alpha)-\epsilon$ approximate
solution in a single pass with at most $O(\kappa/\epsilon^2)$ elements stored at once,
and makes at most $O((\log(k)+\log(1+\alpha))/\epsilon^2)$ queries of $f$.
Their algorithm works by making a pass through $U$ and filtering down the ground set,
where $O(1/\epsilon)$ disjoint sets are maintained, in a related way to \stream.
Finally, they run their $\alpha$-approximation algorithm for cardinality constrained
submodular maximization on the union of the disjoint sets.
\multi, \single, and \bimax are different than the algorithm of
\citeauthor{alaluf2020optimal} in a number of ways:
(1) \multi and \single are for \scp, a different optimization problem;
(2) \citeauthor{alaluf2020optimal} did not have to deal with the same difficulties
as \multi and \single in order to maintain low memory because the cost of
the optimal solution is known
to be $\kappa$;
(3) The final algorithm on the pooled sets for \stream is an
algorithm for unconstrained submodular maximization, not cardinality constrained
submodular maximization as in \citeauthor{alaluf2020optimal};
(4) \citeauthor{alaluf2020optimal} only considers uniform cost;
(5) The conditions \stream uses to store an element is different from what is used
by \citeauthor{alaluf2020optimal}.

\section{Additional Content to Section \ref{section:streamingtheoretical}}
\label{section:appendixtheoretical}
Proofs and helper lemmas
which were omitted from the section Algorithms and Theoretical Guarantees
are presented here. In addition, an example of what makes a single pass algorithm
difficult to develop for \scp is presented.
Finally, a more thorough description of the algorithms
\single and \bimax are included.

\subsection{Omitted Lemmas and Proofs}
The following is Theorem \ref{thm:impossible}, the proof of which was omitted in
Section \ref{section:streamingtheoretical}.
Theorem \ref{thm:impossible} describes
the limitations of how well we can find feasible solutions for \scp in polynomial time,
and is a clear result from the fact that USM
cannot be approximated in polynomial time better than $1/2$ assuming the value query
model \citep{feige2011maximizing}.

\smallskip
\noindent\textbf{Theorem \ref{thm:impossible}.}
\textit{
For any $\epsilon > 0$, there are instances of nonnegative symmetric submodular
cover such that there is no (adaptive, possibly randomized) algorithm
using fewer than $\Omega(\ln(1+\epsilon)e^{\epsilon^2n}/\ln(n))$ queries that always finds a solution of
expected $f$ value at least $(1/2+\epsilon)\tau$.
}
\begin{proof}
Suppose such an algorithm existed, and let it be called $\mathcal{A}$.
Then a new algorithm for unconstrained submodular maximization
is defined as follows: $\mathcal{A}$ is run on instance $\scp (U,f,(1+\epsilon)^i)$
for every $i\in\mathbb{Z}$ such that
$\max_{u\in U}f(\{u\})\leq (1+\epsilon)^i\leq n\max_{u\in U}f(\{u\})$, and the solution with
the highest value of $f$ is returned.
Notice this results in running $\mathcal{A}$ $\ln(n)/\ln(1+\epsilon)$ times.
Because $OPT$ is in the above range, there exists some $i$ such that
$(1+\epsilon)^{i-1}\leq OPT \leq (1+\epsilon)^i$. Once $\mathcal{A}$ is run on
$\scp (U,f,(1+\epsilon)^i)$, by assumption it will return $X$ such that
$\mathbb{E}[f(X)]\geq (1/2+\epsilon)\tau$. This contradicts the result of \citeauthor{feige2011maximizing}.
\end{proof}

Lemma \ref{lemma:streamratio} relied on Claim \ref{claim:mono} in order to proven.
Here, we present a proof of Claim \ref{claim:mono}, but first we will need a
result from \citet{feige2011maximizing}.
\begin{lemma}
  \label{lemma:expected}
  (Lemma 2.2 from \cite{feige2011maximizing}) Let $g:2^U\to\mathbb{R}_{\geq 0}$ be a non-negative submodular
  function. Denote by $A(p)$ a random subset of $A$ where each element appears with
  probability at most $p$ (not necessarily independently). Then
  $\mathbb{E}[g(A(p))]\geq (1-p)g(\emptyset)$.
\end{lemma}
Now we can prove Claim \ref{claim:mono} using Lemma \ref{lemma:expected}.

\smallskip
\noindent\textbf{Claim \ref{claim:mono}.}
\textit{
Let $A_1,...,A_m\subseteq U$ be disjoint, and $B\subseteq U$.
Then there exists $i\in\{1,...,m\}$ such that
$f(A_i\cup B)\geq (1-1/m)f(B)$.
}
\begin{proof}
  Define $g(X)=f(B\cup X)$. Then $g$ is a non-negative submodular function.
  Consider choosing $A$ uniformly randomly from the disjoint sets $A_1,...,A_m$.
  Then any element of $U$ has probability at most $1/m$ of being in $A$. Then
  \begin{align*}
    \frac{1}{m}\sum_{i=1}^m f(B\cup A_i) &= \frac{1}{m}\sum_{i=1}^m g(A_i) \\
    &= \mathbb{E}[g(A)] \\
    &\overset{(a)}{\geq} \left(1-\frac{1}{m}\right)g(\emptyset) \\
    &= \left(1-\frac{1}{m}\right)f(B)
  \end{align*}
  where (a) is a result from \citet{feige2011maximizing}
  which is stated as Lemma \ref{lemma:expected} in the appendix.
  Therefore there must exist some $i\in\{1,...,m\}$ such that
  $f(A_i\cup B)\geq (1-1/m)f(B)$.
\end{proof}

\subsection{Limitations of Single Pass Algorithms for \scp}
To see the difficulty with making a single pass through $U$,
suppose we have some single pass streaming algorithm for \scp
that produces a solution with constraint approximation $\beta$.
Consider two instances of \scp with uniform cost $w$ defined as follows:
(i) \scp$(\{u_1,...,u_n\},f_1,w,\tau)$ where
$f_1$ is modular\footnote{
$f$ is modular if $f(X)=\sum_{x\in X}f(x)$ for all $X\subseteq U$.}
and $f(u_i)=\tau/n$ for all $i$; (ii) \scp$(\{u_1,...,u_n\},f_2,w,\tau)$
where $f_2$ is modular and $f(u_i)=\tau/n$ for all $i\neq n$ and $f(u_n)=\tau$.
Suppose the algorithm receives the universe in order $u_1,...,u_n$. Then because the returned
solution has constraint value at least $\beta n$, in instance (i) the algorithm must store
at least $\beta n - 1$ elements before reading element $u_n$. On the other hand,
instances (i) and (ii) are indistinguishable up to element $u_n$, therefore for
instance (ii) the algorithm also stores at least $\beta n -1$ elements.
However, $OPT=1$ in the latter case, and therefore this stored memory is very large compared
to $OPT$.

\subsection{The Algorithm \single}
\begin{algorithm}[tb]
\caption{\single}\label{alg:single}
\textbf{Input}: Value oracles to $f$ and $w$, $\tau$, $\epsilon$, and $B$\\
\textbf{Output}: $S\subseteq U$
\begin{algorithmic}[1]

 \STATE{$S[(1+\epsilon)^i,j]\gets\emptyset$ $\forall i\in\mathbb{Z}, j\in\jsb$}
 \STATE{$L\gets -1$}
 \FOR {$u$ received from stream}\label{line:univs}
  \IF {$f(u)/w(u)>\epsilon\tau/(2 L)$}\label{line:decreaseopts}
   \STATE {$L\gets \epsilon\tau w(u)/(2f(u))$}
  \ENDIF
  \FOR {$\guesssingle$ in $\{(1+\epsilon)^i: i\in\mathbb{Z}, L\leq (1+\epsilon)^i\leq B\}$} \label{line:lazy}
   \IF {$w(S[\guesssingle,i])<2\guesssingle/\epsilon$ $\forall i$ and $w(u)\leq\sigma$}
    \IF {$\exists i$ s.t. $\Delta f(S[\guesssingle,i], u)\geq w(u)\epsilon \tau/(2\guesssingle)$}
     \STATE {$S[\guesssingle,i]\gets S[\guesssingle,i] \cup \{u\}$}
    \ENDIF
   \ENDIF
   \STATE {$S[\guesssingle, 0]\gets$\uncons$(\cup_{i=1}^{\smallpb}S_{\guesssingle,i})$}
   \IF {$\max\{f(S[\guesssingle,i]): i\in\{0,...,\smallpb\}\} \geq \gamma(1-\epsilon)\tau\}$}\label{line:test}
    \STATE {$B\gets\guesssingle$}
    %\STATE {$S[(1+\epsilon)^i,j]\gets\emptyset$ $\forall i,j$ $s.t. (1+\epsilon)^i > B$}
   \ENDIF
  \ENDFOR
 \ENDFOR
 \STATE \textbf{return} $\argmax\{f(S[B,i]): i\in\jsb\}$

\end{algorithmic}
\end{algorithm}

\single works in a similar way to \multi, except: (i) The instances of \stream
corresponding to each guess of $OPT$ are running in parallel rather than
sequentially; (ii) Instead of each instance of \stream running \uncons at the
end, \uncons is run periodically as elements are received.
A lower bound on the guesses of $OPT$, $L$, is updated lazily as \single runs.
An upper bound on the guesses of $OPT$, $B$, is initially given as an input
and then is updated by running
\uncons over stored elements periodically as \single runs.
Guesses of $OPT$ are
$\{(1+\epsilon)^i: i\in\mathbb{Z}, L\leq (1+\epsilon)^i \leq B\}$, and each guess of $OPT$
corresponds to an instance of \stream running in parallel.
The sets for the instance of \stream
corresponding to guess $\sigma$ are $S[\sigma,0],...,S[\sigma,\smallpb]$.
In particular, let $U_i=\{u_1,...,u_i\}$ be the elements received at a certain
point of \stream.
Then the lower bound $L$ is $L=\min\{\epsilon\tau w(u)/(2f(u)): u\in U_i\}$.
$B$ is maintained such that there does not exist a guess $\sigma$
such that $\sigma < B$ and
$\max\{f(S[\sigma,i]): i\in\smallpb\}\geq \gamma(1-\epsilon)\tau$,
i.e. only the biggest guess of $OPT$ has found a solution with $f$ value at least
$\gamma(1-\epsilon)\tau$, if such a solution has been found at all.
Any instances
of \stream corresponding to guesses of $OPT$ above the upper bound are
assumed to be discarded.
Pseudocode for \single is given in Algorithm \ref{alg:single}.

\smallskip
\noindent\textbf{Theorem \ref{theorem:single}.}
\textit{
Suppose that \single is run for an instance of \scp,
and input $B\geq OPT$.
Define the following two functions:
$$m(x)=(1+\epsilon)(4/\epsilon^2+1)\frac{x\ln\left(2x/(\epsilon\tau\xi)\right)}{\ln(1+\epsilon)}$$
$$q(x)=\frac{\ln\left(2x/(\epsilon\tau\xi)\right)}{\ln(1+\epsilon)}\left(\frac{2}{\epsilon}+\frac{\mathcal{T}((4/\epsilon^2+1)x)}{w_{min}}\right)$$
where $\xi = \min_{u\in U}w(u)/f(\{u\})$, and
$\mathcal{T}(m)$ is the number of queries of \uncons on an input set of size $m$.
Then, \single:
\vspace{-1em}
\begin{itemize}[noitemsep]
  \item[(i)] Returns a set $S$ such that
    $f(S)\geq\gamma(1-\epsilon)\tau$ and $w(S) \leq \ratio OPT$;
  \item[(ii)] The total cost of all elements stored at once is at
    most $m(B)$;
  \item[(iii)] Makes at most $q(B)$
  queries of $f$ per arriving element of the stream.
\end{itemize}
Let $u_1,...,u_n$ be the order that the elements of $U$ arrive in, and
$U_i=\{u_1,...,u_i\}$.
If the instance $SC(U_i,f,w,\tau)$ is feasible and has optimal cost $OPT_i$,
then from the end of the $i$th iteration of the loop in \single onwards:
\begin{itemize}[noitemsep]
  \item[(iv)] The total cost of all elements stored at once is
    at most $m(OPT_i)$;
  \item[(v)] At most $q(OPT_i)$
  queries of $f$ are made per arriving element of the stream.
\end{itemize}
}
\begin{proof}
  Consider an alternate version of \stream where instead of running \uncons
  on $\cup S_i$ after
  receiving all elements in the stream (Line \ref{line:uncons}), \uncons is run at the end
  of each iteration of the loop on Line \ref{line:loop} of \stream.
  %I.e., \uncons is run after reading in each element of $U$.
  Notice that this does not change
  any of the properties of \stream detailed in Lemmas \ref{lemma:streamweight} and \ref{lemma:streamratio}
  except the number of queries to $f$.
  From this point on in the proof, we will consider this alternative version of \stream.

  Consider the value of $B$ at the end of some iteration of the for loop on
  Line \ref{line:univs} of \single. It is now shown that without loss of generality, one
  can assume that up to this point \single is running
  \stream in parallel with guesses of $OPT$
  %$\{(1+\epsilon)^i: i\in\mathbb{Z}, \epsilon\tau /(2m)\leq (1+\epsilon)^i\leq B\}$
  $\{(1+\epsilon)^i: i\in\mathbb{Z}, (1+\epsilon)^i\leq B\}$.
  $B$ is only decreasing throughout \stream, and so changes in $B$ only result in
  removing instances of \stream, not adding them.
  Therefore we only need to show that
  guesses of $OPT$ smaller than $B$ are w.l.o.g. running in parallel.

  Consider any $(1+\epsilon)^i\leq B$.
  Consider any previous iteration of the loop on Line \ref{line:univs}
  such that for the first time an $u$ has arrived such that
  $\Delta f(\emptyset, u)\geq w(u)\epsilon\tau/(2(1+\epsilon)^i)$
  (i.e. the first time an element should be added to $S[(1+\epsilon)^i,j]$ for some
  $j\in\js$), and we are at the beginning of the loop on Line \ref{line:univs}.
  If $L > (1+\epsilon)^i$, then
  \begin{align*}
    f(u)/w(u) &\geq \Delta f(\emptyset, u)/w(u) \\
    &\geq \epsilon\tau/(2(1+\epsilon)^i) \\
    &> \epsilon\tau/(2L).
  \end{align*}
  Therefore the if statement on Line \ref{line:decreaseopts} will be true,
  $L$ will be reset to $\epsilon\tau w(u)/(2f(u))$, and $(1+\epsilon)^i$ added to
  the guesses of $OPT$ since
  \begin{align*}
    (1+\epsilon)^i &\geq w(u)\epsilon\tau/(2\Delta f(\emptyset, x)) \\
    &\geq L.
  \end{align*}

  Item (i) is now proven. By Lemma \ref{lemma:streamratio}, if there
  exists a run of \stream with a guess of $OPT$ that is at least as big, then
  the set returned by \stream has $f$ value at least $\gamma(1-\epsilon)\tau$.
  Therefore by the end of \single, any run of \stream corresponding to a guess
  of $OPT$ that is at least as big as $OPT$ must have triggered the if
  statement on Line \ref{line:test}.
  Initially $B\geq OPT$, and only decreases if the if statement on
  Line \ref{line:test} is true, it must be that the solution $S$ of \single
  has $f(S)\geq\gamma(1-\epsilon)\tau$.
  In addition, the above discussion implies that $B$ is no greater than
  $(1+\epsilon)OPT$ at the end of \single, then Lemma \ref{lemma:streamweight}
  implies the remaining part of item (i).

  Item (ii) is now proven.
  By Lemma \ref{lemma:streamweight}, the total cost of all elements stored by each
  run of \stream with input $(\epsilon,\sigma)$ is $(4/\epsilon^2+1) \sigma$, which is bounded
  above by $(4/\epsilon^2+1)B$.
  In addition, $L \geq \epsilon\tau\xi/2$, and therefore there are at most
  $\ln(2B/(\epsilon\tau\xi))/\ln(1+\epsilon)$ parallel instances of \stream running
  in \single. This proves item (iii).

  Item (iii) is now proven.
  Since $B$ is the biggest guess of $OPT$,
  the alternative versions of \stream that \single is running makes at most
  $2/\epsilon+\mathcal{T}((4/\epsilon^2+1)\guess/w_{min})\leq 2/\epsilon+\mathcal{T}((4/\epsilon^2+1)B/w_{min})$
  queries to $f$ per element, which can be proven using the same argument as in
  Lemma \ref{lemma:streamweight}.
  Combining this with the fact that there are at most
  $\ln(2B/(\epsilon\tau\xi))/\ln(1+\epsilon)$ parallel instances of \stream running
  proves item (iii).

  Finally, item (iv) and (v) are proven.
  Suppose the iteration of the for loop on Line \ref{line:univs} corresponding to
  element $u_i$ is complete.
  By a nearly identical argument to that used for item (i), one
  can see that the largest guess of $OPT$ is no bigger than $(1+\epsilon)OPT_i$
  from this point on. Therefore the largest memory for any run of \stream from this
  point on is
  $\ratio OPT_i$,
  and any run will make at most $2/\epsilon+\mathcal{T}((4/\epsilon^2+1)OPT_i/w_{min})$
  queries per element received, which can be proven using the same argument as
  in Lemma \ref{lemma:streamweight}.
  There are at most $\ln(2OPT_i/(\epsilon\tau\xi))$
  parallel instances of \stream running
  in \single. Altogether this implies items (iv) and (v).

\end{proof}

\subsubsection{The Algorithm \bimax}
The algorithm \bimax was presented for the problem
\smp in Section \ref{section:bimax}.
In \smp, the $f$ value of the optimal solution is unknown (in contrast, in
\scp it is known to be $\tau$).
\smp runs versions of \stream in parallel where instead of input $\guess$, $\guess$ is
fixed at $\kappa$, and instead the value of $\tau$ is guessed.
In particular, the set
$\{(1+\epsilon)^i: i\in\mathbb{Z}, m\leq (1+\epsilon)^i\leq 2m\kappa/\epsilon\}$
are the guesses of $OPT$
Because the cost of the optimal solution is known to be at most
$\kappa$ in \smp, the total stored cost
at once for every instance of \stream is bounded by $(4/\epsilon^2+1)\kappa$
(see Lemma \ref{lemma:streamweight}).
For this reason, we avoid difficulties of having too high
total stored cost as we did in \single.
\bimax lazily keeps track of an upper and lower bound
for the $f$ value of the optimal solution as elements arrive from the stream in
a similar manner as the lower bound $L$
was updated in \single.
Pseudocode for \bimax is presented in Algorithm \ref{alg:bimax}.
In addition, the theoretical guarantees of \bimax, stated in Theorem \ref{theorem:bimax}
of the main content, are proven below.

\begin{algorithm}[tb]
\caption{\bimax}\label{alg:bimax}
\textbf{Input}: Value oracles to $f$ and $w$, $\kappa$, and $\epsilon$\\
\textbf{Output}: $S\subseteq U$
\begin{algorithmic}[1]
 \STATE{$S[(1+\epsilon)^i,j]\gets\emptyset$ $\forall i\in\mathbb{Z}, j\in\jsb$}
 \FOR {$x$ received from stream}
  \IF{$f(\{x\})/w(x)>m$}
   \STATE{$m\gets f(\{x\})/w(x)$}
   %\STATE{$S_{(1+\epsilon)^i,j}\gets\emptyset$ $\forall i\in\mathbb{Z}, (1+\epsilon)^i < f(\{x\}), j\in\jsb$}\label{line:lazysm}
  \ENDIF
  \FOR {$\tau$ in $\{(1+\epsilon)^i: i\in\mathbb{Z}, m\leq (1+\epsilon)^i \leq 2m\kappa/\epsilon\}$}
   \IF {$w(S[\tau,i])<2\kappa/\epsilon$ $\forall i$ and $w(x)\leq\kappa$}
    \IF {$\exists i$ s.t. $\Delta f(S[\tau,i], x)\geq w(x)\epsilon \tau/(2\kappa)$}
     \STATE {$S[\tau,i]\gets S[\tau,i] \cup \{x\}$}
    \ENDIF
   \ENDIF
  \ENDFOR
 \ENDFOR
 \FOR {$\tau$ in $T$}
  \STATE{$S[\tau, 0]\gets$\uncons$(\cup_{i=1}^{\smallpb}S[\tau,i])$}
 \ENDFOR
 \STATE \textbf{return} $\argmax\{f(S[\tau,i]): \tau\in T, i\in\jsb\}$

\end{algorithmic}
\end{algorithm}

\smallskip
\noindent\textbf{Theorem \ref{theorem:bimax}.}
\textit{
Suppose that \bimax is run for an instance of \smp:
Then:
\vspace{-1em}
\begin{itemize}[noitemsep]
  \item[(i)] The set $S$ returned by \bimax satisfies $f(S)\geq\gamma(1-\epsilon)OPT$
  and $w(S)\leq \ratio\kappa$;
  \item[(ii)] The total cost of all elements needing to be stored at once is at most
  $(4/\epsilon^2+1)\ln(2\kappa/(w_{min}\epsilon))/\ln(1+\epsilon)\kappa$;
  \item[(iii)] And at most
  $\ln(2\kappa/(w_{min}\epsilon))(2n/\epsilon + \mathcal{T}((4/\epsilon^2+1)\kappa))/\ln(1+\epsilon)$
  queries of $f$ are made in total
  where $\mathcal{T}(m)$ is the number of queries of \uncons on an input set of size $m$.
\end{itemize}
}
\begin{proof}
  In order to prove Theorem \ref{theorem:bimax}, a new version of Lemma \ref{lemma:streamratio}
  is needed. The following Lemma is proved in as essentially identical way to Lemma \ref{lemma:streamratio}:
  \begin{lemma}
    \label{lemma:streamratio2}
    Suppose that \stream is run with input \defineepsilon,
    and $\tau\geq OPT$.
    Let $S$ be the set returned by \stream.
    Then $f(S)\geq\gamma (1-\epsilon)OPT$.
  \end{lemma}
  Similar to \single, \bimax is essentially running
  a bunch of instances of \stream in parallel as $U$ is read in.
  In particular, the set
  $\{(1+\epsilon)^i: i\in\mathbb{Z}, m\leq (1+\epsilon)^i\leq 2m\kappa/\epsilon\}$
  are the guesses of $OPT$, and there is an instance of \stream corresponding to
  each guess.
  For each guess $\tau$, $S_{\tau,0},...,S_{\tau,2/\epsilon}$ in \single
  correspond to the sets $S_0,...,S_{2/\epsilon}$ in \stream.

  Define $q\in\mathbb{Z}$ to be the unique value such that
  \begin{align*}
    (1+\epsilon)^q \leq OPT < (1+\epsilon)^{q+1}.
  \end{align*}
  Then we may assume without loss of generality that there is an instance of \stream
  corresponding to $(1+\epsilon)^q$ as a guess of $OPT$ for the duration of
  \bimax, as explained as follows.
  First of all, clearly $(1+\epsilon)^q\geq \max\{f(\{x\}): x\in U\}$ and therefore is
  at least the smallest guess throughout the duration of \bimax.
  On the other hand, suppose that for the first time we have received from the stream an
  element $x$ such that $\Delta f(\emptyset,x)\geq \epsilon w(x) (1+\epsilon)^q/(2\kappa)$
  (i.e. the first time an element $x$ should be added to $S_{(1+\epsilon)^q, i}$ for some $i\in\js$).
  If $(1+\epsilon)^q>2m\kappa/\epsilon$ at the beginning of the for loop then
  \begin{align*}
    f(\{x\})/w(x) &\overset{(a)}{\geq} \Delta f(\emptyset,x)/w(x) \\
    &\geq \epsilon(1+\epsilon)^q/(2\kappa) \\
    &> m
  \end{align*}
  where (a) is because $f(\emptyset)\geq 0$.
  Therefore the if statement will be true, $m$ will be re-assigned
  as $f(\{x\})/w(x)$, and $(1+\epsilon)^q$ added to the guess of $OPT$ since
  \begin{align*}
    (1+\epsilon)^q &\leq 2\Delta f(\emptyset,x)\kappa/(w(x)\epsilon) \\
    &\leq 2f(\{x\})\kappa/(w(x)\epsilon) \\
    &= 2m\kappa/\epsilon
  \end{align*}
  and will remain in the guesses until the end.

  In light of the above, items (i), (ii), and (iii) follow by an analogous
  argument as in Theorem \ref{theorem:single}.
  \end{proof}

\section{Additional Content to Section \ref{section:streamexperiments}}
\label{section:appendixexp}
The experimental results here are a superset of those included in the main paper.
In addition, additional details about the applications and setup are included here.

\subsection{Applications of \scp}
  In sections \ref{section:streamexperiments}, the algorithms \single and \multi are evaluated
  on instances of non-monotone submodular cover involving graph cut \citep{balkanski2018non}
  and diverse summarization \citep{tschiatschek2014learning} functions.
  Definitions of both of these applications are now provided.

  The first application considered is \scp where $f$ is a
  graph cut function, which is a submodular but not necessarily monotone function.
  Graph cut functions have frequently been used as applications of non-monotone
  submodular maximization.
  A graph cut function takes in a set of vertices in a graph $X$ and computes the
  total number of edges between $X$ and $U\setminus X$.
  The problem definition is defined as follows.
  \begin{definition}[Graph cut]
    Let $G=(V,E)$ be a graph, and $w:E\to\mathbb{R}_{\geq 0}$ be a function that gives a
    weight for every edge in the graph. Define \definef to be a function that takes
    $X\subseteq V$ to the total weight of edges between $X$ and $V\setminus X$, i.e.
    \begin{align*}
      f(X) = \sum_{x\in X, y\notin X}w(x,y).
    \end{align*}
    Then $f$ is submodular and non-negative, but is not necessarily monotone.
  \end{definition}

  The second application considered is \scp where $f$ is a diverse data summarization
  function, which is also a submodular but not necessarily monotone function.
  A diverse data summarization function takes in a subset $X$ of a data set $U$
  and returns a score of how effective $X$ summarizes $U$, while penalizing
  for similarity between the elements of $X$.
  Variants of diverse data summarization are also a popular application for non-monotone
  submodular maximization.
  The particular formulation used here is based on \textit{tagged data}, and is defined
  as follows.
  \begin{definition}[Diverse summarization of a tagged data set]
    Suppose the data points in $U$ are each tagged by a subset of tags $T$ via the
    function $t:U\to 2^T$. E.g. if $U$ is a set of images then $T$ may be words
    describing each image.
    Given parameter $\gamma\geq 0$, define \definef to be
    \begin{align*}
      f(X)=|\cup_{x\in X}t(X)| - \gamma \sum_{x\in X, y\in X}\frac{|t(x)\cup t(y)|}{|t(x)\cap t(y)|}.
    \end{align*}
    The first term in $f(X)$ is the total number of tags covered by a summary $X$,
    while the second term is a penalty to encourage diversity in the summary
    (using the Jaccard similarity).
    $f$ is submodular, but not necessarily monotone or non-negative.
    If $\gamma$ is sufficiently small, then $f$ is non-negative.
  \end{definition}

  For the experiments in this paper, we set
  $$\gamma = |\cup_{x\in U}t(X)|/\sum_{x\in U, y\in U}\frac{|t(x)\cup t(y)|}{|t(x)\cap t(y)|}.$$

  \subsection{Experimental Setup}
  The graph cut instances presented here are on the
  ca-AstroPh ($n=18772$, 198110 edges), com-Amazon ($n=334863$, 925872 edges), and
  email-Enron ($n=36692$, 183831 edges)
  networks from the SNAP large network collection \citep{snapnets}.
  In all of the cut instances the cost of each element is uniformly set as 1.
  The diverse summarization instances are on the
  Corel5k set of images \citep{duygulu2002object} ($n=5000$),
  and a subset of tagged webpages from the delicious.com website
  \citep{soleimani2016semi} ($n=5000$ or $n=50000$ depending on the instance)
  The smaller instances are used in some experiments because some the
  comparison algorithms cannot run on the
  larger datasets within a couple of hours. Specifically, when the local search
  algorithm of \cite{feige2011maximizing} is used as a subroutine for USM then
  \single takes too long to run.
  In addition, the cover algorithm by using the
  stochastic greedy algorithm of \cite{buchbinder2017comparing} for submodular
  maximization
  along with the approach of \cite{iyer2013} takes too long to run.
  The Corel5k images are each losslessly compressed, and their cost is
  assigned to be their size in kB after compression.
  The costs of the websites from the delicious.com website are uniform.

  Several USM algorithms are run as subroutines of \single and \multi:
  (i) Repeated runs of the randomized double greedy algorithm
  of \citet{buchbinder2015tight} (``DG'');
  (ii) The local search
  algorithm of \citet{feige2011maximizing} with parameter 0.25 (``LS''),
  which is a deterministic $1/3$ approximation for USM, and on a set of size $r$
  makes $O(r^3\ln(r))$ queries to $f$;
  (iii) Repeatedly returning a random set, as described by
  \citet{feige2011maximizing} (``RS'').
  The randomized algorithms for USM (i and iii)
  are repeated 50 times and the best solution is chosen.

  The comparison algorithms using the method of \cite{iyer2013} described in the main
  text are only for uniform cost. For the Corel5k dataset, which has non-uniform cost,
  we use a modified version of each algorithm
  where the marginal gain function $\Delta f(X,x)$ is replaced
  by $\Delta f(X,x)/w(x)$. This modified version does not have any proven approximation
  guarantee.

  %Any parameters that are not otherwise defined are given in Table \ref{table:parameters}.
  %\input{sections/parameters}

  \subsection{Additional Experimental Results}
  \label{section:resultsapp}
\begin{figure*}[t!]
  \centering
  \hspace{-1em}
  \subfigure[enron, cut] {
    \includegraphics[width=0.24\textwidth]{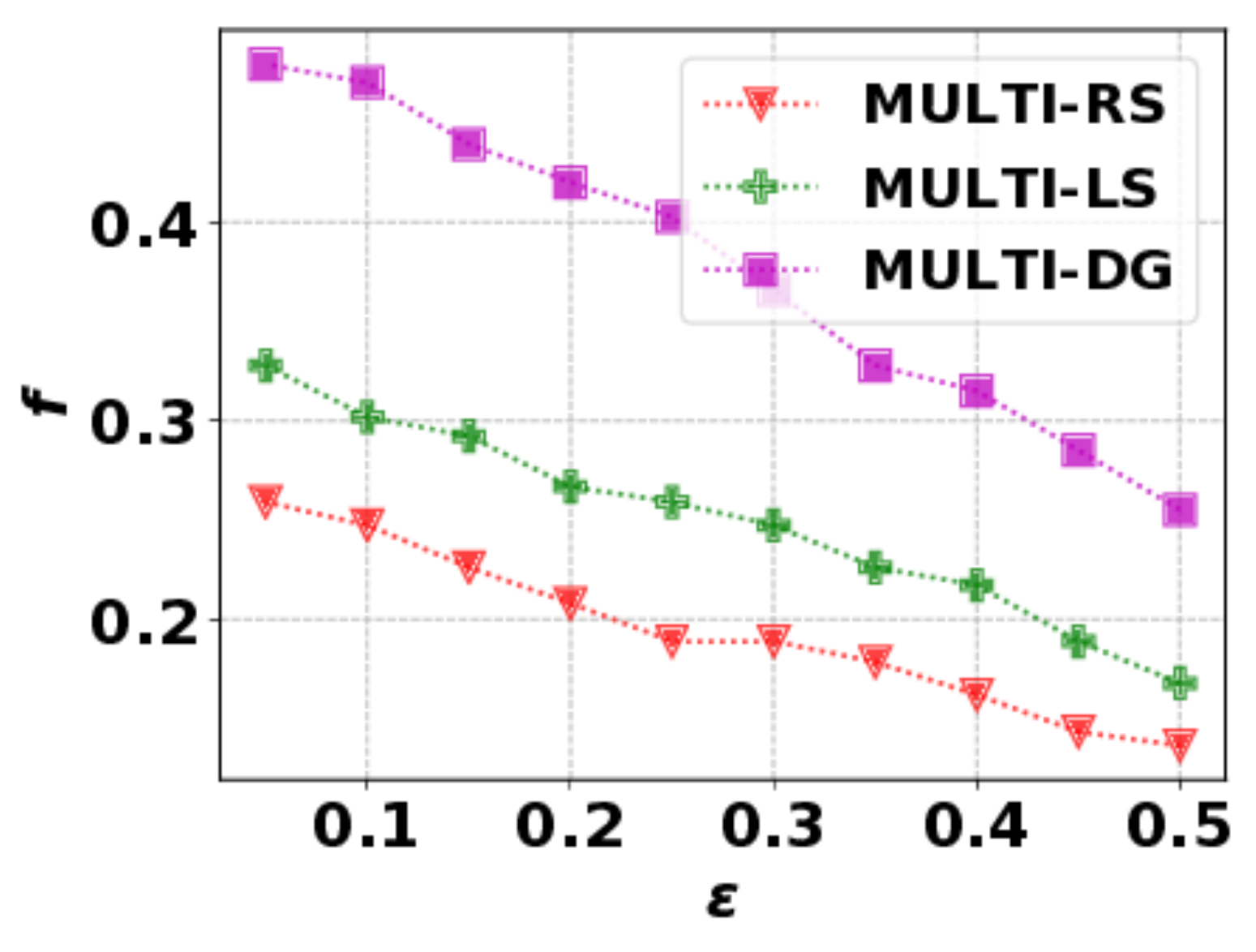}
  }
  \hspace{-1em}
  \subfigure[astro, cut] {
    \includegraphics[width=0.24\textwidth]{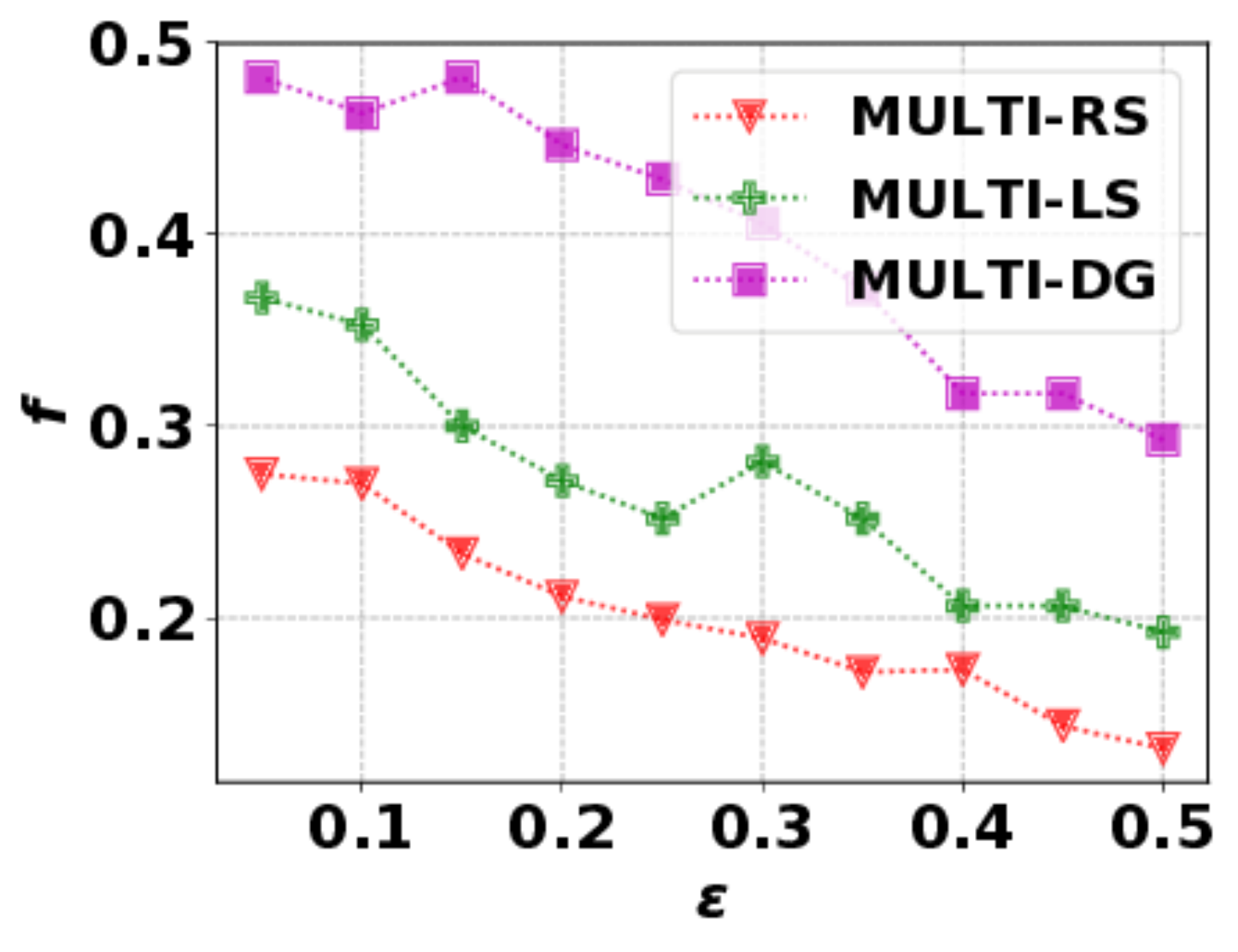}
  }
  \hspace{-1em}
  \subfigure[corel, cover] {
    \includegraphics[width=0.24\textwidth]{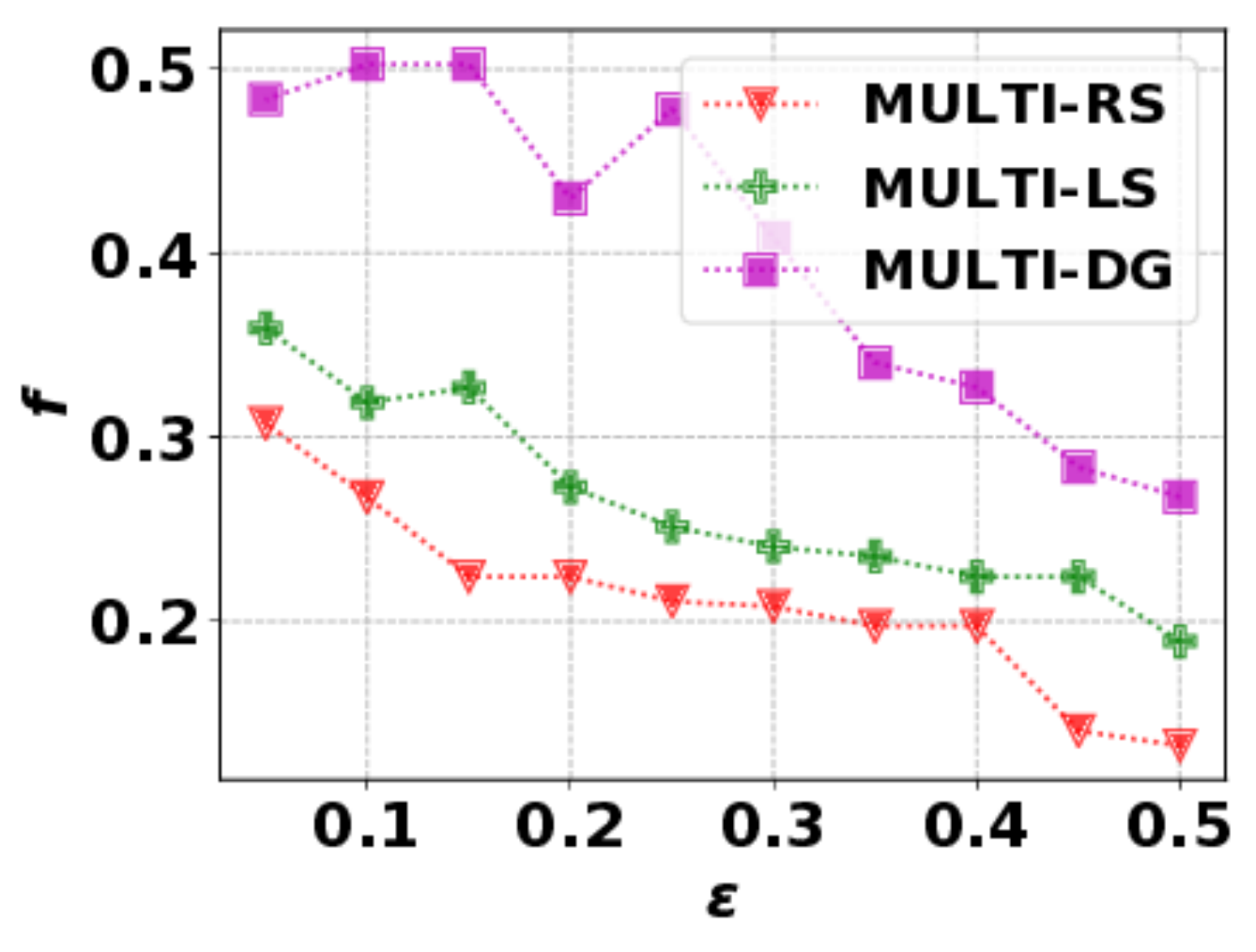}
  }
  \hspace{-1em}
  \subfigure[delicious5k, cover] {
    \includegraphics[width=0.24\textwidth]{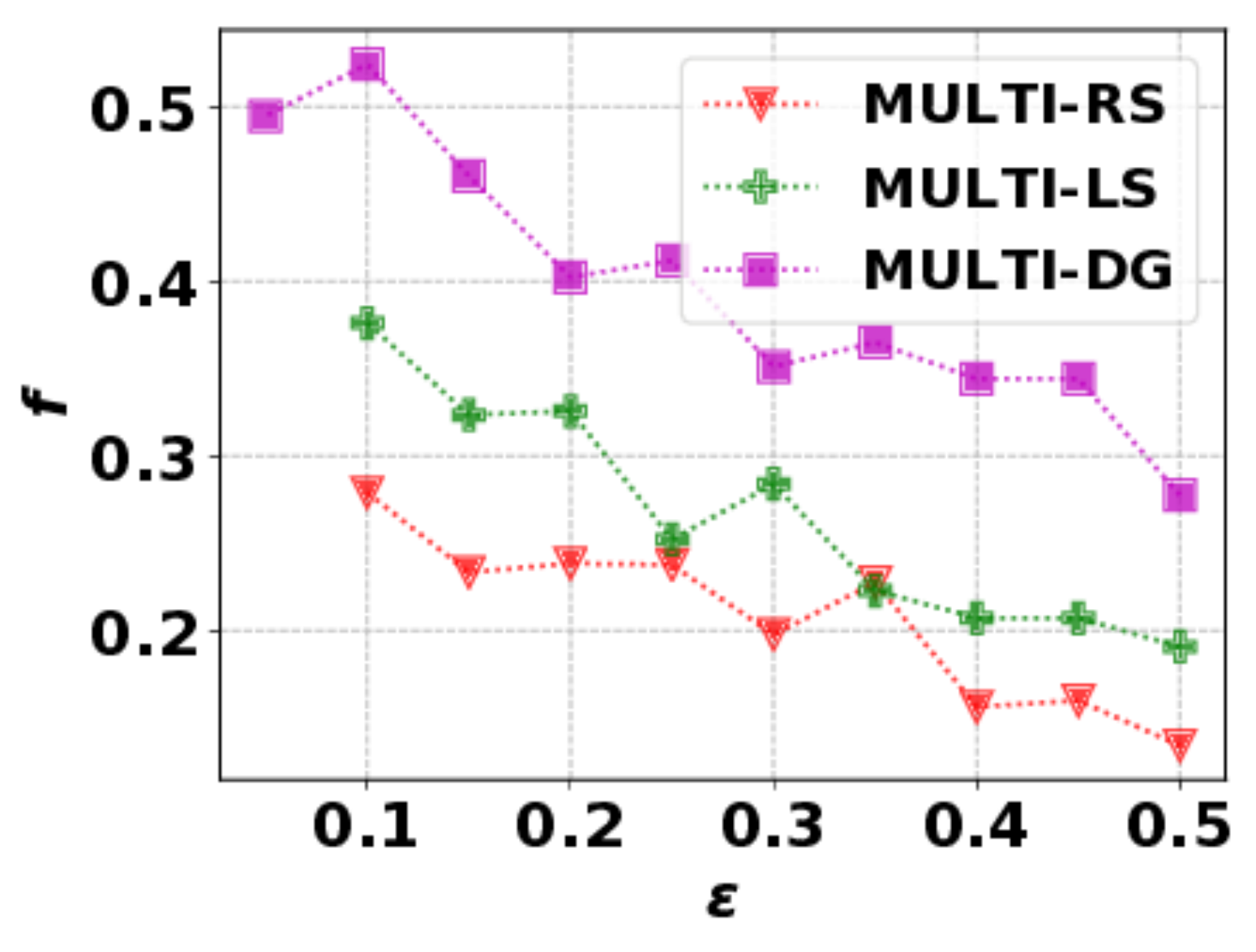}
  }
  \hspace{-1em}
  \subfigure[enron, cut] {
    \includegraphics[width=0.24\textwidth]{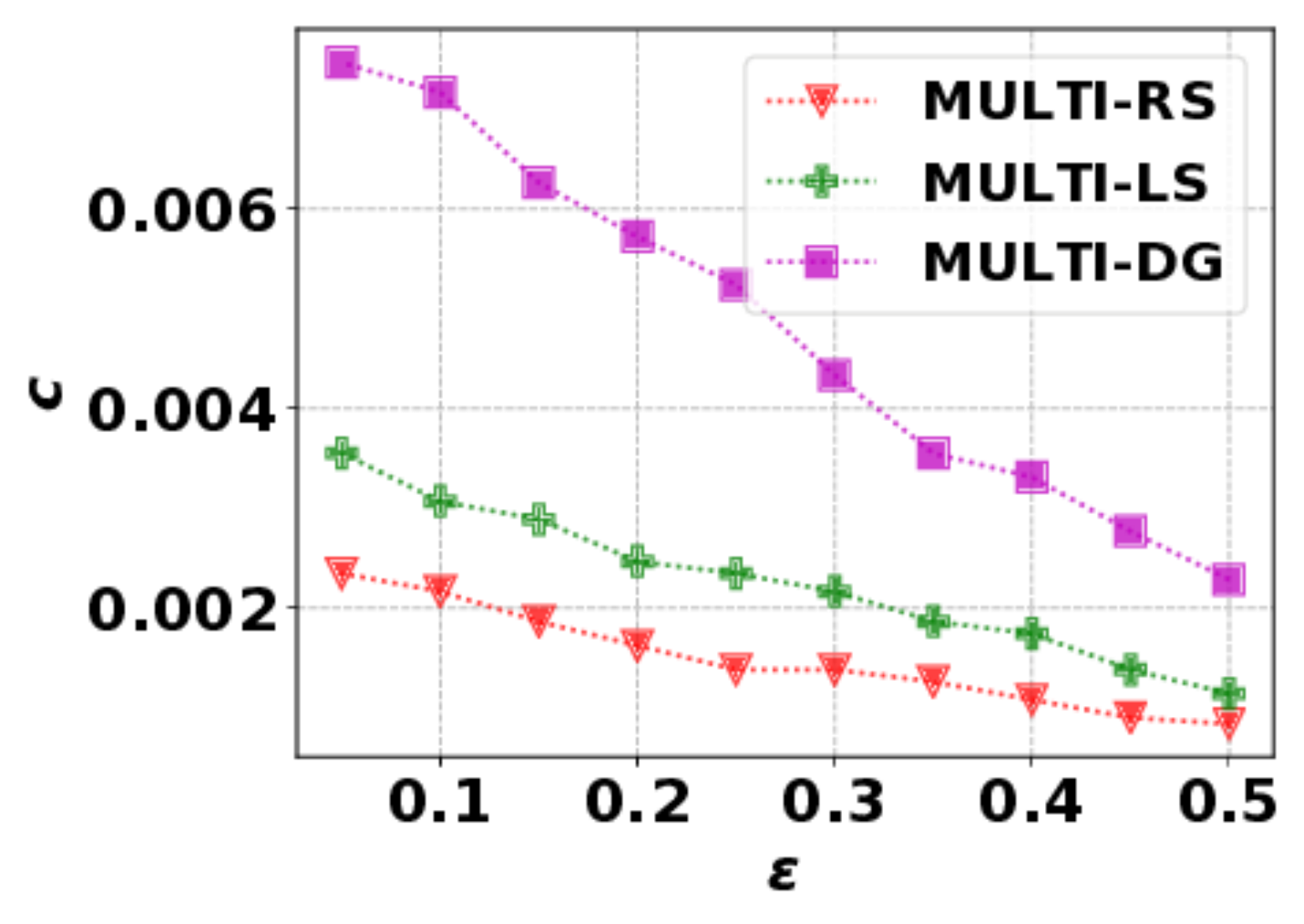}
  }
  \hspace{-1em}
  \subfigure[astro, cut] {
    \includegraphics[width=0.24\textwidth]{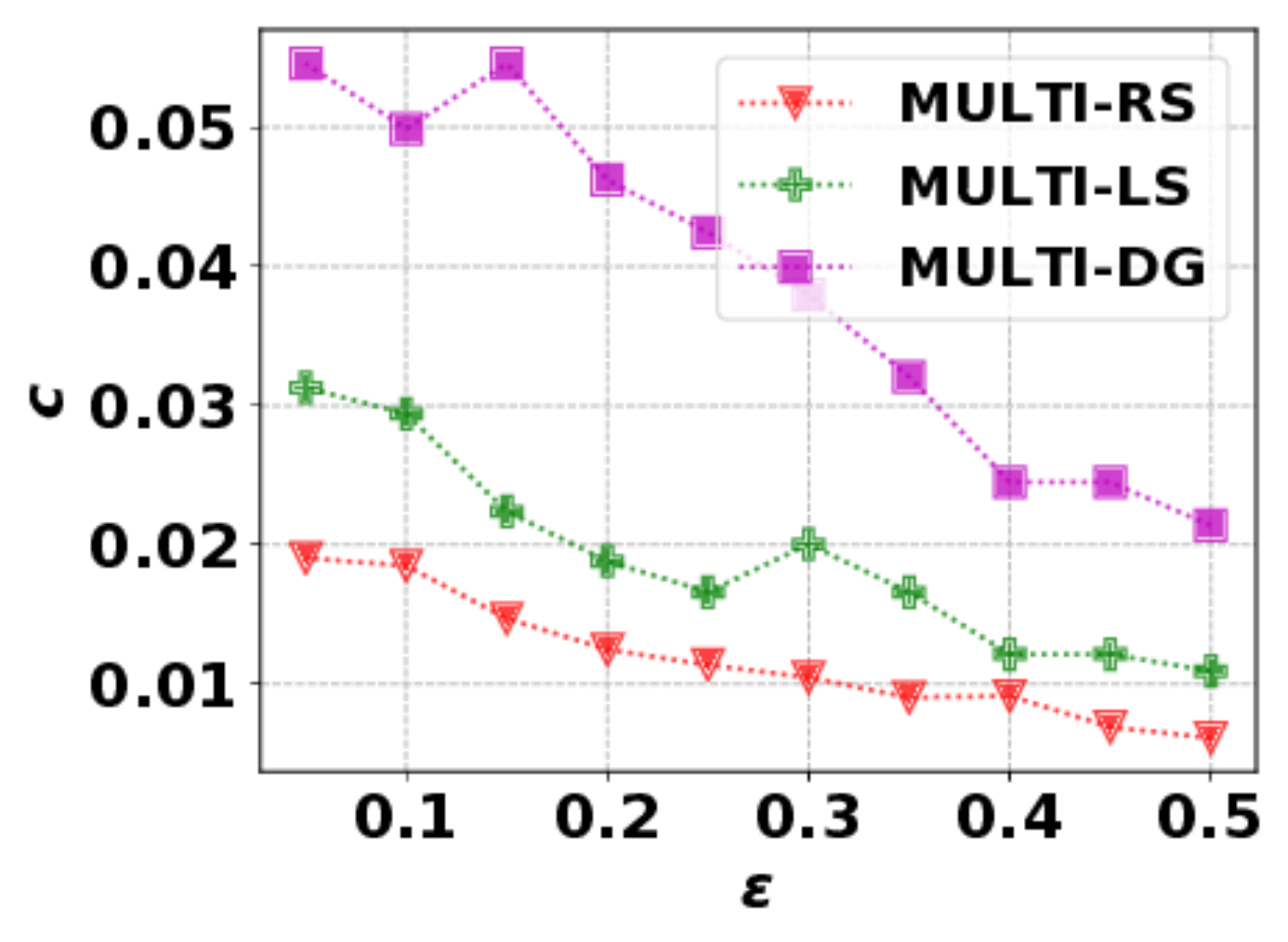}
  }
  \hspace{-1em}
  \subfigure[corel, cover] {
    \includegraphics[width=0.24\textwidth]{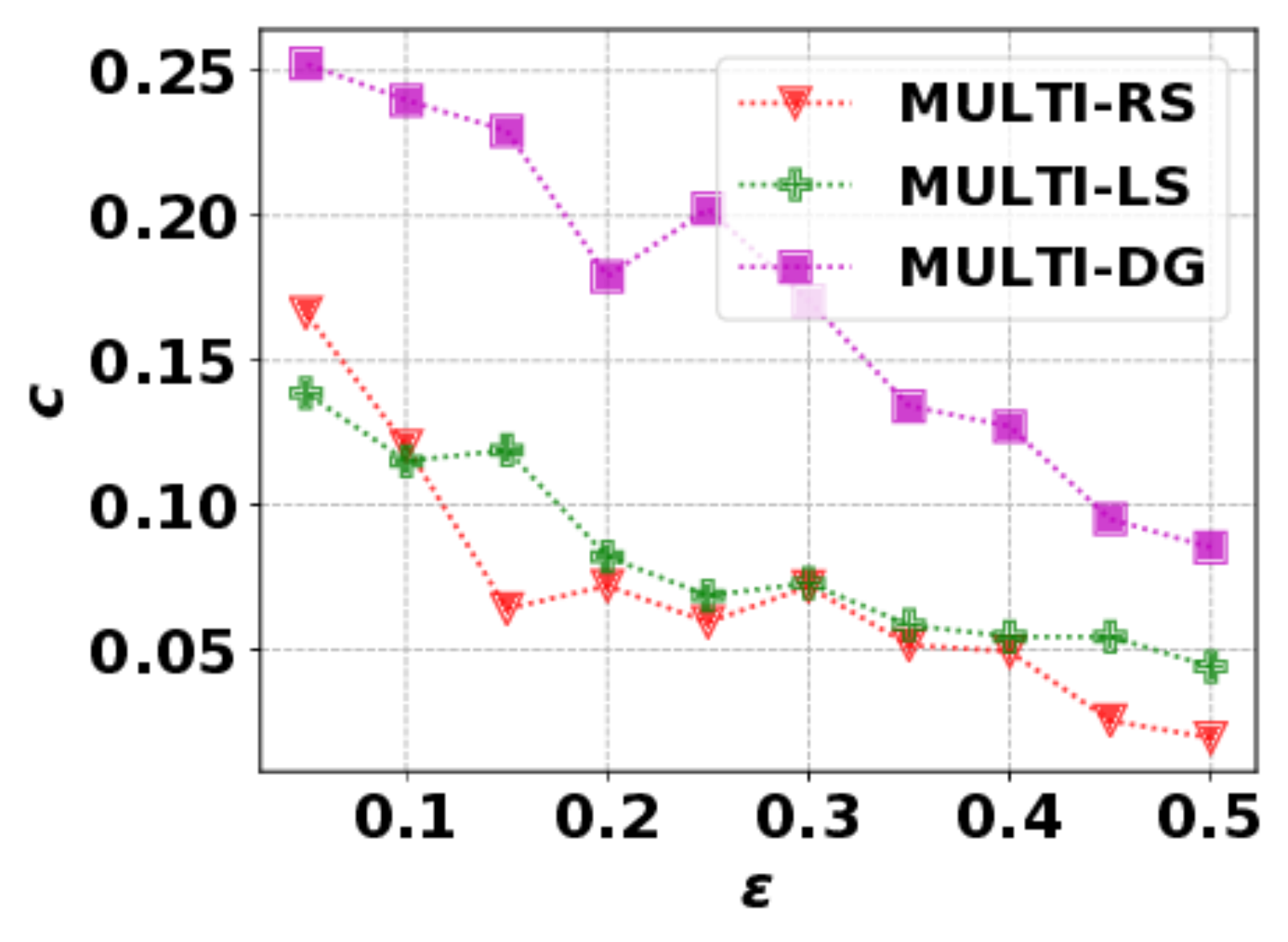}
  }
  \hspace{-1em}
  \subfigure[delicious5k, cover] {
    \includegraphics[width=0.24\textwidth]{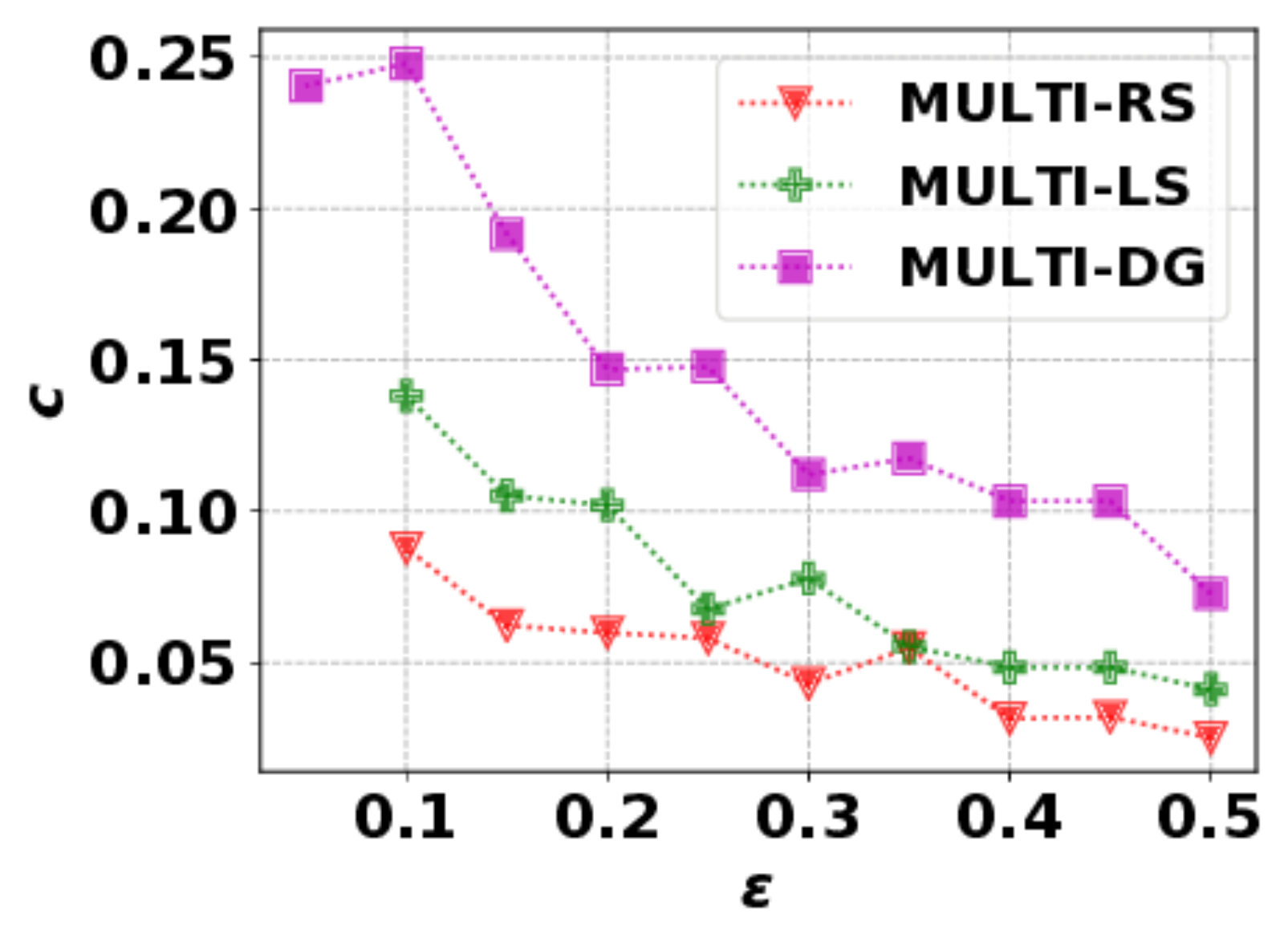}
  }
  \hspace{-1em}
  \subfigure[enron, cut] {
    \includegraphics[width=0.24\textwidth]{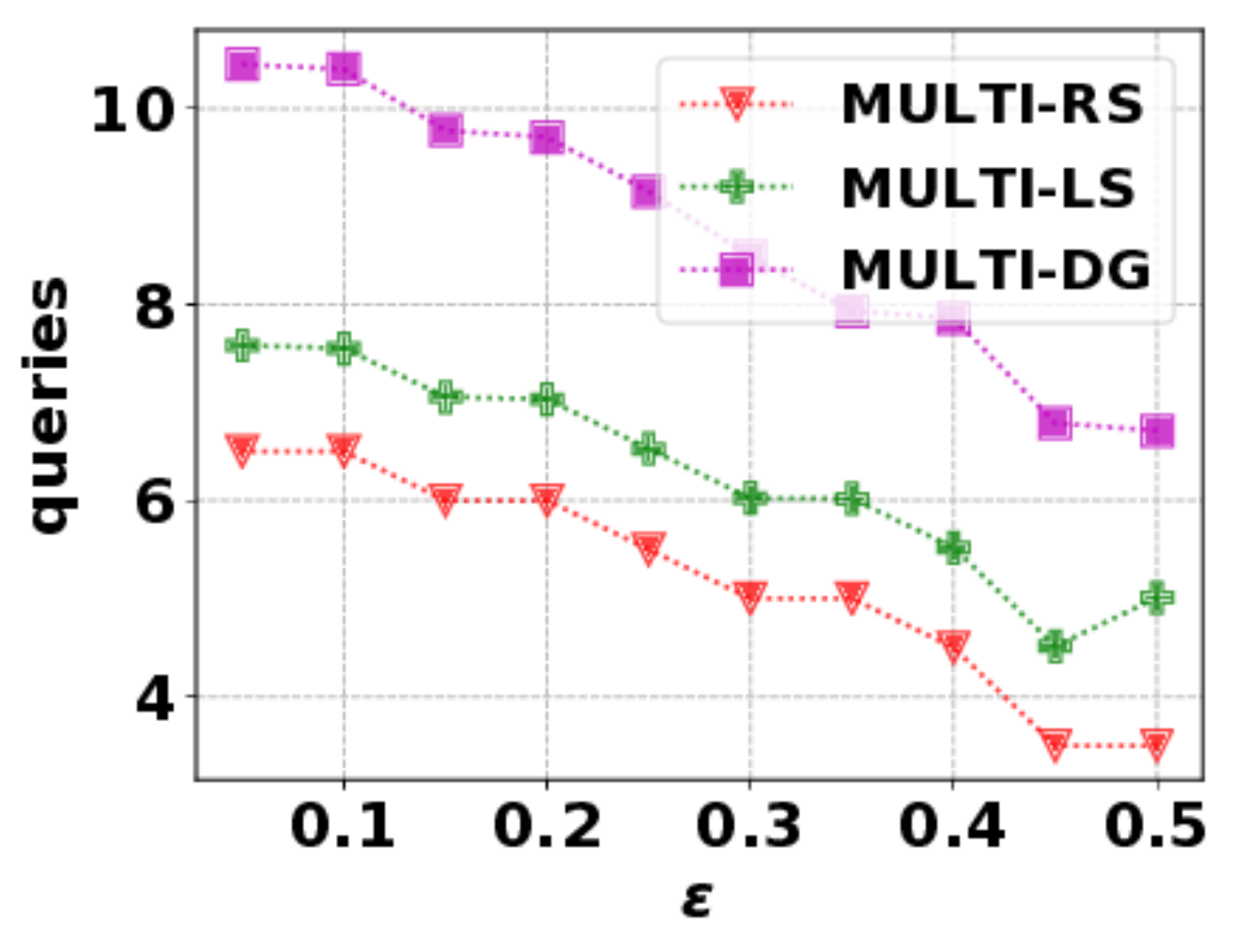}
  }
  \hspace{-1em}
  \subfigure[astro, cut] {
    \includegraphics[width=0.24\textwidth]{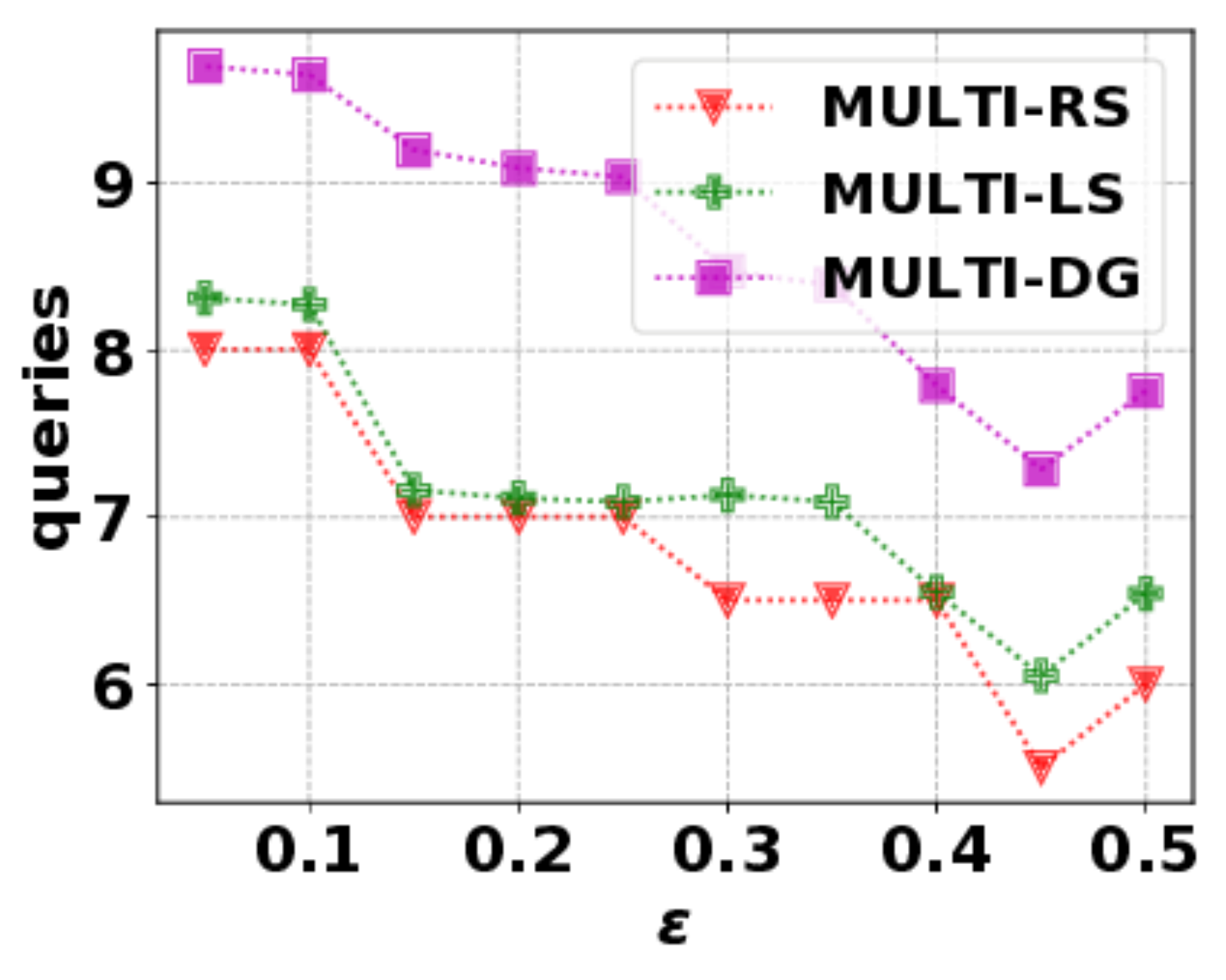}
  }
  \hspace{-1em}
  \subfigure[corel, cover] {
    \includegraphics[width=0.24\textwidth]{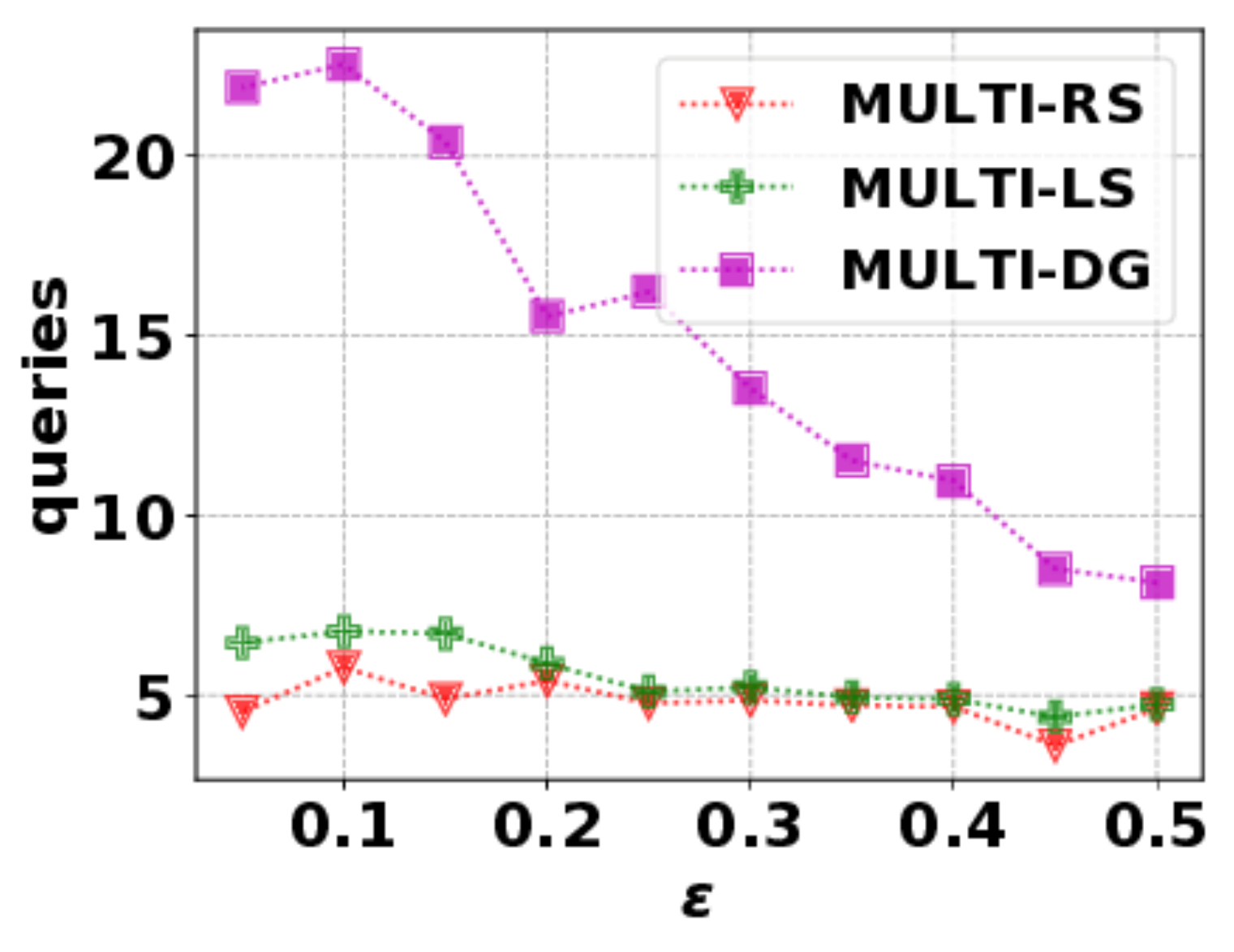}
  }
  \hspace{-1em}
  \subfigure[delicious5k, cover] {
    \includegraphics[width=0.24\textwidth]{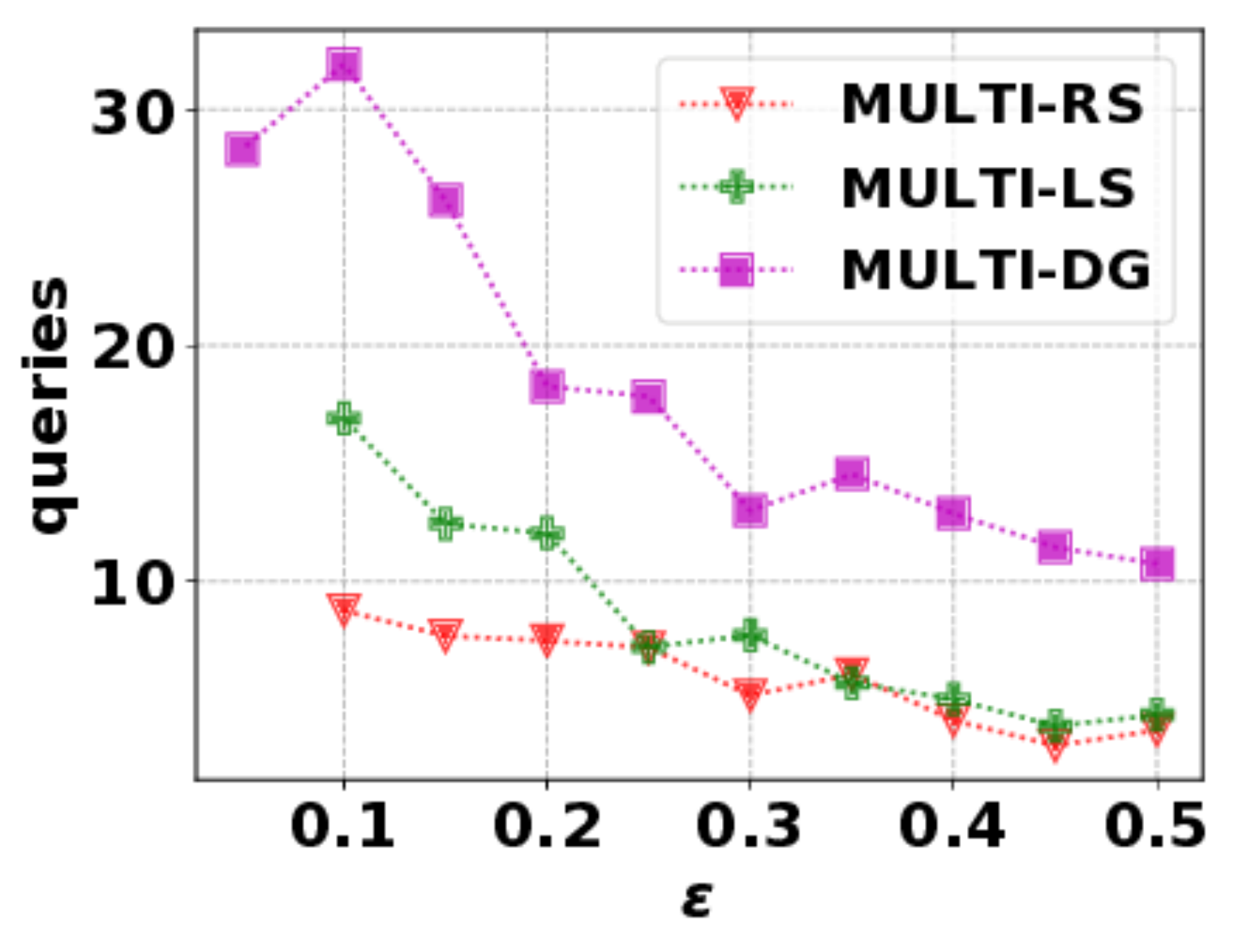}
  }
  \hspace{-1em}
  \subfigure[enron, cut] {
    \includegraphics[width=0.24\textwidth]{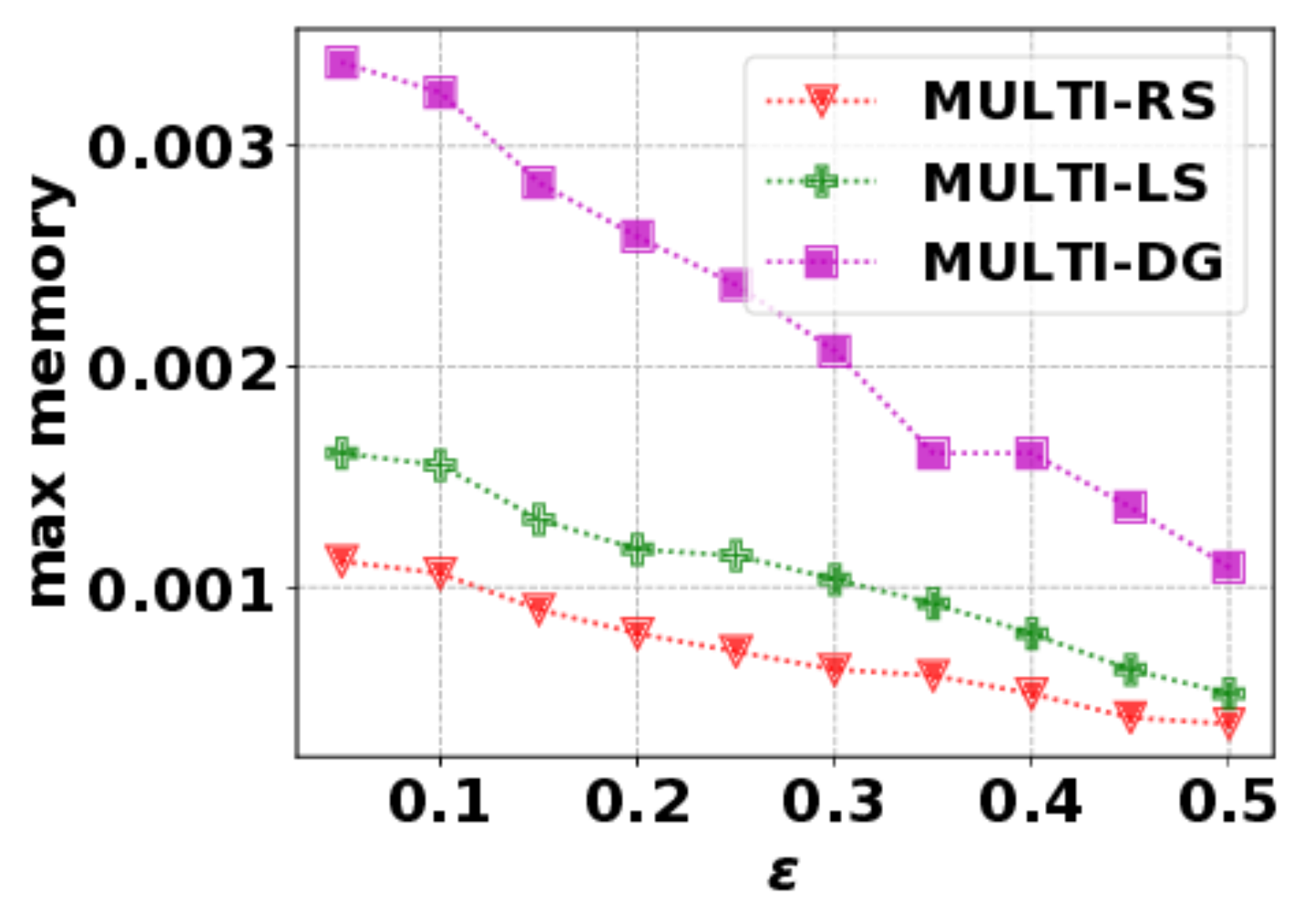}
  }
  \hspace{-1em}
  \subfigure[astro, cut] {
    \includegraphics[width=0.24\textwidth]{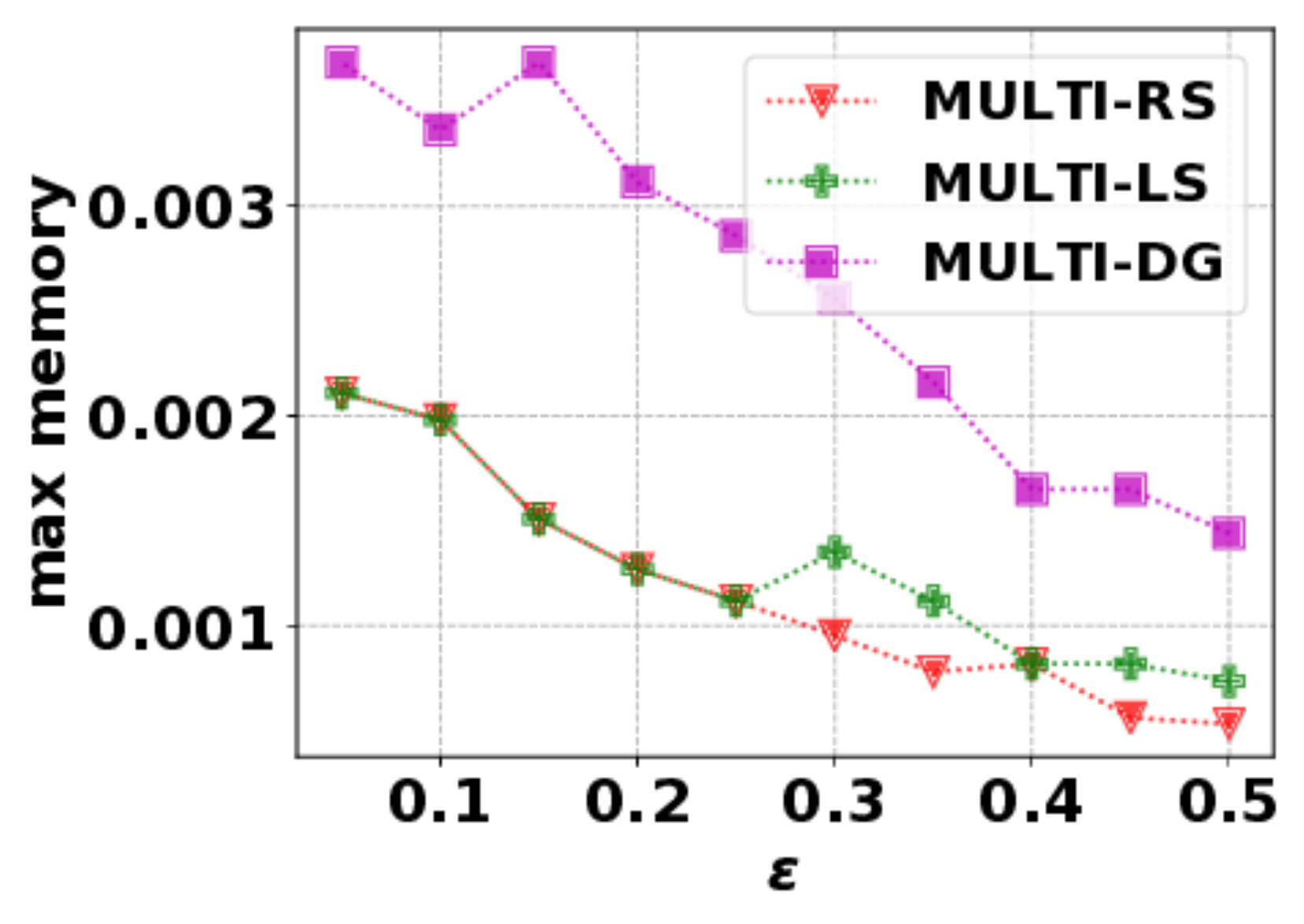}
  }
  \hspace{-1em}
  \subfigure[corel, cover] {
    \includegraphics[width=0.24\textwidth]{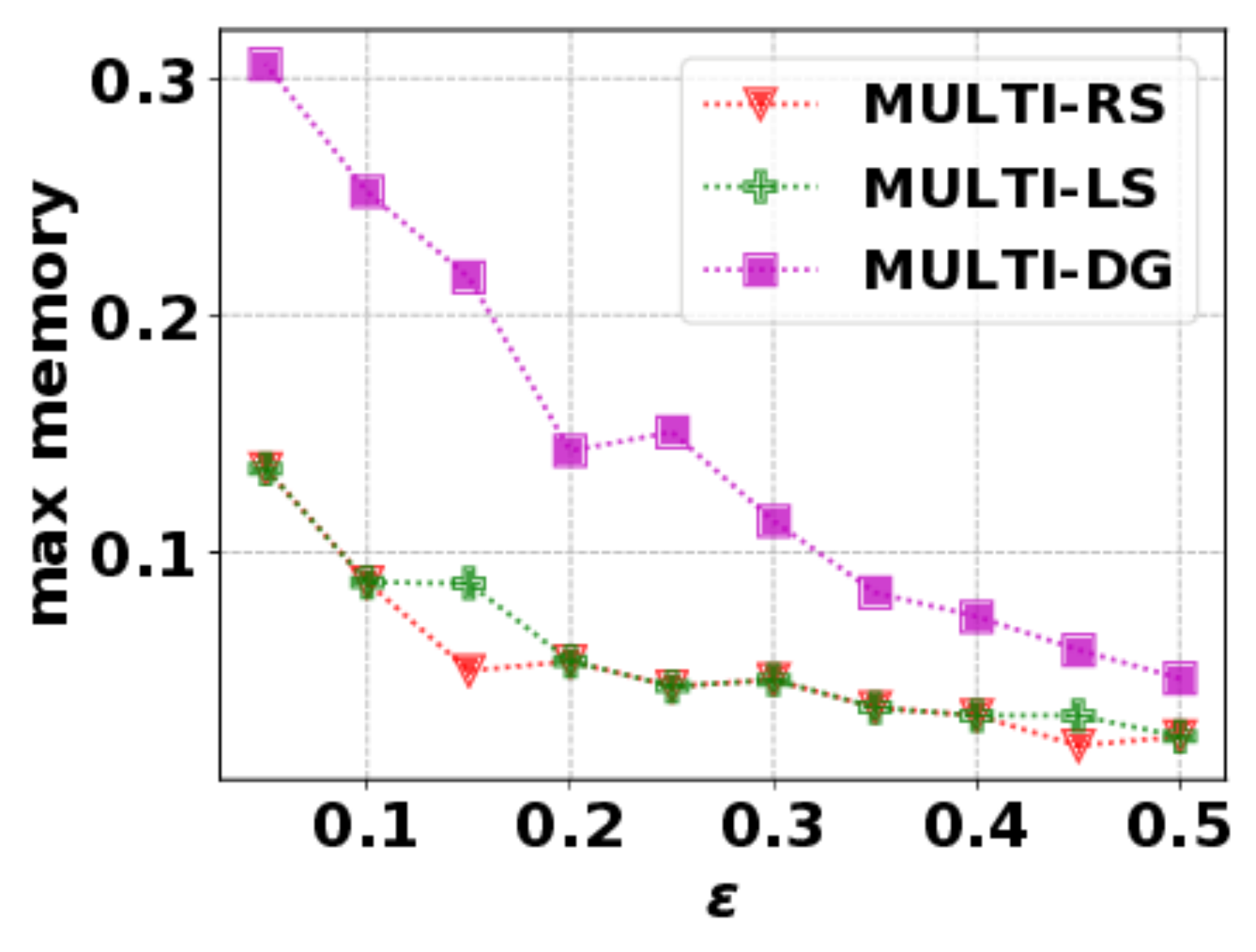}
  }
  \hspace{-1em}
  \subfigure[delicious5k, cover] {
    \includegraphics[width=0.24\textwidth]{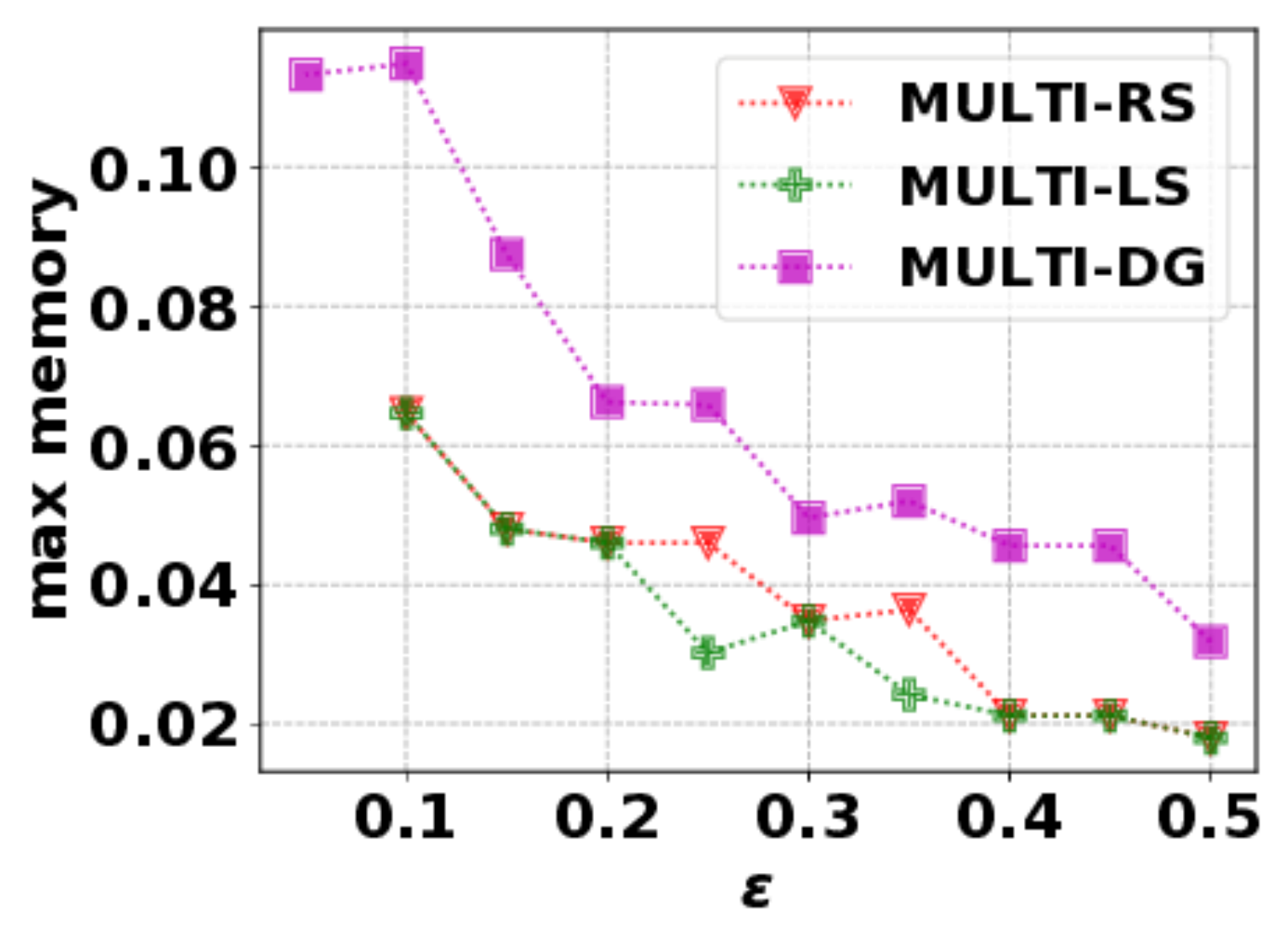}
  }
  \hspace{-1em}
  \subfigure[enron, cut] {
    \includegraphics[width=0.24\textwidth]{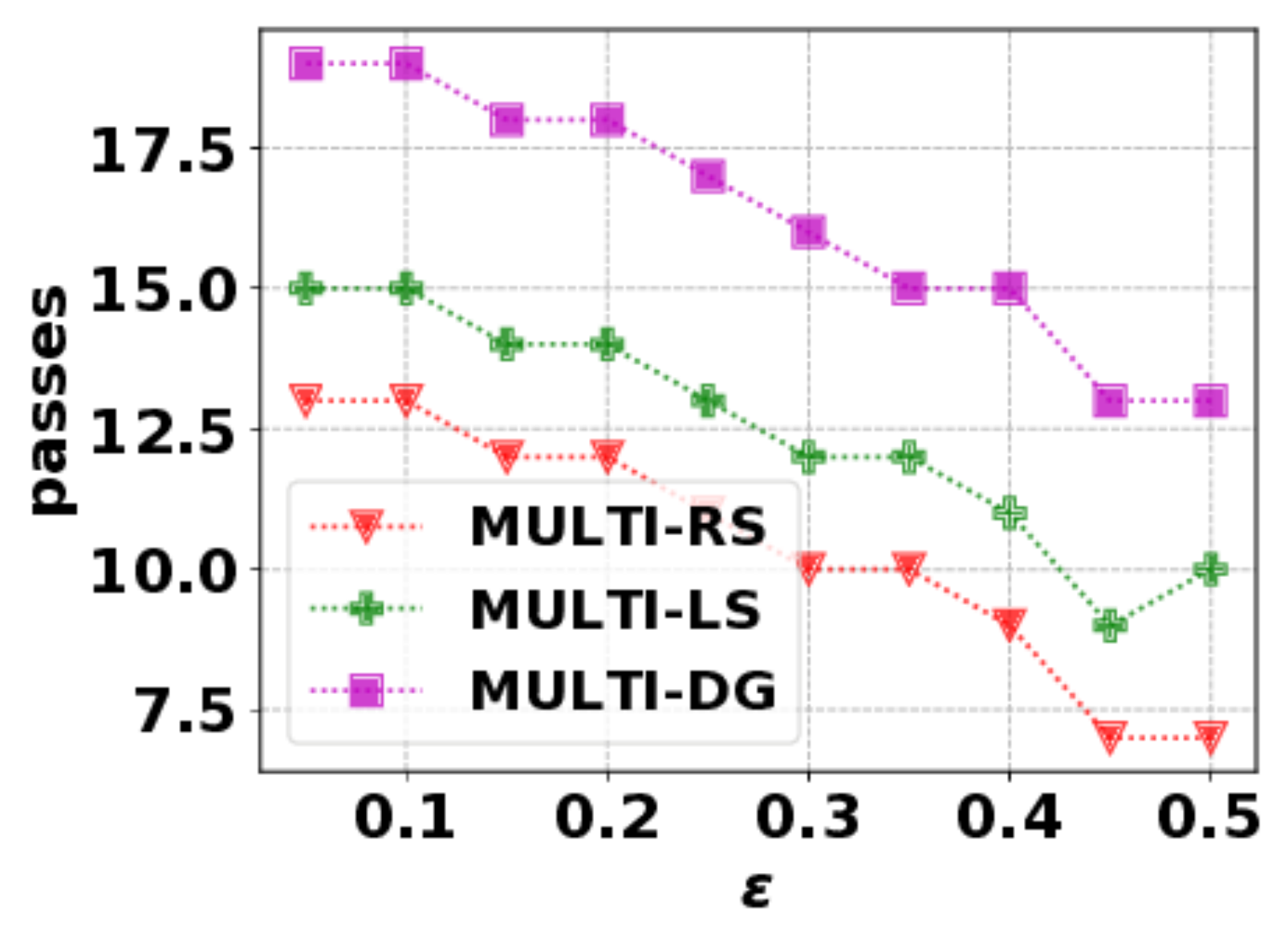}
  }
  \hspace{-1em}
  \subfigure[astro, cut] {
    \includegraphics[width=0.24\textwidth]{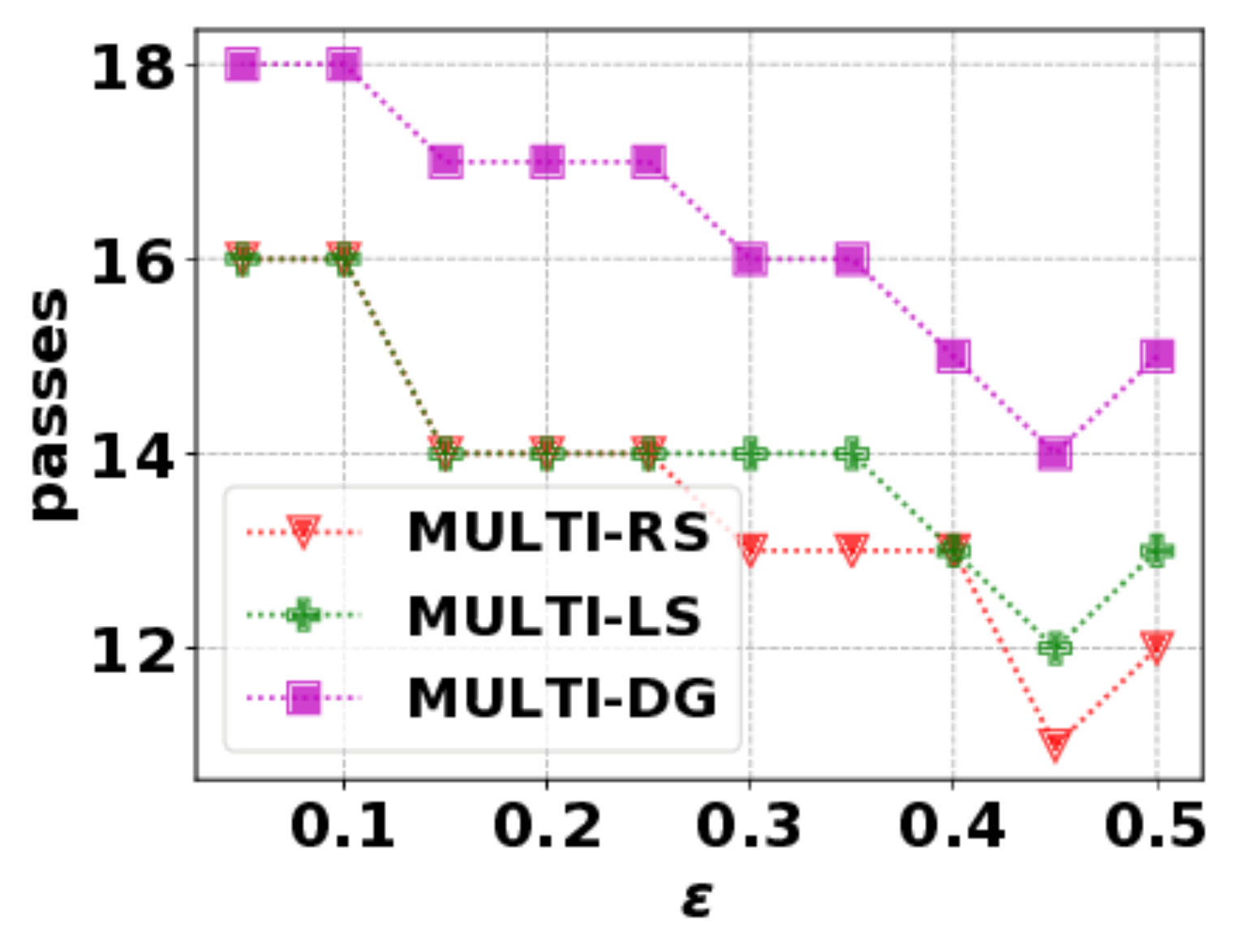}
  }
  \hspace{-1em}
  \subfigure[corel, cover] {
    \includegraphics[width=0.24\textwidth]{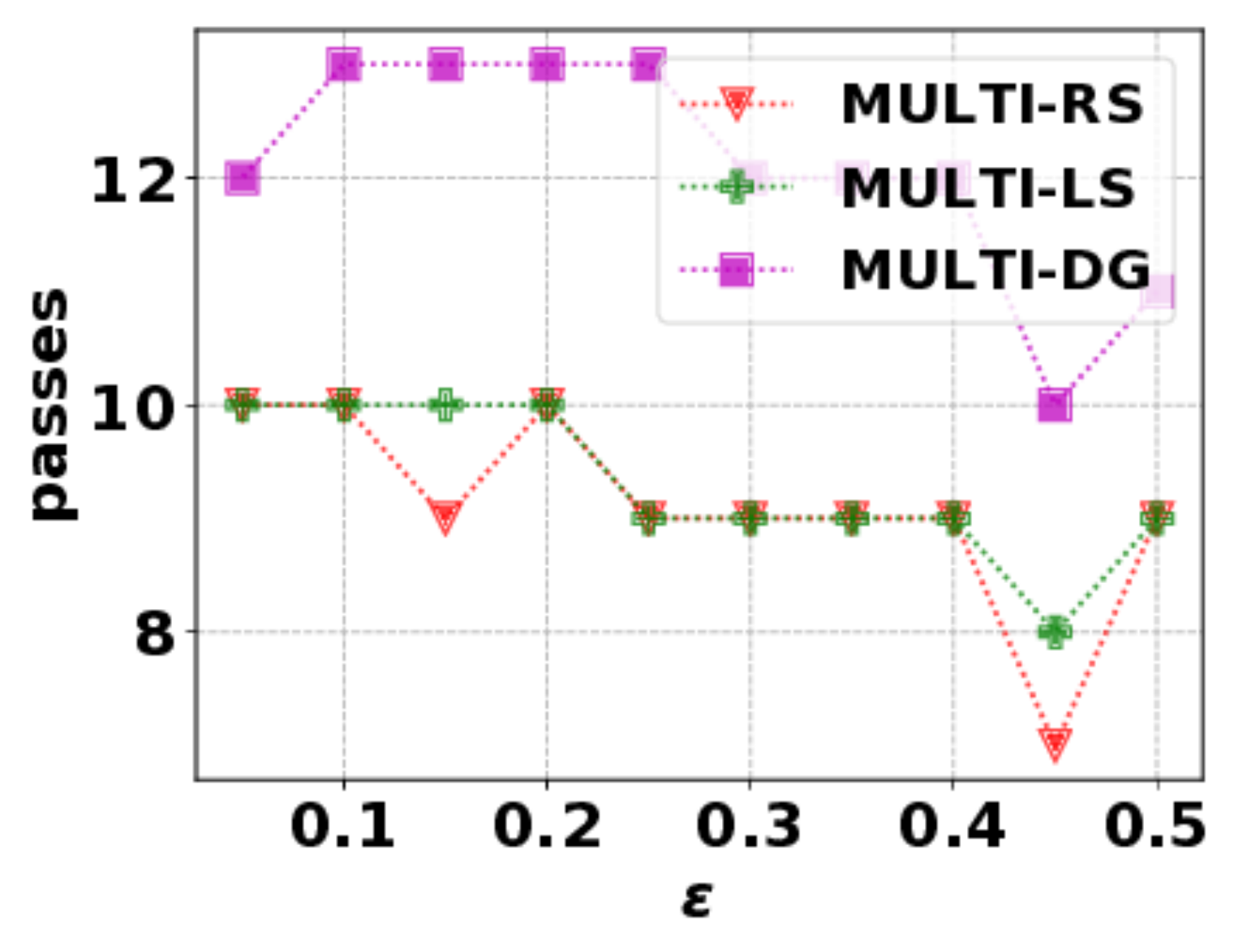}
  }
  \hspace{-1em}
  \subfigure[delicious5k, cover] {
    \includegraphics[width=0.24\textwidth]{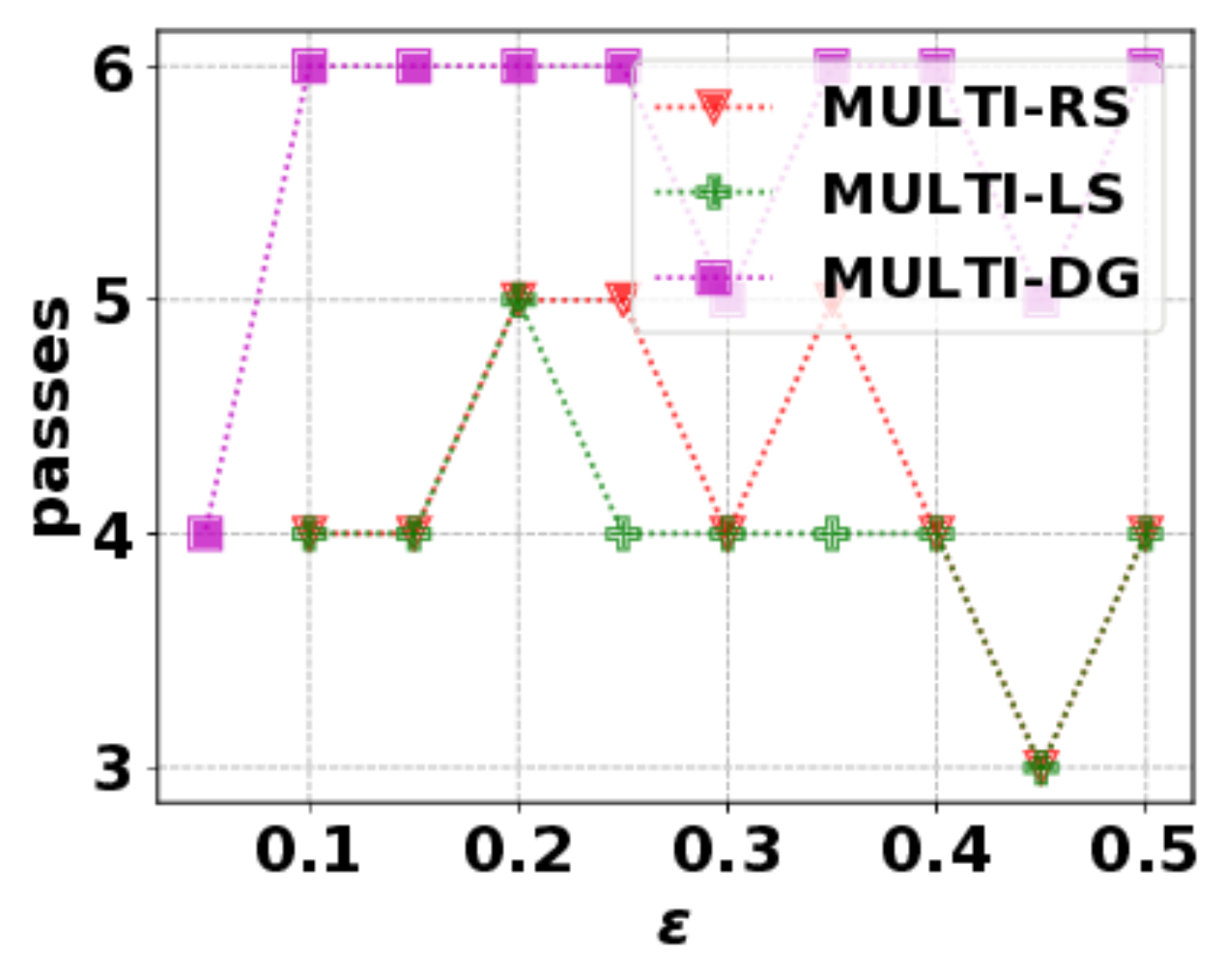}
  }
  \caption[]{
  A comparison of the performance of \multi using different algorithms for USM as
  a subroutine.
  $f$ and $c$ refer to the $f$ and cost values of the returned solution.
  Queries is the total number of queries to $f$ that the algorithm took.
  Max memory is the maximum cost of all elements stored at once
  over the duration of the algorithm.
  The instances of
  graph cut and diverse data summarization are on the
  Corel5k (``corel'') and
  delicious (``delicious5k'') dataset,
  and instances of graph cut are on the ca-AstroPh (``astro'')
  and email-Enron (``amazon'') datasets.
  \multi using the double greedy algorithm of \citet{buchbinder2015tight}, the local
  search algorithm of \citet{feige2011maximizing}, and the random set algorithm are
  referred to as ``MULTI-DG'', ``MULTI-LS'', and ``MULTI-RS'' respectively.
  All $x$ and $y$ axes are normalized as described in Section \ref{section:resultsapp}.
  }
  \label{fig:mssubroutines}
\end{figure*}

\begin{figure*}[t!]
  \centering
  \hspace{-1em}
  \subfigure[enron, cut] {
    \includegraphics[width=0.24\textwidth]{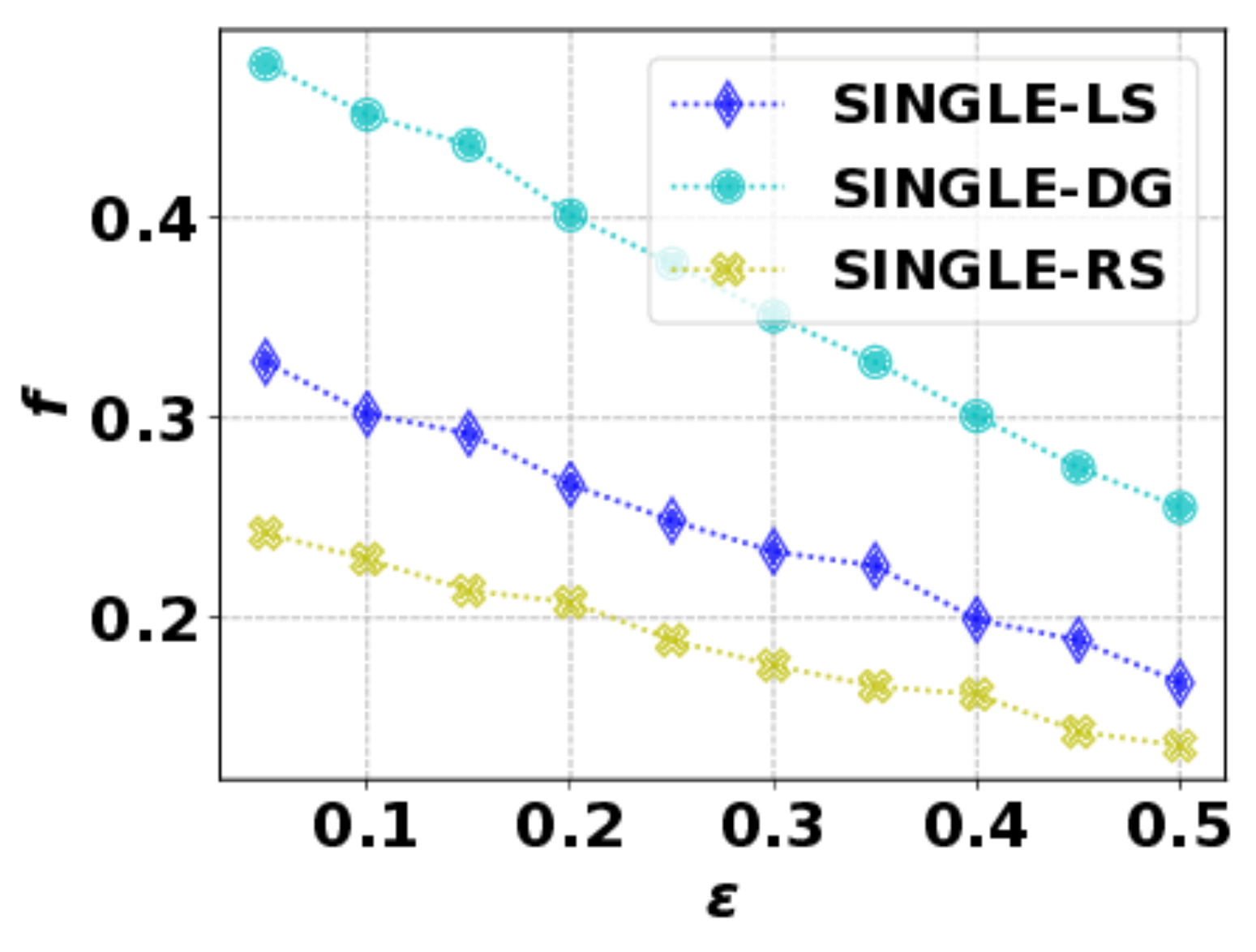}
  }
  \hspace{-1em}
  \subfigure[astro, cut] {
    \includegraphics[width=0.24\textwidth]{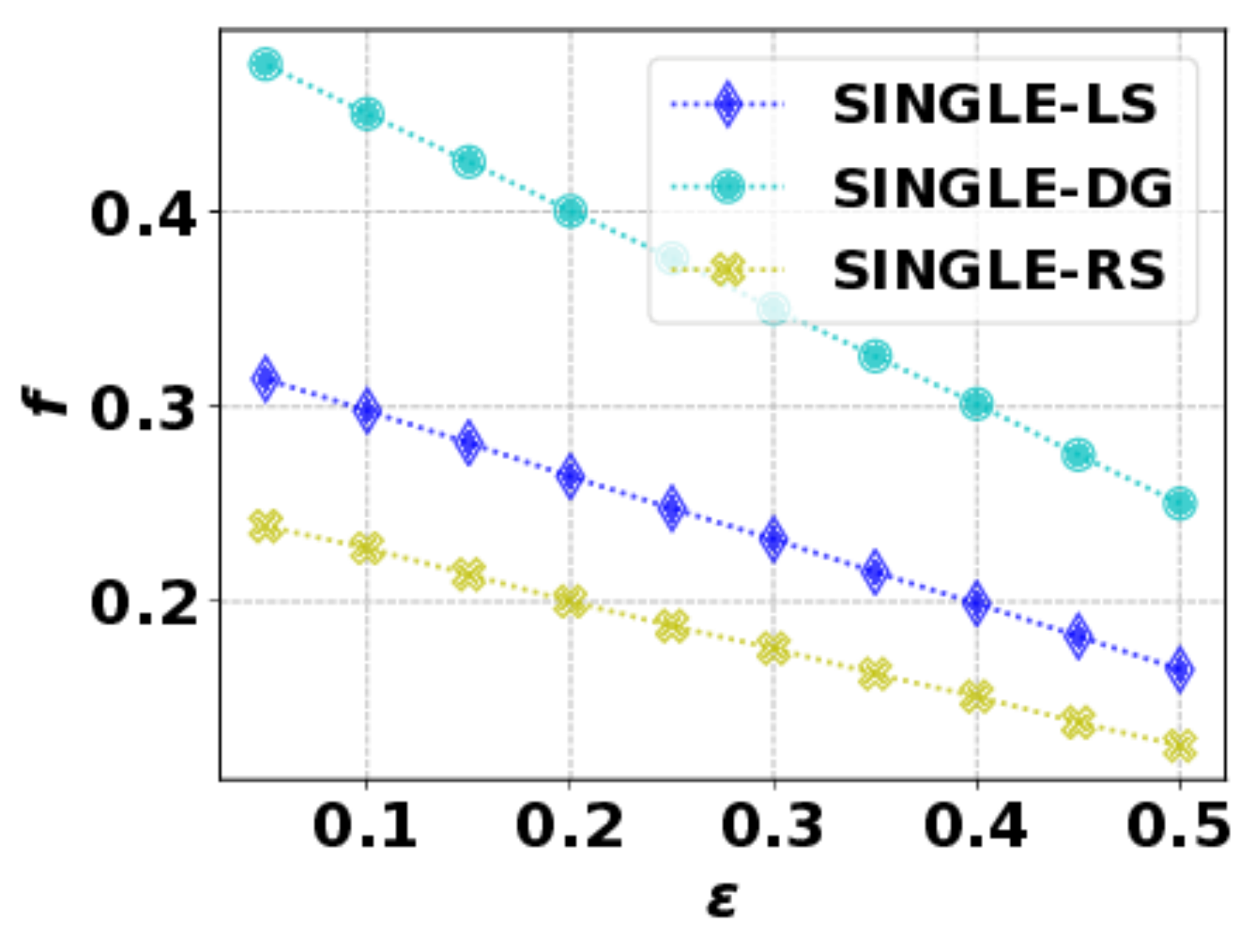}
  }
  \hspace{-1em}
  \subfigure[corel, cover] {
    \includegraphics[width=0.24\textwidth]{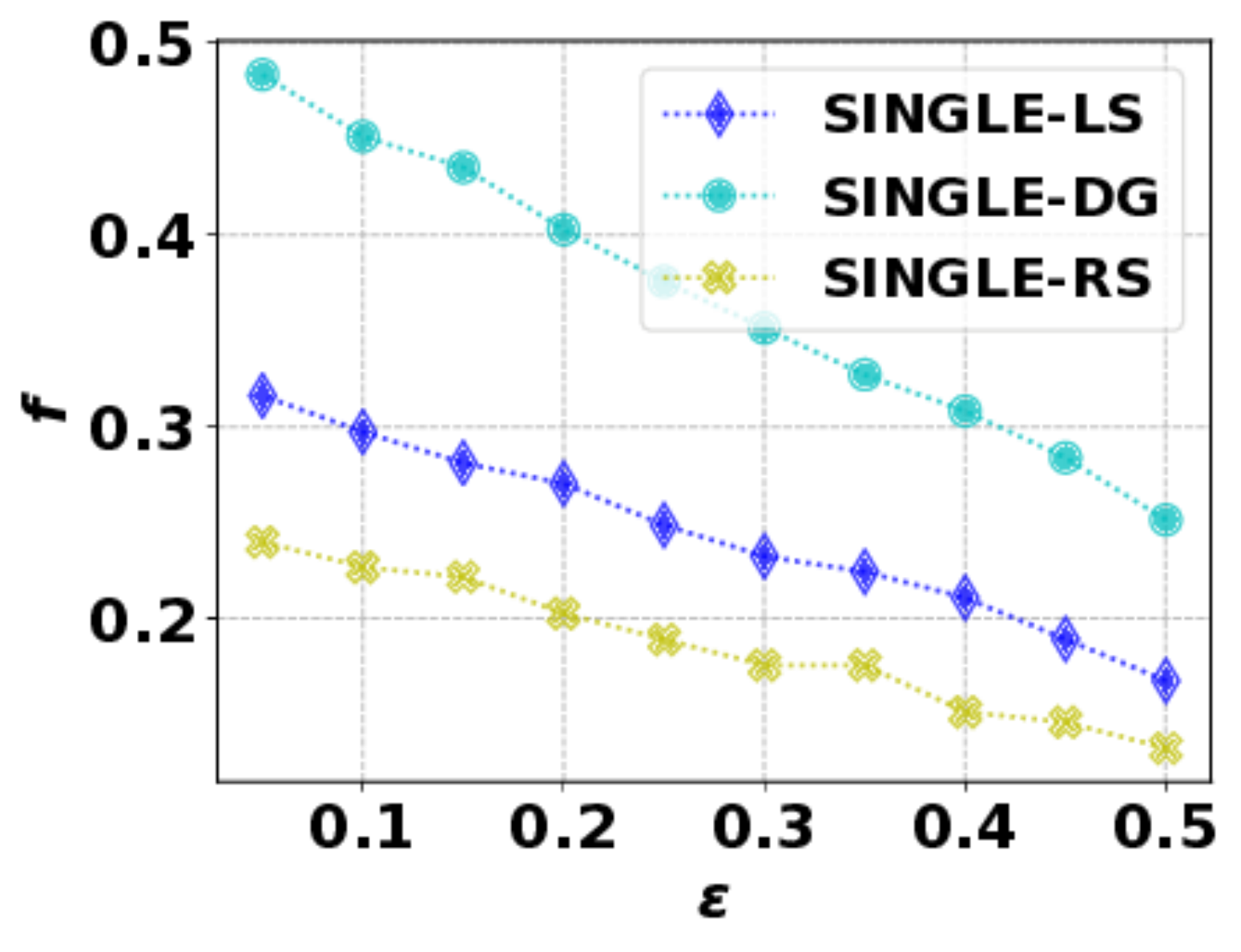}
  }
  \hspace{-1em}
  \subfigure[delicious5k, cover] {
    \includegraphics[width=0.24\textwidth]{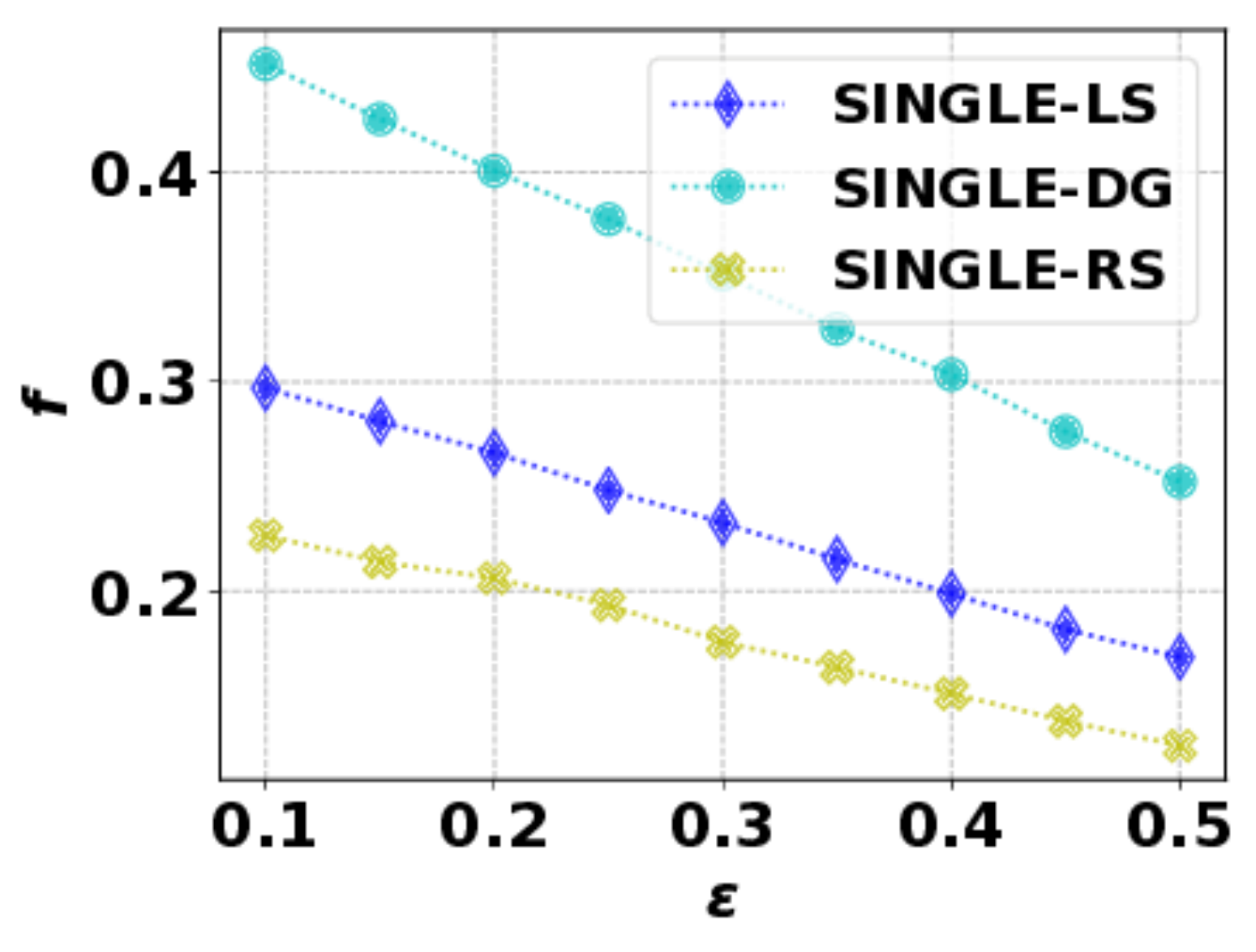}
  }
  \hspace{-1em}
  \subfigure[enron, cut] {
    \includegraphics[width=0.24\textwidth]{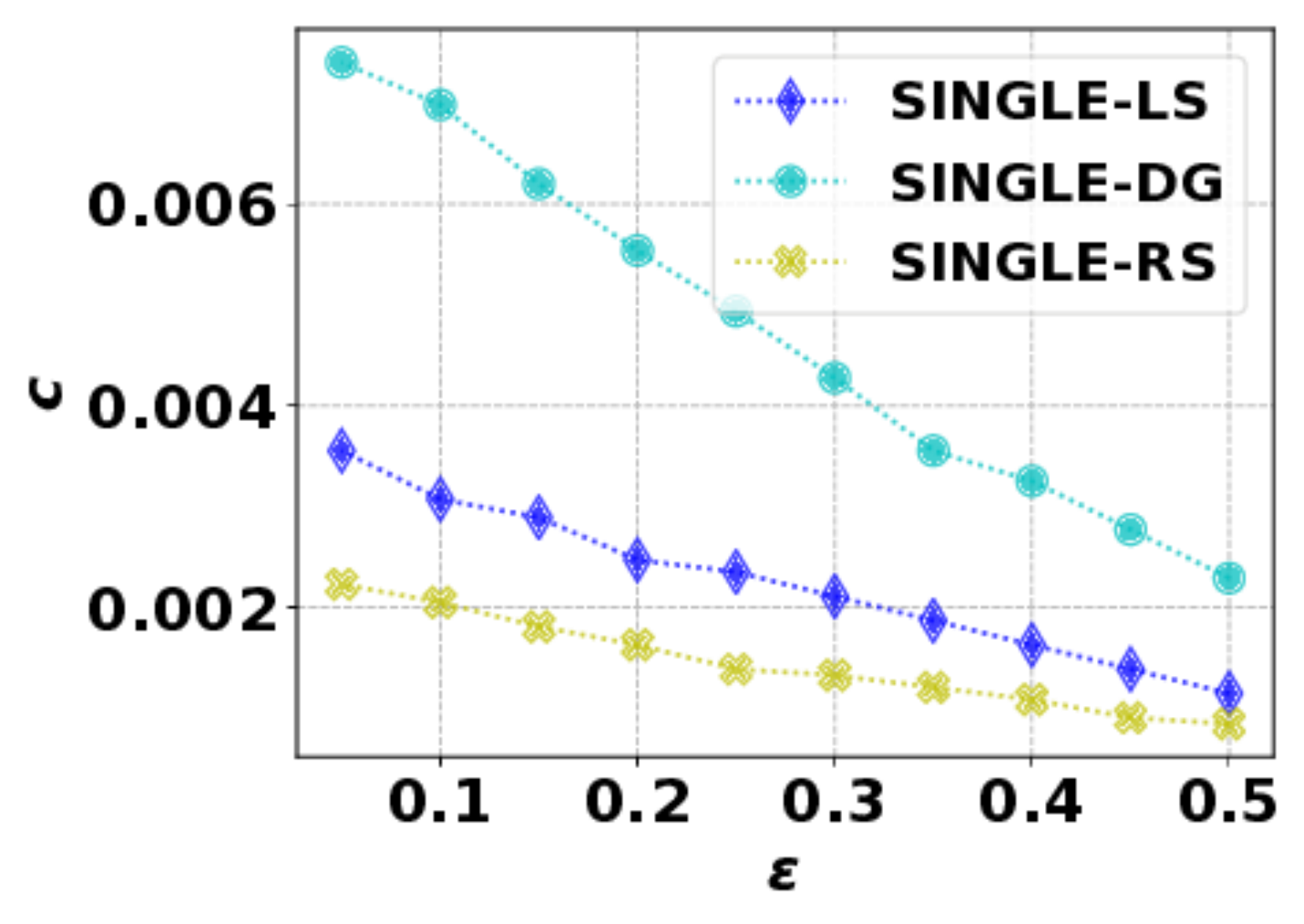}
  }
  \hspace{-1em}
  \subfigure[astro, cut] {
    \includegraphics[width=0.24\textwidth]{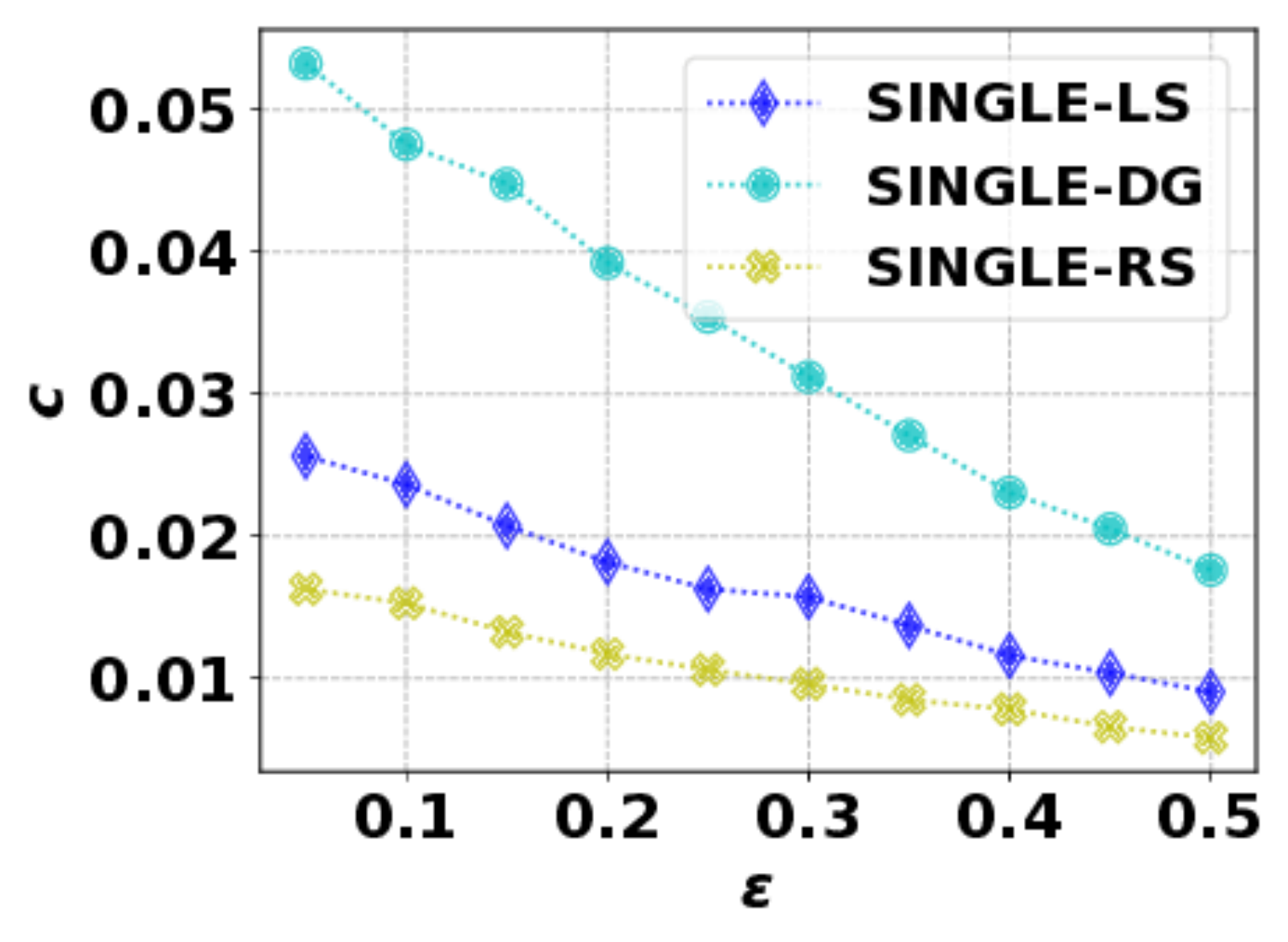}
  }
  \hspace{-1em}
  \subfigure[corel, cover] {
    \includegraphics[width=0.24\textwidth]{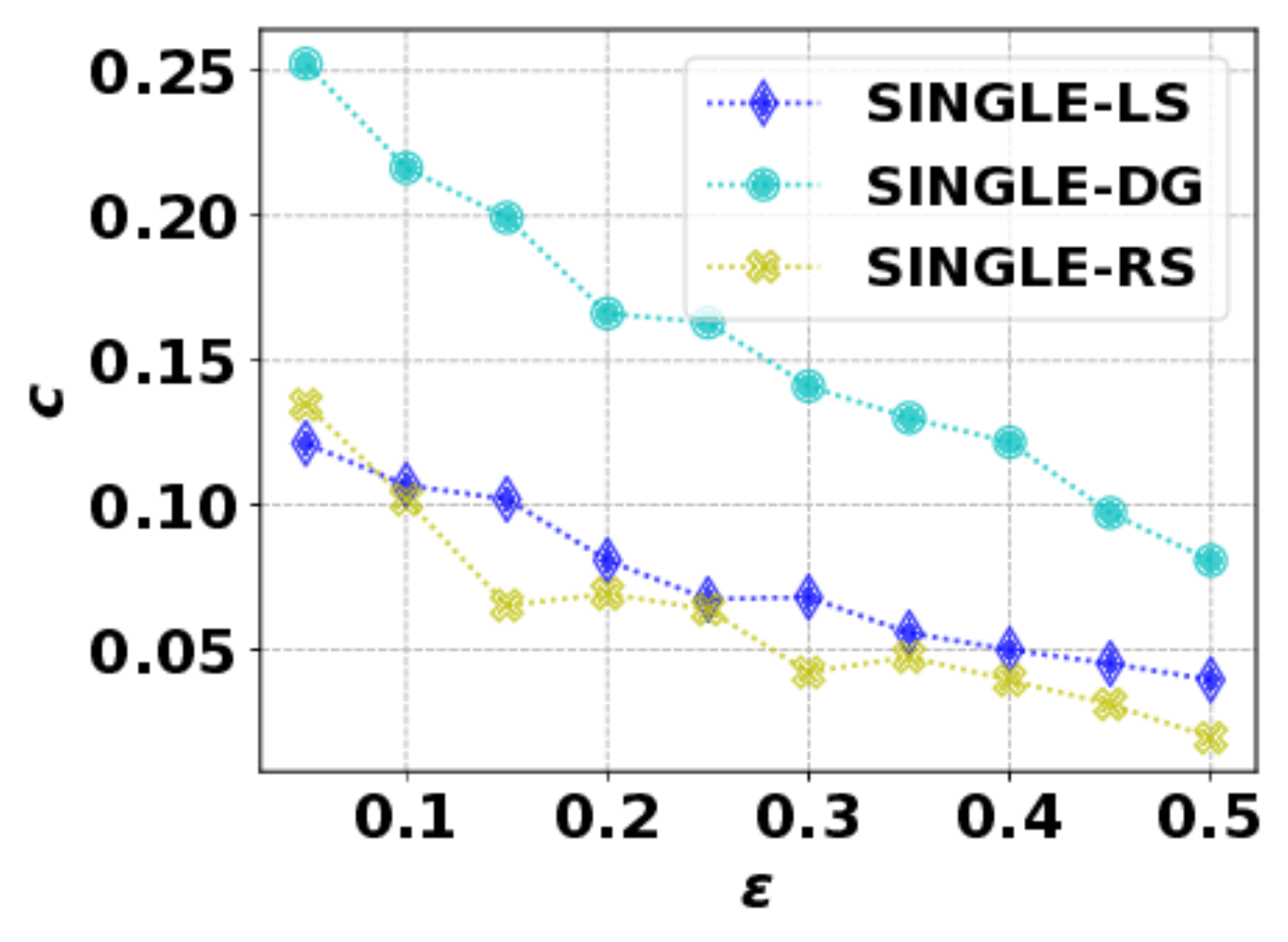}
  }
  \hspace{-1em}
  \subfigure[delicious5k, cover] {
    \includegraphics[width=0.24\textwidth]{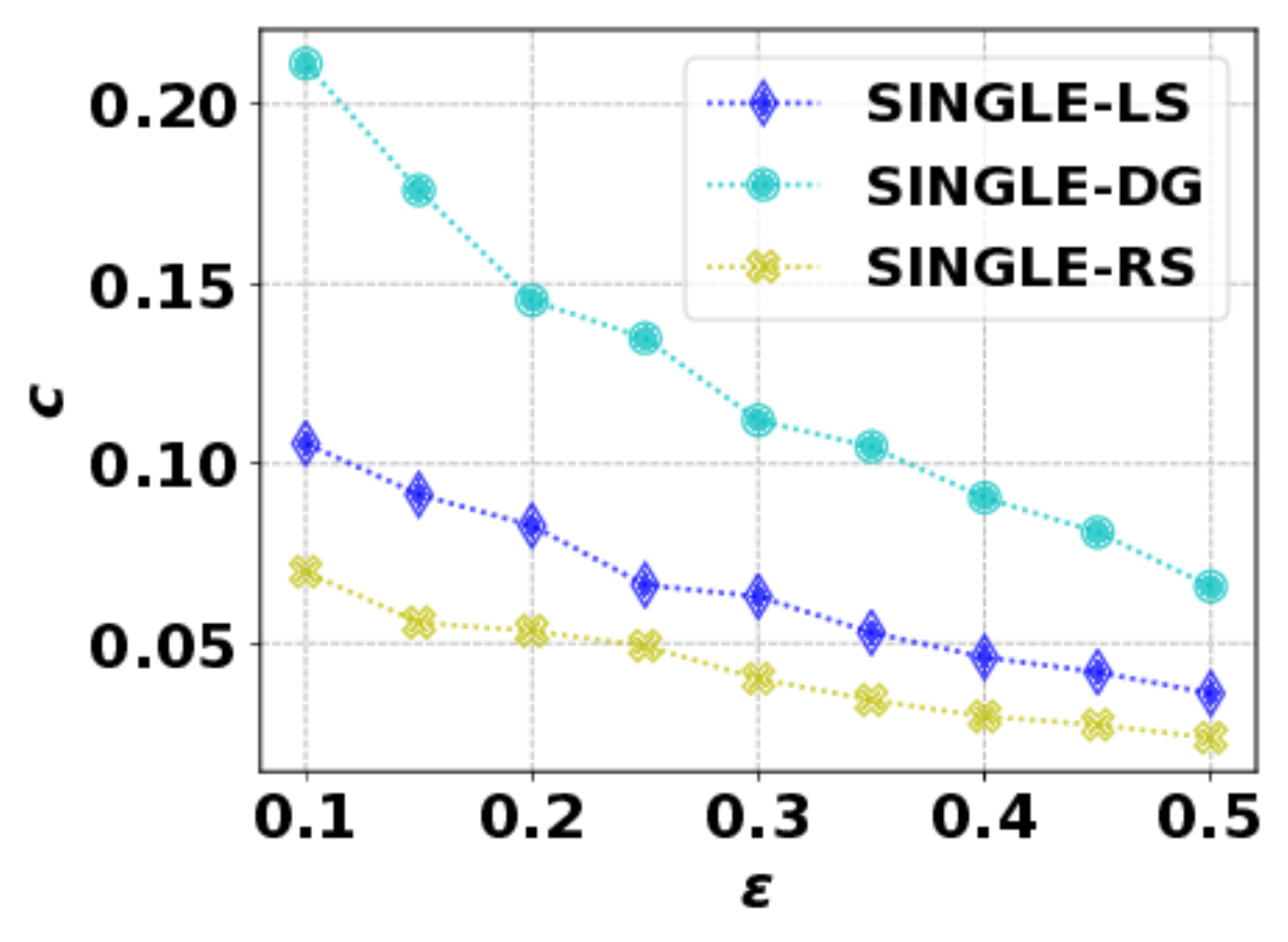}
  }
  \hspace{-1em}
  \subfigure[enron, cut] {
    \includegraphics[width=0.24\textwidth]{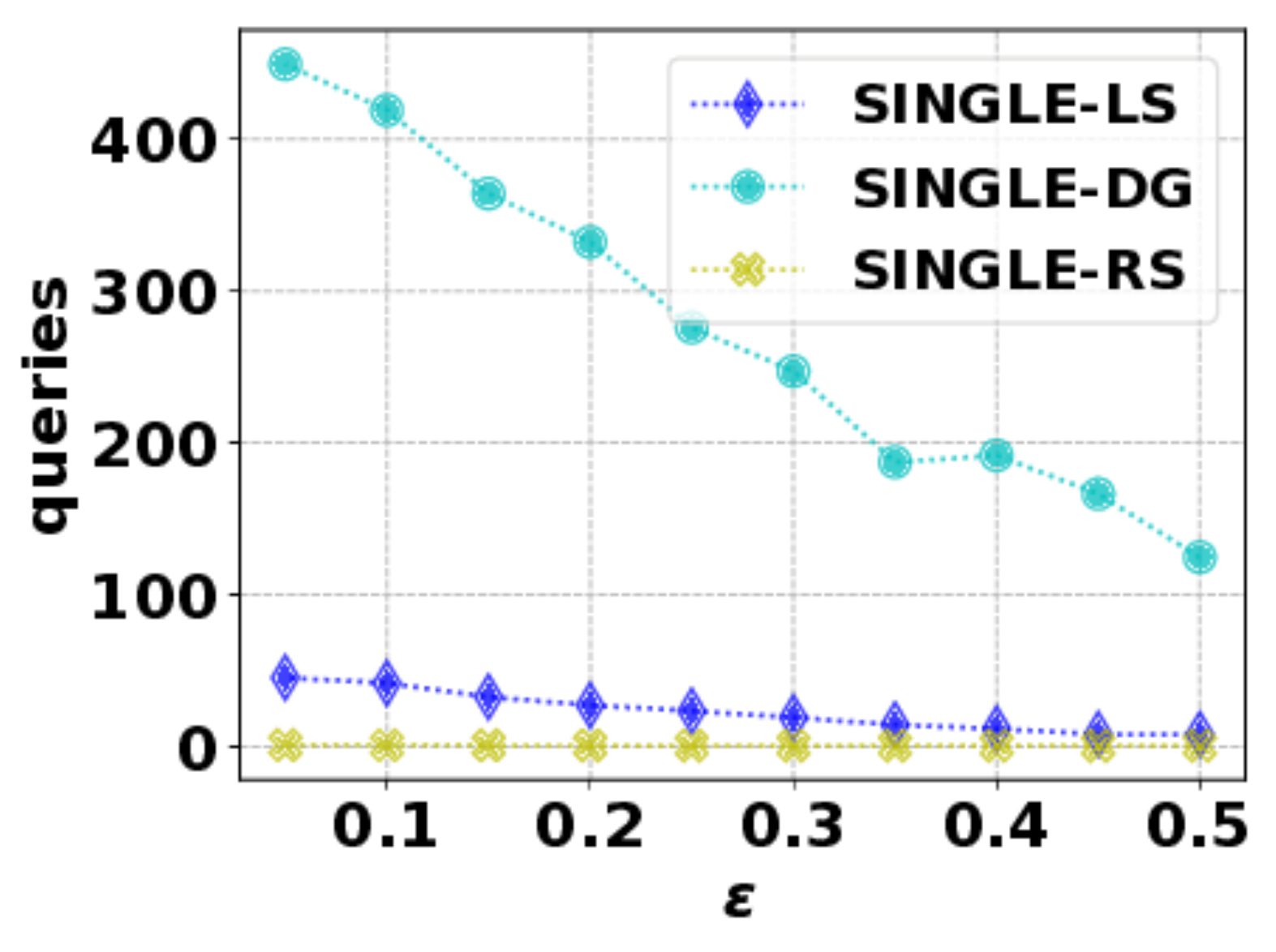}
  }
  \hspace{-1em}
  \subfigure[astro, cut] {
    \includegraphics[width=0.24\textwidth]{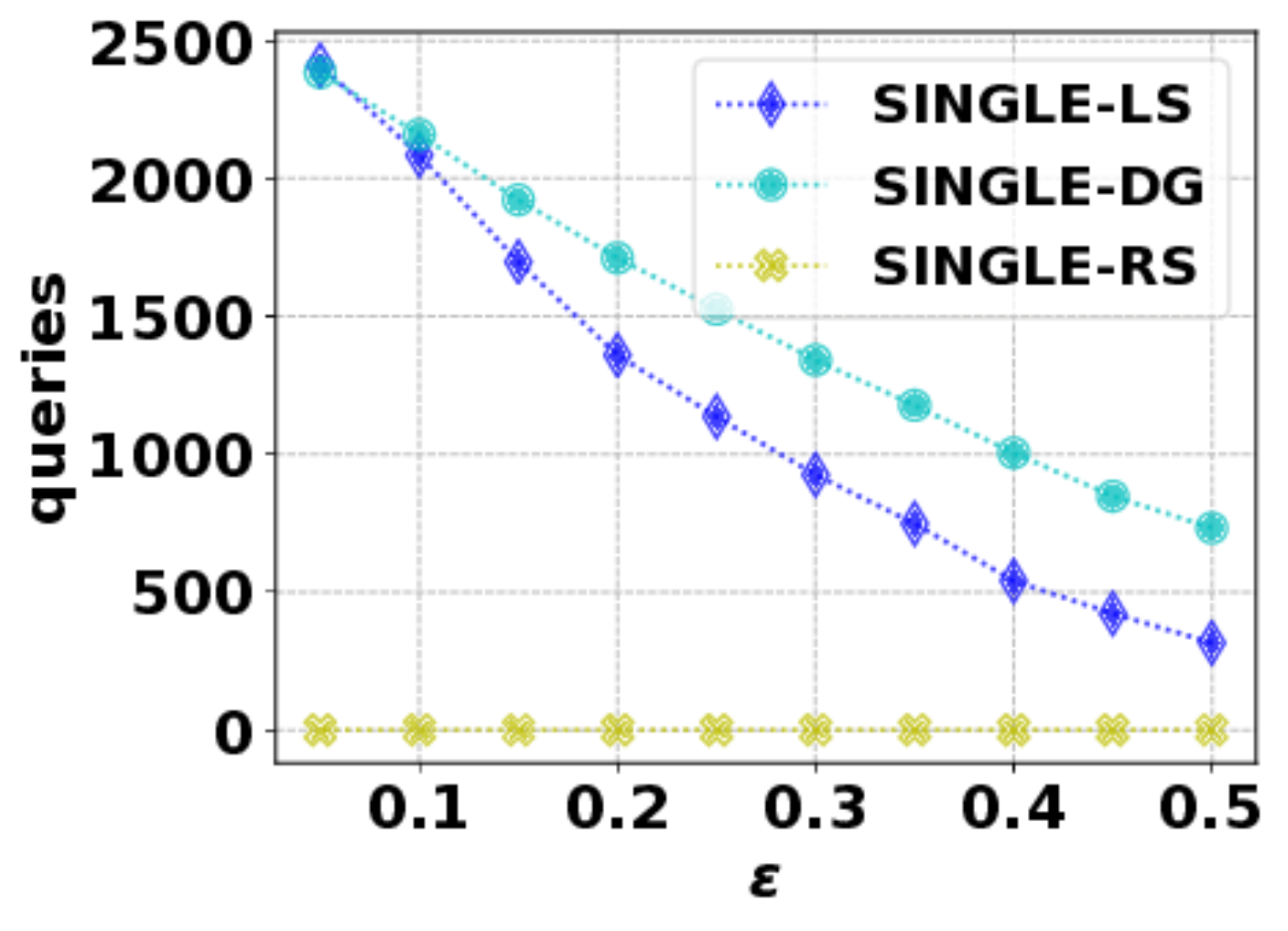}
  }
  \hspace{-1em}
  \subfigure[corel, cover] {
    \includegraphics[width=0.24\textwidth]{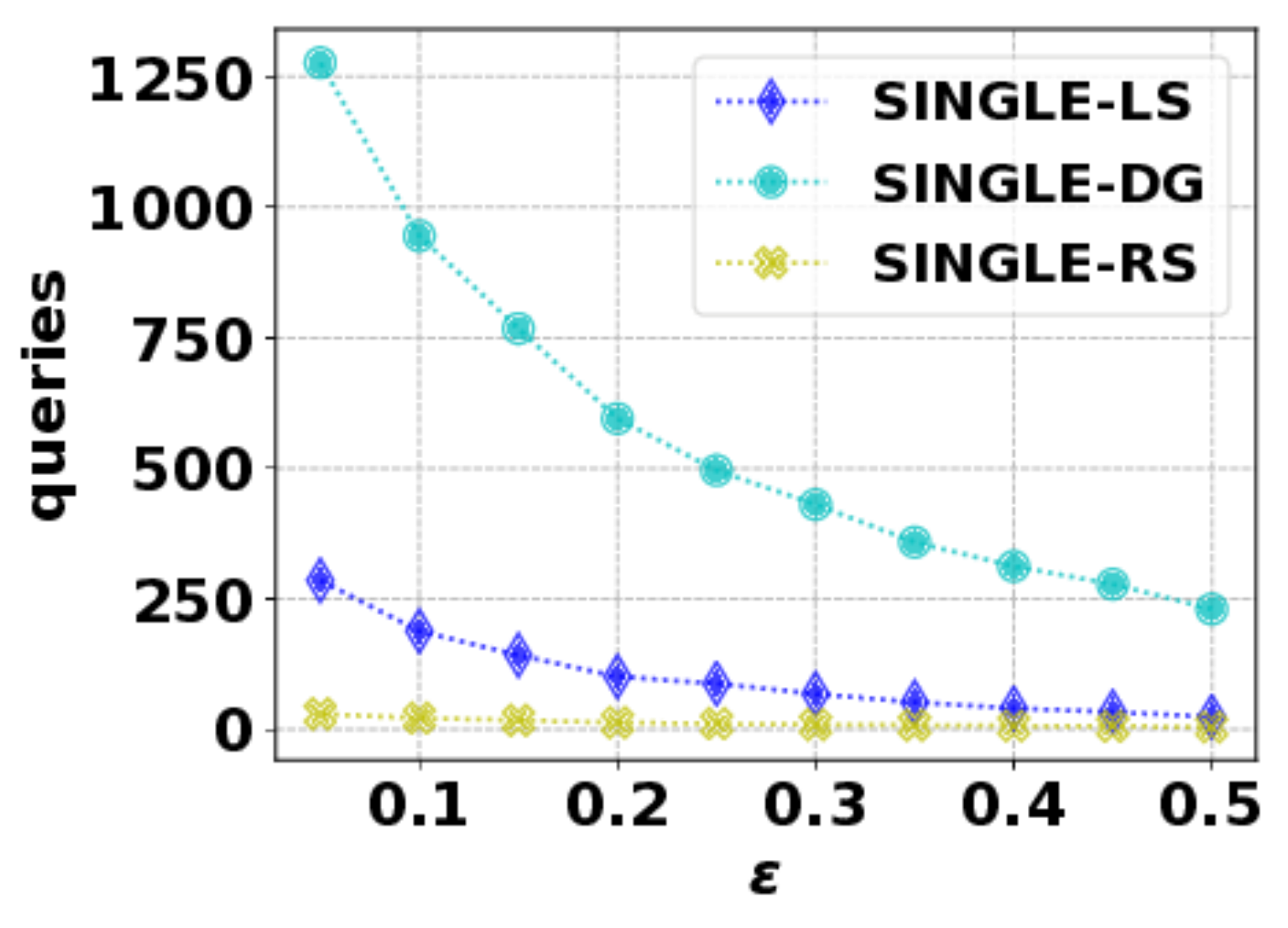}
  }
  \hspace{-1em}
  \subfigure[delicious5k, cover] {
    \includegraphics[width=0.24\textwidth]{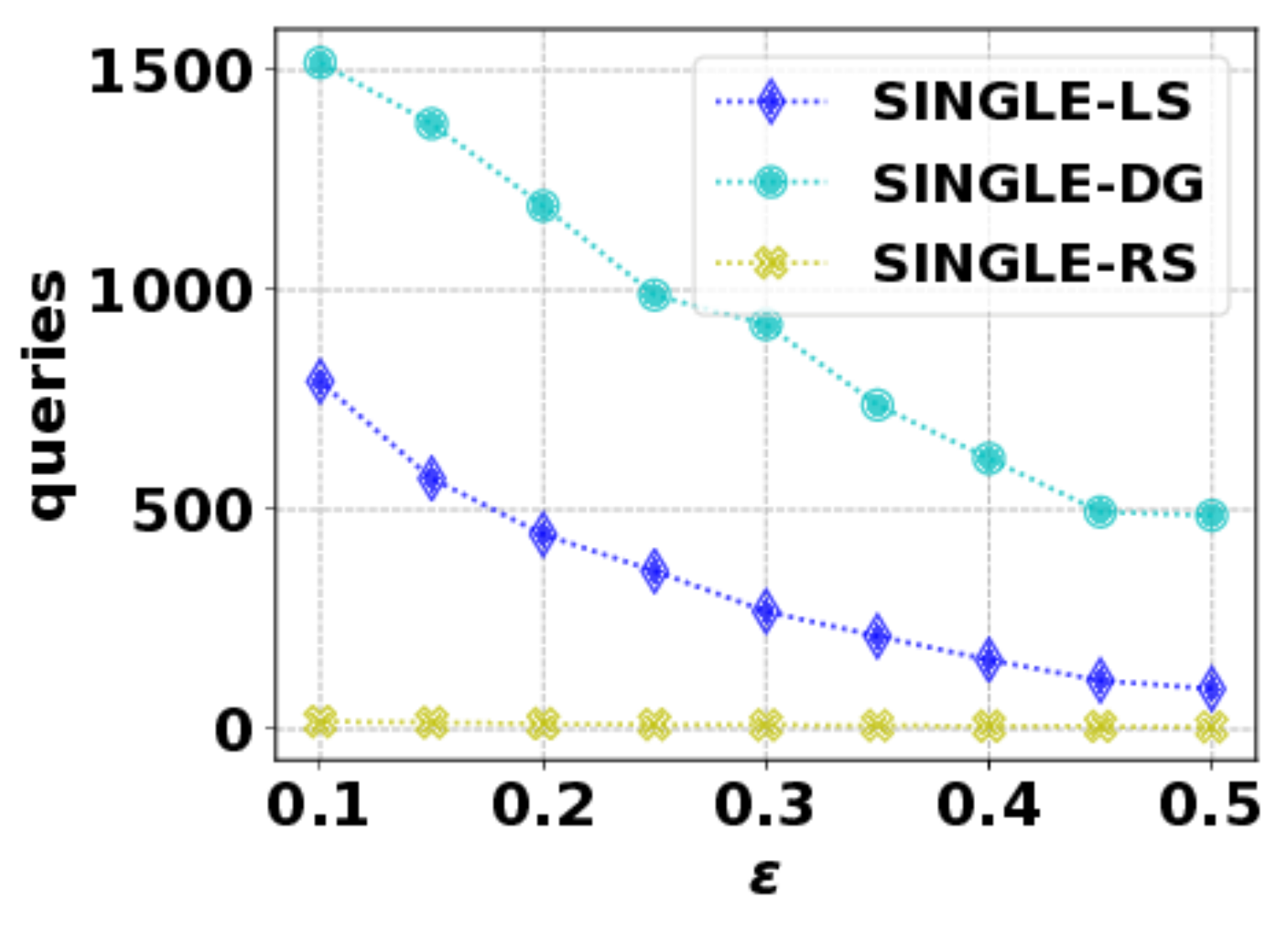}
  }
  \hspace{-1em}
  \subfigure[enron, cut] {
    \includegraphics[width=0.24\textwidth]{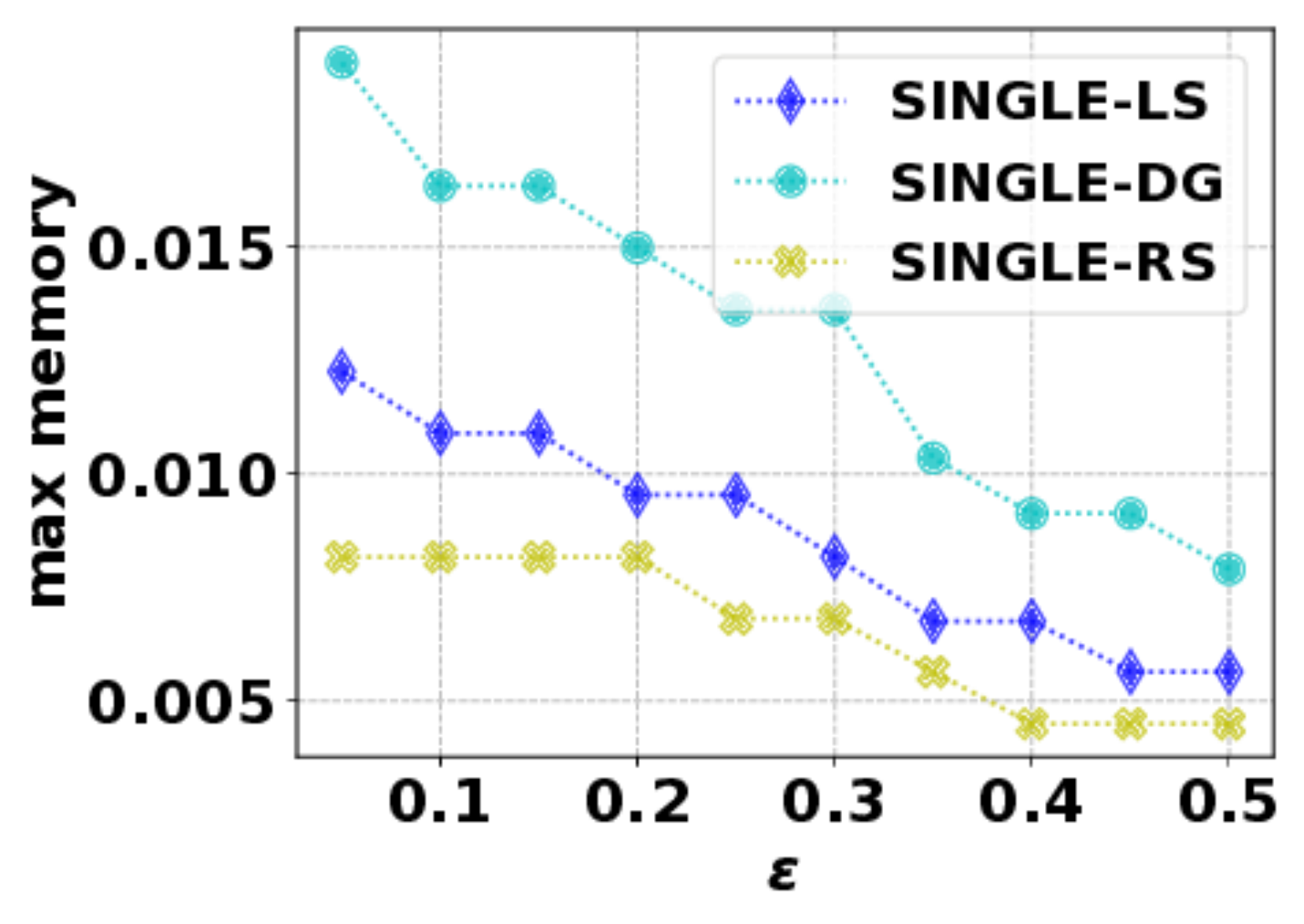}
  }
  \hspace{-1em}
  \subfigure[astro, cut] {
    \includegraphics[width=0.24\textwidth]{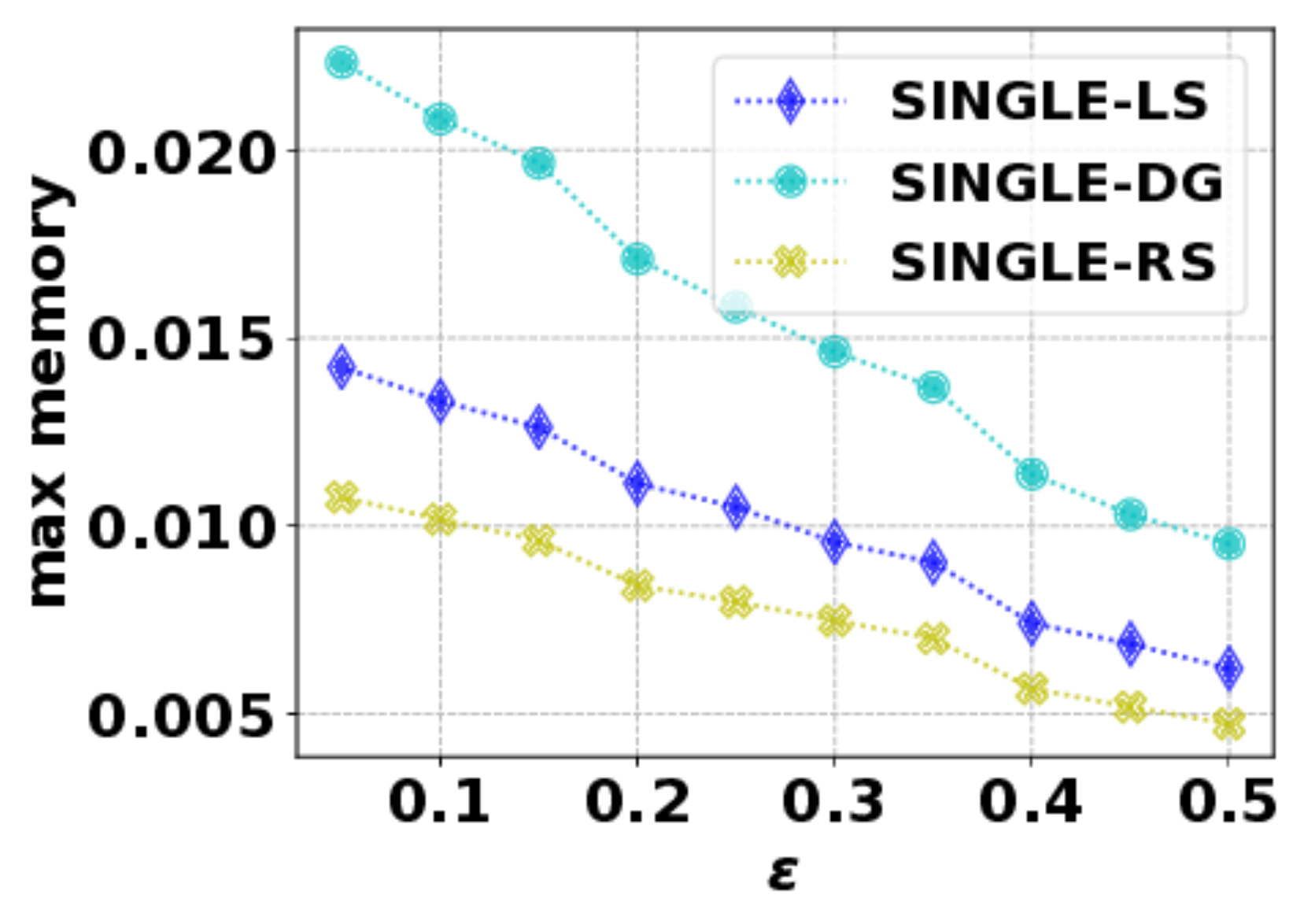}
  }
  \hspace{-1em}
  \subfigure[corel, cover] {
    \includegraphics[width=0.24\textwidth]{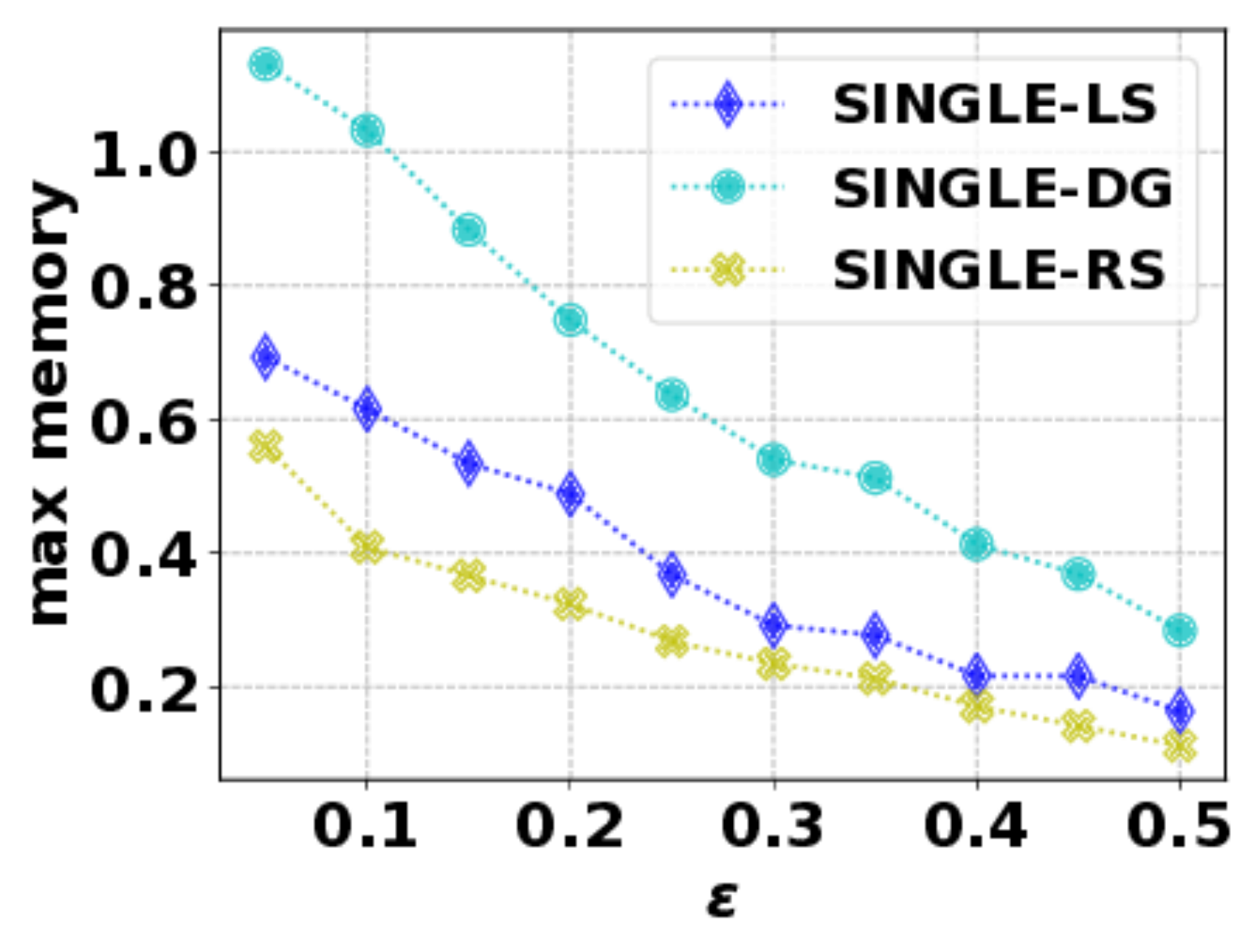}
  }
  \hspace{-1em}
  \subfigure[delicious5k, cover] {
    \includegraphics[width=0.24\textwidth]{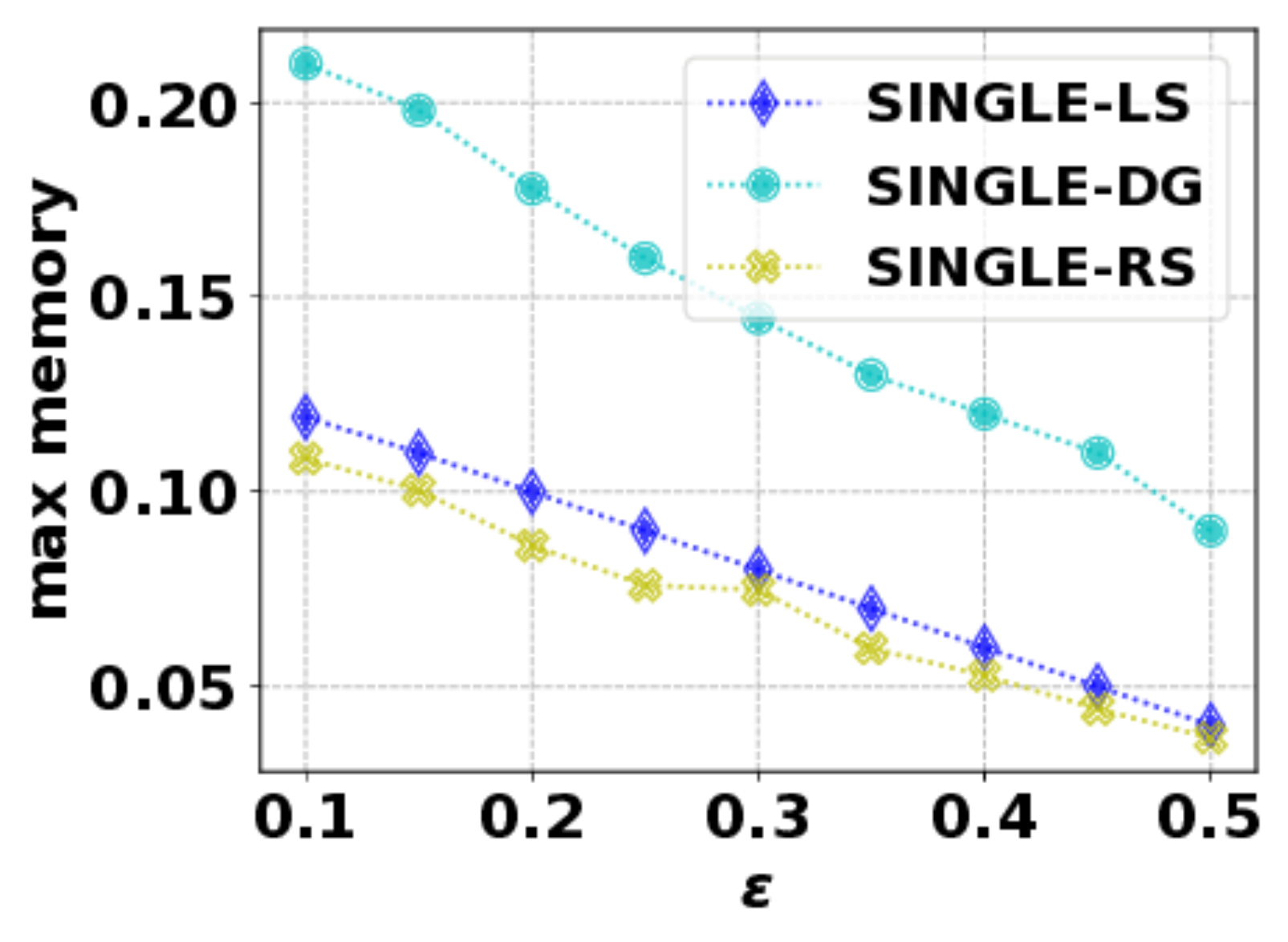}
  }
  \caption[]{
  A comparison of the performance of \single using different algorithms for USM as
  a subroutine.
  %$f$ and $c$ refer to the $f$ and cost values of the returned solution.
  %Queries is the total number of queries to $f$ that the algorithm took.
  %Max memory is the maximum cost of all elements stored at once
  %over the duration of the algorithm.
  %The instances of
  %graph cut and diverse data summarization are on the
  %Corel5k (``corel'') and
  %delicious (``delicious5k'') dataset,
  %and instances of graph cut are on the ca-AstroPh (``astro'')
  %and email-Enron (``amazon'') datasets.
  \single using the double greedy algorithm of \citet{buchbinder2015tight}, the local
  search algorithm of \citet{feige2011maximizing}, and the random set algorithm are
  referred to as ``SINGLE-DG'', ``SINGLE-LS'', and ``SINGLE-RS'' respectively.
  %All $x$ and $y$ axes are normalized as described in Section \ref{section:resultsapp}.
  }
  \label{fig:ssubroutines}
\end{figure*}

\begin{figure*}[t!]
  \centering
  \hspace{-1em}
  \subfigure[amazon, cut] {
    \includegraphics[width=0.24\textwidth]{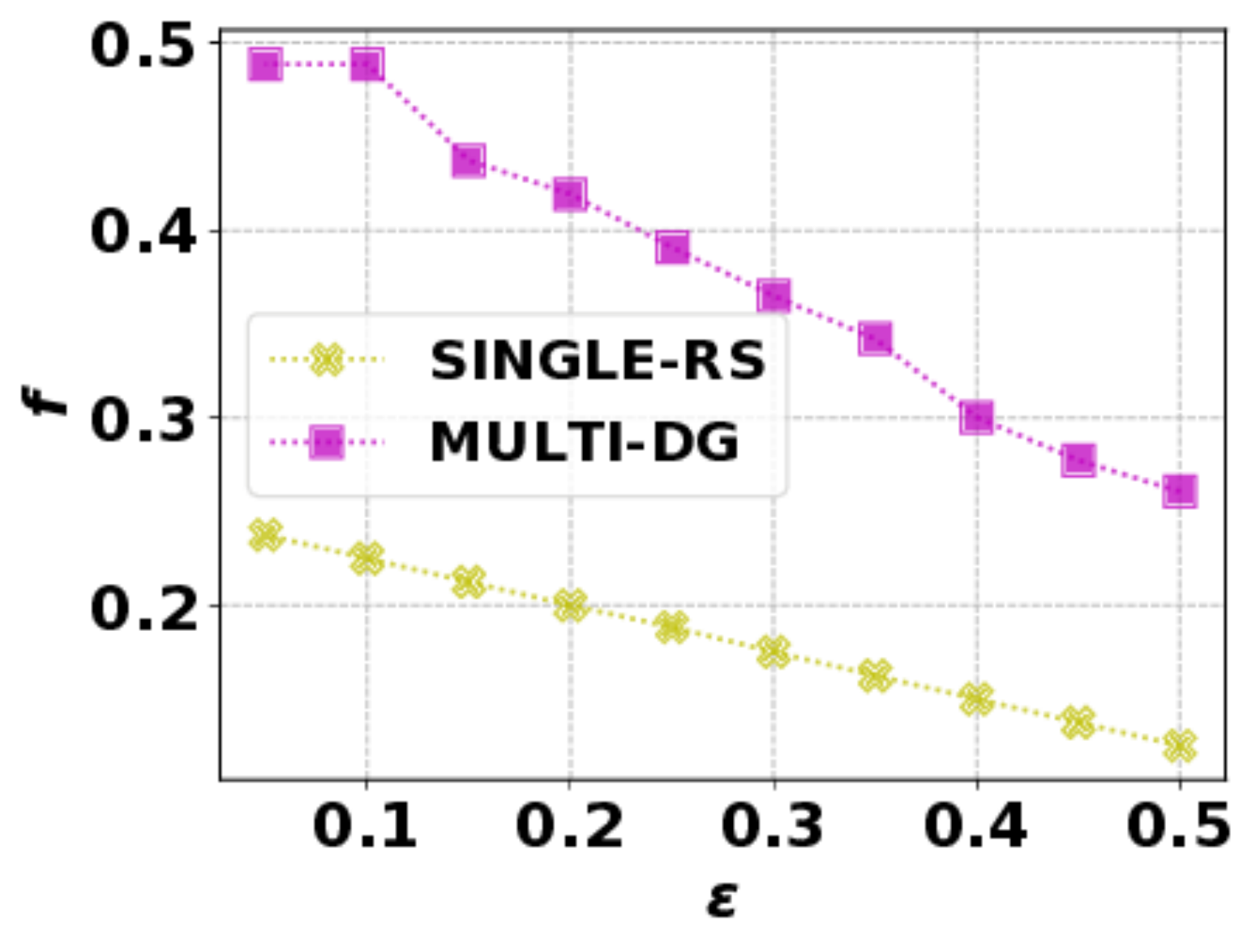}
  }
  \hspace{-1em}
  \subfigure[astro, cut] {
    \includegraphics[width=0.24\textwidth]{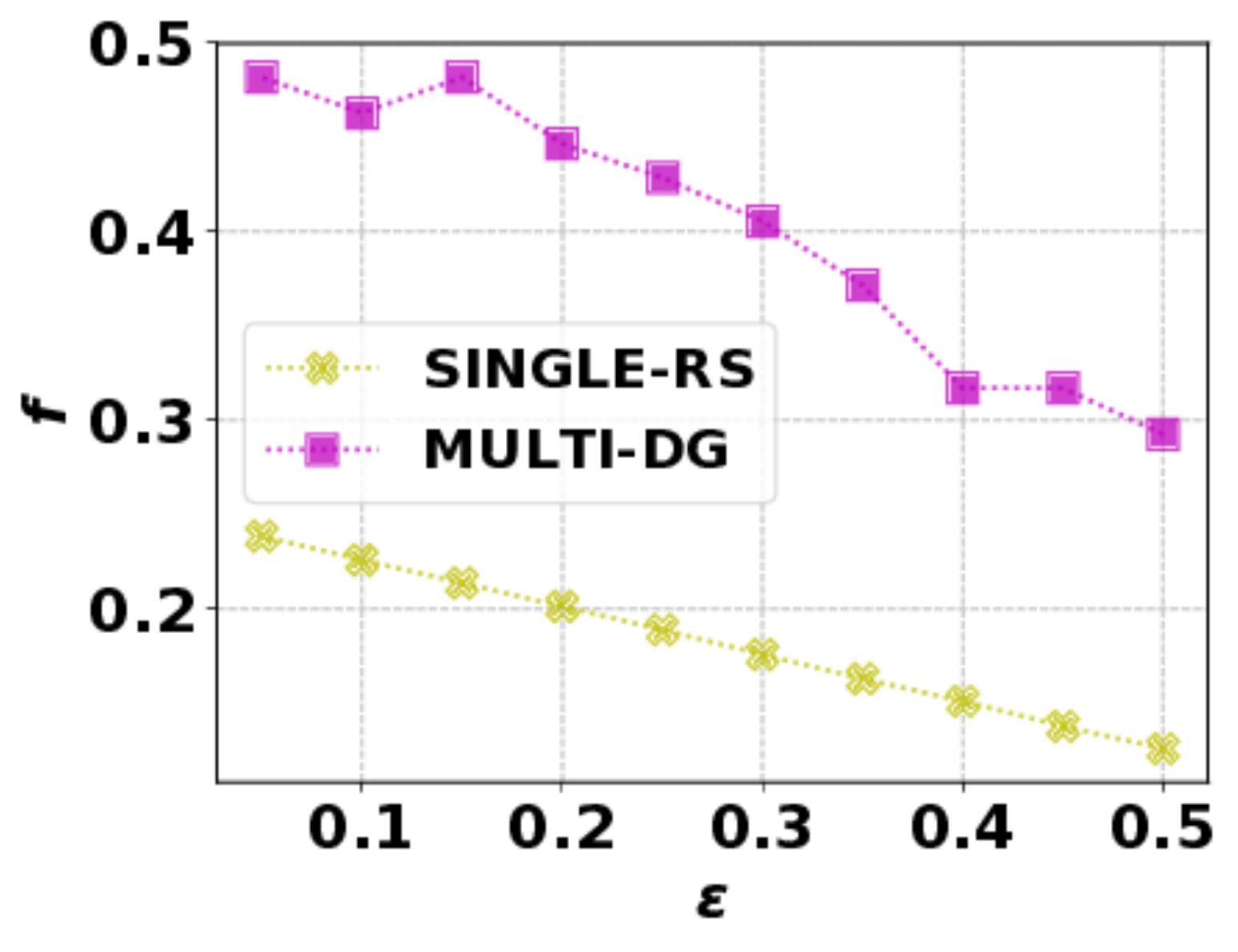}
  }
  \hspace{-1em}
  \subfigure[corel, cover] {
    \includegraphics[width=0.24\textwidth]{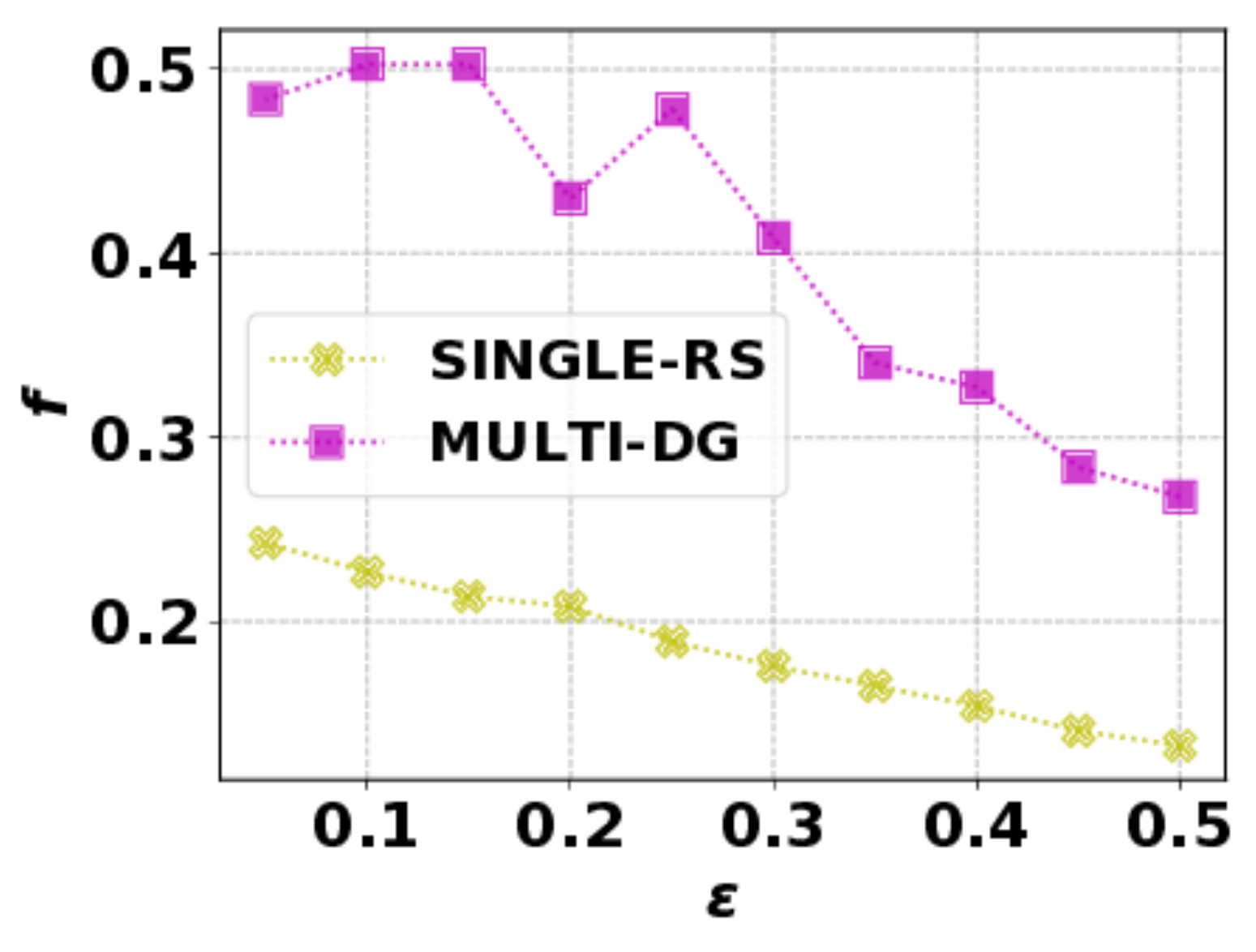}
  }
  \hspace{-1em}
  \subfigure[delicious50k, cover] {
    \includegraphics[width=0.24\textwidth]{sections/figures/cover_n50000_epsf.pdf}
  }
  \hspace{-1em}
  \subfigure[amazon, cut] {
    \includegraphics[width=0.24\textwidth]{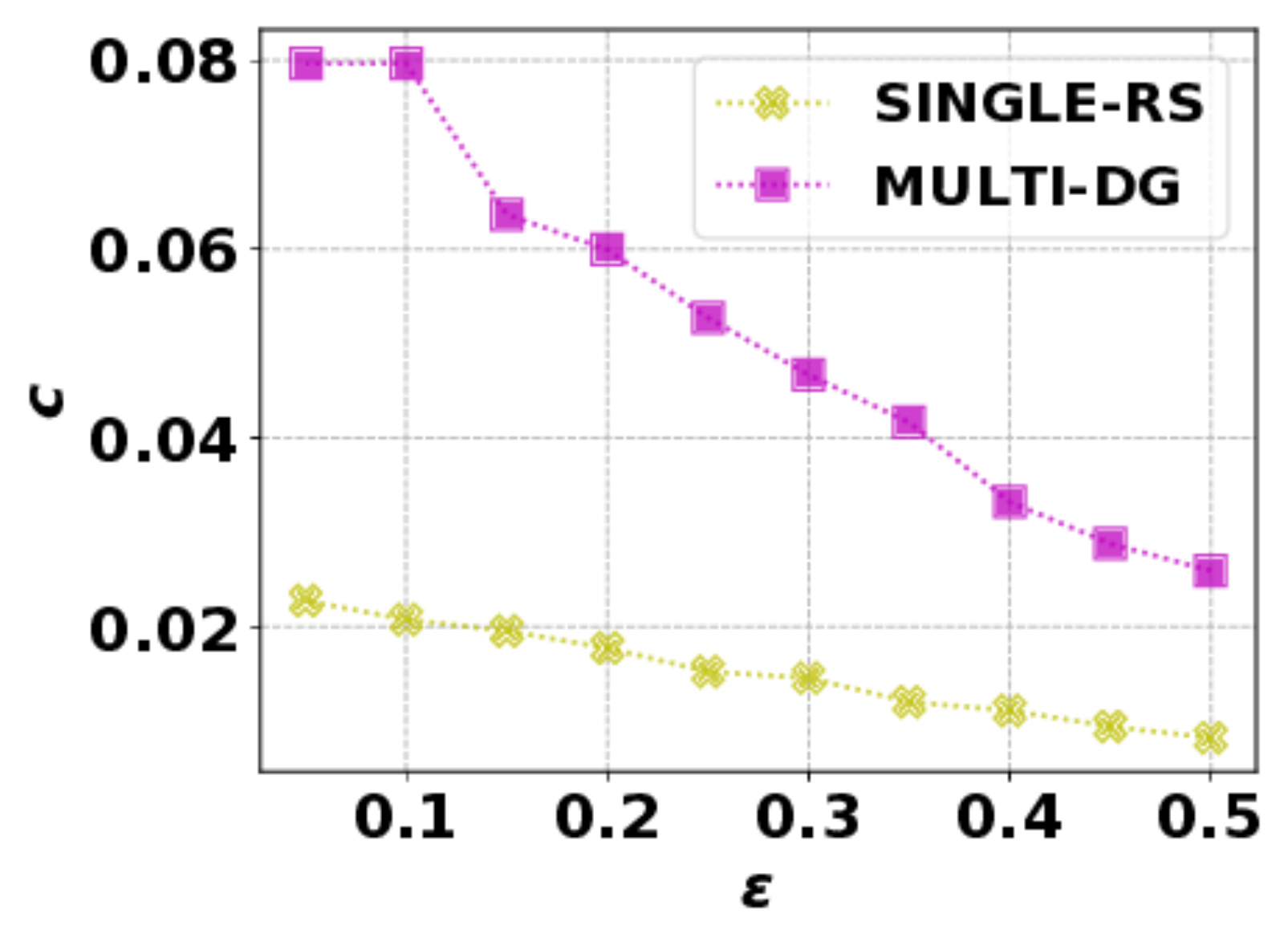}
  }
  \hspace{-1em}
  \subfigure[astro, cut] {
    \includegraphics[width=0.24\textwidth]{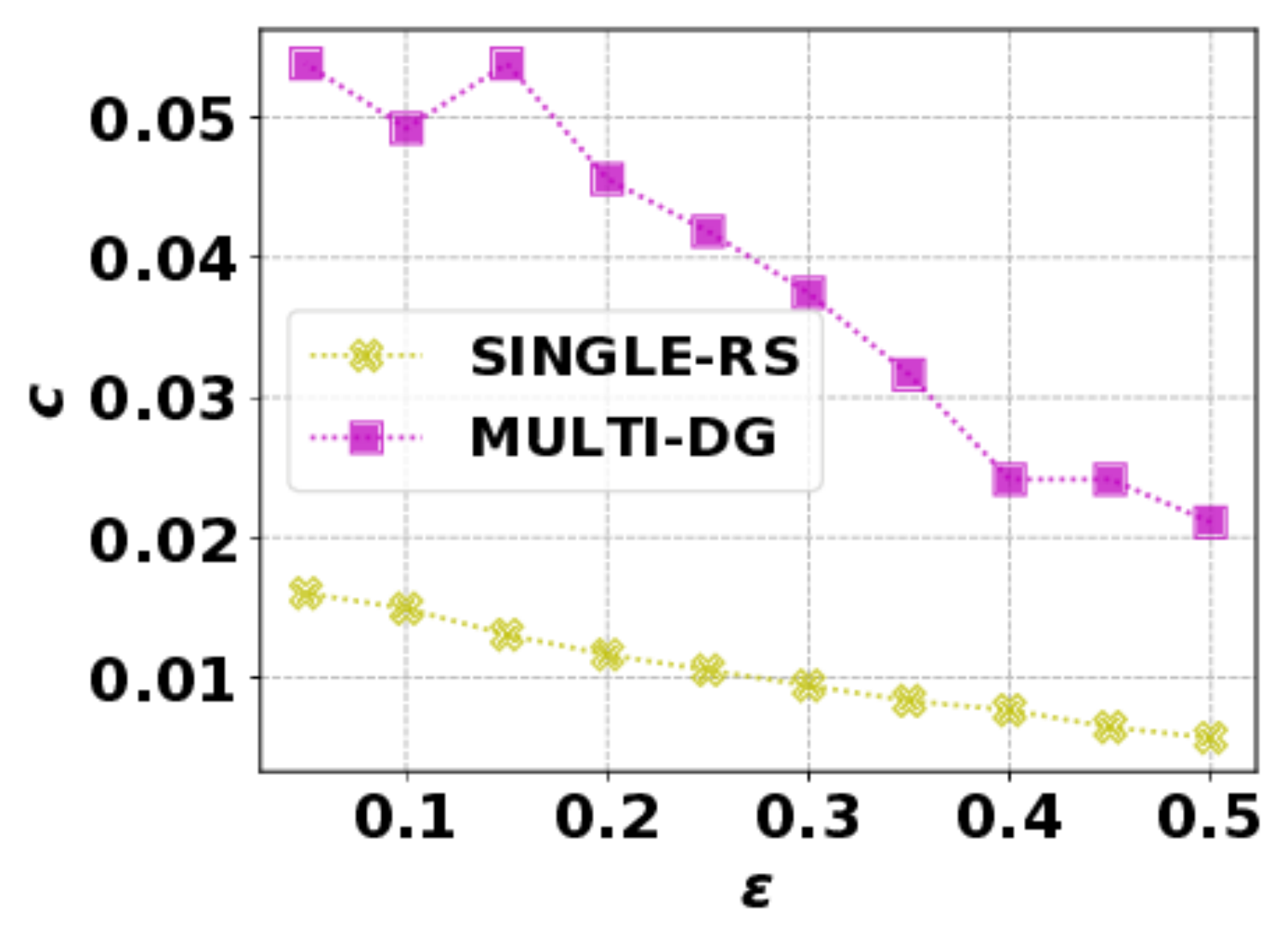}
  }
  \hspace{-1em}
  \subfigure[corel, cover] {
    \includegraphics[width=0.24\textwidth]{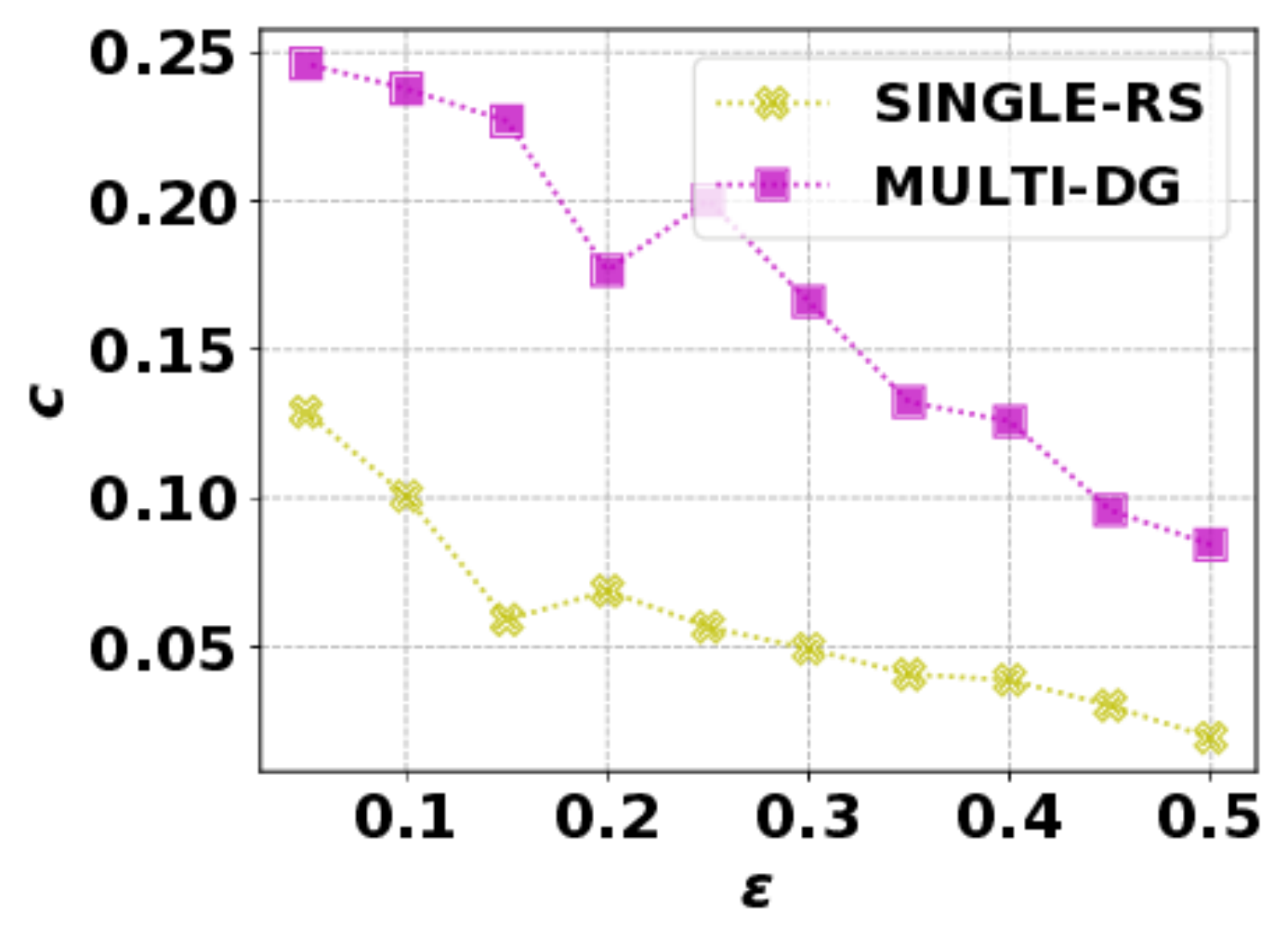}
  }
  \hspace{-1em}
  \subfigure[delicious50k, cover] {
    \includegraphics[width=0.24\textwidth]{sections/figures/cover_n50000_epsc.pdf}
  }
  \hspace{-1em}
  \subfigure[amazon, cut] {
    \includegraphics[width=0.24\textwidth]{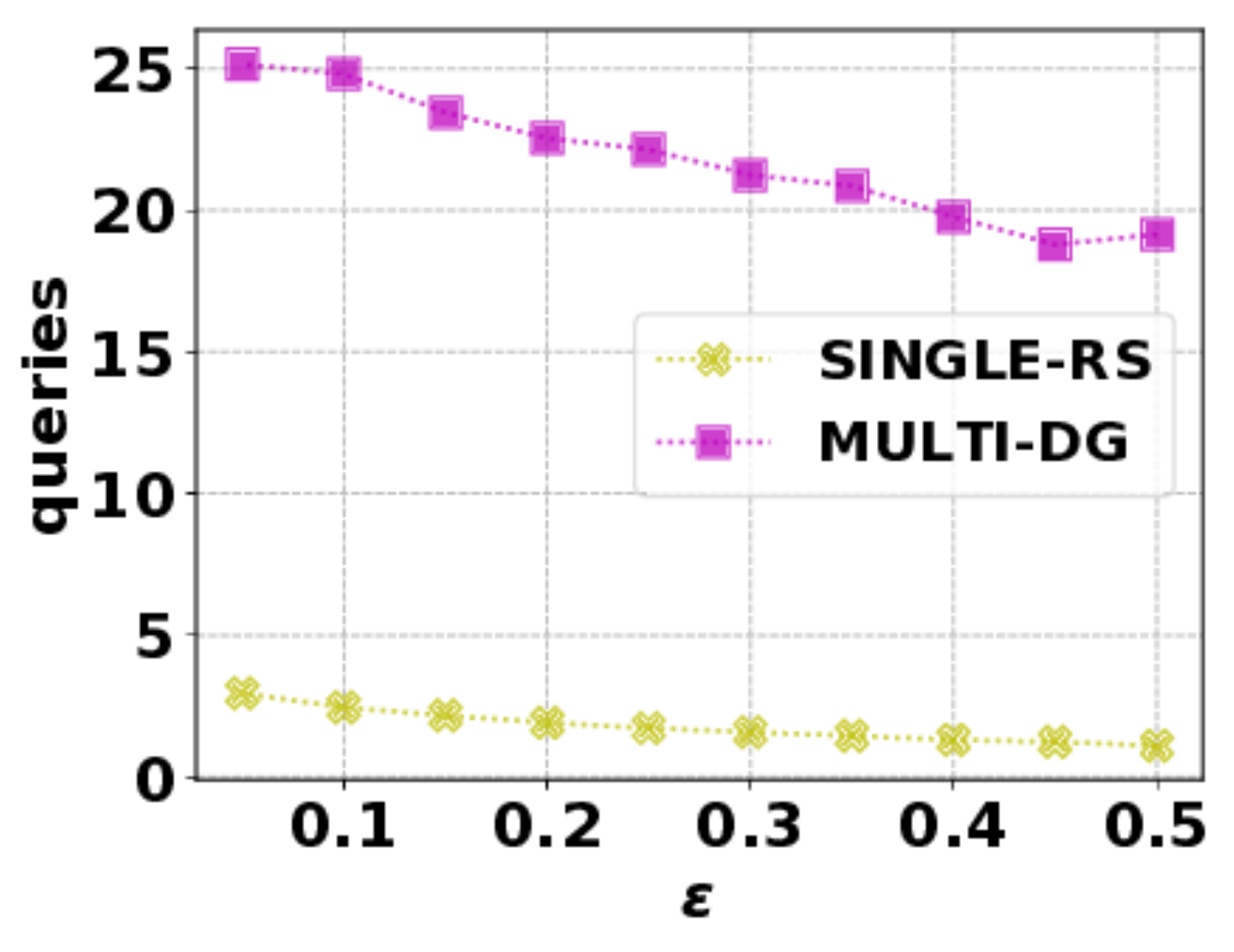}
  }
  \hspace{-1em}
  \subfigure[astro, cut] {
    \includegraphics[width=0.24\textwidth]{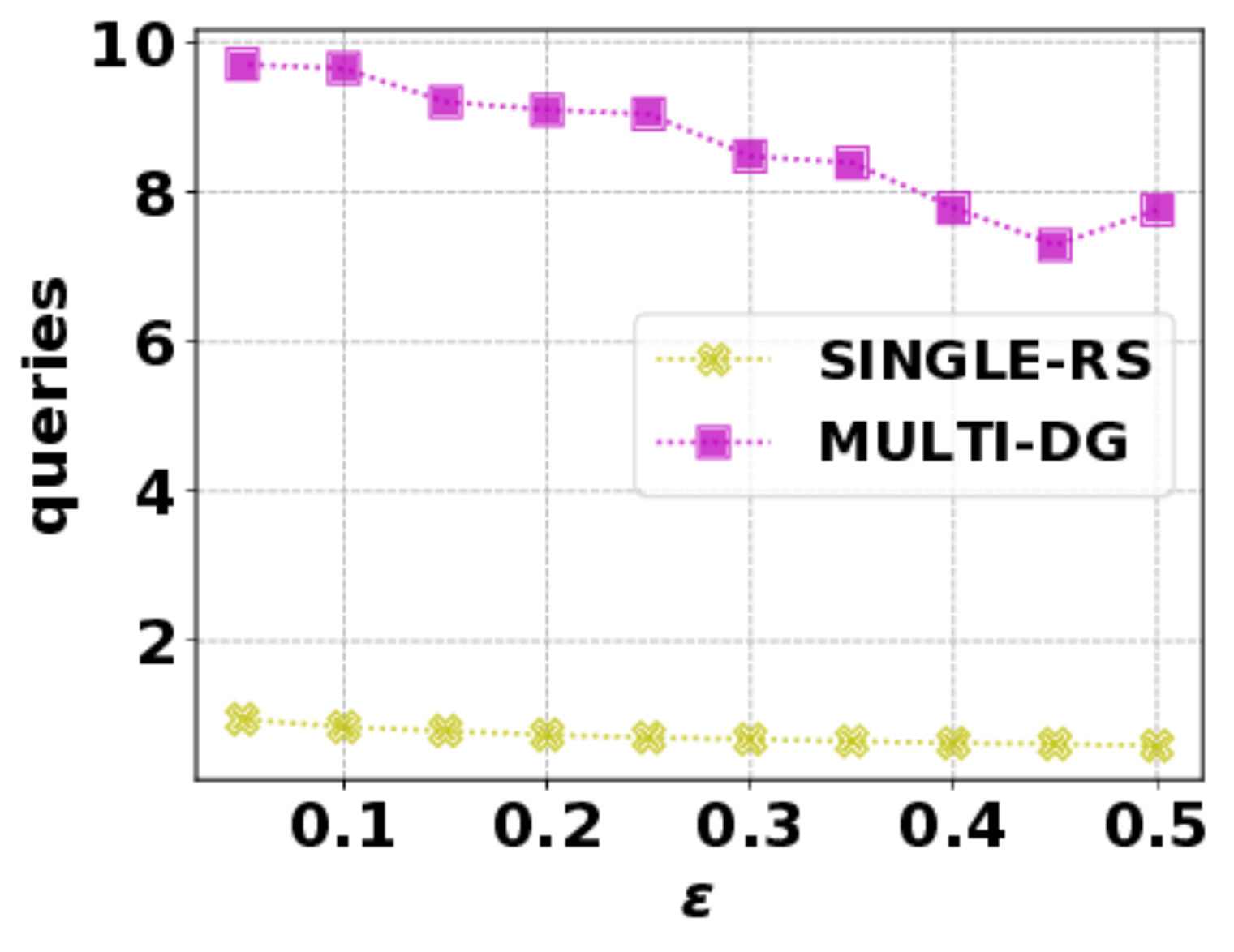}
  }
  \hspace{-1em}
  \subfigure[corel, cover] {
    \includegraphics[width=0.24\textwidth]{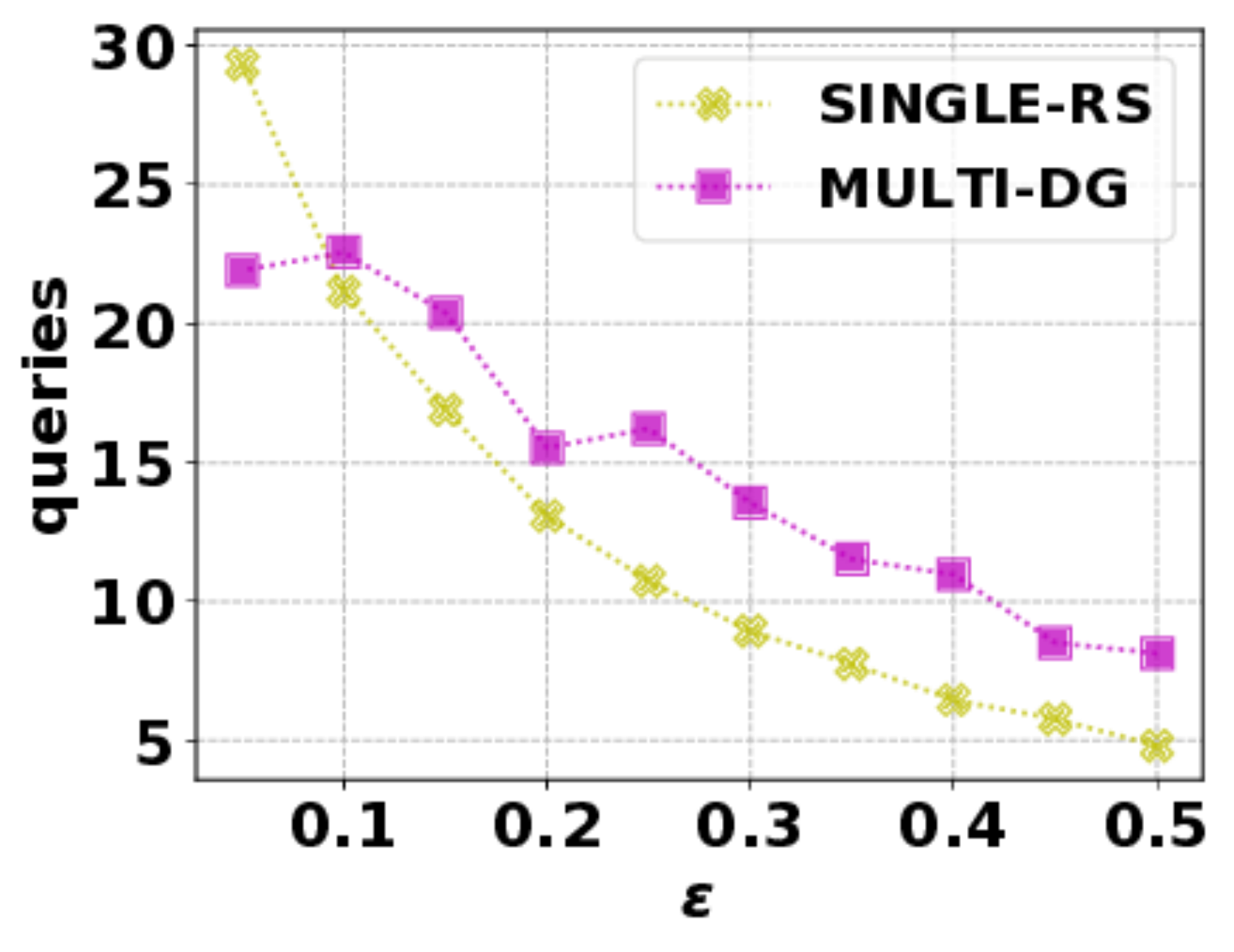}
  }
  \hspace{-1em}
  \subfigure[delicious50k, cover] {
    \includegraphics[width=0.24\textwidth]{sections/figures/cover_n50000_epsq.pdf}
  }
  \hspace{-1em}
  \subfigure[amazon, cut] {
    \includegraphics[width=0.24\textwidth]{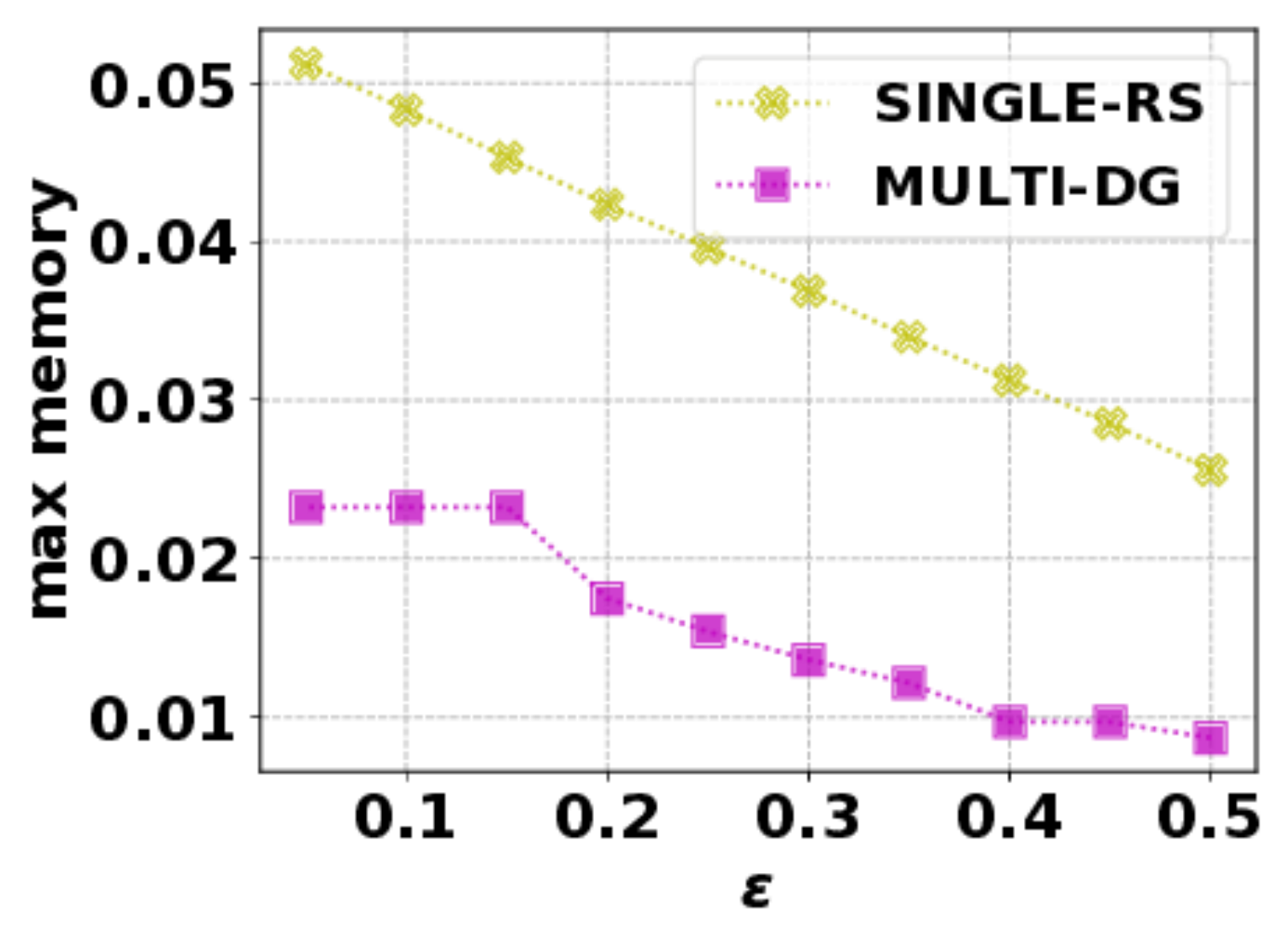}
  }
  \hspace{-1em}
  \subfigure[astro, cut] {
    \includegraphics[width=0.24\textwidth]{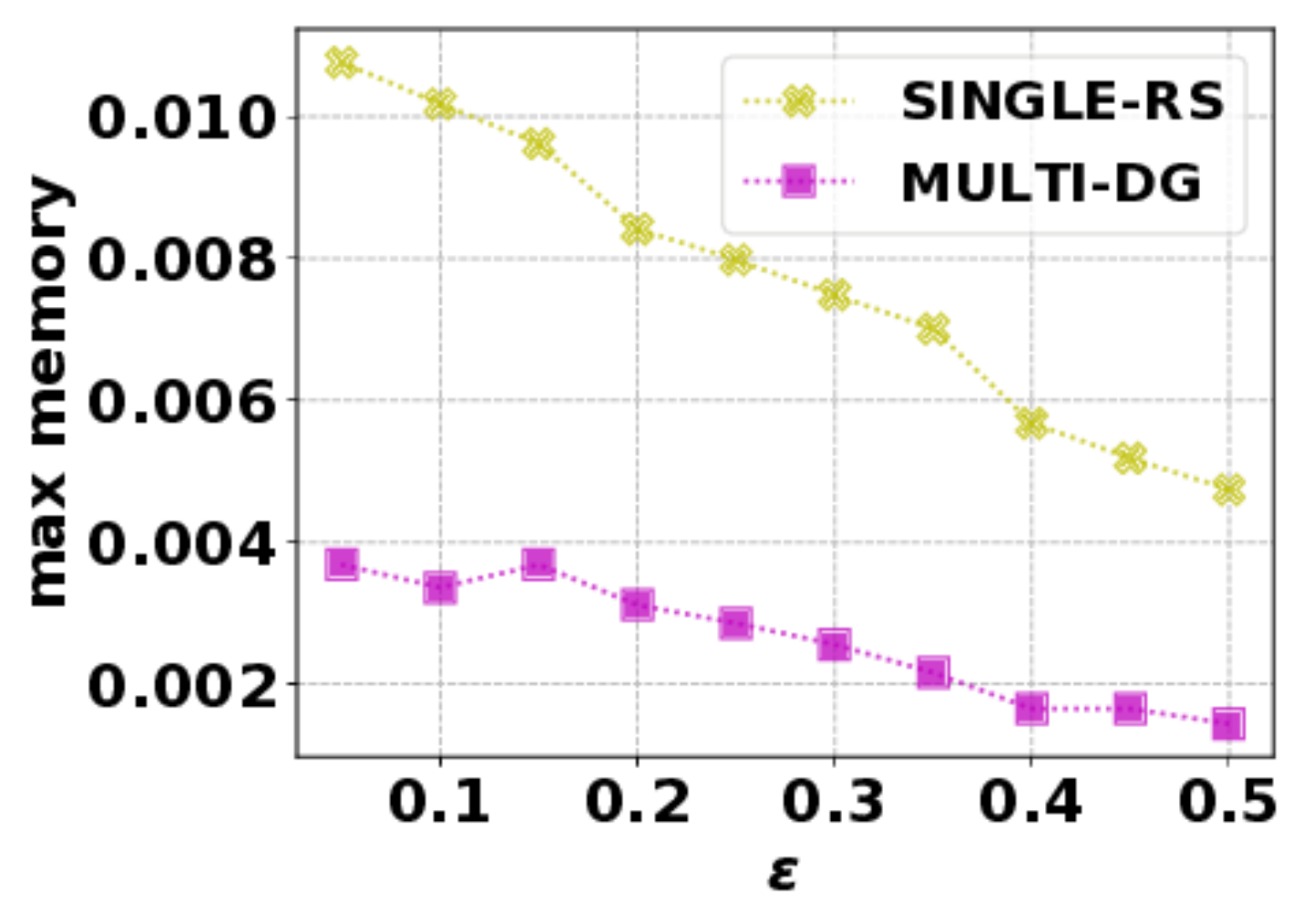}
  }
  \hspace{-1em}
  \subfigure[corel, cover] {
    \includegraphics[width=0.24\textwidth]{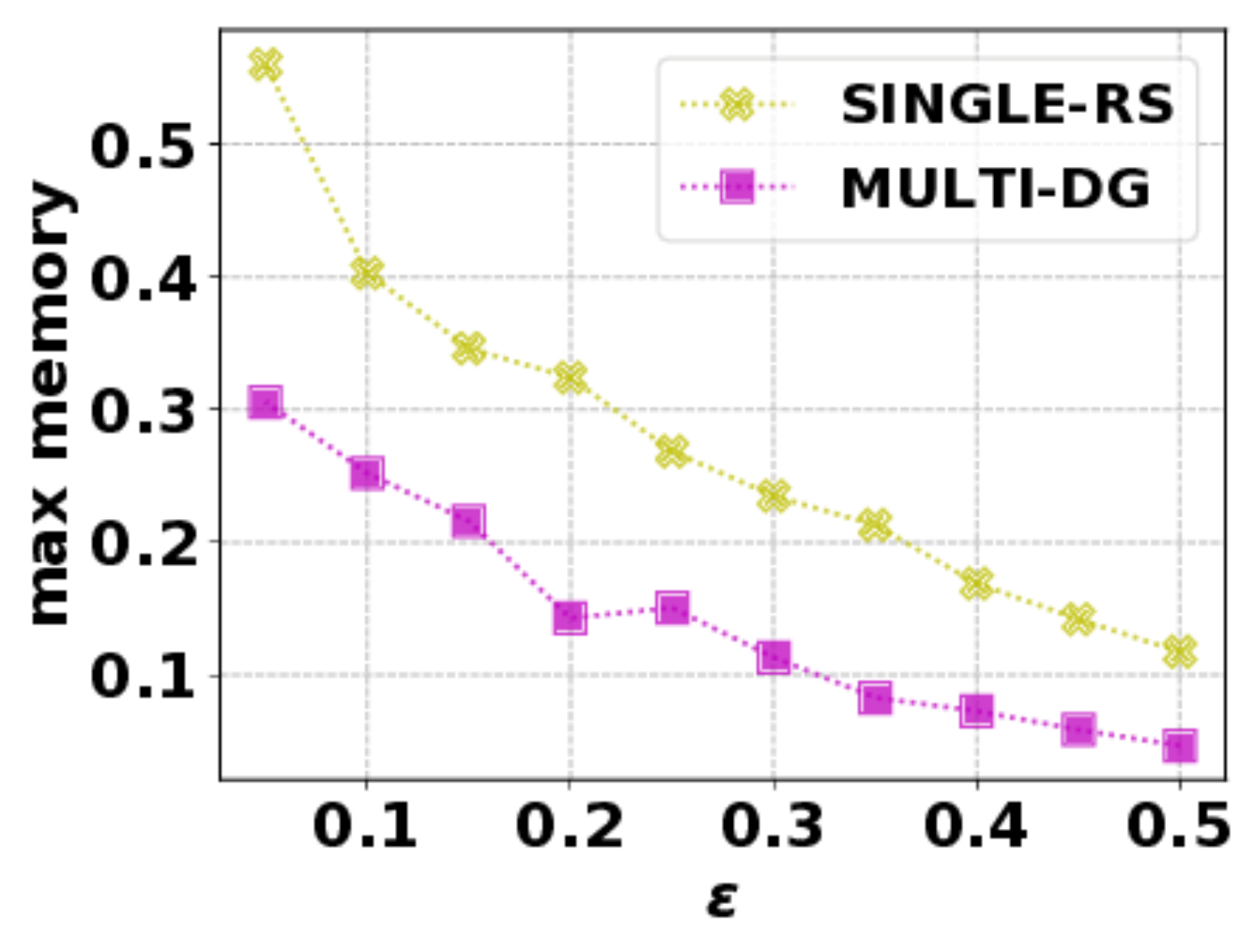}
  }
  \hspace{-1em}
  \subfigure[delicious50k, cover] {
    \includegraphics[width=0.24\textwidth]{sections/figures/cover_n50000_epsmem.pdf}
  }
  \caption{
  The outcome of running \multi and \single
  for varying $\epsilon$
  on instances of
  diverse data summarization on the
  Corel5k (``corel'') and
  delicious (``delicious50k'') dataset with $n=50000$,
  and instances of graph cut on the ca-AstroPh (``astro'')
  and com-Amazon (``amazon'')
  datasets.
  %$f$ and $c$ refer to the $f$ and cost values of the returned solution.
  %Queries is the total number of queries to $f$ that the algorithm took.
  %Max memory is the maximum cost of all elements stored at once
  %over the duration of the algorithm.
  %\multi using the double greedy algorithm of \citet{buchbinder2015tight}
  %is referred to as ``MULTI-DG''. \single using random sets as described by
  %\citet{feige2011maximizing} is referred to as ``SINGLE-RS''.
  %All $x$ and $y$ axes are normalized as described in Section \ref{section:resultsapp}.
  }
  \label{fig:epsilons}
\end{figure*}

\begin{figure*}[t!]
  \centering
  \hspace{-1em}
  \subfigure[enron, cut] {
    \includegraphics[width=0.24\textwidth]{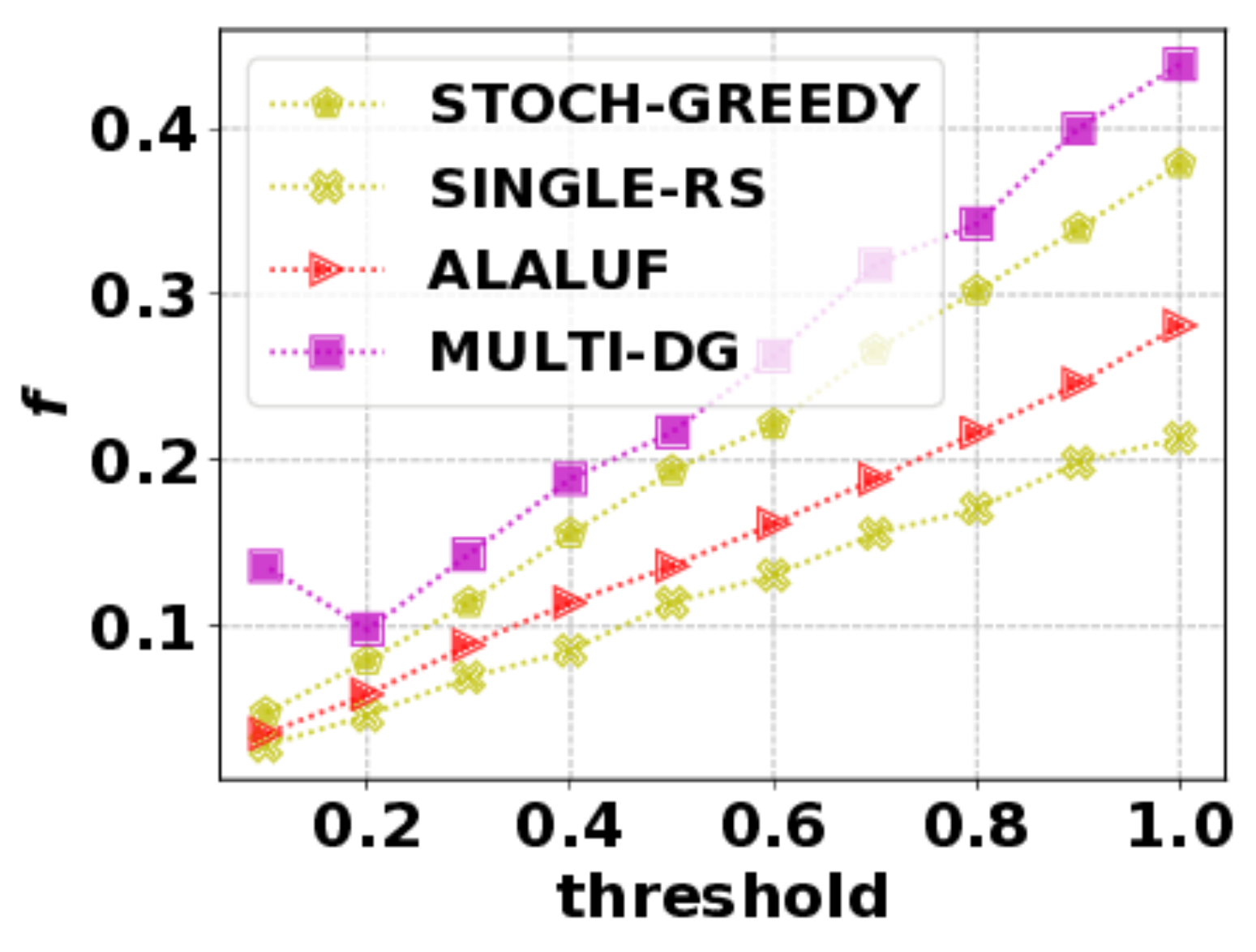}
  }
  \hspace{-1em}
  \subfigure[astro, cut] {
    \includegraphics[width=0.24\textwidth]{sections/figures/astro_tauf.pdf}
  }
  \hspace{-1em}
  \subfigure[corel, cover] {
    \includegraphics[width=0.24\textwidth]{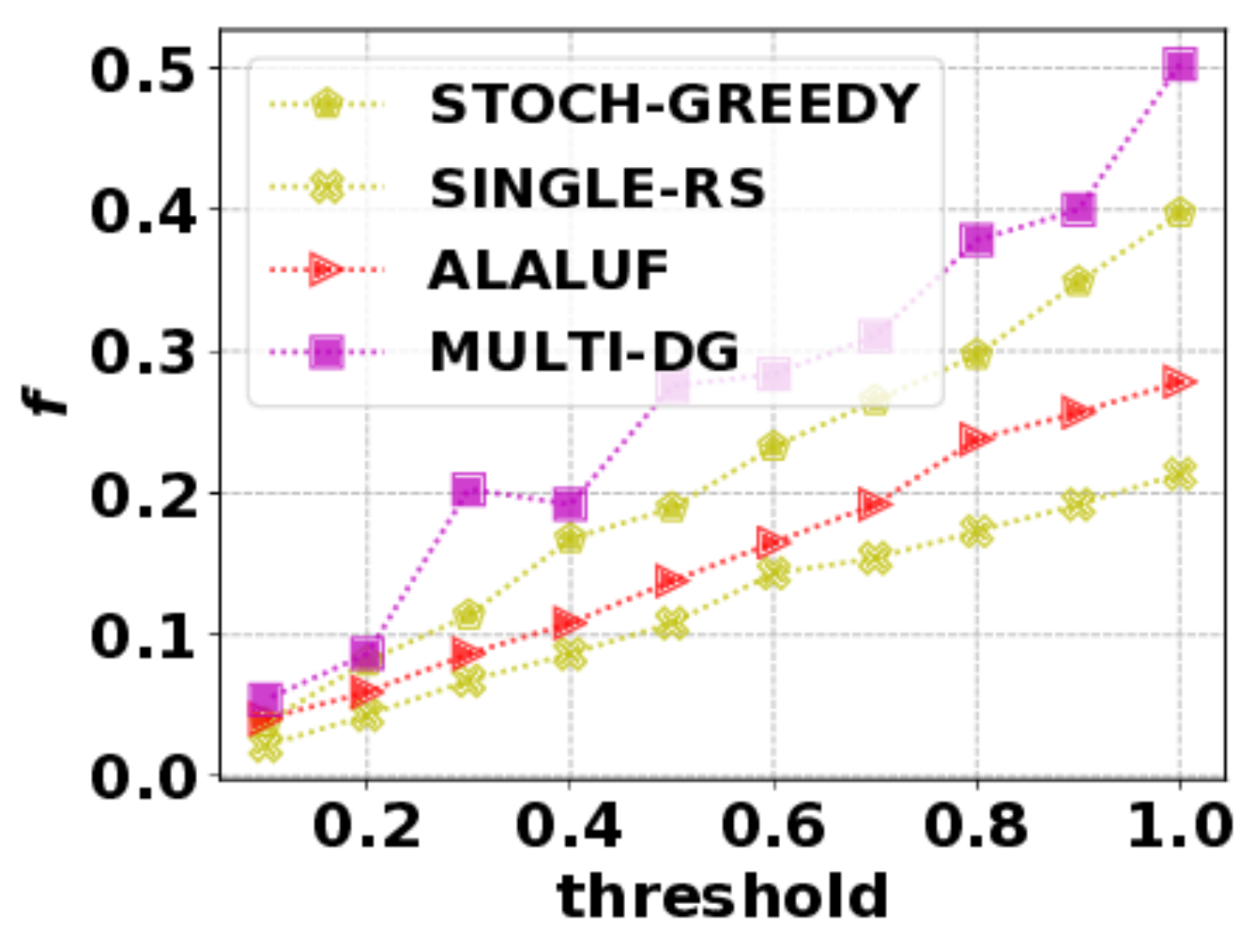}
  }
  \hspace{-1em}
  \subfigure[delicious, cover] {
    \includegraphics[width=0.24\textwidth]{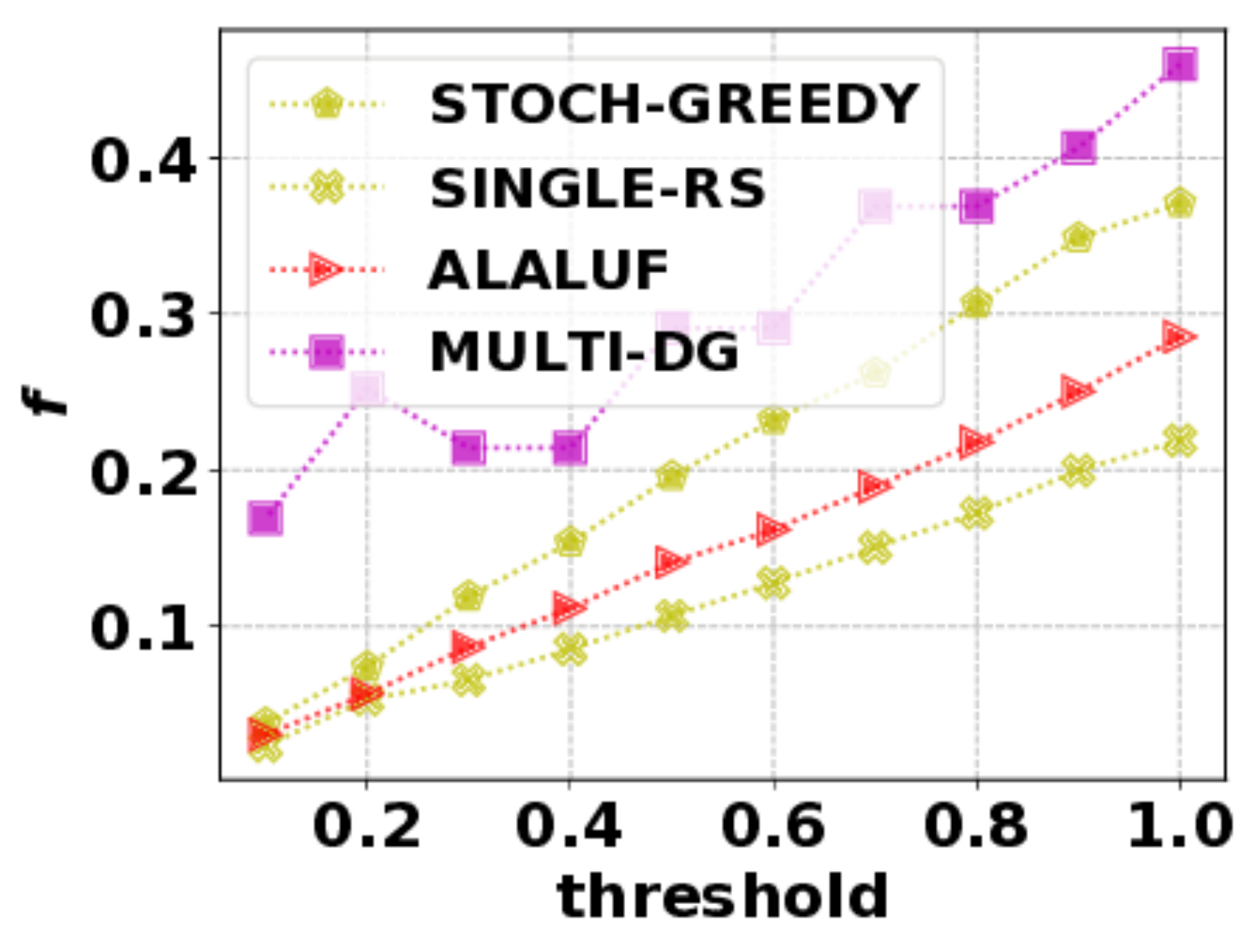}
  }
  \hspace{-1em}
  \subfigure[enron, cut] {
    \includegraphics[width=0.24\textwidth]{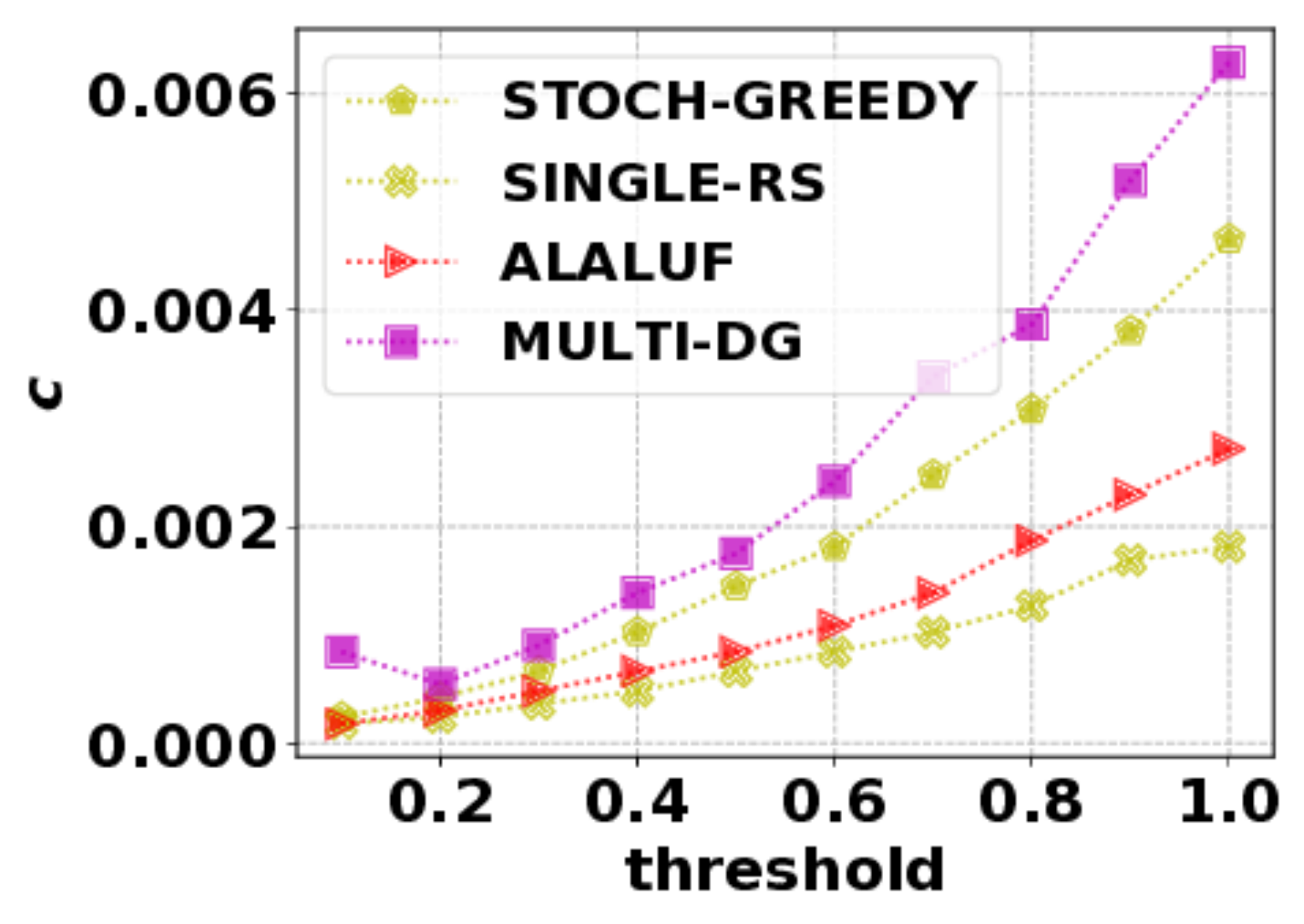}
  }
  \hspace{-1em}
  \subfigure[astro, cut] {
    \includegraphics[width=0.24\textwidth]{sections/figures/astro_tauc.pdf}
  }
  \hspace{-1em}
  \subfigure[corel, cover] {
    \includegraphics[width=0.24\textwidth]{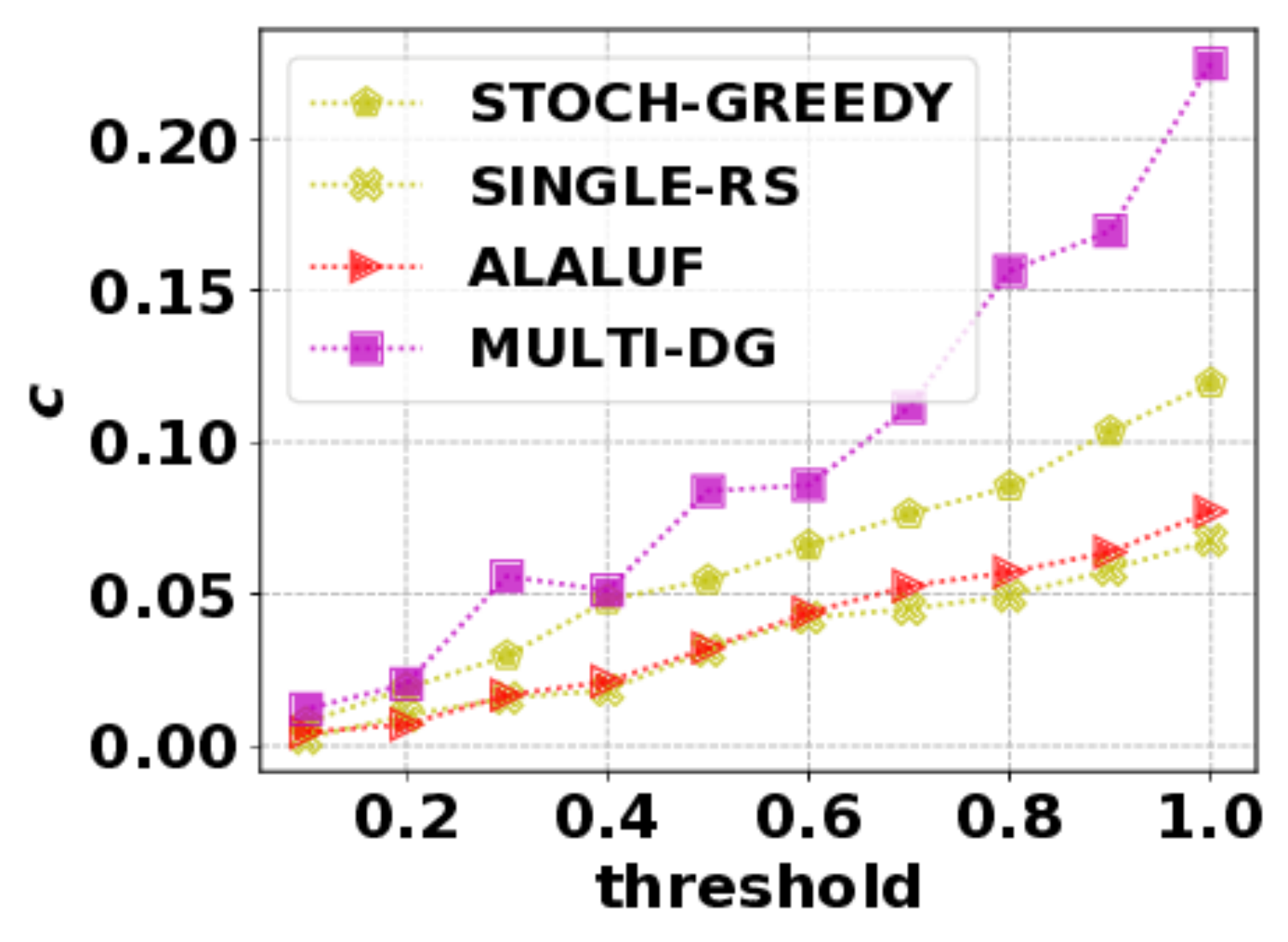}
  }
  \hspace{-1em}
  \subfigure[delicious, cover] {
    \includegraphics[width=0.24\textwidth]{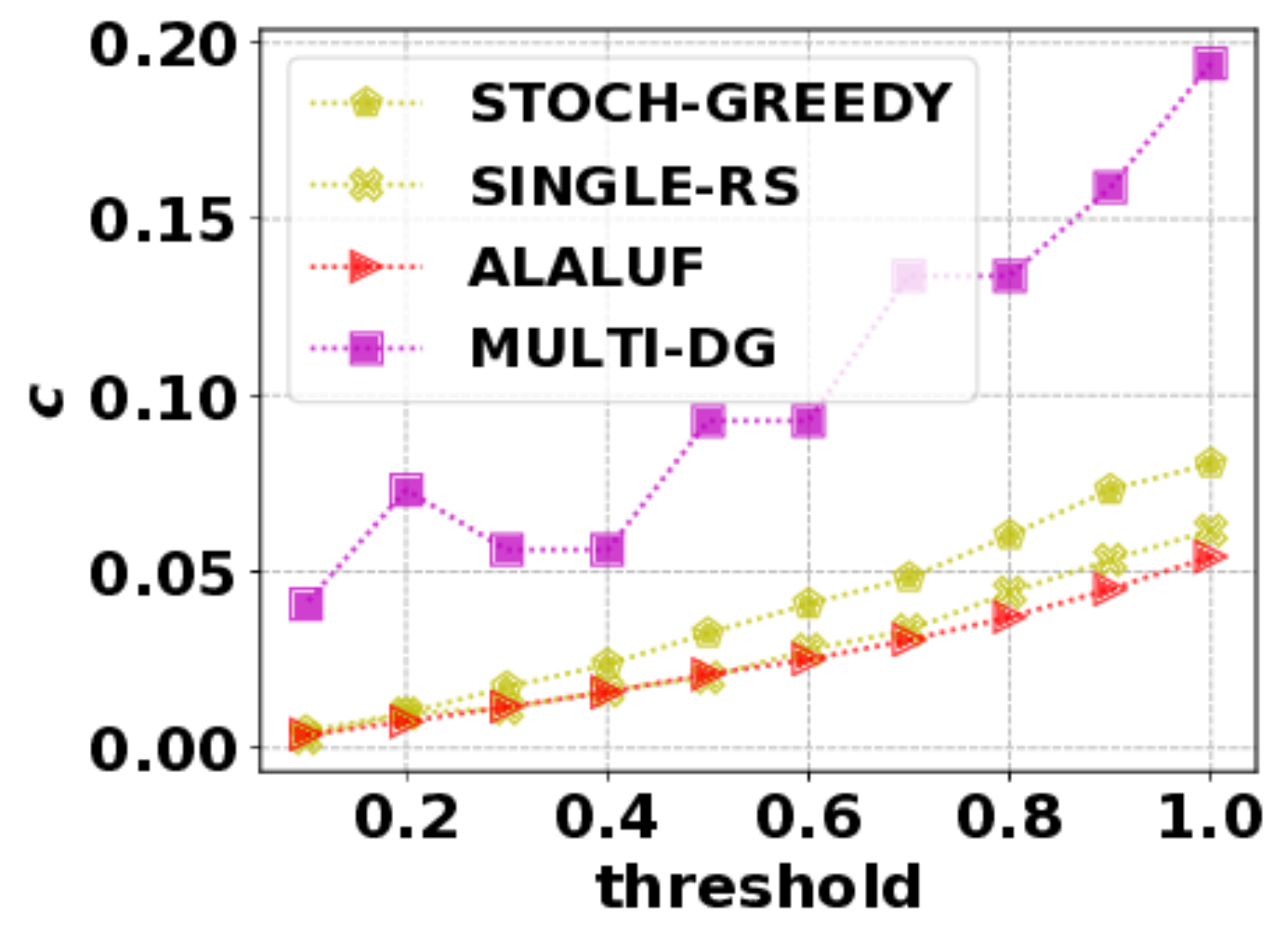}
  }
  \hspace{-1em}
  \subfigure[enron, cut] {
    \includegraphics[width=0.24\textwidth]{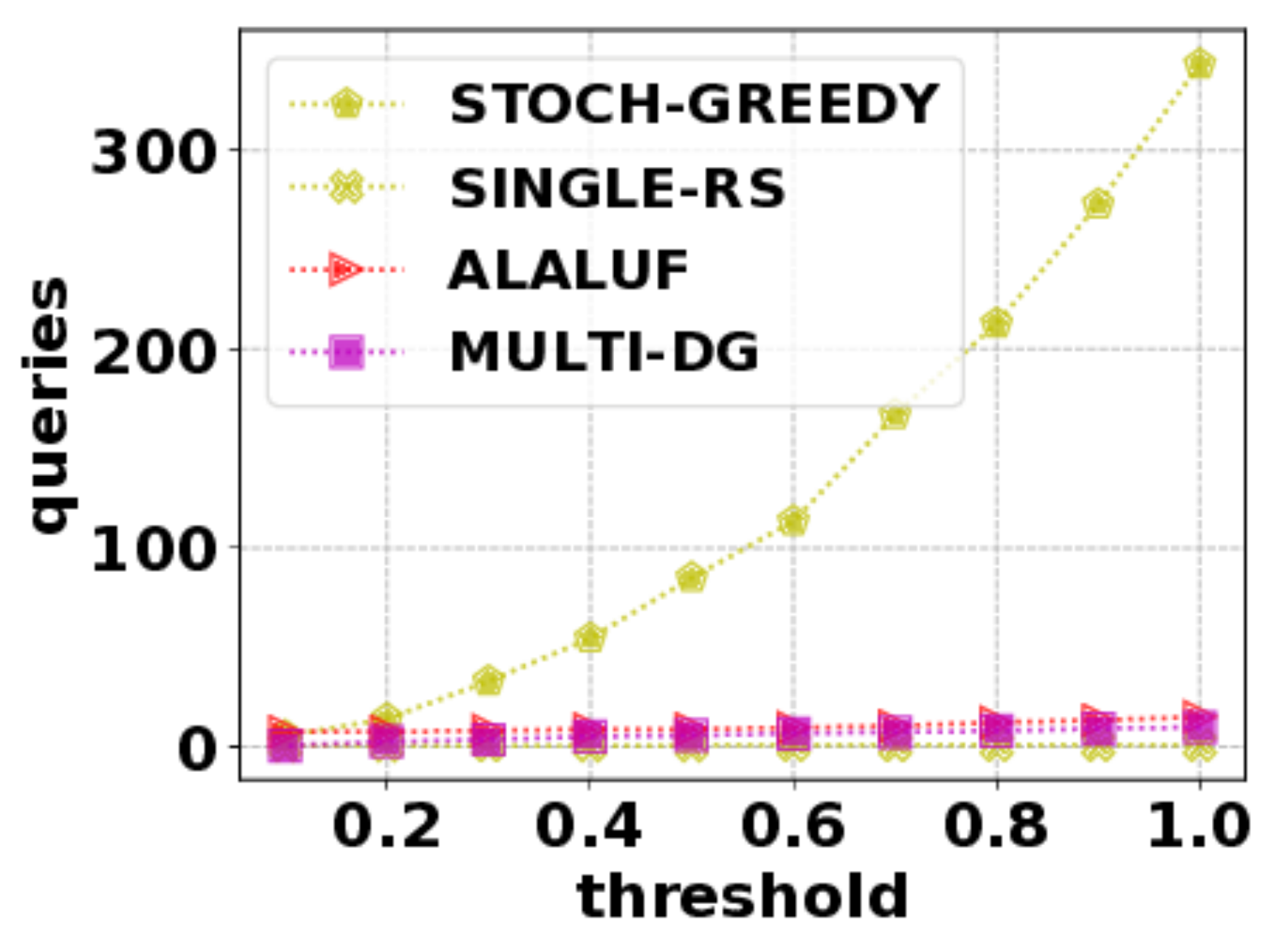}
  }
  \hspace{-1em}
  \subfigure[astro, cut] {
    \includegraphics[width=0.24\textwidth]{sections/figures/astro_tauq.pdf}
  }
  \hspace{-1em}
  \subfigure[corel, cover] {
    \includegraphics[width=0.24\textwidth]{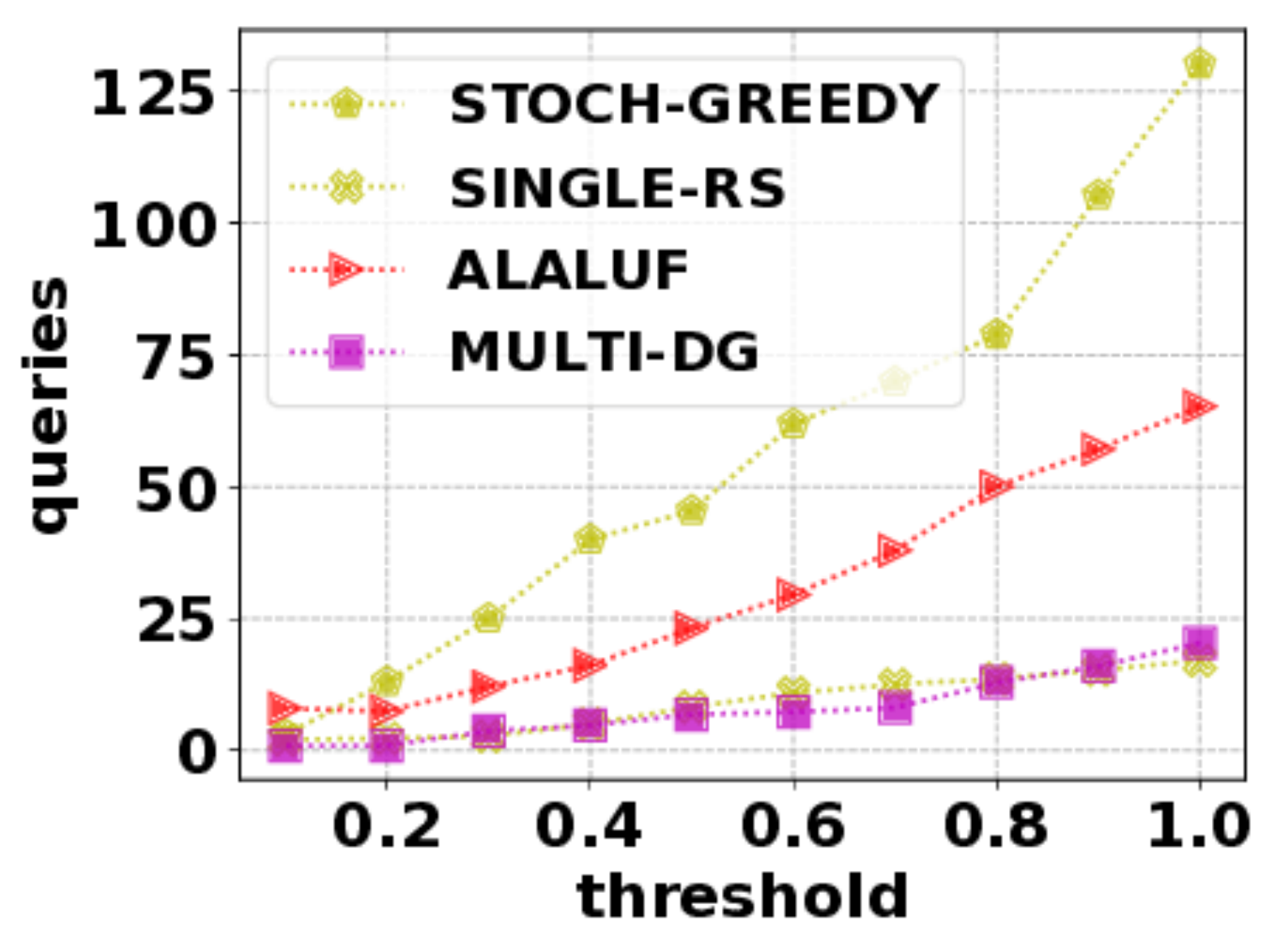}
  }
  \hspace{-1em}
  \subfigure[delicious, cover] {
    \includegraphics[width=0.24\textwidth]{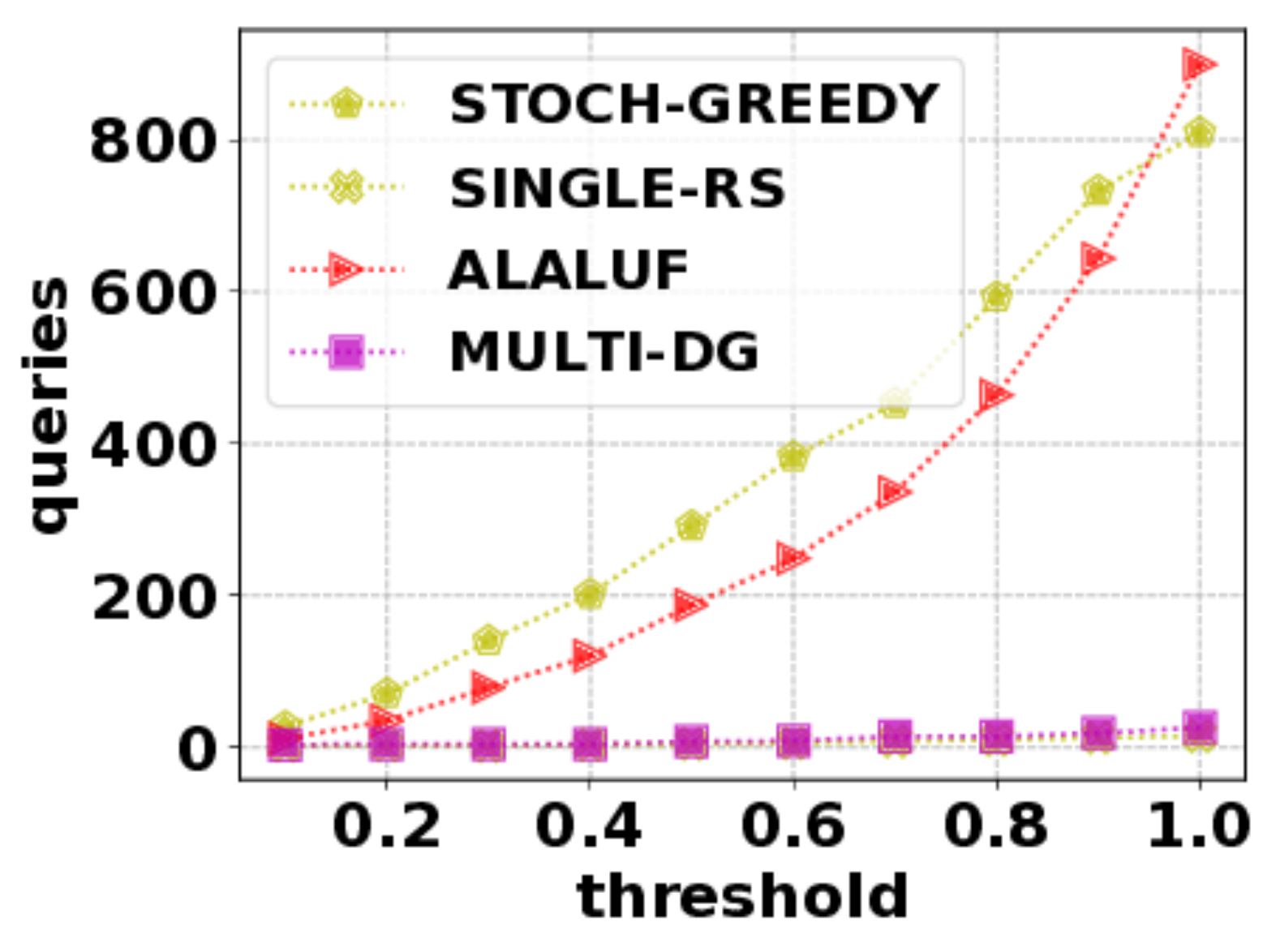}
  }
  \hspace{-1em}
  \subfigure[enron, cut] {
    \includegraphics[width=0.24\textwidth]{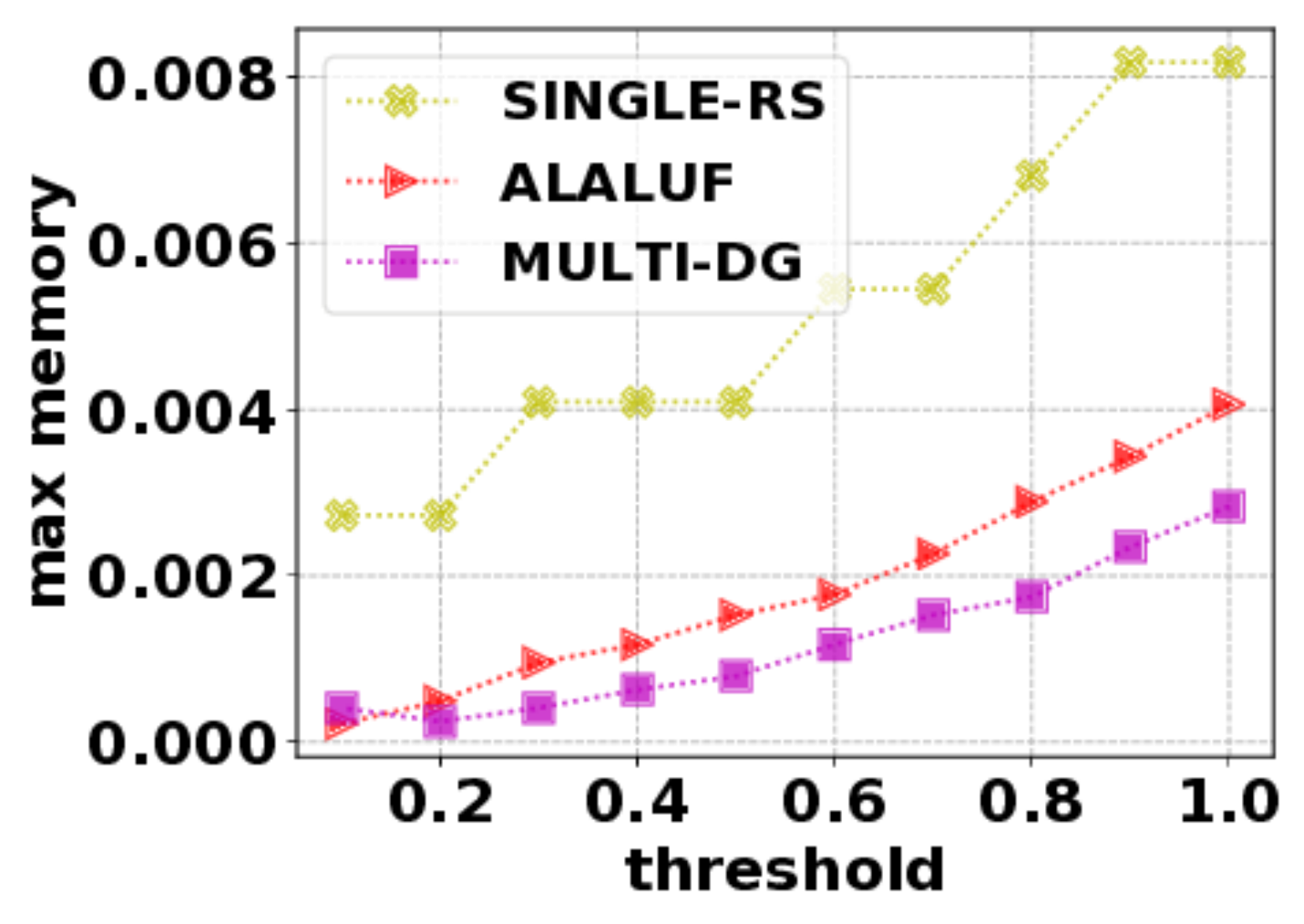}
  }
  \hspace{-1em}
  \subfigure[astro, cut] {
    \includegraphics[width=0.24\textwidth]{sections/figures/astro_taumem.pdf}
  }
  \hspace{-1em}
  \subfigure[corel, cover] {
    \includegraphics[width=0.24\textwidth]{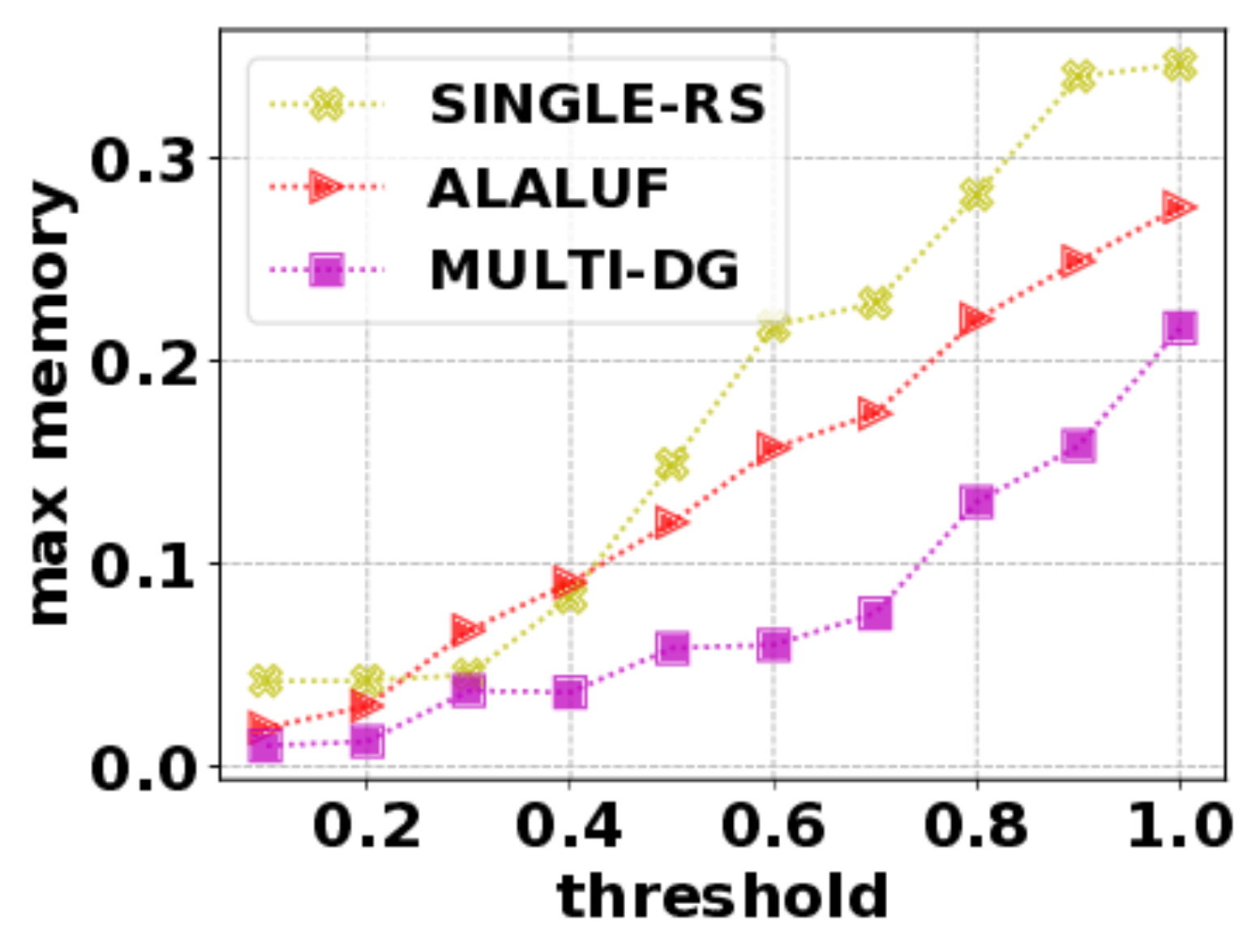}
  }
  \hspace{-1em}
  \subfigure[delicious, cover] {
   \includegraphics[width=0.24\textwidth]{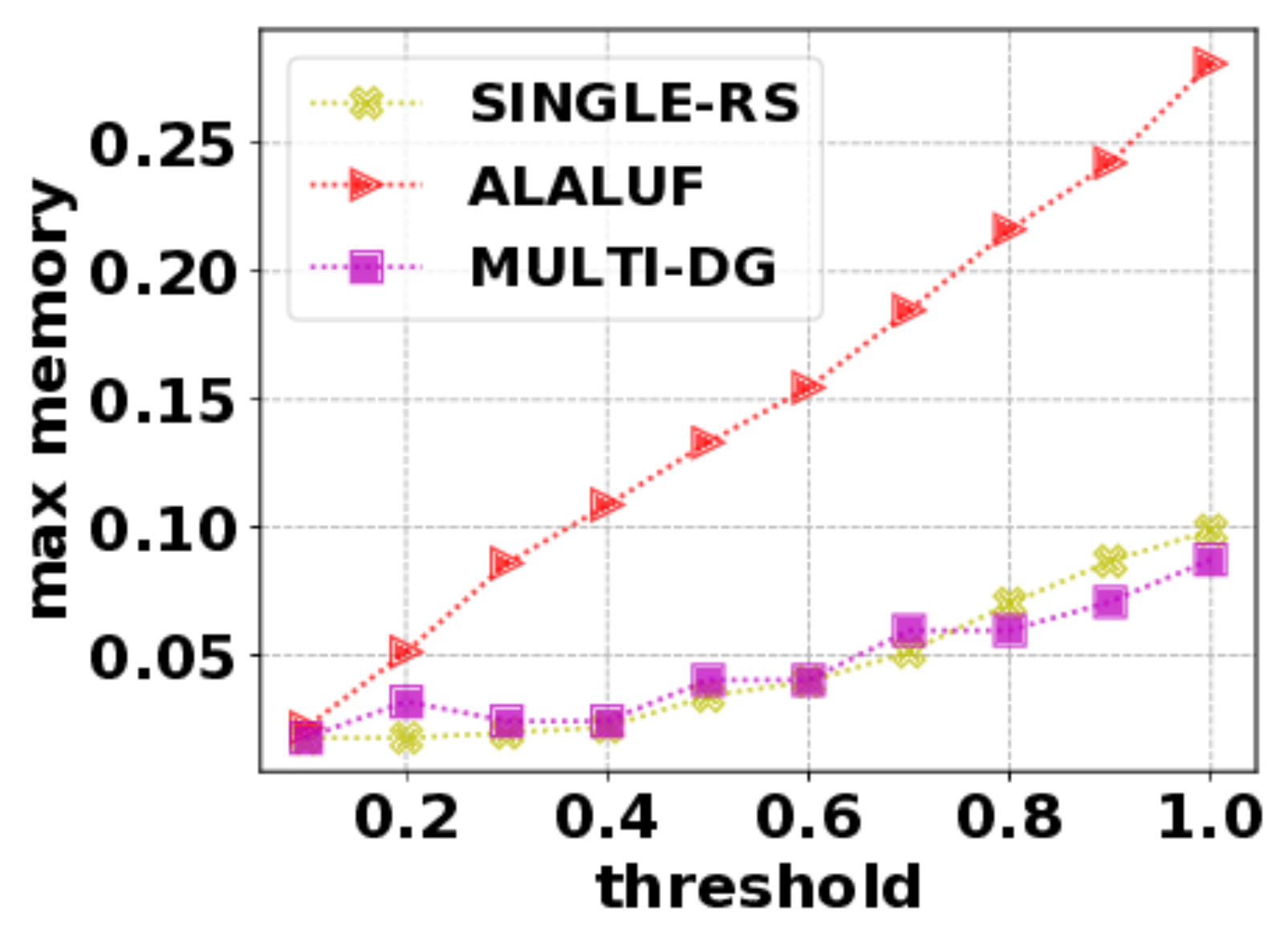}
  }
  \hspace{-1em}
  \subfigure[enron, cut] {
    \includegraphics[width=0.24\textwidth]{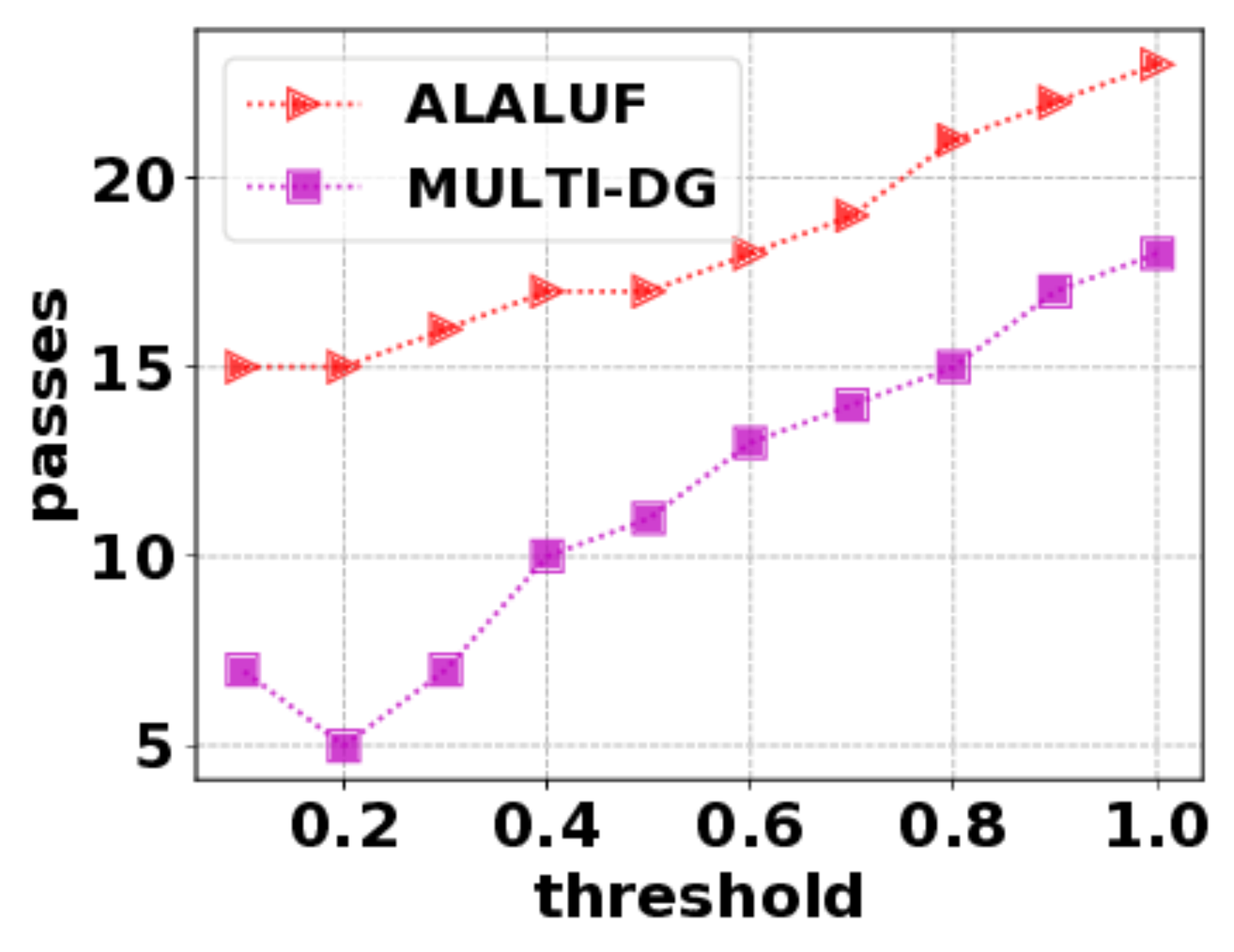}
  }
  \hspace{-1em}
  \subfigure[astro, cut] {
    \includegraphics[width=0.24\textwidth]{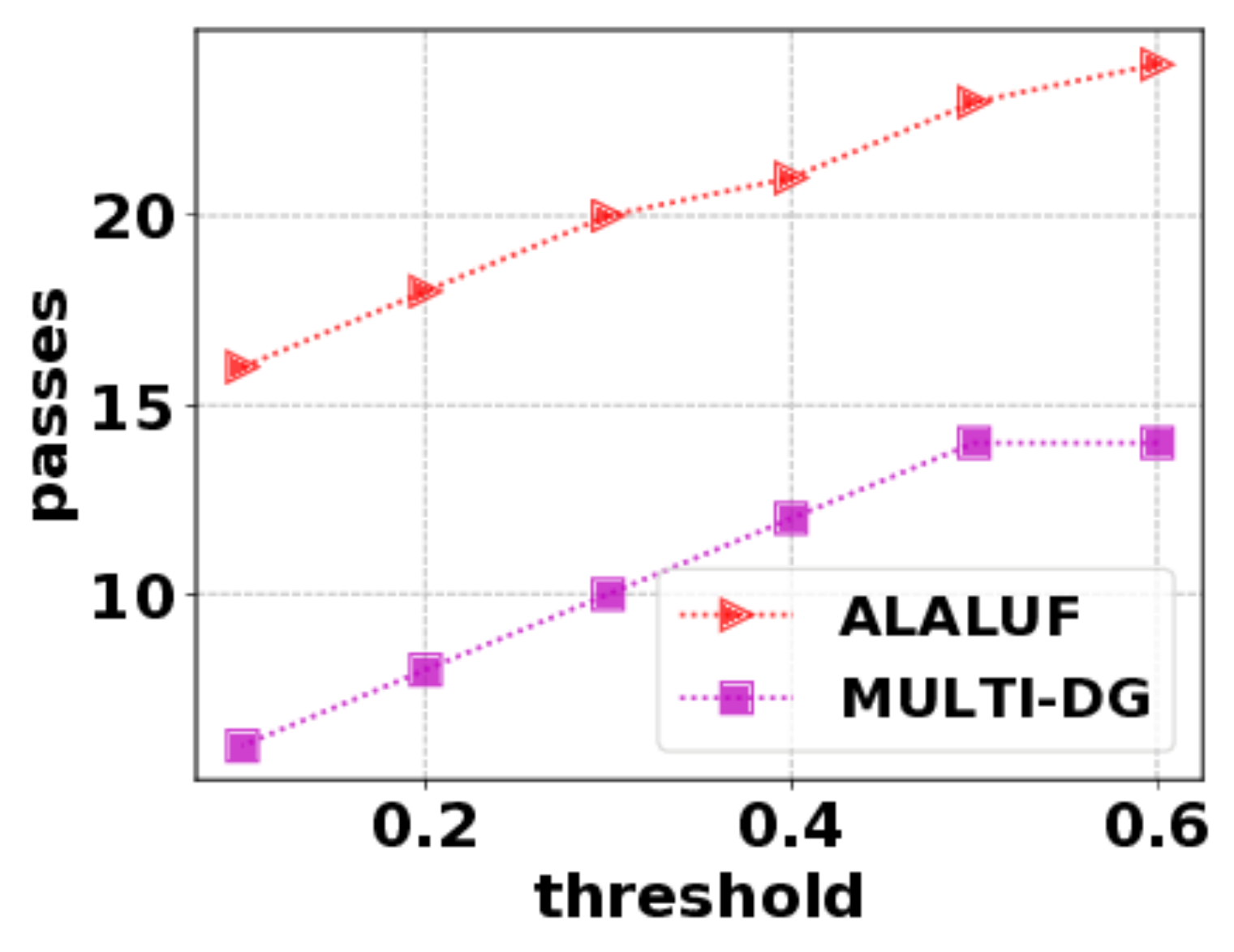}
  }
  \hspace{-1em}
  \subfigure[corel, cover] {
    \includegraphics[width=0.24\textwidth]{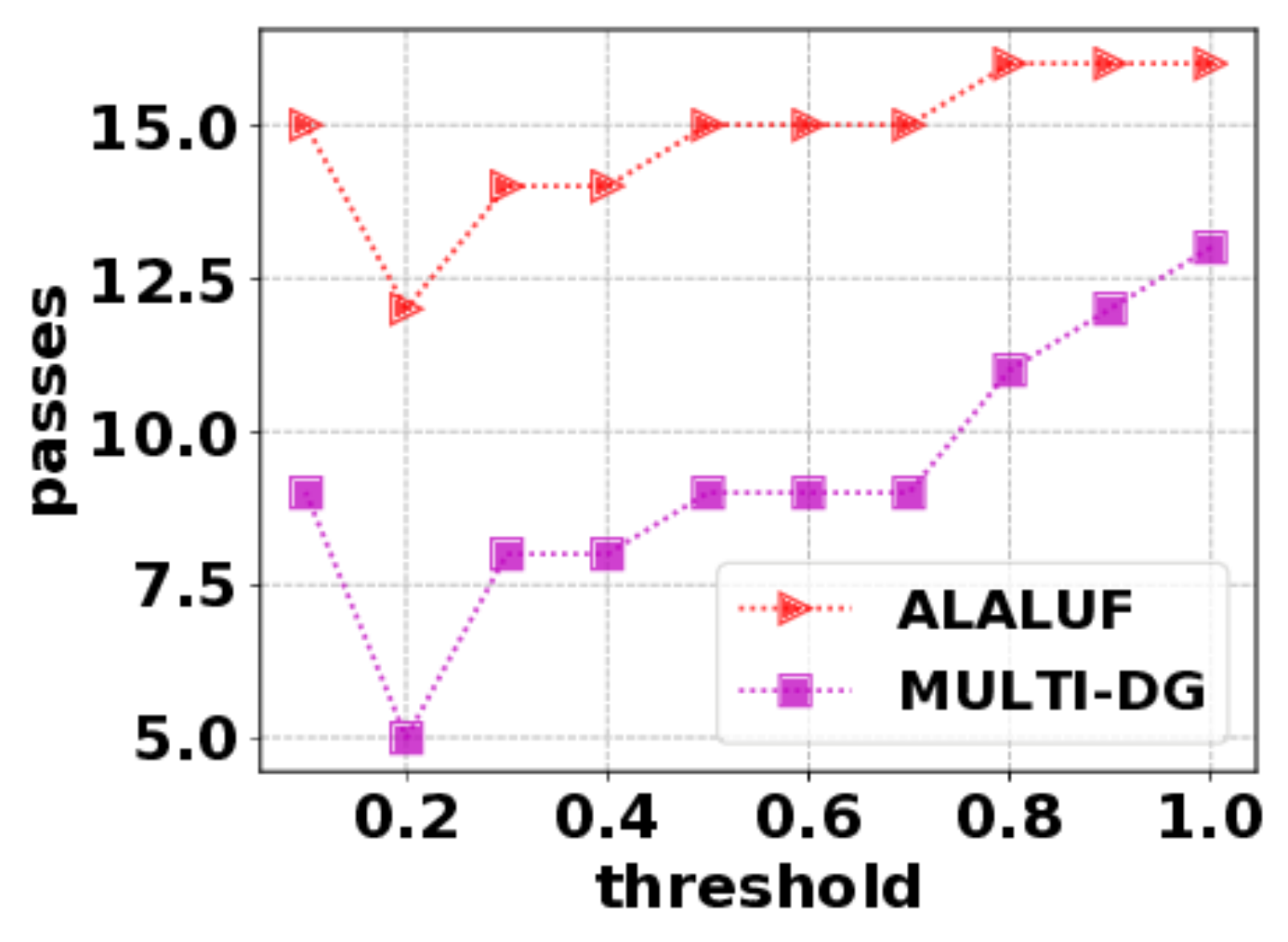}
  }
  \hspace{-1em}
  \subfigure[delicious, cover] {
    \includegraphics[width=0.24\textwidth]{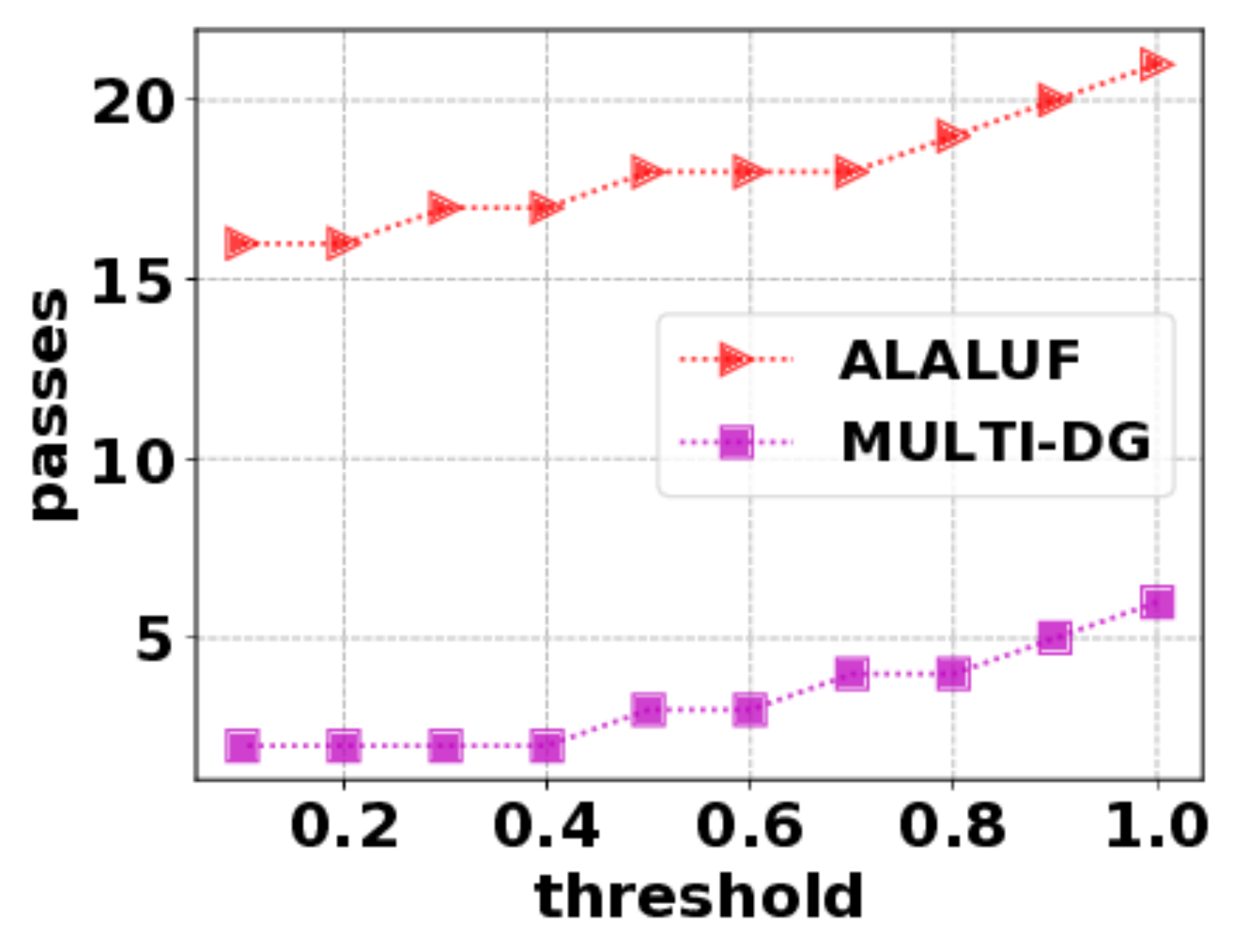}
  }
  \caption[]{
  A comparison of algorithms for \scp using different thresholds $\tau$
  (normalized as described in Section \ref{section:resultsapp}).
  STOCH-GREEDY is converting the stochastic greedy algorithm of \cite{buchbinder2017comparing}
  to an algorithm for \scp by using the method of \cite{iyer2013}.
  ALALUF is the same, but using the algorithm of \cite{alaluf2019optimal}.
  }
  \label{fig:cover}
\end{figure*}

\begin{figure*}[t!]
  \centering
  \hspace{-1em}
  \subfigure[enron, cut] {
    \includegraphics[width=0.24\textwidth]{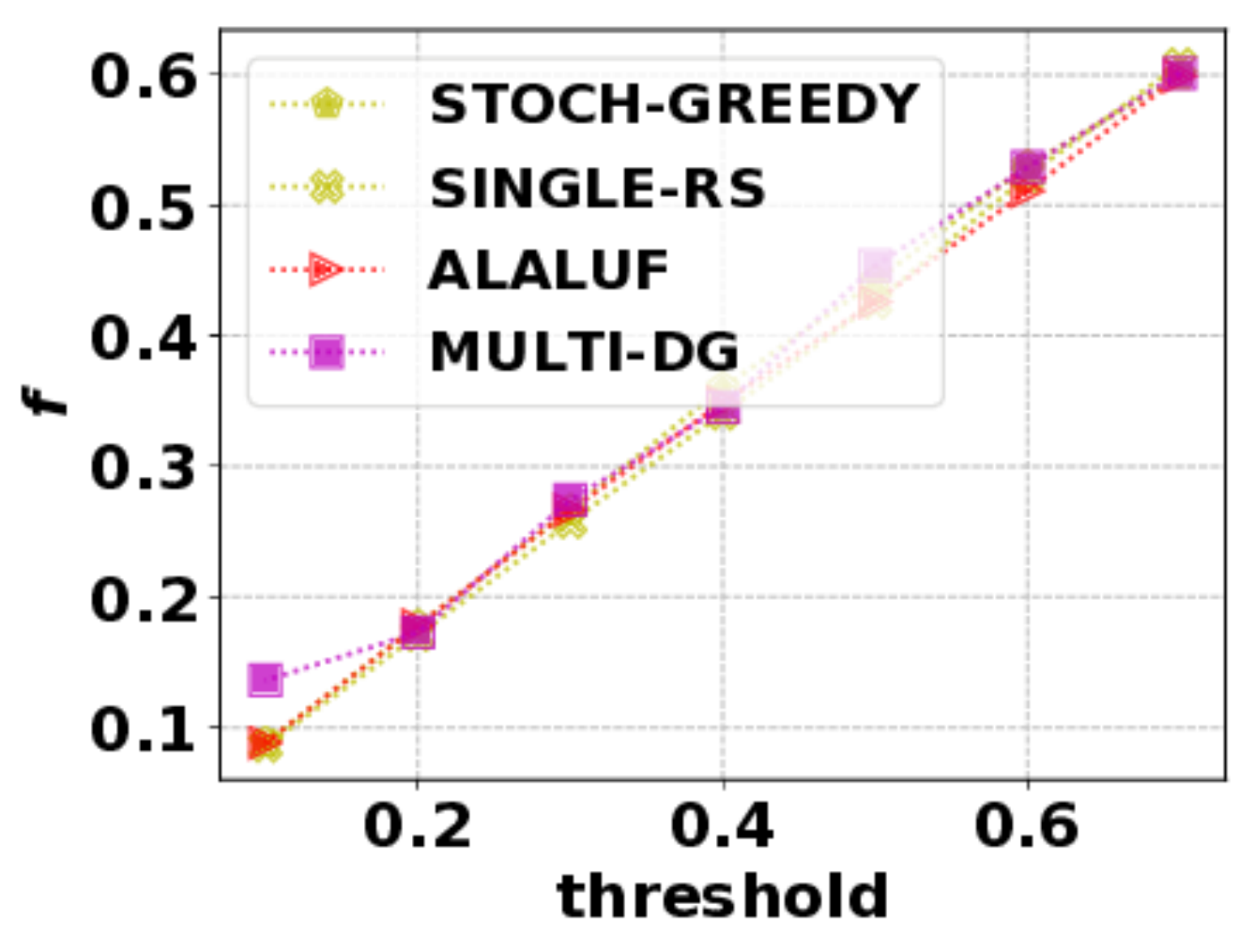}
  }
  \hspace{-1em}
  \subfigure[astro, cut] {
    \includegraphics[width=0.24\textwidth]{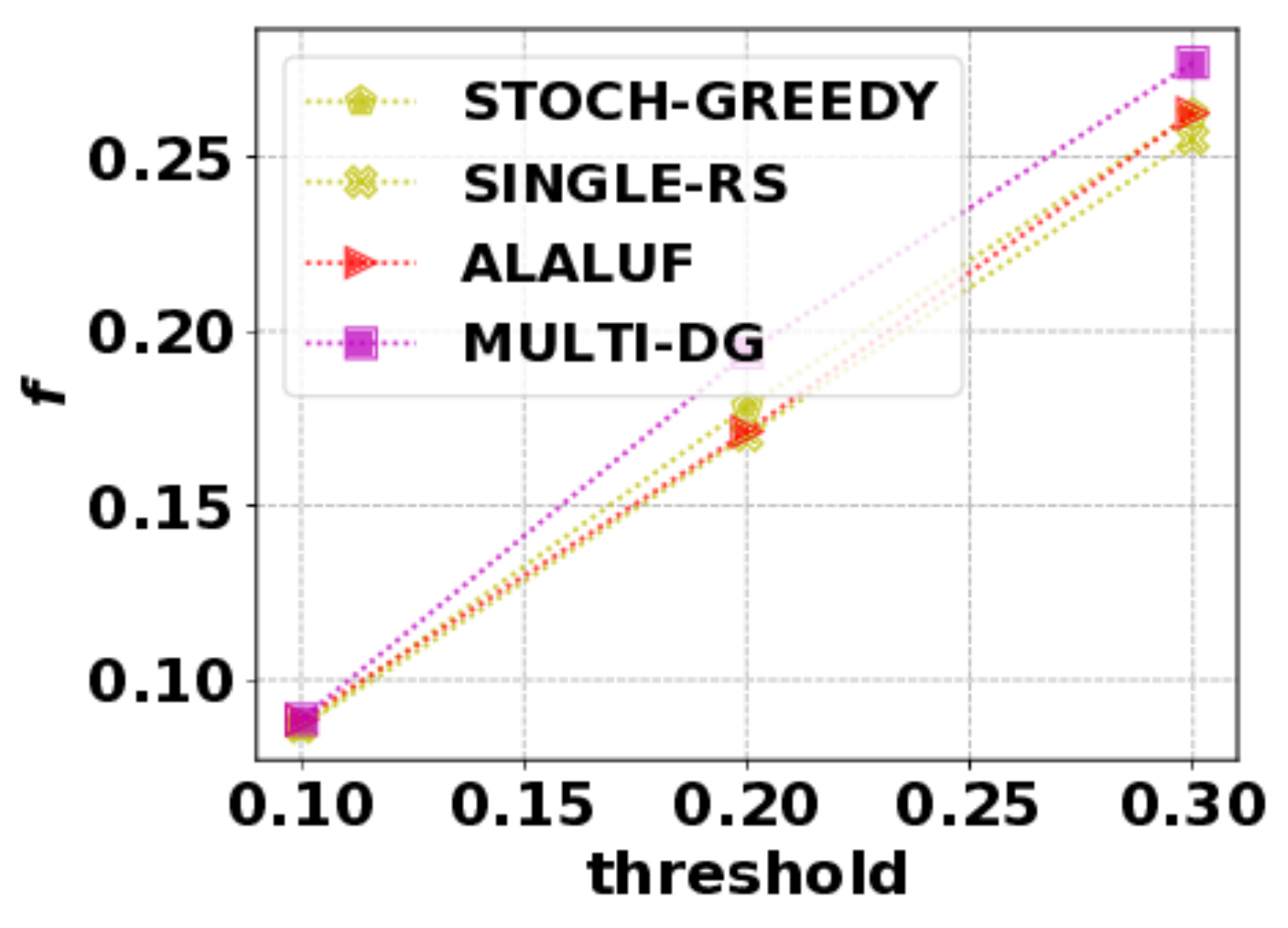}
  }
  \hspace{-1em}
  \subfigure[corel, cover] {
    \includegraphics[width=0.24\textwidth]{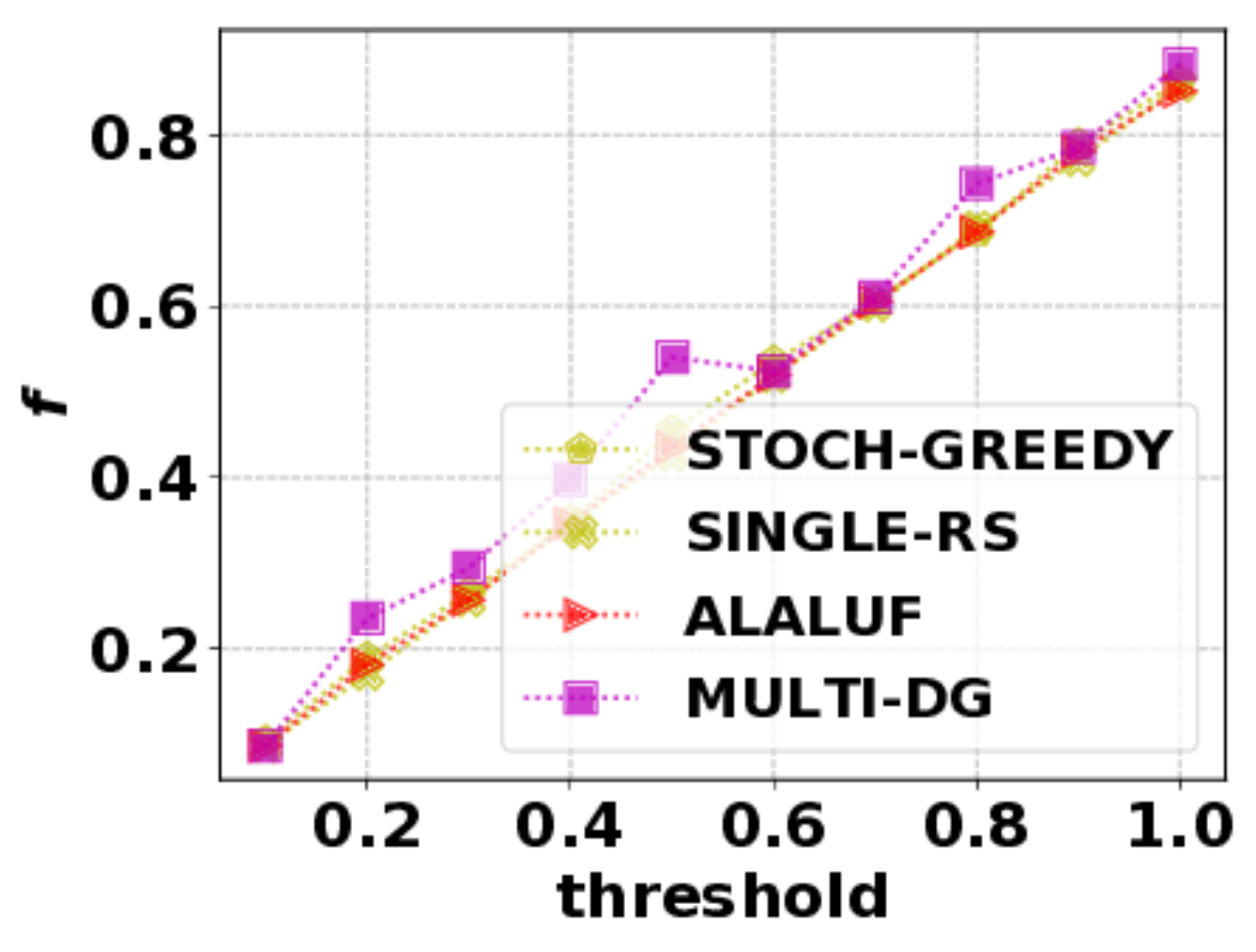}
  }
  \hspace{-1em}
  \subfigure[delicious, cover] {
    \includegraphics[width=0.24\textwidth]{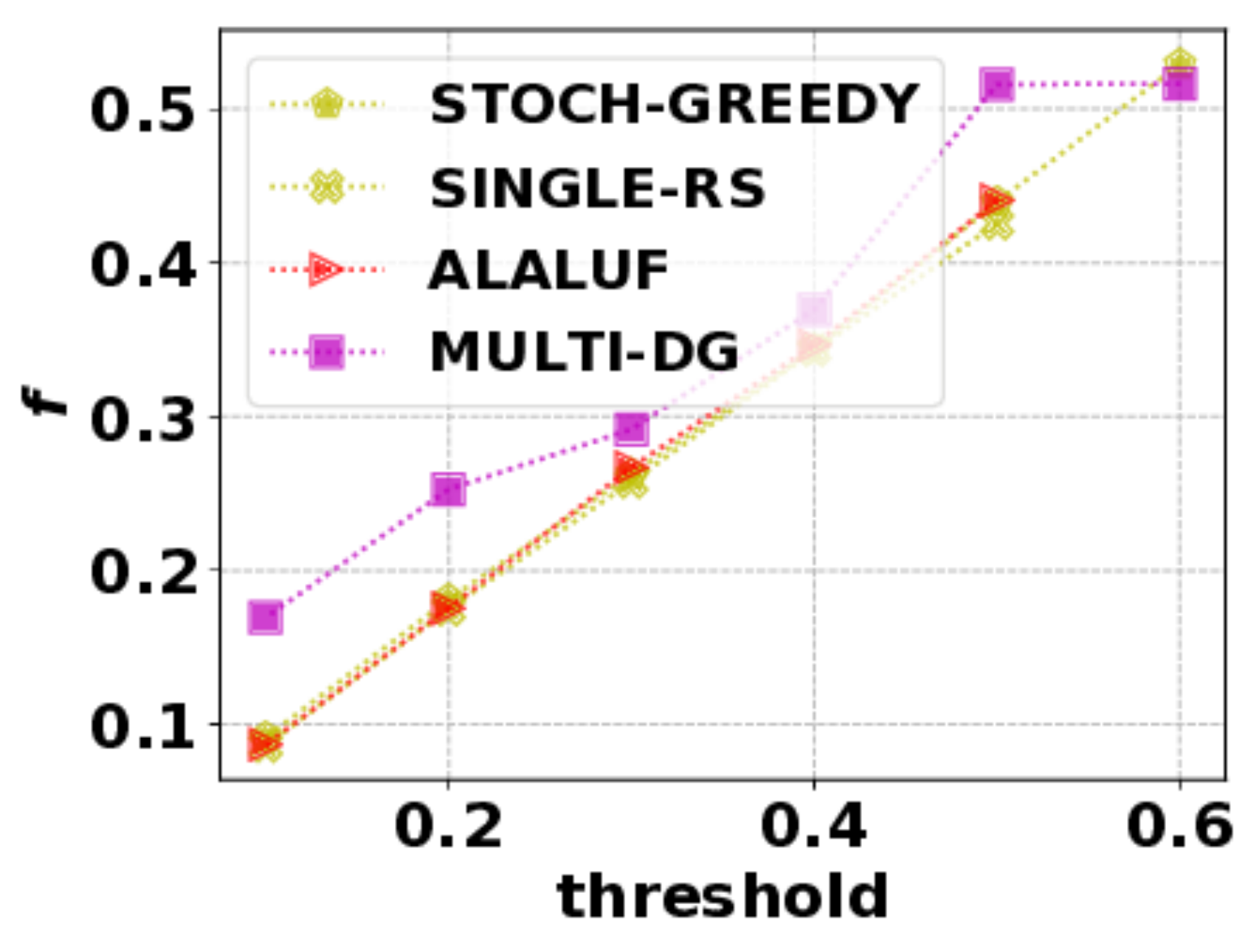}
  }
  \hspace{-1em}
  \subfigure[enron, cut] {
    \includegraphics[width=0.24\textwidth]{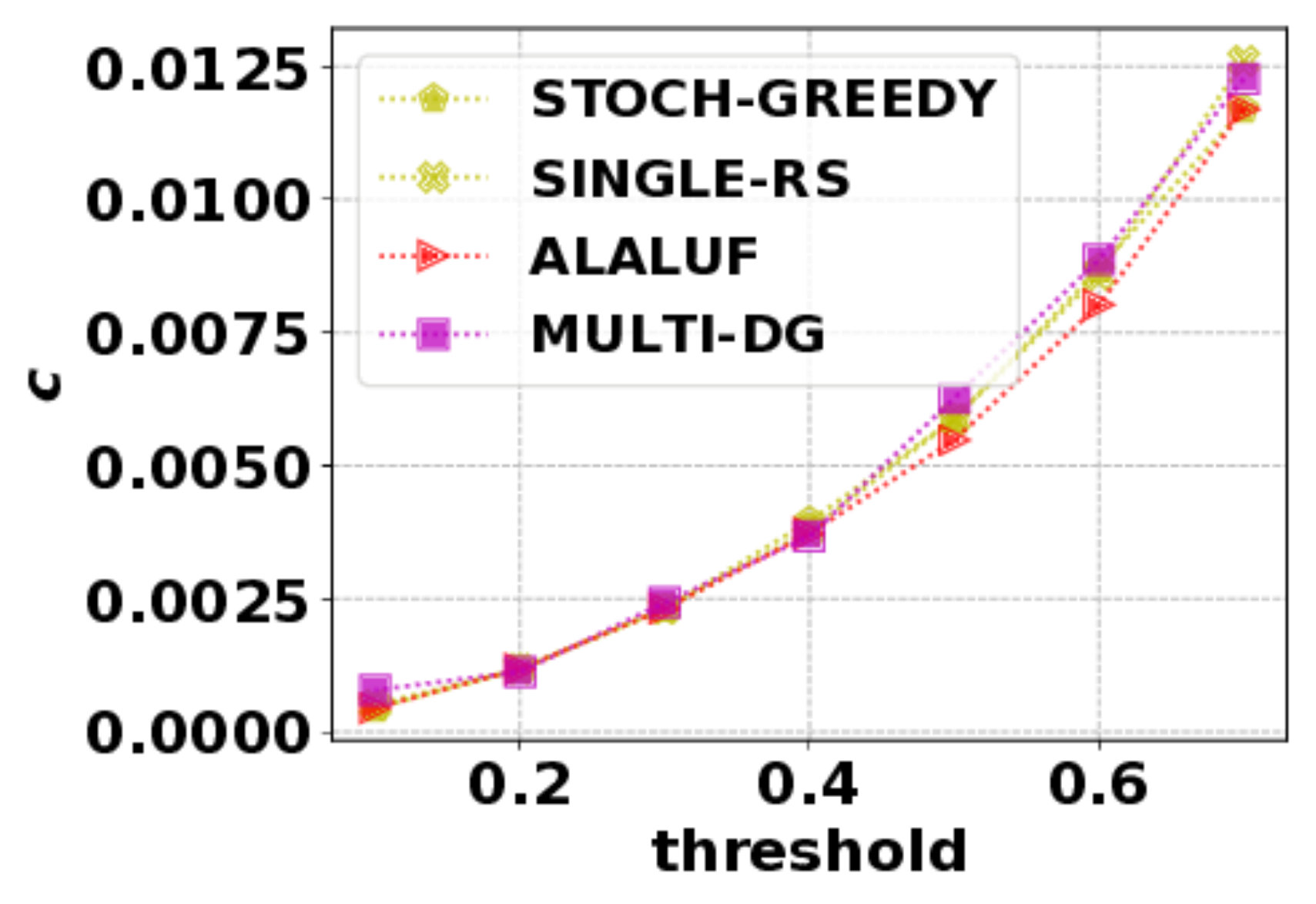}
  }
  \hspace{-1em}
  \subfigure[astro, cut] {
    \includegraphics[width=0.24\textwidth]{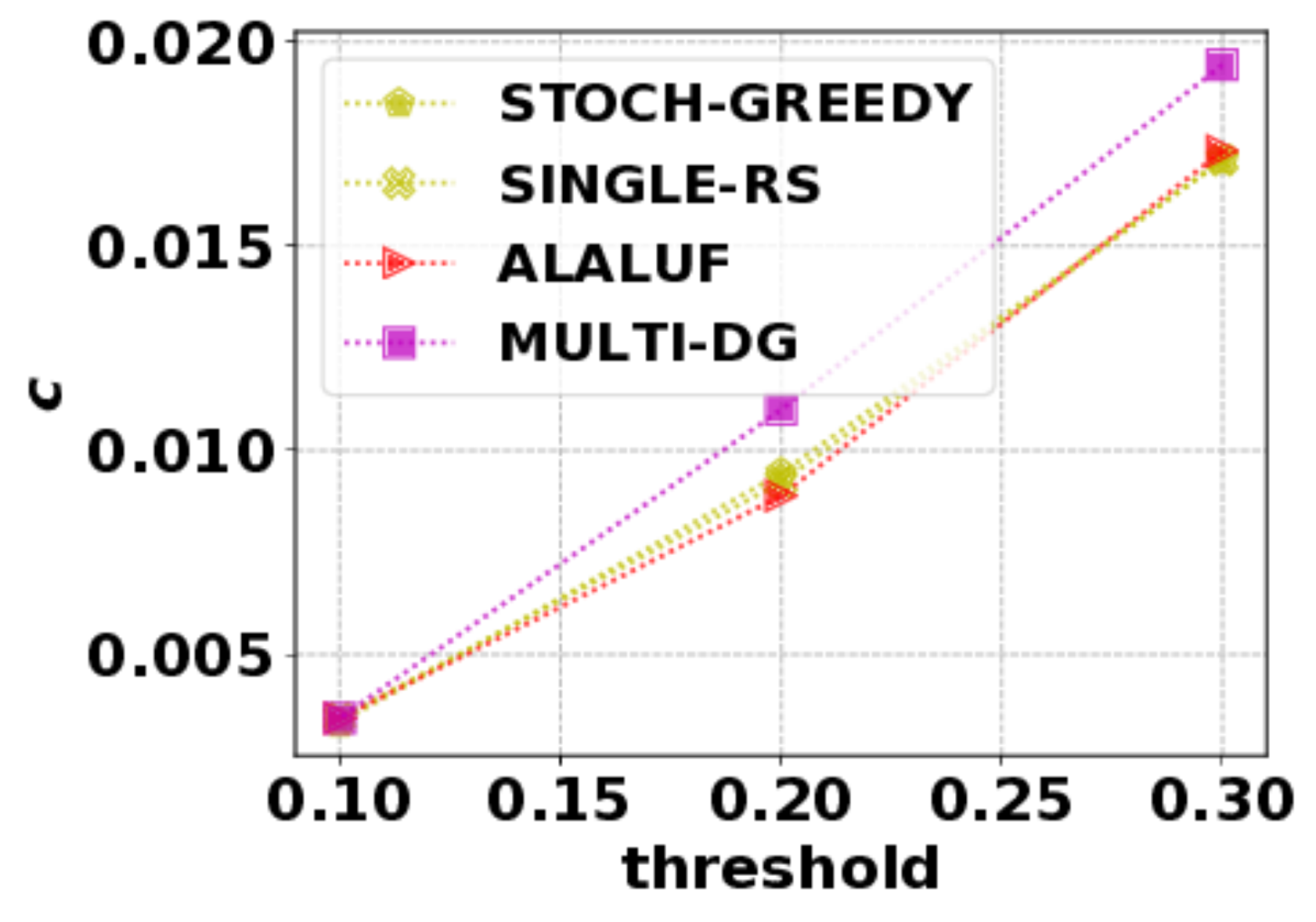}
  }
  \hspace{-1em}
  \subfigure[corel, cover] {
    \includegraphics[width=0.24\textwidth]{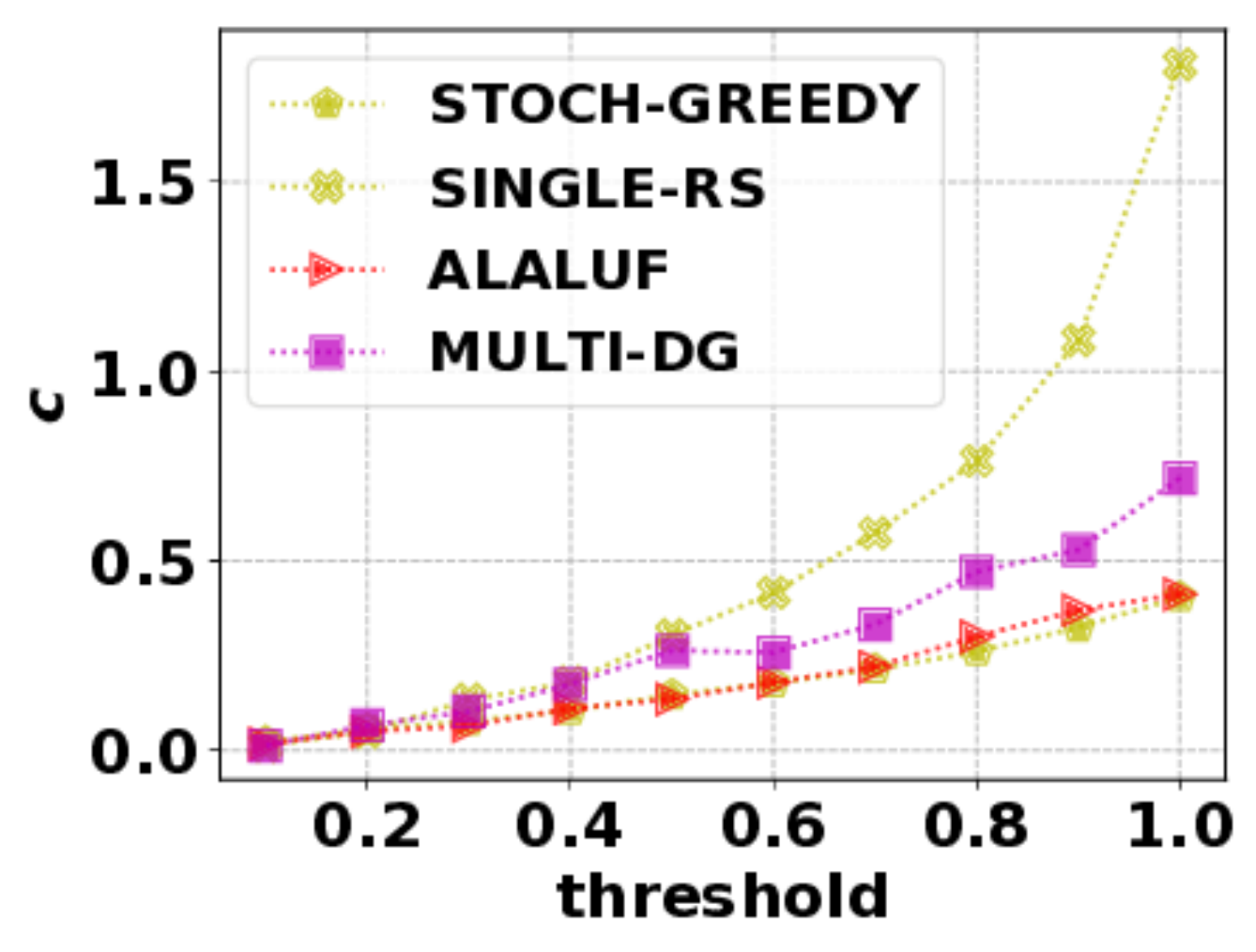}
  }
  \hspace{-1em}
  \subfigure[delicious, cover] {
    \includegraphics[width=0.24\textwidth]{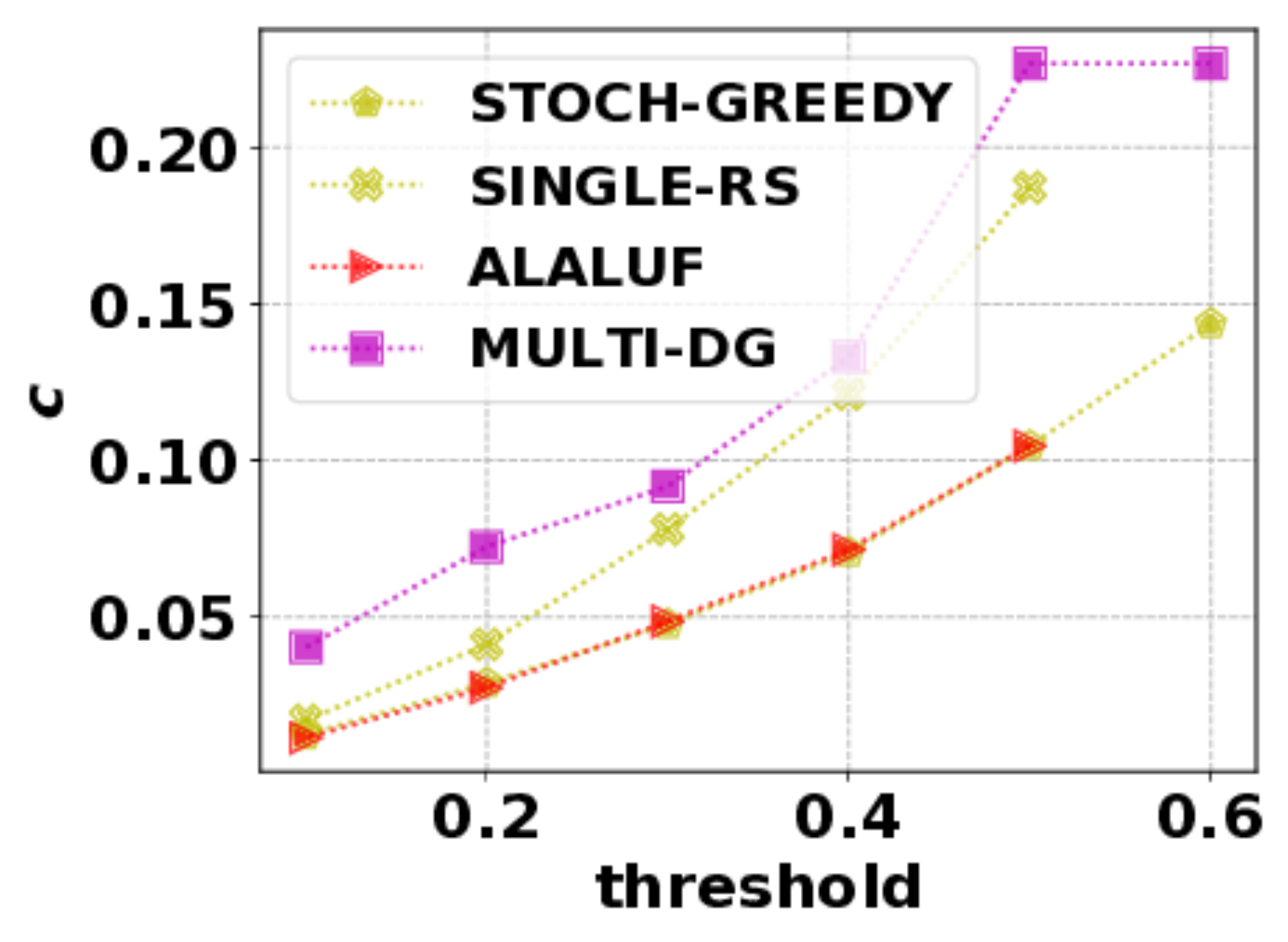}
  }
  \hspace{-1em}
  \subfigure[enron, cut] {
    \includegraphics[width=0.24\textwidth]{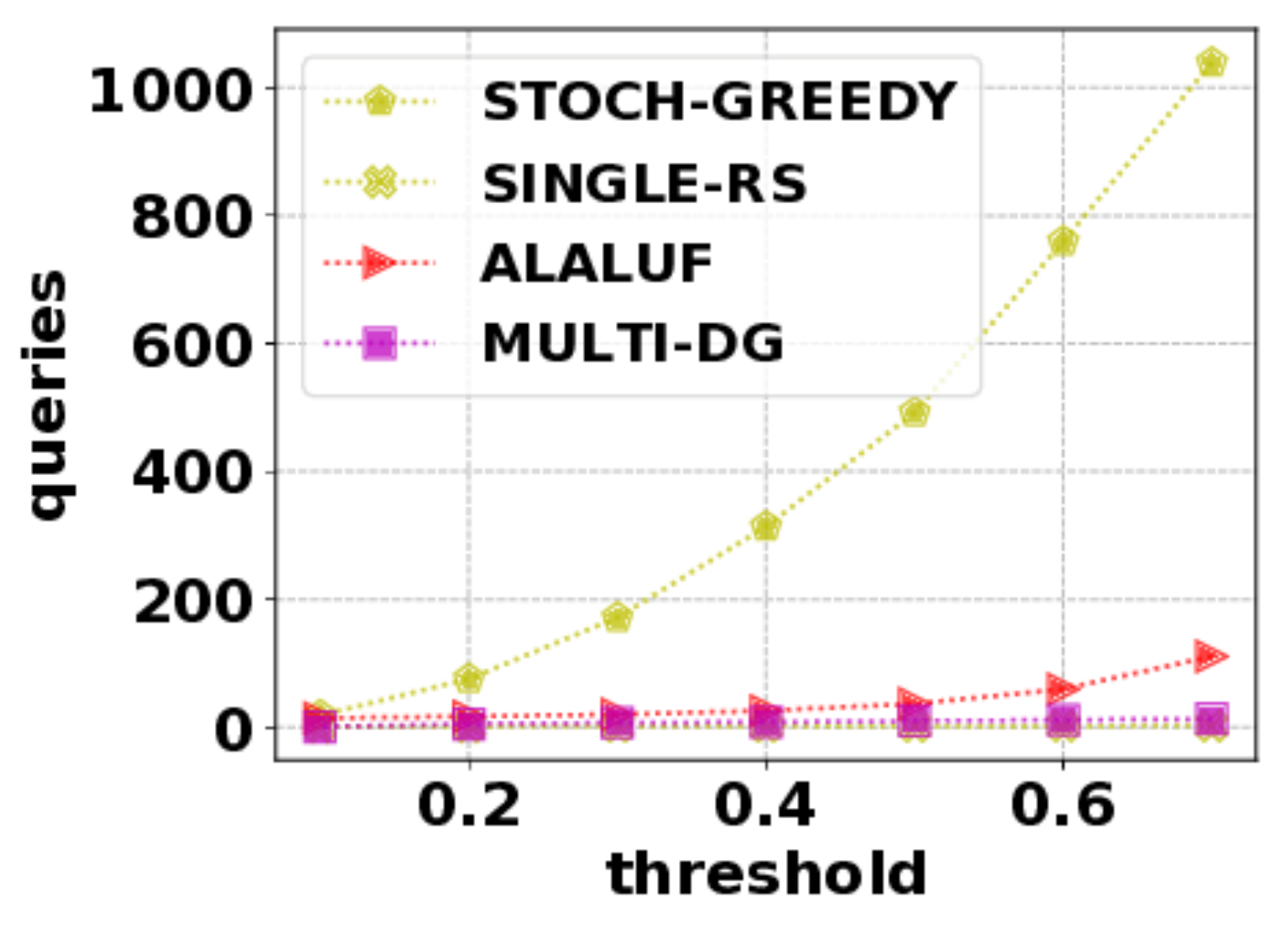}
  }
  \hspace{-1em}
  \subfigure[astro, cut] {
    \includegraphics[width=0.24\textwidth]{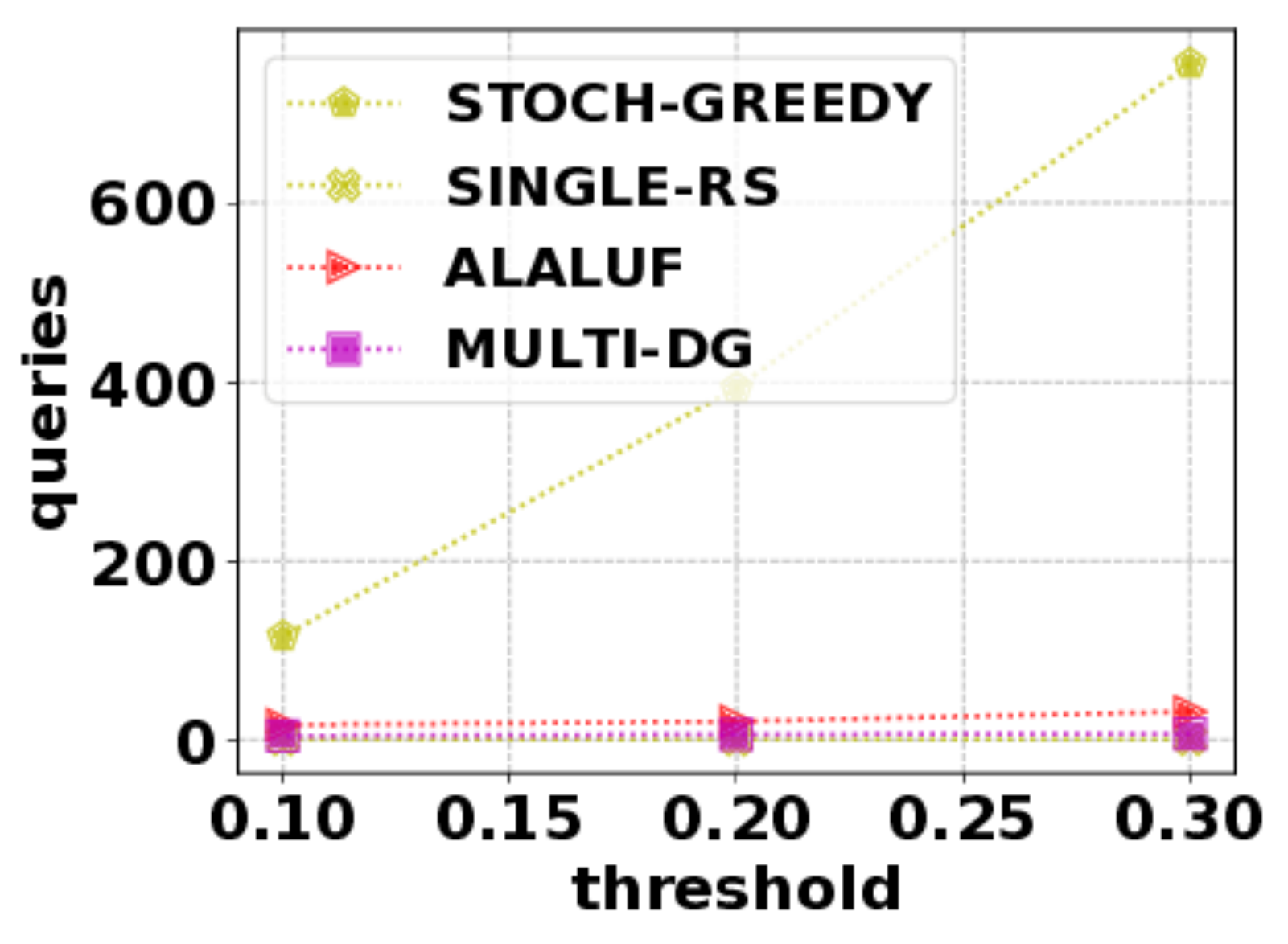}
  }
  \hspace{-1em}
  \subfigure[corel, cover] {
    \includegraphics[width=0.24\textwidth]{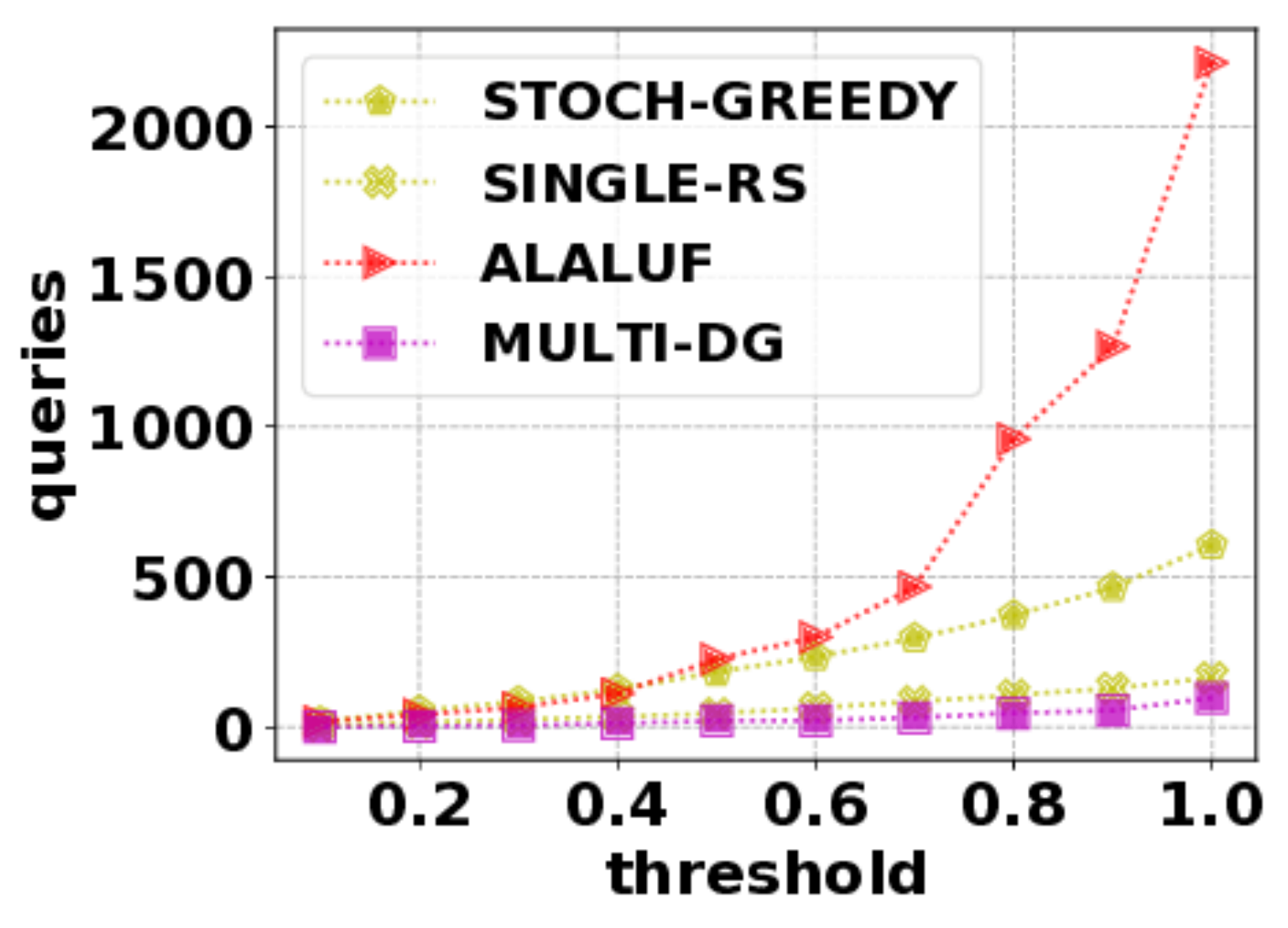}
  }
  \hspace{-1em}
  \subfigure[delicious, cover] {
    \includegraphics[width=0.24\textwidth]{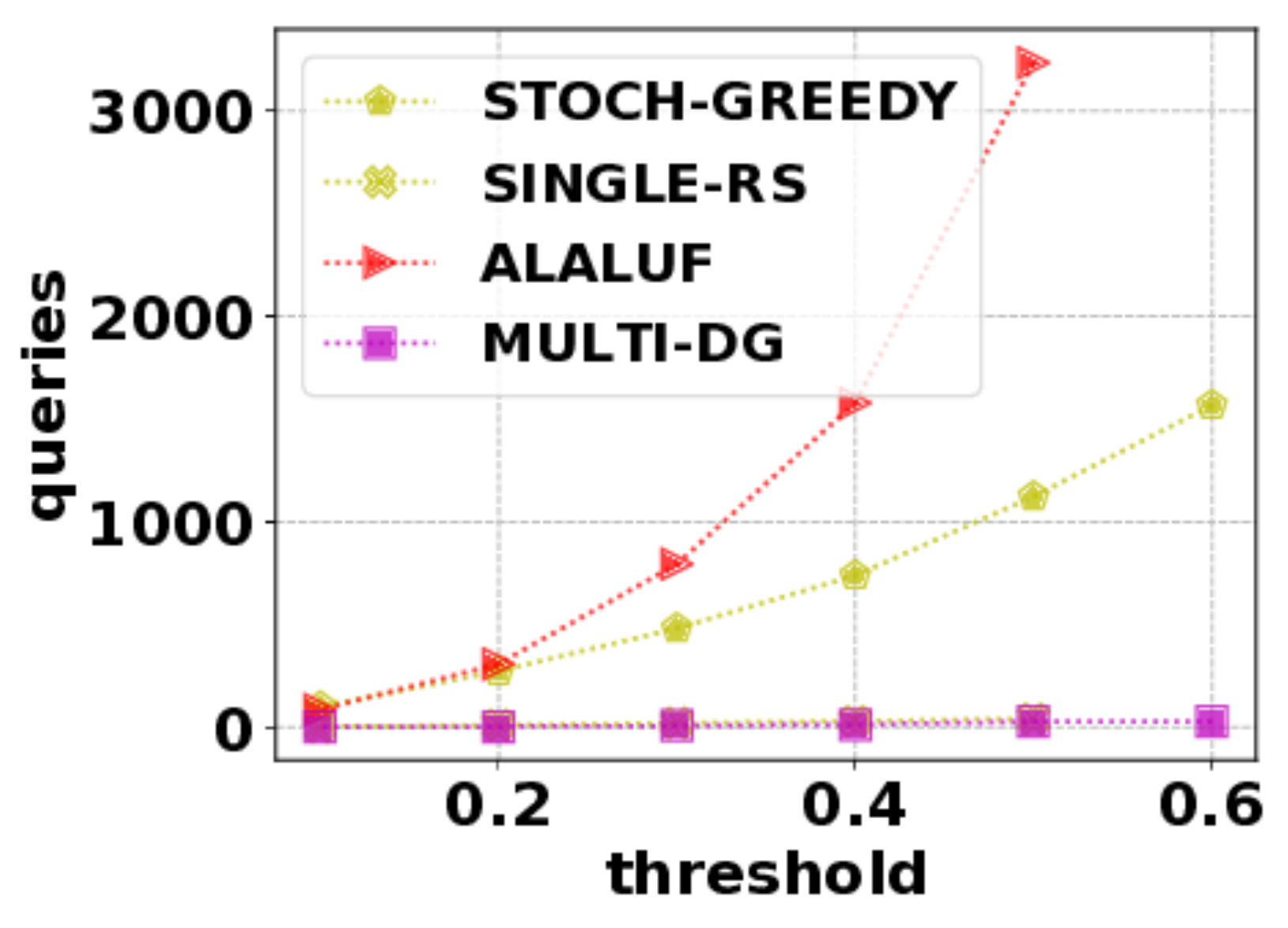}
  }
  \hspace{-1em}
  \subfigure[enron, cut] {
    \includegraphics[width=0.24\textwidth]{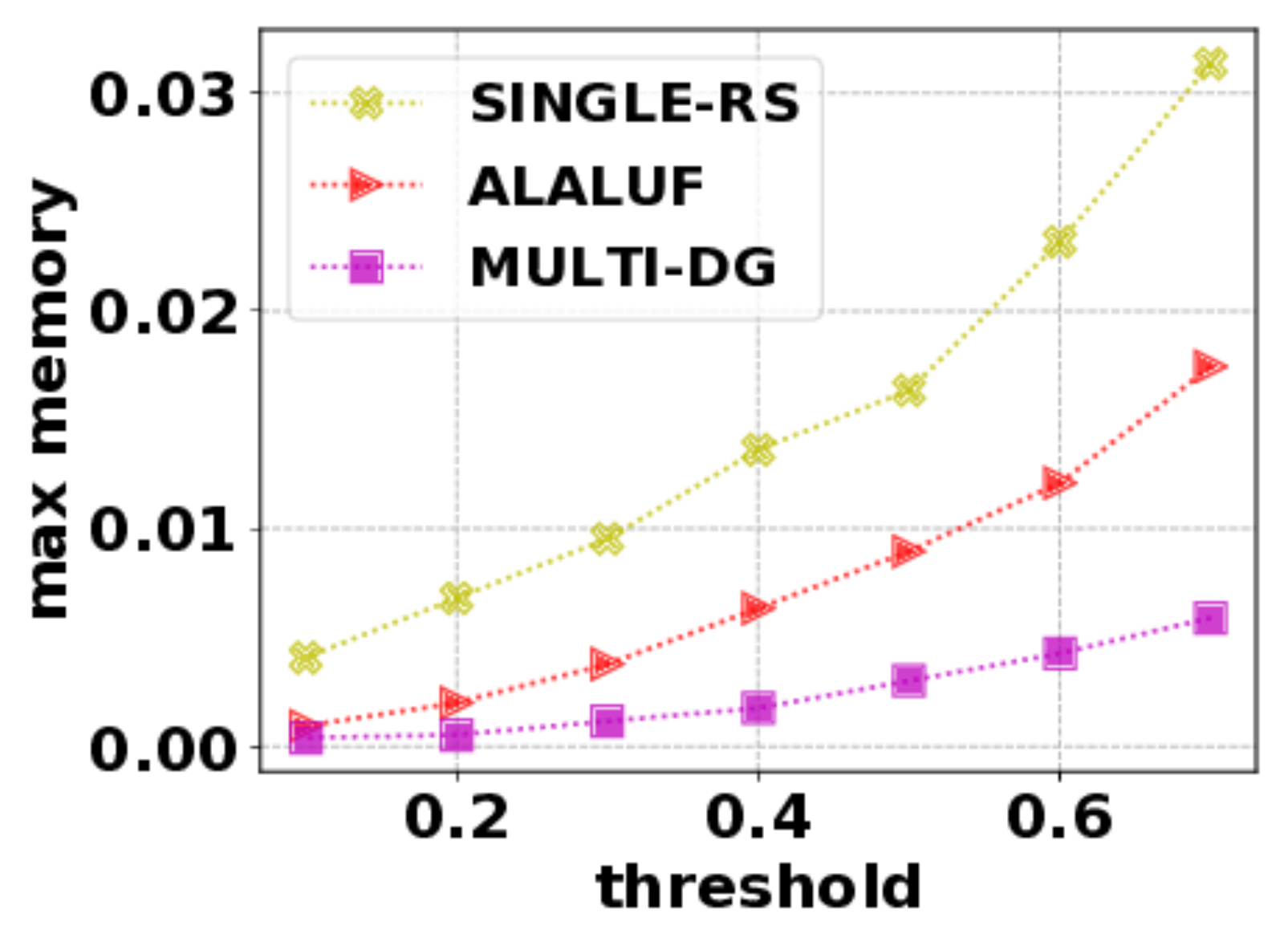}
  }
  \hspace{-1em}
  \subfigure[astro, cut] {
    \includegraphics[width=0.24\textwidth]{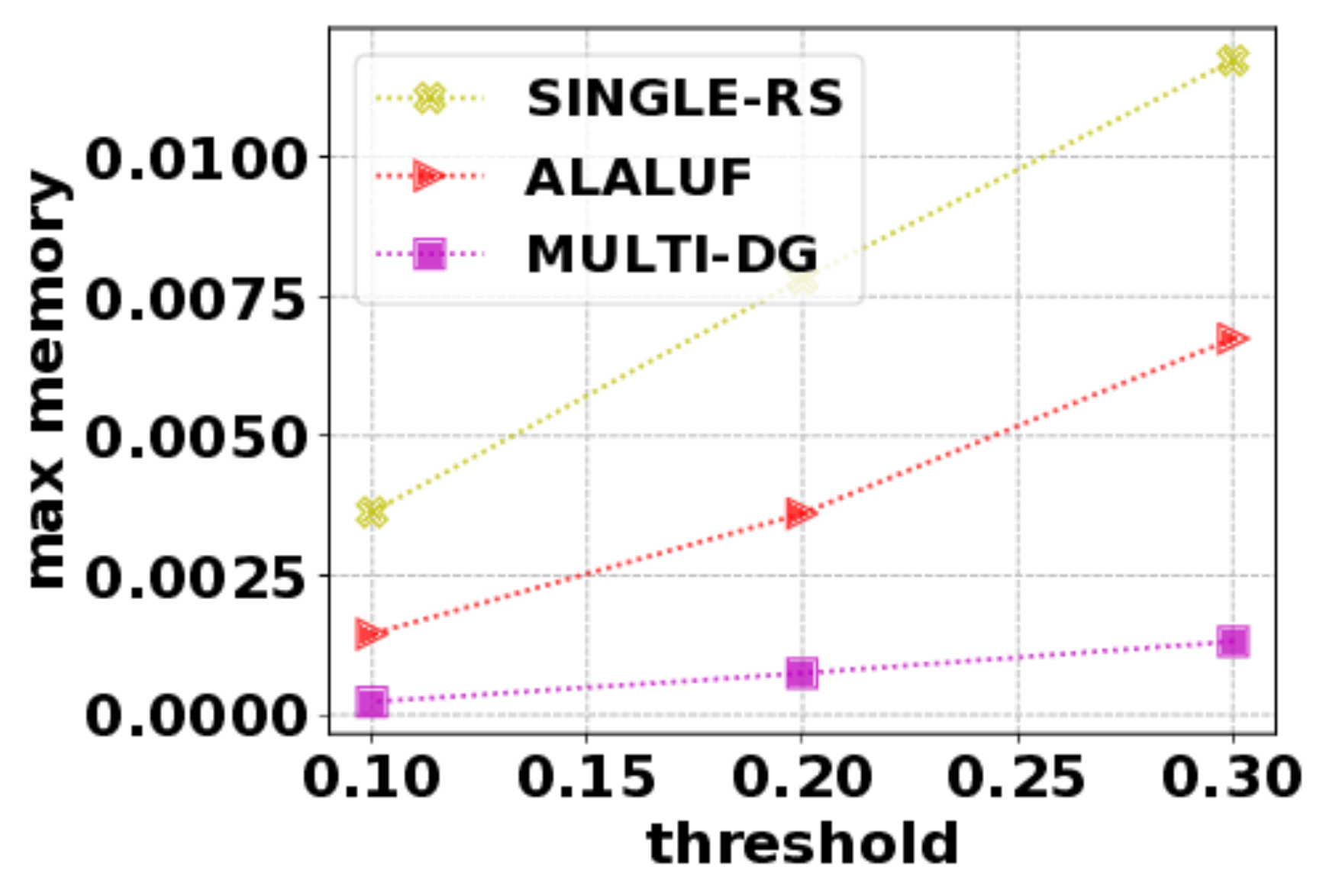}
  }
  \hspace{-1em}
  \subfigure[corel, cover] {
    \includegraphics[width=0.24\textwidth]{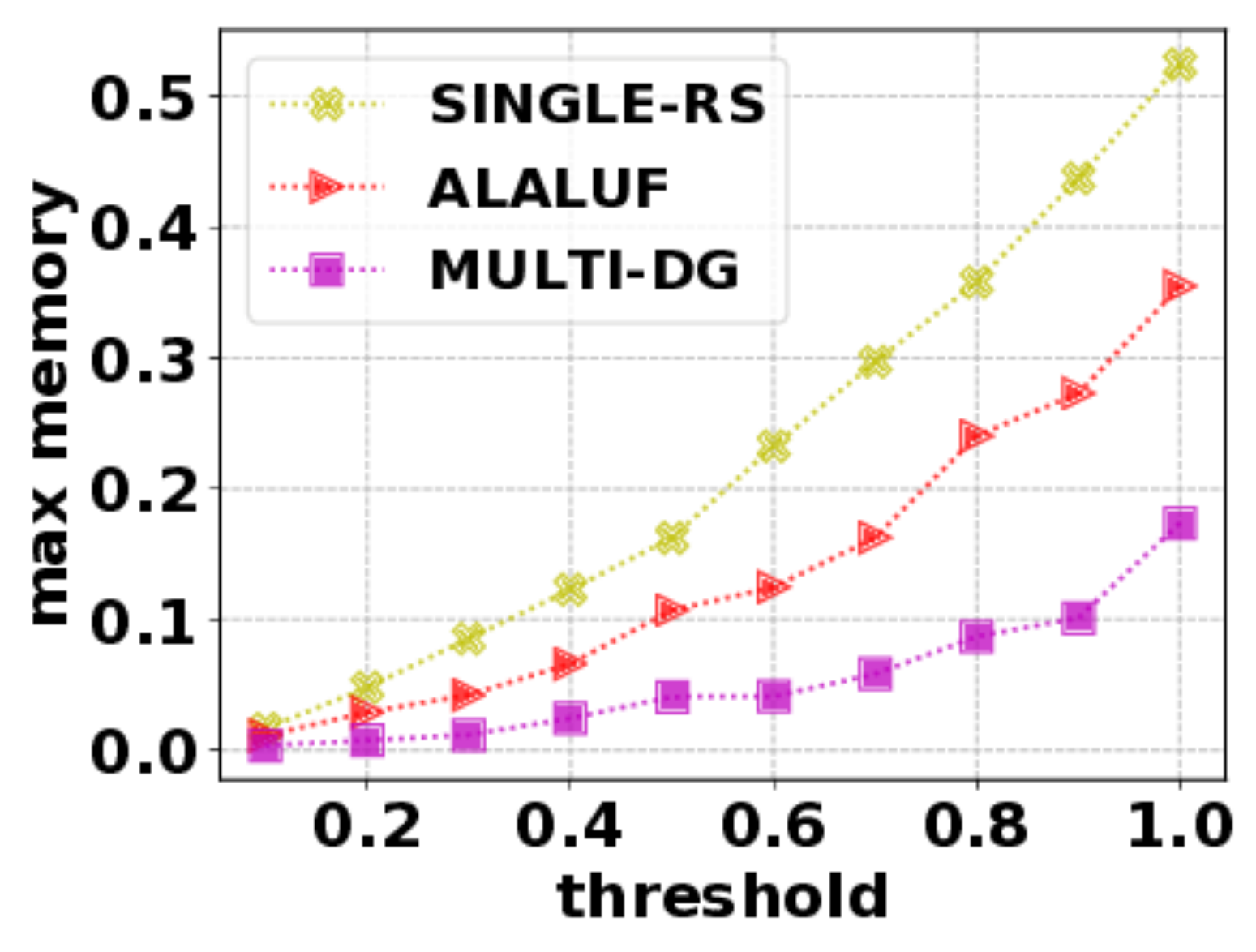}
  }
  \hspace{-1em}
  \subfigure[delicious, cover] {
   \includegraphics[width=0.24\textwidth]{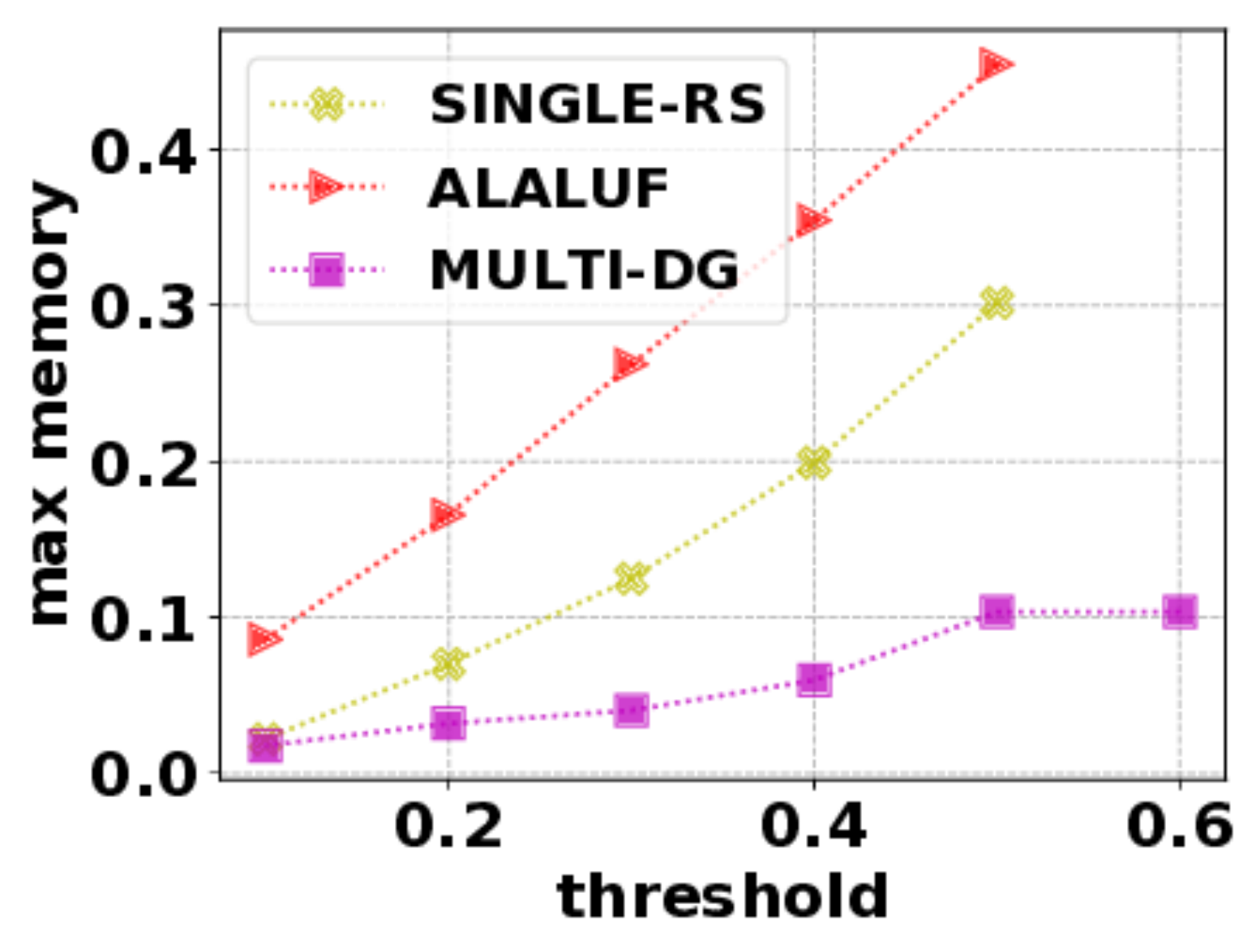}
  }
  \hspace{-1em}
  \subfigure[enron, cut] {
    \includegraphics[width=0.24\textwidth]{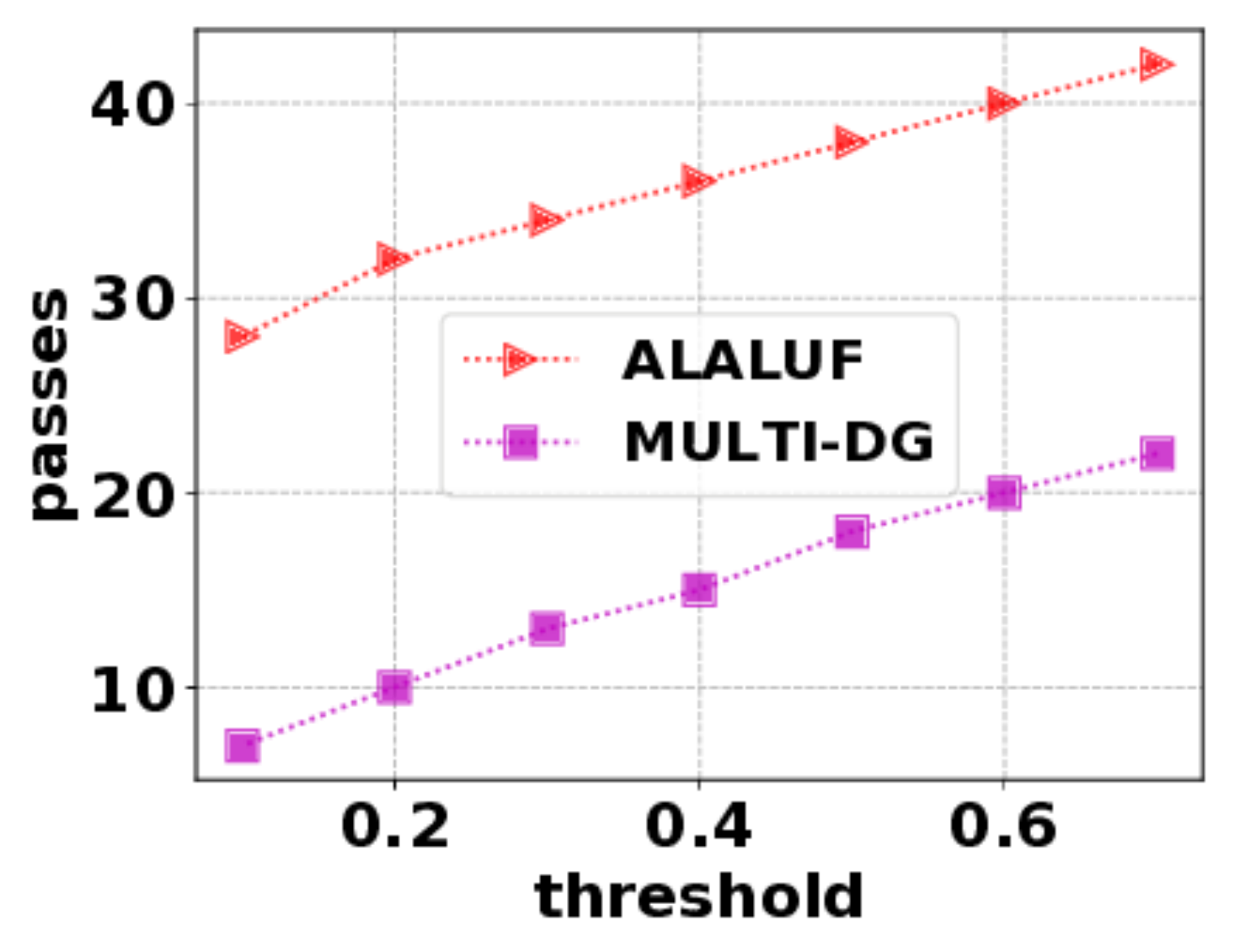}
  }
  \hspace{-1em}
  \subfigure[astro, cut] {
    \includegraphics[width=0.24\textwidth]{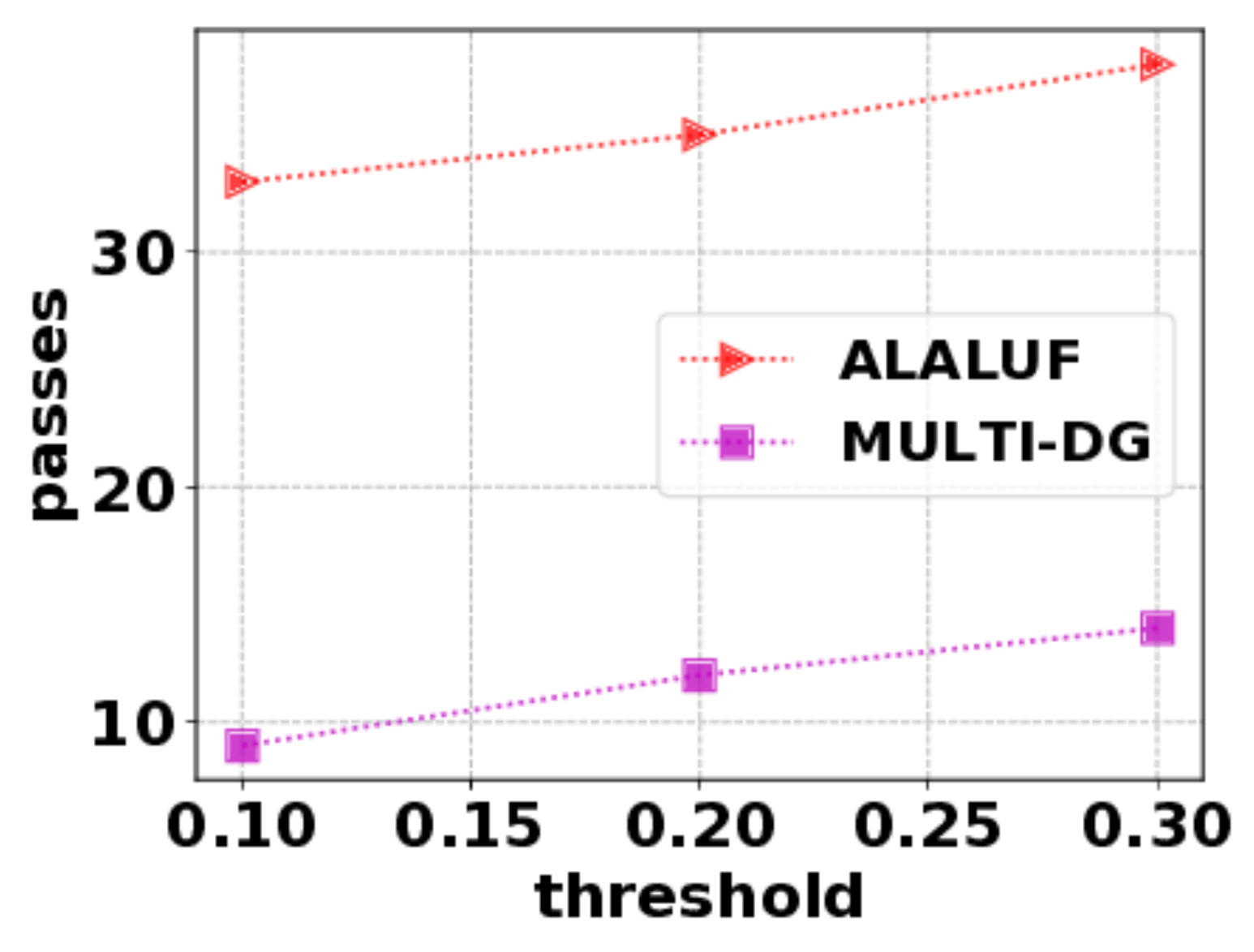}
  }
  \hspace{-1em}
  \subfigure[corel, cover] {
    \includegraphics[width=0.24\textwidth]{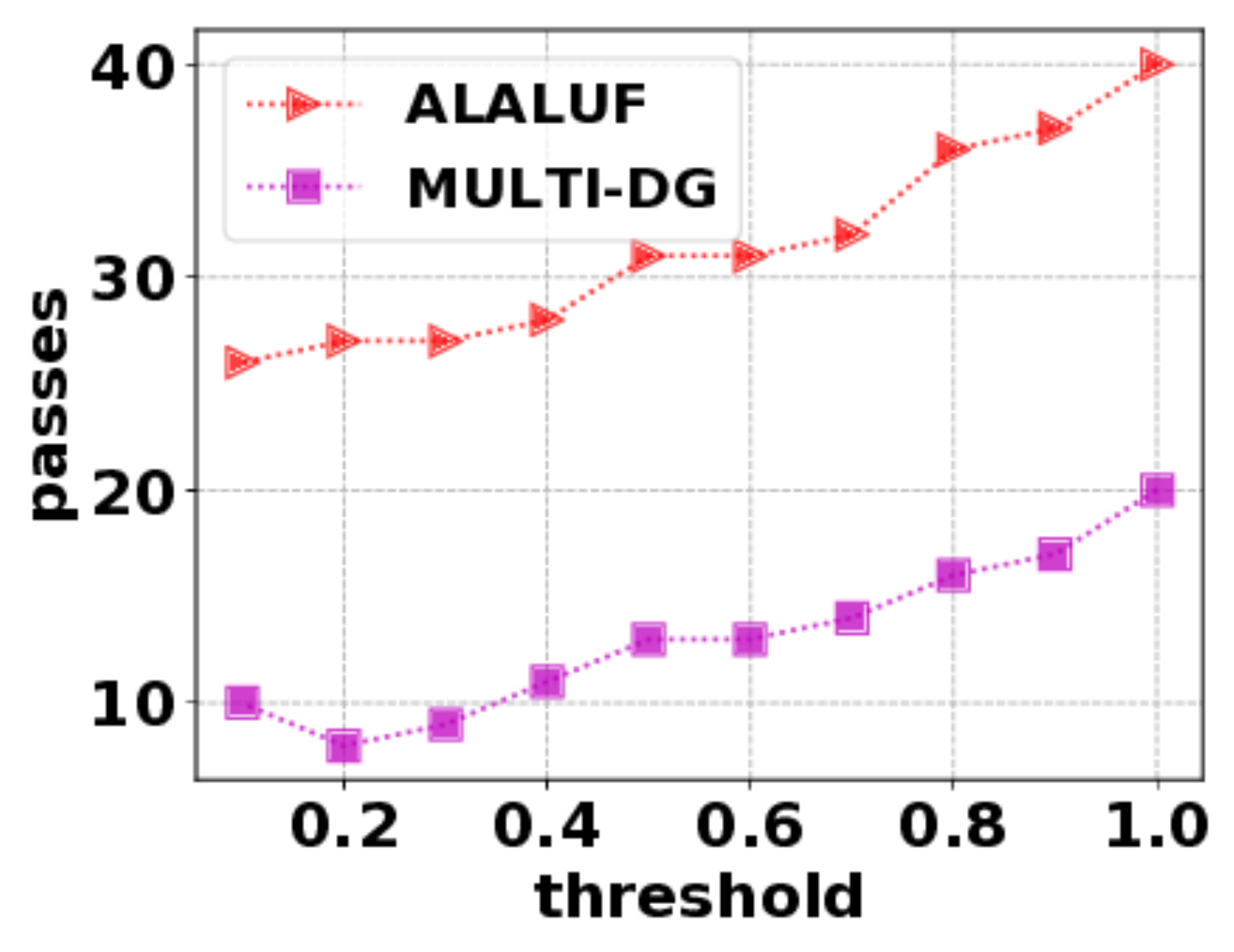}
  }
  \hspace{-1em}
  \subfigure[delicious, cover] {
    \includegraphics[width=0.24\textwidth]{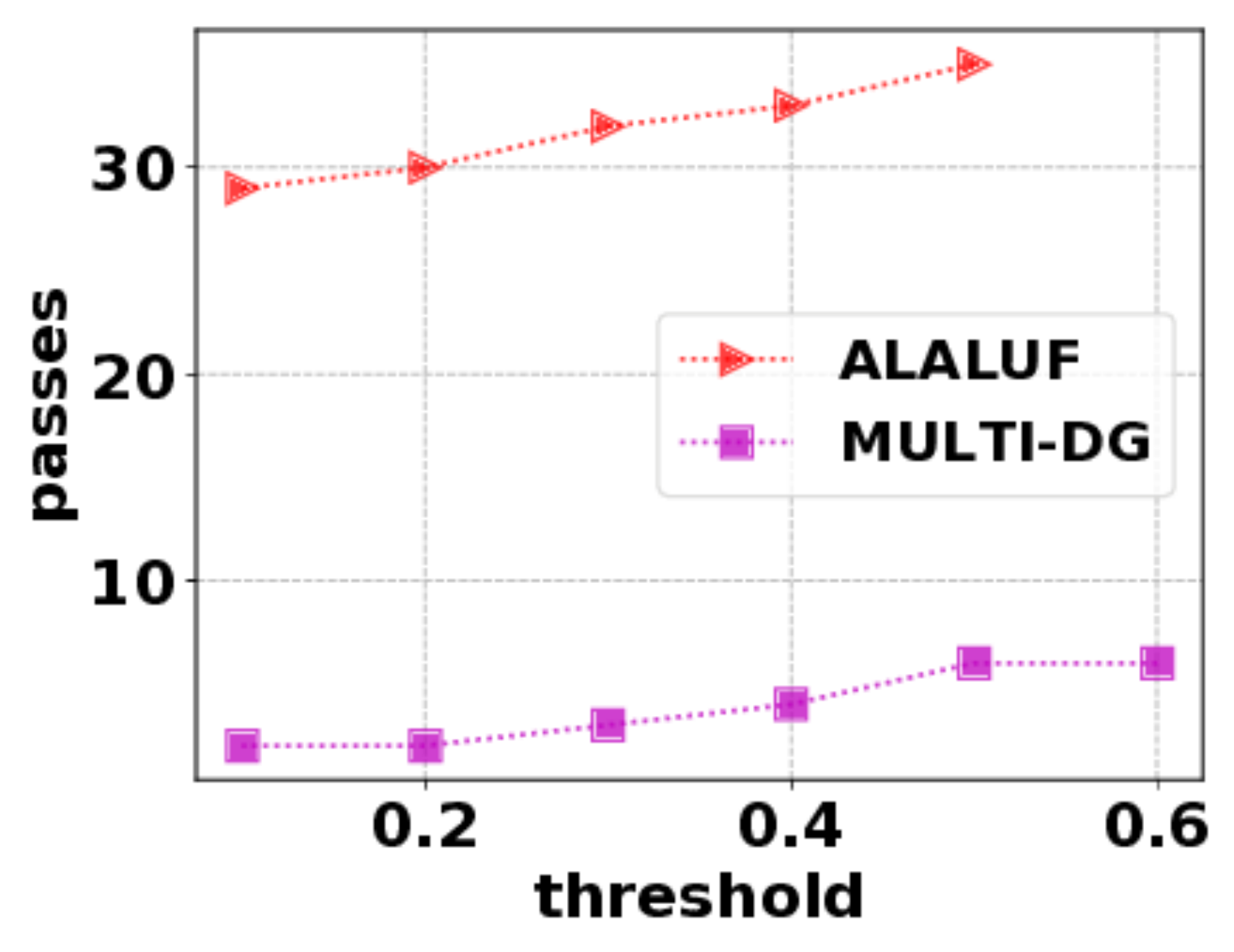}
  }
  \caption[]{
  A comparison of algorithms for \scp using different thresholds $\tau$.
  However, the algorithms are run as heuristics where all approximation guarantees are
  assumed to be 1. See Section \ref{section:resultsapp} to see further description.
  }
  \label{fig:heuristic}
\end{figure*}

  The additional experimental results are presented in Figures \ref{fig:mssubroutines} to
  \ref{fig:heuristic}.
  In every experiment, the double greedy
  USM algorithm of \citet{buchbinder2015tight} (``DG'') is
  initially run as a baseline comparison.
  Let the cost, $f$ value, and number of queries of DG
  be $c_0$, $f_0$, and $q_0$ respectively.
  For all of the plots,
  the $f$ values on the y-axis are normalized by $f_0$, the $c$ values by $c_0$,
  the threshold $\tau$ by $f_0$,
  the number of queries by $q_0$, and the max memory by $n$.
  Notice that the total cost in memory at one time of the algorithm is $n$.

  The first set of experiments compare different USM algorithms used as a subroutine
  in \stream. The results for \multi are in Figure \ref{fig:mssubroutines}, and the
  results for \single are in Figure \ref{fig:ssubroutines}.
  One can see that no matter the algorithm used for USM, \multi and \single can
  substantially improve on the unconstrained algorithm DG when it comes to cost
  ($c$) while reaching reasonably high values of $f$.
  As observed in Section \ref{section:streamexperiments} in the main paper,
  The $f$ values of the
  solutions returned are close to their theoretical bounds.
  When DG is the subroutine for USM then the highest $f$
  value is returned, and RS returns the lowest $f$ value, as expected based on their
  approximation guarantees.
  Surprisingly, for \multi the DG subroutine results in the highest total number of
  queries to $f$ despite LS being the algorithm with the highest theoretical run time.
  This is because
  \multi is making a lot more passes to reach the higher approximation guarantee of
  DG relative to LS. This results in more total queries to $f$ since each pass is
  $O(n/\epsilon)$ queries even before any USM subroutine is used.
  Any of the subroutines are practical for \multi in terms of number of queries.
  In all other experiments,
  we set DG as the subroutine for \multi since it gives the highest $f$ value.
  In contrast, both DG and LS are relatively
  impractical in terms of the number of queries for \single since \single needs to
  run USM so many times, especially as $\epsilon$ decreases.
  Therefore in all other experiments we use
  RS as the subroutine for \single.

  In the second set of experiments, \multi and \single are run on the
  instances of
  diverse data summarization on the
  Corel5k (``corel'') and
  delicious (``delicious50k'') dataset with $n=50000$,
  and instances of graph cut on the ca-AstroPh (``astro'')
  and com-Amazon (``amazon'') datasets.
  with input $\tau=f_0$ and varying $\epsilon$
  (Figure \ref{fig:epsilons}).
  These experiments are just like those in
  Figures \ref{delicious_epsf} to \ref{delicious_epsmem}
  of the main paper, and are presented here to show that the same patterns
  hold for additional datasets.
  Similarly, in Figure \ref{fig:cover},
  \multi and \single are run with varying input $\tau$
  and the results are
  similar to those featured in Figures \ref{astro_tauf} to \ref{astro_taumem}
  in the main paper.

  Finally, in Figure \ref{fig:heuristic} the same experiments as in Figure
  \ref{fig:cover} are run except the algorithms are run as heuristics where the
  approximation guarantees are assumed to be 1.
  In particular, \multi and \single both run \stream where $\gamma=1$,
  and when the approach of \cite{iyer2013} is used the submodular maximization
  subroutines are run until a solution is found with $f$ value at least $(1-\epsilon)\tau$.
  The purpose of these experiments is to compare the algorithms without explicitly
  using their known approximation guarantees, since that was found to heavily influence
  the $f$ values of the solutions in Figure \ref{fig:cover} as discussed in the
  main text.
  The differences in $f$ values between the algorithms are now practically
  eliminated, and the difference in cost values greatly lessened although ALALUF
  and SG still find
  relatively lower cost solutions. However, the differences in memory and queries
  to $f$ between the algorithms are increased.
  This is because the approach of \cite{iyer2013} requires running
  ALALUF and SG many more times to reach the higher threshold for $f$.
  In this setting, ALALUF and SG are not practical
  because of their large numbers of queries compared to \multi and \single.

\end{document}